\documentclass[a4paper,oneside,11pt]{book}

\newcommand{\betabb}{\boldsymbol{\beta}}

\newcommand{\ubb}{\boldsymbol{u}}
\newcommand{\xbb}{\boldsymbol{x}}
\newcommand{\hbb}{\boldsymbol{h}}

\newcommand{\sx}{\sigma^x}
\newcommand{\sy}{\sigma^y}
\newcommand{\sz}{\sigma^z}

\newcommand{\dlambda}{\partial_{\lambda}}
\newcommand{\AGP}[1]{\mathcal{A}_{#1}}
\newcommand{\adj}[1]{#1^{\dagger}}
\newcommand{\dotlambda}{\dot{\lambda}}
\newcommand{\approxAGP}{\mathbb{A}_{\lambda}}
\newcommand{\gammabar}{\Bar{\gamma}}
\newcommand{\zetabar}{\Bar{\zeta}}

\newcommand{\HCD}{H_{\rm CD}}
\newcommand{\R}{\mathbb{R}}

\newcommand{\acrref}[1]{\hyperref[acr:#1]{#1}}

\makeatletter
\def\BState{\State\hskip-\ALG@thistlm}
\makeatother

\usepackage{amsbsy}
\usepackage{amsmath}
\usepackage{amsfonts}
\usepackage{graphicx}
\usepackage{multirow}
\usepackage{mathrsfs}
\usepackage{color}
\usepackage{xurl}
\usepackage[hidelinks]{hyperref}
\hypersetup{breaklinks=true}
\usepackage{cite}
\usepackage{enumitem}
\usepackage{epsfig}
\usepackage{caption}
\usepackage{subcaption}
\usepackage{physics}
\usepackage[strict]{changepage}
\usepackage{epigraph}
\usepackage{lipsum}
\usepackage{wrapfig}
\usepackage{algorithm}
\usepackage{algorithmicx}
\usepackage{algpseudocode}
\usepackage{mdframed}
\usepackage{cleveref}
\usepackage{dsfont}

\usepackage[left=4cm,right=2.5cm,top=2cm,bottom=4cm,includehead,includefoot,headheight=15pt]{geometry}

\usepackage{fancyhdr}
\usepackage{setspace}
\usepackage[most]{tcolorbox}

\newtcolorbox[auto counter,number within=section,crefname={box}{boxes}]{mycolorbox}[2][]{%
title=Box 1 $\mid$ Timeline, 
enhanced,
colback=PineGreen!10!white,
colframe=PineGreen!40!black,
colbacktitle=PineGreen!95!white,
fonttitle=\bfseries\upshape, 
code={\singlespacing},
subtitle style={boxrule=0.4pt},title=Box ~\thetcbcounter $\mid$ #2, label={#1}}

\newtcbtheorem[auto counter,number within=section]{theo}%
  {Theorem}{fonttitle=\bfseries\upshape, fontupper=\slshape,
     arc=0mm, colback=Periwinkle!10!white,colframe=Periwinkle!40!black,colbacktitle=Periwinkle!95!white}{theorem}

\usepackage{titling}
\pretitle{%
  \begin{center}
  \huge
  \includegraphics[width=6cm]{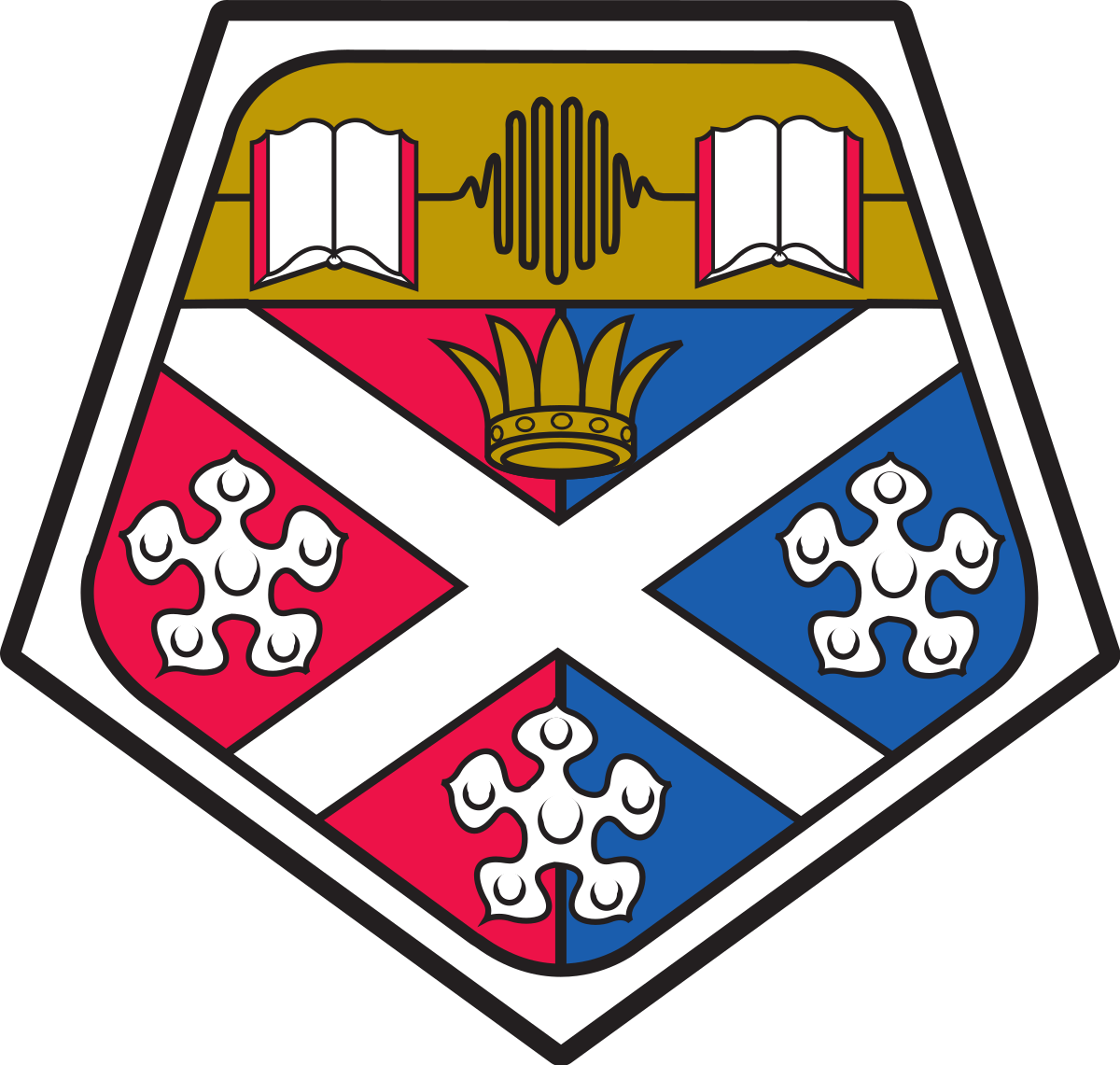}\\[\bigskipamount]
}
\posttitle{\end{center}}

\title{Counterdiabatic, Better, Faster, Stronger: \\
\Large{Optimal control for approximate counterdiabatic driving} \\ PhD Thesis}
\author{Ieva \v{C}epait\.{e}
\\ \small Quantum Optics and Quantum Many-Body Physics\\[-0.8ex]
\small Department of Physics\\[-0.8ex]
\small University of Strathclyde, Glasgow\\
}

\begin{document}

\maketitle

\frontmatter

\topskip0pt
\vspace*{\fill}
\noindent
\begin{quote}
	\centering
	This thesis is the result of the author's original research. It has been composed by the author and has not been previously submitted for examination which has led to the award of a degree. \\[5pt]
	The copyright of this thesis belongs to the author under the terms of the United Kingdom Copyright Acts as qualified by University of Strathclyde Regulation 3.50. Due acknowledgement must always be made of the use of any material contained in, or derived from, this thesis. \\[5pt]
	%
\end{quote}
\vspace*{\fill}
\chapter{Abstract}\label{chap:abstract}

Adiabatic protocols are employed across a variety of quantum technologies, from implementing state preparation and individual operations that are building blocks of larger devices, to higher-level protocols in quantum annealing and adiabatic quantum computation. The main drawback of adiabatic processes, however, is that they require prohibitively long timescales. This generally leads to losses due to decoherence and heating processes. The problem of speeding up system dynamics while retaining the adiabatic condition has garnered a large amount of interest, resulting in a whole host of diverse methods and approaches made for this purpose. Most of these methodologies are encompassed by the fields of quantum optimal control and shortcuts to adiabaticity (\acrref{STA}), which are in themselves complementary approaches. Optimal control often concerns itself with the design of control fields for steering system dynamics while minimising the use of some resource, like time, while the goal of \acrref{STA} is to retain the adiabatic condition upon speed-up.

This thesis is dedicated to the discovery of new ways to combine optimal control techniques with a universal method from \acrref{STA}: counterdiabatic driving (\acrref{CD}). The \acrref{CD} approach offers perfect suppression of all non-adiabatic effects experienced by a system driven by a time-dependent Hamiltonian regardless of how fast the process occurs. In practice, however, exact \acrref{CD} is difficult to derive often even more difficult to implement. The main result presented in the thesis is thus the development of a new method called counterdiabatic optimized local driving (\acrref{COLD}), which implements optimal control techniques in tandem with \emph{approximations} of exact \acrref{CD} in a way that maximises suppression of non-adiabatic effects. We show, using numerical methods, that using \acrref{COLD} results in a substantial improvement over optimal control or approximate \acrref{CD} techniques when applied to annealing protocols, state preparation schemes, entanglement generation, and population transfer on a synthetic lattice. We explore how \acrref{COLD} can be enhanced with existing advanced optimal control methods and we show this by using the chopped randomized basis method and gradient ascent pulse engineering. Furthermore, we demonstrate a new approach for the optimization of control fields that does not require access to the wave function or the computation of system dynamics. In their stead, we use components of the approximate counterdiabatic drive to inform the optimisation, owing to the fact that \acrref{CD} encodes information about non-adiabatic effects of a system for a given dynamical Hamiltonian. 

\tableofcontents

\chapter{Lay Summary}
\epigraph{“With magic, you can turn a frog into a prince. With science, you can turn a frog into a Ph.D and you still have the frog you started with.”}{Terry Pratchett}

\begin{wrapfigure}{r}{0.4\textwidth}
\centering
\includegraphics[width=0.9\linewidth]{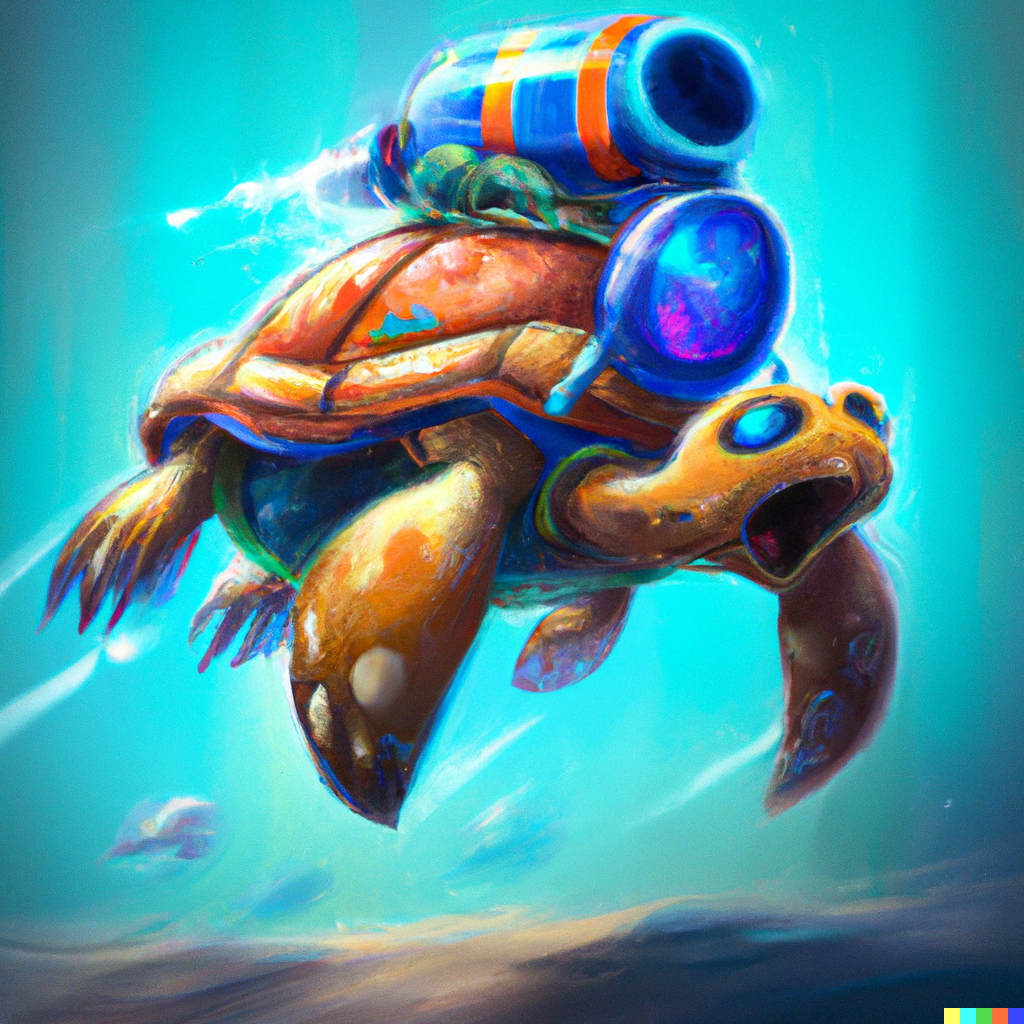} \caption[DALL$\cdot$E turtle illustration.]{A turtle with a jetpack strapped to its back, illustrating the speed-up of what is canonically a slow (adiabatic) process. This image was created with the assistance of DALL$\cdot$E 2 \cite{noauthor_dalle_nodate}.}\label{fig:COLD_TURTLE}
\vspace{-5pt}
\end{wrapfigure}

Quantum systems are notoriously volatile creatures. In our quest to build better quantum technologies, we must first learn the art of \emph{controlling} them with very high precision in a way that produces useful information or work. This must be done while protecting the information such systems contain from an environment that is often hell-bent on making this job as difficult as possible \footnote{This anthropomorphisation of quantum systems and the environment is for literary effect - I do not believe that the environment has much in the way of a political agenda to inflict decoherence upon quantum systems}.

A particularly useful type of controlled process that we would like to be able to perform is an \emph{adiabatic} process, which involves slowly changing some parameter affecting a quantum system, \@e.g.~the strength or direction of an electromagnetic field. The `slow' part here is required to keep the system from getting excited out of the `instantaneous' energy level that it starts in. Think of a magnetic field slowly rotating through some angle such that a bar magnet placed in said field always stays aligned with it. If the rotation happens too quickly, the magnet overshoots in the direction of the changing field. An analogous process happens in the quantum case, where the quantum state `jumps' out of its energy level. For many applications of quantum technologies, we would like to avoid such jumps, hence we perform adiabatic (slow) transformations.

Unfortunately, the volatility of quantum systems does not often allow us to abide by this slow condition. The longer a quantum process takes to complete, the more time it spends exposed to the environment, leaking information and absorbing heat. In order to combat this lossiness, entire fields of study have been developed with the sole aim of imitating the results of adiabatic processes on shorter timescales. The techniques used to achieve this vary vastly and achieve various levels of success: some suppress the losses that come with fast processes, others try to avoid them entirely with increasingly complex protocols.

In this thesis, we present a method which aims to speed up adiabatic processes in a way that caters to the practical constraints of quantum experiments. We assume that we are given a limited set of operations that we can actually perform in order to suppress some of the jumps that occur during fast driving. We then optimise the path through which the system travels in a way that helps this very restricted set of operations perform as best as they could. This approach follows the fact that the losses depend on the path that the system parameter takes: for example, the magnetic field can rotate from its starting direction to the final one while including detours and oscillations along the way. If you get the set of rotations just right, it is possible to mitigate or suppress many of the effects of a fast change quite efficiently in many cases. We demonstrate this in some of the later chapters with simulations of such optimised counterdiabatic protocols for different systems and different rates of change in the system parameters. 

For more details, I invite you to read my blog post on the topic given by Ref. \cite{cepaite_cold_2023}, which is slightly more technical and detailed, but brief and full of animations to explain the concepts involved. 

\chapter{Acknowledgements}

The last few years have been as exciting as they were tough, but they were perpetually made better by the many wonderful people around me. It would be absolutely impossible to include everyone whose company enriched my mind and spirit during this PhD, although I would absolutely love to. Instead, I will strive to mention as many as I possibly can, as without them this journey would definitely not have been possible.

First, there are a multitude of people who supported me directly in my academic endeavours and beyond. I want to thank my supervisor Andrew Daley for giving me all the opportunities to learn, travel and engage with fascinating, cutting-edge scientific endeavours as well as for mentoring me in both how to be a better researcher and a better person. I would also like to thank Anatoli Polkovnikov and Pieter Claeys for all the great help, interesting discussions and mentoring, which helped me get through a number of barriers in understanding things. Thank you to Callum Duncan for guiding me through the most difficult parts of the project.

Secondly, in a similar vein, I want to thank all of the wonderful physicists at Strathclyde, former and current, for their friendship and support and for the company in complaining about things. Thank you to Sridevi, Sebastian, Tomas, Ewen, Ryan, Sebastian, Gerard, Johannes, Pablo, Natalie, Emmanuel, Emanuele, Rosaria, Jorge, Stewart, Tom, Grant, Jonathan and many others who will remain unnamed only for the sake of keeping this to less than ten pages. I am grateful to all of you, from the bottom of my heart. 

I would also like to thank my friends at the Mathematically Structured Programming group at Strathclyde and those adjacent, who provided me with shelter, sanity, pints and fantastic advice. In particular, thank you to Jules Hedges for the healthy cynicism and to Conor McBride for making me arguably far more sensible. To everyone else: Alasdair, Joe, Matteo, Giorgi, Sean, Dylan, Riu, Zanzi, Ezra, Fred, Bob, Clemens, Malin, Toby, André and others - you were the best of friends and I learned as much from you as I did while reading papers.

Finally, I would like to thank my family: my mother Silvija and my father Darius, for their unconditional love and the knowledge that I am safe and cared for, my brothers Džiugas, Joris and Stepas, for the joy and company they have brought into my life, my uncle Evaldas and his family, for their support and companionship, and my grandparents Vytautas, Milda, Viktoras and Virginija, although only one of you gets to see me complete this PhD. You are all remembered and loved. Most importantly, in the last few years, I have met my best friend and partner, someone whom I love and cherish and who, I daresay, helped me the most to become both a great person and a good one: my deepest gratitude and love goes to you, Bruno Gavranović.

\begin{flushright}
\emph{Ieva \v{C}epait\.{e}, 13\textsuperscript{th} July, 2023}
\end{flushright}

\chapter{Acronyms and abbreviations}

\begin{table}[h]
      \begin{tabular}{p{3cm}  p{8cm}}

        \textbf{AGP}\label{acr:AGP} & Adiabatic Gauge Potential \\ [7pt]
        \textbf{ARP}\label{acr:ARP} & Adiabatic Rapid Passage \\ [7pt]
        \textbf{BDA}\label{acr:BDA} & Bare Dual Annealing \\[7pt]
        \textbf{BPO}\label{acr:BPO} & Bare Powell Optimisation \\[7pt]
        \textbf{CD}\label{acr:CD} & Counterdiabatic driving \\[7pt]
        \textbf{COLD}\label{acr:COLD} & Counterdiabatic Optimised Local Driving \\[7pt]
        \textbf{CRAB}\label{acr:CRAB} & Chopped Randomised Basis \\ [7pt]
        \textbf{FO}\label{acr:FO} & First order \\ [7pt]
        \textbf{GRAPE}\label{acr:GRAPE} & Gradient Ascent Pulse Engineering \\ [7pt]
        \textbf{GSA}\label{acr:GSA} & Generalized Simulated Annealing \\ [7pt]
        \textbf{LCD}\label{acr:LCD} & Local Counterdiabatic Driving \\[7pt]
        \textbf{PMP}\label{acr:PMP} & Pontryagin Maximum Principle \\[7pt]
        \textbf{QOCT}\label{acr:QOCT} & Quantum Optimal Control Theory \\[7pt]
        \textbf{SO}\label{acr:SO} & Second order \\ [7pt]
        \textbf{STA}\label{acr:STA} & Shortcuts to Adiabaticity \\[7pt]

    \end{tabular}

\end{table}\label{table}

\addcontentsline{toc}{chapter}{List of Figures}
\listoffigures

\makeatletter
\@mainmattertrue
\pagenumbering{arabic}
\makeatother

\chapter{Introduction}

\epigraph{Everything starts somewhere, although many physicists disagree.}{Terry Pratchett, \emph{Hogfather (1996)}}

Despite the fact that quantum mechanics has been established for around a century, only recently have we begun to harness the unique features found in the quantum domain, a development spurred by and further proliferating the rapid progress of experimental advances for quantum systems. It is often control that turns scientific knowledge into technology. Thus control, or the precise manipulation of and interaction with quantum systems, is a fundamental goal of quantum technologies. This may be for the purpose of gaining insight into the physics governing quantum systems, in order to build better devices or in order to solve complex computational problems. We are currently on the cusp of a new age of quantum technologies and control of quantum systems, driven by the methodical exploitation of phenomena such as coherence and entanglement, allowing us to probe and predict the behaviour of quantum systems in ways that could never be done before. 

With this development in experimental capabilities, the demand for theoretical techniques for the time-dependent manipulation of quantum systems has increased considerably. Such techniques are imperative for the development of efficient transformations of quantum states, like in the case of quantum gate design \cite{pelegri_high-fidelity_2022}, quantum computing \cite{albash_adiabatic_2018} or state preparation for the study of condensed matter physics \cite{dimitrova_many-body_2023}, among many other examples. Simultaneously, there has been a rise in demand for techniques which refine and enhance existing protocols with the aim of reducing or mitigating decoherence and unwanted losses, whether through information-theoretic techniques like quantum error-correction \cite{roffe_quantum_2019}, or via approaches for designing driving pulses like in the case of quantum optimal control methods \cite{glaser_training_2015, koch_quantum_2022}. 

\paragraph*{Non-adiabatic losses}

An important example of control imperfections experienced by a system driven in a time-dependent manner is that of losses in the form of undesired transitions that can occur between instantaneous eigenstates of a dynamical Hamiltonian \cite{berry_transitionless_2009, kolodrubetz_geometry_2017}. There are many processes where one might want to end up in \@e.g.~the ground state of a given Hamiltonian whose parameters have been modified in a time-dependent manner. This holds true in the case of state-preparation \cite{dimitrova_many-body_2023}, population transfer \cite{meier_counterdiabatic_2020} or in the case of solutions to combinatorics problems encoded in ground states of Hamiltonians \cite{pichler_quantum_2018, ebadi_quantum_2022}. This is why many quantum driving protocols rely on adiabatic dynamics, where the system follows the instantaneous eigenstates of time-dependent Hamiltonians and transitions are naturally suppressed\cite{born_beweis_1928, kato_adiabatic_1950}. Ideal adiabatic processes are reversible, making them, in principle, highly robust \cite{jarzynski_geometric_1995, kolodrubetz_geometry_2017}. Ideal adiabatic processes, however, require very slow system dynamics and one must make compromises on the timescales of competing heating and decoherence processes. This has led to a rise in the development of methods which aim to speed up adiabatic dynamics while minimising the undesired transitions associated with fast driving, either by entirely removing or by suppressing them. These types of methods are collectively referred to as `shortcuts to adiabaticity' or \acrref{STA} \cite{guery-odelin_shortcuts_2019, torrontegui_chapter_2013}. 

\paragraph*{Shortcuts to adiabaticity} 

The field of \acrref{STA} concerns itself with fast routes to the final results of slow, adiabatic changes of the time-dependent parameters of a system. Such routes are generally designed via a set of analytical and numerical methods for different systems and conditions. Speeding up adiabatic protocols to enable their completion within the system’s coherence time is important
for the development of any quantum technologies relying on such protocols. Thus, \acrref{STA} methods have become instrumental in preparing and driving internal and motional states in atomic, molecular, and solid-state physics. Some \acrref{STA} techniques rely on specific formalisms like invariants and scaling \cite{deffner_classical_2014, deng_superadiabatic_2018, chen_fast_2010}, which exploit symmetries in the physical systems in order to simplify models of non-adiabatic effects, or fast-forward \cite{masuda_fast-forward_2009, masuda_fast-forward_2008}, which adds an external phase to the system wavefunction in order to allow for fast transport. These methods, within specific domains, can be related to each other and potentially be made equivalent because of underlying common structures. A universal \acrref{STA} approach like this is counterdiabatic driving or \acrref{CD}, which will be a focal point of this thesis. 

\paragraph*{Counterdiabatic driving}

The idea of \acrref{CD} was first introduced by Demirplak and Rice in the context of physical chemistry \cite{demirplak_adiabatic_2003} and independently developed by Berry \cite{berry_transitionless_2009}, where it was referred to as `transitionless' driving. The aim of \acrref{CD} is the complete suppression of non-adiabatic effects experienced by a system driven at finite time via the application of an external `counterdiabatic' driving pulse. This is generally not possible, however, due to the fact that the exact counterdiabatic drive is often difficult to compute in the case of complex systems and may be near-impossible to implement in most experimental settings, as well as being undefined for \@e.g.~chaotic systems \cite{kolodrubetz_geometry_2017, pandey_adiabatic_2020, sugiura_adiabatic_2021}. This has led to the development of several approximate \acrref{CD} methods, like the variational approach first introduced by Sels and Polkovnikov in \cite{sels_minimizing_2017} as well as the nested-commutator method of Claeys et al \cite{claeys_floquet-engineering_2019}. Such approaches allow for some suppression of non-adiabatic effects, but their efficacy is highly variable between different systems and the Hamiltonians driving them. Discrete, quantum gate-based versions of \acrref{CD} and its approximations have also been developed, under the moniker of `digitized counterdiabatic quantum optimization' (DCQO) \cite{hegade_digitized_2022}, as well as within the context of the quantum approximate optimisation algorithm or QAOA \cite{wurtz_counterdiabaticity_2022}, although this is a relatively new line of research.

\paragraph*{Quantum optimal control}

A different but complementary approach to achieving the target state of adiabatic dynamics more rapidly is that of quantum optimal control theory or \acrref{QOCT} \cite{glaser_training_2015, koch_quantum_2022}. \acrref{QOCT} is primarily concerned with the development of driving schedules for quantum systems which satisfy specific constraints and behave optimally with respect to a given metric. Links between optimal control and \acrref{STA} have existed throughout the development of both approaches \cite{stefanatos_frictionless_2010, stefanatos_shortcut_2021, zhang_connection_2021}. This has included the realisation of \acrref{CD} through fast oscillations of the Hamiltonian \cite{petiziol_accelerated_2020, petiziol_fast_2018} as well as a fusion of machine learning methods and \acrref{STA}, demonstrating significant improvements for optimizing quantum protocols through machine learning with the inclusion of concepts from \acrref{CD} \cite{bukov_reinforcement_2018, yao_reinforcement_2021, khait_optimal_2022}. While \acrref{QOCT} methods certainly play a part in many aspects of \acrref{STA}, however, they are not applied uniquely to the problem of speeding up adiabatic dynamics. \acrref{QOCT} techniques are often implemented with the goal of driving a system to some desired target state, as in the case of much of \acrref{STA}, however they can also be implemented in determining protocols which satisfy criteria that are unrelated to some target state, like minimising the magnitude of energy expenditure. Due to the versatility of optimal control techniques, they can often be incorporated into many aspects of quantum technologies in order to improve them. Examples include the design of quantum computing gates \cite{pelegri_high-fidelity_2022} as well as improving measurement techniques \cite{wiseman_quantum_2009}, along with the aforementioned applications to speeding up adiabatic dynamics \cite{guery-odelin_shortcuts_2019}.


\paragraph*{Goals and contributions of the thesis}

Speeding up adiabatic processes while suppressing non-adiabatic losses remains an open problem in most practical settings. In the case of \acrref{CD}, issues generally arise at the point of implemention, with the counterdiabatic term requiring operators that are simply not available in an experimental setting, even if the exact counterdiabatic term could be theoretically obtained. The variational approach of Sels and Polkovnikov \cite{sels_minimizing_2017}, which we refer to as `local counterdiabatic driving' or \acrref{LCD}, has attempted to circumvent this by constructing approximations which allow one to choose an ansatz set of operators rather than requiring them to have full support over the exact counterdiabatic drive. Such an approach makes for a far more accessible method, however it is also one which has no guarantees of performance due to the restrictions placed on the operators by \acrref{CD} theory. Optimal control methods, on the other hand, while far more flexible also generally offer very little insight into the way an optimal pulse should be constructed in order to suppress non-adiabatic effects. Thus pure \acrref{QOCT} approaches are often even more ineffective than approximate \acrref{CD} for this purpose. In this thesis, we present a new combination of \acrref{LCD} and optimal control methods which aims to improve upon both of the existing approaches while retaining their advantages. The method, which we will call `counterdiabatic optimised local driving' or \acrref{COLD} \cite{cepaite_counterdiabatic_2023}, is based on the observation that the effectiveness of a given \acrref{LCD} approximation depends on the path of the dynamical Hamiltonian and furthermore, that this path can be optimised using \acrref{QOCT} methods. We will also show that the optimal control component of \acrref{COLD} can be extended by using an optimisation metric constructed using information about the counterdiabatic drive. We will demonstrate the effectiveness and flexibility of \acrref{COLD} and its extensions via numerical analysis, comparing it to both of its components, \acrref{LCD} and quantum optimal control.

\section{Thesis overview}

The thesis is divided into four parts, prefaced by this introduction. \textbf{Part \ref{part:background}} introduces key background concepts relevant to the new results discussed later in the thesis: quantum adiabaticity and quantum optimal control or \acrref{QOCT}. First, we discuss the concept of an adiabatic quantum process, with particular focus in Sec.~\ref{sec:2.1.2_adiabatic_condition} on what it means for a change in the Hamiltonian parameters to be `slow enough' to be adiabatic. We cover how non-adiabatic effects are generated by an operator known as the adiabatic gauge potential (or \acrref{AGP}) and subsequently introduce the concept of a counterdiabatic drive. We then discuss the difficulties of obtaining an exact counterdiabatic drive for a given Hamiltonian and introduce several existing approximations of \acrref{CD}. This includes \acrref{LCD}, which plays a large part in the rest of the thesis. We introduce \acrref{QOCT}, beginning with the mathematical foundations of optimal control as well as several popular numerical optimisation methods. We discuss how optimal control techniques can be applied specifically to quantum systems and describe several \acrref{QOCT} methods that are implemented in order to acquire the results presented later in the thesis.

\begin{figure}[t!]
    \centering
    \includegraphics[width=\linewidth]{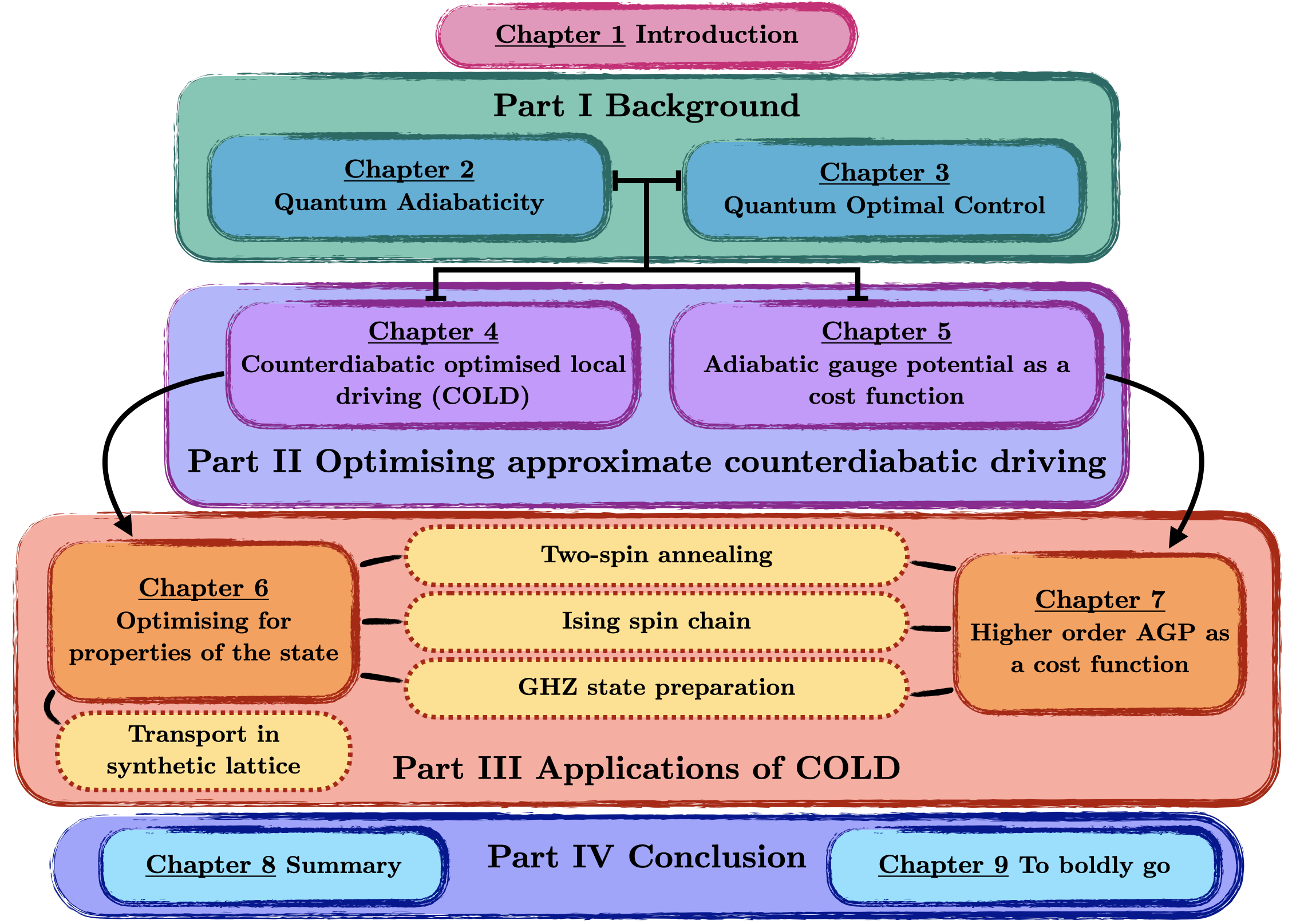} \caption[Thesis outline.]{Thesis outline. In Part~\ref{part:background}, we will introduce the concept of quantum adiabaticity, the counterdiabatic driving (\acrref{CD}) method and its approximations as well as quantum optimal control theory (\acrref{QOCT}) and several optimisation techniques that we will apply later in the thesis. Then, in Part~\ref{part:COLD}, we will combine ideas from \acrref{CD} and \acrref{QOCT} in order to develop a new method for speeding up adiabatic dynamics and the focal point of this thesis: ``Counterdiabatic Optimised Local Driving" or \acrref{COLD}. We will then extend the method with a new optimisation metric based on information about non-adiabatic effects experienced by the system in fast driving. In Part~\ref{part:applications} we will numerically implement the \acrref{COLD} method and its extensions in several different quantum systems to evaluate their performance and compare it to existing techniques. Then, finally, in Part~\ref{part:conclusion}, we will conclude with a summary of the thesis and a look towards the future and several open questions that arise from the work presented here.}\label{fig:thesis_overview}
\end{figure}

In \textbf{Part~\ref{part:COLD}}, we introduce the main new material of the thesis: the \acrref{COLD} method and its extension using several \acrref{AGP}-inspired cost functions. First, we discuss the ways in which \acrref{LCD} and quantum optimal control methods can be combined to obtain better results than either approach alone, and how that follows from the dependence of the counterdiabatic drive on the path of the Hamiltonian in parameter space. We expand on the optimal control methods used for \acrref{COLD} and introduce the idea of using information about the counterdiabatic drive itself, like its total power across the driving time, as a metric for optimising the control pulse in \acrref{COLD} and for the case where no \acrref{LCD} is applied. 

In \textbf{Part.~\ref{part:applications}} we demonstrate implementations of the new methods in numerical simulations of several example quantum systems. First, in Ch.~\ref{chap:6_Applications_fidelity} we present and discuss results obtained when applying \acrref{COLD} to a simple two-spin annealing protocol, the Ising spin chain of varying lengths, the case of population transfer in a synthetic lattice, and finally for the preparation of maximally entangled GHZ states in the setting of frustrated spin systems. We compare the results obtained with \acrref{COLD} to those obtained using un-optimised \acrref{LCD} as well as different optimal control pulses with no counterdiabatic component. In Ch.~\ref{chap:7_higher_order_agp} we do the same but implement \acrref{CD}-inspired cost functions in the optimisation of \acrref{COLD} and plain optimal control instead of using fidelity or (as in the case of GHZ state preparation) entanglement as optimisation metrics. We present results for the two-spin annealing case, the Ising spin chain, and finally for the GHZ state preparation protocol in a system of frustrated spins, to compare and contrast to the case where optimisation is based on final state fidelity. We discuss when such optimisation metrics may be better than those used in Ch.~\ref{chap:6_Applications_fidelity} and in which cases they might fail.

Finally, in \textbf{Part~\ref{part:conclusion}} we conclude with a summary of the thesis and an outlook into future research directions that are left to be explored. A diagram of the thesis structure can be found in Fig.~\ref{fig:thesis_overview}, linking the relevant parts together.

\section{Publications and manuscripts}

The majority of this work is based on the following publications and manuscripts:

\begin{enumerate}
    \item \textbf{Counterdiabatic Optimised Local Driving}, \textit{Ieva Čepaitė, Anatoli Polkovnikov, Andrew J. Daley, Callum W. Duncan. PRX Quantum \textbf{4}, 010309, 2023.} Eprint arxiv:2203.01948. \cite{cepaite_counterdiabatic_2023}
    \item \textbf{Many-body spin rotation by adiabatic passage in
    spin-1/2 XXZ chains of ultracold atoms}, \textit{Ivana Dimitrova, Stuart Flannigan, Yoo Kyung Lee, Hanzhen Lin,  Jesse Amato-Grill, Niklas Jepsen, Ieva Čepaitė, Andrew J. Daley, Wolfgang Ketterle. Quantum Sci. Technol. \textbf{8} 035018, 2023} Eprint arxiv:2301.00218.\cite{dimitrova_many-body_2023}.
    \item \textbf{A numerical approach for calculating exact non-adiabatic terms in quantum dynamics}, \textit{Ewen D. C. Lawrence, Sebastian Schmid, Ieva Čepaitė, Peter Kirton, Callum W. Duncan}, Eprint arxiv: 2401.10985 \cite{lawrence_numerical_2024}.
\end{enumerate}

My contributions to (1) include theoretical work, numerical analysis and writing of the manuscript. In the case of (2) I contributed to some discussions and some numerical analysis relating to the results. In the case of (3), my contribution was confined to theoretical discussions and the writing of the introduction and theoretical component of the manuscript.

\section{Talks and presentations}

Throughout my PhD I gave several talks on my work, including on topics that are not covered in this thesis. Here I list most of them.

\begin{itemize}
    \item \textit{``Solving Partial Differential Equations (PDEs) with Quantum Computers"}, AWE, (March 2020)
    \item \textit{``A Continuous Variable Born Machine"}, \href{https://www.youtube.com/live/ImQeEs0BcQs?feature=share&t=1996}{Pittsburgh Quantum Institute Virtual Poster Session}, Online (April 2020)
    \item \textit{``A Continuous Variable Born Machine"}, \href{https://www.youtube.com/watch?v=6v1IiXRToPU&t=3685s}{Quantum Techniques in Machine Learning}, Online (November 2020)
    \item \textit{``Variational Counterdiabatic Driving"}, University of Strathclyde and University of Waterloo Joint Virtual Research Colloquium on Quantum Technologies, Online (November 2020)
    \item \textit{``A Continuous Variable Born Machine"}, Bristol QIT Online Seminar Series, Online (March 2021)
    \item \textit{``Optimised counderdiabatic driving with additional terms"}, \href{https://meetings.aps.org/Meeting/MAR21/Session/S21.8}{APS March Meeting}, Online (March 2021)
    \item \textit{``Counterdiabatic Optimised Local Driving"}, \href{https://www.youtube.com/watch?v=YkoCPIlFl70}{DAMOP}, Orlando (May 2022)
    \item \textit{``Counterdiabatic Optimised Local Driving"}, QCS Hub Project Forum, Oxford (January 2023)
    \item \textit{``Counterdiabatic Optimised Local Driving"}, \href{https://meetings.aps.org/Meeting/MAR23/Session/Q71.8}{APS March Meeting}, Las Vegas (March 2023)
    \item \textit{``Counterdiabatic Optimised Local Driving"}, \href{https://youtu.be/-btmXDNaQX4}{INQA Seminar}, Online (March 2023)
\end{itemize}
\part{Background}\label{part:background}
\chapter{Quantum Adiabaticity}\label{chap:2_adiabaticity}

\epigraph{``I saw this movie about a bus that had to SPEED around a city, keeping its SPEED over fifty, and if its SPEED dropped, it would explode! I think it was called `The Bus That Couldn’t Slow Down'."}{Homer Simpson, \emph{The Simpsons (S7E10)}}

The concept of quantum adiabaticity is the central starting point of the work presented in this thesis. In classical thermodynamics, an adiabatic process is one where no heat is transferred between a system and its environment. On a microscopic quantum mechanical level, this means not changing the occupation/population of Hamiltonian eigenstates. The quantum adiabatic theorem then describes how slowly changes to the Hamiltonian and therefore the eigenstates have to be made so as not to change the distribution. To illustrate, imagine a system that starts in some eigenstate of a Hamiltonian. If a parameter of the Hamiltonian is varied slowly enough, then the system is expected to stay in the corresponding eigenstate of the time-independent `snapshot' Hamiltonian throughout the change and the process is `adiabatic'. In Sec.~\ref{sec:2.1_adiabatic_theorem} we will derive the adiabatic condition and explore what happens when the rate of change in the Hamiltonian parameters is too fast for adiabaticity. As we will find, the non-adiabatic effects that result from fast driving have a geometric interpretation, relating to the Berry connection \cite{berry_quantal_1984} and an operator known as the adiabatic gauge potential \cite{kolodrubetz_geometry_2017, jarzynski_geometric_1995} or \acrref{AGP}. We will describe the \acrref{AGP} in detail in Sec.~\ref{sec:2.2_AGP} and proceed to use it in order to define the concept of counterdiabatic driving \cite{berry_transitionless_2009, demirplak_adiabatic_2003} (\acrref{CD}) in Sec.~\ref{sec:2.3_CD}. \acrref{CD} is a method under the more general umbrella of Shortcuts to Adiabaticity \cite{guery-odelin_shortcuts_2019} (\acrref{STA}), which aim to suppress the non-adiabatic eigenstate deformations that occur when the Hamiltonian parameters are changed too fast, in order to achieve pseudo-adiabatic processes at shorter timescales. In Sec.~\ref{sec:2.4_approximate_CD}, we will demonstrate that exact suppression of non-adiabatic effects in the general case turns out to be impractical (if not impossible) and discuss how one can construct approximate \acrref{CD} protocols which are physically implementable and can mitigate some level of the losses brought about by fast driving.
        
    \section{The quantum adiabatic theorem}\label{sec:2.1_adiabatic_theorem}
    
    Imagine a quantum system that finds itself in the ground state of a time-dependent Hamiltonian at some given point in time. According to the quantum adiabatic theorem, it will \emph{remain} in the instantaneous ground state provided the Hamiltonian changes sufficiently slowly or `adiabatically' (where the meaning of `slow' will become clearer as this section progresses). We note that the quantum adiabatic theorem is often presented in the literature as being valid only when the instantaneous eigenstates of the Hamiltonian are non-degenerate throughout the system evolution. However, more general versions of the quantum adiabatic theorem do not impose this restriction \cite{katanaev_adiabatic_2011}, \@e.g.~defining it with respect to a system remaining in particular eigenspaces rather than eigenstates as it evolves. Here we will always work with the simpler version, wherein the instantaneous eigenstates are non-degenerate.
    
\begin{figure}[t]
    \centering
    \includegraphics[width=0.9\linewidth]{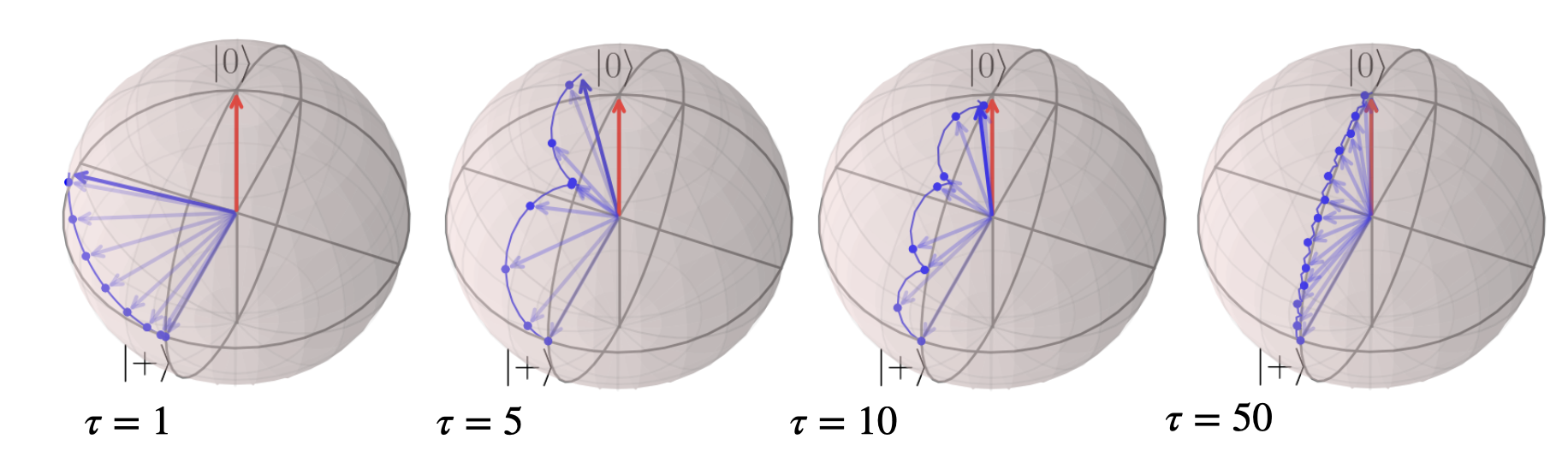} \caption[Rotating spin Bloch sphere illustrations]{Bloch sphere illustration of the single-spin system driven by the Hamiltonian of Eq.~\eqref{eq:rotating_spin_H} for different total driving times $\tau$. The red arrow indicates the ground state of the Hamiltonian at $t = \tau$ while the blue path is that taken by the spin during the evolution.}\label{fig:bloch_rotating_spin}
    \end{figure}

    To take an intuitive example, we can consider a spin in a magnetic field that is rotated from the $x$ direction to the $z$ direction during some total time $\tau$. The Hamiltonian might be written in a chosen basis as:
    \begin{equation}\label{eq:rotating_spin_H}
        H(t) = -\cos\Big(\frac{\pi t}{2 \tau}\Big)\sx - \sin \Big(\frac{\pi t}{2 \tau}\Big)\sz,
    \end{equation}
    with the Pauli matrices defined as:
    \begin{equation}
        \sx = \mqty(0 & 1 \\ 1 & 0), \quad \sy = \mqty(0 & -i \\ i & 0), \quad \sz = \mqty(1 & 0 \\ 0 & -1).  
    \end{equation}
    If the spin starts in the ground state of $H(0)$,\@i.e.~pointing in the $x$ direction such that $\ket{\psi(0)} = \ket{+}$, then as the magnetic field is rotated, the spin starts precessing about the new direction of the field. This moves the spin toward the $z$ axis but also produces a component out of the $x-z$ plane. As the total time for the rotation becomes longer (\@i.e.~the rotation gets slower compared to the precession), the state maintains a tighter and tighter orbit around the field direction. In the limit of $\tau \rightarrow \infty$, the state of the spin tracks the magnetic field perfectly, and is always in the ground state of $H(t)$ for all $t$. This is illustrated in Fig.~\ref{fig:bloch_rotating_spin}, which shows the evolution of the system for increasing $\tau$ (and thus decreasing speed). At very fast times, \@e.g.~when $\tau = 1$, the state of the spin veers away from the instantaneous ground state completely, while for $\tau = 50$, the evolution tracks the instantaneous ground state quite closely.
    
    \subsection{Proof of the adiabatic theorem}\label{sec:2.1.1_proof_adiabatic_theorem}

    The above example gives some intuition for the behaviour of quantum systems as the time of evolution is slowed down, but it doesn't quite answer the question of what it means to be `slow enough' in the general case, \@i.e.~what one would refer to as the \emph{adiabatic condition}. In order to characterise this regime, we first imagine a state $\ket{\psi(t)}$ which evolves under some time-dependent Hamiltonian $H(t)$. For convenience, we redefine time through the parameter $\lambda = \frac{t}{\tau} \in [0,1]$, such that $\psi(t), H(t) \rightarrow \psi(\lambda), H(\lambda)$ vary smoothly as a function of $\lambda$. This is often done to capture the fact that there may be a natural parameterisation of the changing Hamiltonian such as, for example, two different angles describing a varying magnetic field. The parameter space we build generally has some geometric properties that relate to non-adiabatic effects, so it becomes important to talk about abstract parameters like $\lambda$ instead of time.
    
    For each value of $\lambda$ throughout the evolution, we have a time-independent `instantaneous' Hamiltonian which can be diagonalised:
    \begin{equation}\label{eq:instantaneous_schroedinger}
        H(\lambda)\ket{n(\lambda)} = E_n(\lambda)\ket{n(\lambda)},
    \end{equation}
    where $E_n(\lambda)$ are the eigenenergies and $\ket{n(\lambda)}$ are the eigenstates. The time-evolution of a system is given by the Schr\"{o}dinger equation $i\hbar \partial_t \ket{\psi(\lambda)} = H(\lambda) \ket{\psi(\lambda)}$ and since the family of eigenvectors $\ket{n(\lambda)}$ constitute a basis at every value of $\lambda$, we can expand the system state as:
    \begin{equation}\label{eq:adiabatic_basis_expansion}
        \ket{\psi(\lambda)} = \sum_n c_n(\lambda)e^{i \dotlambda^{-1} \theta_n(\lambda)}\ket{n(\lambda)},
    \end{equation}
    where $c(\lambda)$ are time-dependent coefficients through the parameter $\lambda$, $\dotlambda = \frac{d\lambda}{dt}$ and
    \begin{equation}\label{eq:dynamical_phase}
        \theta_n(\lambda) = -\frac{1}{\hbar} \int_0^{\lambda} E_n(\lambda') d\lambda'
    \end{equation}
    is commonly referred to as the dynamic (or dynamical) phase. 
    
    Thus, the task is now to solve the time-dependent Schr\"{o}dinger equation:
    \begin{equation}\label{eq:td-schroedinger}
        i\hbar \dotlambda \ket{\dlambda \psi(\lambda)} = H(\lambda) \ket{\psi(\lambda)},
    \end{equation}
    where $\dlambda$ is the partial derivative with respect to the parameter $\lambda$. We can use the expansion Eq.~\eqref{eq:adiabatic_basis_expansion}, differentiate and take the inner product with some eigenstate $\bra{m(\lambda)}$ to obtain:
    \begin{equation}\label{eq:adiabatic_derivation}
        \begin{aligned}
         i\hbar \dotlambda \dlambda \sum_n c_n e^{i \dotlambda^{-1} \theta_n} \ket{n} &= H \sum_n c_n e^{i \dotlambda^{-1} \theta_n} \ket{n} \\
        \sum_n \Big( \dlambda c_n \ket{n} + c_n \ket{\dlambda n} + i \dotlambda^{-1} \dlambda\theta_n c_n \ket{n} \Big)e^{i \dotlambda^{-1} \theta_n} &= -\frac{i}{\hbar \dotlambda} \sum_n E_n c_n e^{i \dotlambda^{-1} \theta_n} \ket{n} \\
        \sum_n \Big( \dlambda c_n \ket{n} + c_n \ket{\dlambda n} \Big)e^{i \dotlambda^{-1} \theta_n} &= 0 \\
        \dlambda c_m  &= - \sum_n c_n \braket{m}{\dlambda n}e^{i\dotlambda^{-1}(\theta_m-\theta_n),}
        \end{aligned}
    \end{equation}
    where the last two lines are a consequence of the fact that $i \dotlambda^{-1} \dlambda \theta_n(\lambda) = -\frac{i}{\hbar \dotlambda} E_n(\lambda)$ and the orthogonality of $\ket{m}$ and $\ket{n}$ when $m \neq n$. Note that we have removed the explicit dependence on $\lambda$ for the sake of readability and to make the writing more compact and will continue with this convention for the rest of the chapter unless otherwise stated. 
    
    The above differential equation is exact and describes the evolution of the coefficients $c_n$,  but it does not give much of a clue as to what `slow' time evolution means with respect to the changes in the Hamiltonian. For that, we can express the term $\braket{m}{\dlambda n}$ in terms of the changing Hamiltonian. This is done by differentiating Eq.~\eqref{eq:instantaneous_schroedinger} with respect to time and then again taking the inner product with $\bra{m}$ to get:
    \begin{equation}\label{eq:hamiltonian_derivative}
        \begin{aligned}
            \dotlambda \Big(\dlambda{H}\ket{n} + H \ket{\dlambda n}\Big)  &= \dotlambda \Big(\dlambda{E_n}\ket{n} + E_n \ket{\dlambda n}\Big) \\
            \mel{m}{\dlambda H}{n} + \mel{m}{H}{\dlambda n} &= \dlambda E_n\braket{m}{n} + E_n \braket{m}{\dlambda n} \\
            E_m \braket{m}{\dlambda n} - E_n \braket{m}{\dlambda n} &= - \mel{m}{\dlambda H}{n}, \quad m \neq n \\
            \braket{m}{\dlambda n} &= - \frac{\mel{m}{\dlambda H}{n}}{E_m - E_n}, \quad m \neq n
        \end{aligned}
    \end{equation}
    Inserting this into the final line of Eq.~\eqref{eq:adiabatic_derivation}, we find that:
    \begin{equation}\label{eq:coefficient_exact}
            \dlambda c_m + c_m \braket{m}{\dlambda m} = \sum_{n \neq m} c_n  \frac{\mel{m}{\dlambda H}{n}}{E_m - E_n}e^{i \dotlambda^{-1} (\theta_m-\theta_n)}.
    \end{equation}
    
    When the term on the RHS is small, which is a condition that will be discussed in more detail in the next section, we can neglect it and the solution for the remaining differential equation of $c_m$ is just:
    \begin{equation}\label{eq:c_adiabatic}
        c_m(\lambda) = c_m(0)e^{i \gamma_m(\lambda)},
    \end{equation}
    where
    \begin{equation}\label{eq:geometric_phase}
        \gamma_m(\lambda) = i \int_0^{\lambda} \braket{m}{\partial_{\lambda'} m} d \lambda' 
    \end{equation}
    is the geometric (or Berry) phase \cite{pancharatnam_generalized_1956, longuet-higgins_studies_1958, berry_quantal_1984}. It arises from the fact that if the Hamiltonian varies according to $\lambda$ in a closed loop way, \@i.e.~it returns to its starting point at the end of the evolution, the wavefunction might not. Think of Foucault's pendulum, which changes its plane of swinging due to the Earth's rotation around its own axis and does not necessarily return to its initial state after a full rotation. Both the appearance of the geometric phase in Eq.~\eqref{eq:c_adiabatic} and the changing plane of Foucault's pendulum are consequences of the geometry or `curvature' of the parameter space in which the dynamics occur and are related to concepts like parallel transport. To illustrate this, we can absorb the geometric phase into the adiabatic eigenstates via the transformation
    \begin{equation}\label{eq:parallel_transport_state}
        \ket{m'} = e^{i \gamma_m(\lambda)}\ket{m} = e^{- \int_0^{\lambda} \braket{m}{\partial_{\lambda'} m} d \lambda'}\ket{m},
    \end{equation}
    and then take the derivative of the above expression with $\lambda$ followed by taking the inner product with $\bra{m'}$. This gives:
    \begin{equation}\label{eq:geometric_phase_cancellation}
        \braket{m'}{\dlambda m'} = 0,
    \end{equation}
    which in other words just means that some change in the parameter $\lambda$ produces an eigenvector that is orthogonal to the unchanged eigenstate. This turns out to be the condition which defines parallel transport along a curve in a curved space, as analogous to the classical example of Foucault's pendulum. The choice of phases in Eq.~\eqref{eq:parallel_transport_state} is generally referred to as the \emph{parallel transport gauge}\cite{nakahara_geometry_2003}.
    
    The constraint that the RHS of Eq.~\eqref{eq:coefficient_exact} be negligible is exactly the adiabatic condition, which can be seen by checking that $\abs{c_m(\lambda)}^2 = \abs{c_m(0)}^2$ in Eq.~\eqref{eq:c_adiabatic}. What this means is that a state starting in a particular eigenstate $\ket{m(\lambda)}$ will remain in that state under these circumstances, \@e.g.~for $c_m(0) = 1$ and $c_{m \neq n}(0) = 0$:
    \begin{equation}\label{eq:adiabatic_states}
        \ket{\psi(\lambda)} = e^{i \dotlambda^{-1} \theta_m(\lambda)}e^{i \gamma_m(\lambda)} \ket{m(\lambda)}
    \end{equation}
    the $m^{\text{th}}$ eigenstate stays in the $m^{\text{th}}$ eigenstate.
    
    So to understand adiabaticity, we need to understand what conditions lead to the case where the additional term in Eq.~\eqref{eq:coefficient_exact} is small enough to be neglected, or:
    \begin{equation}\label{eq:adiabaticity_condition}
        \sum_{n \neq m} c_n \frac{\mel{m}{\dlambda H}{n}}{E_m - E_n}e^{i \dotlambda^{-1} (\theta_m-\theta_n)} \ll 1,
    \end{equation}
    which is exactly what the next section sets out to do.
    
    \subsection{The adiabatic condition: how slow is \emph{slow}?}\label{sec:2.1.2_adiabatic_condition}

    The condition given by Eq.~\eqref{eq:adiabaticity_condition} contains terms relating both to the rate of change of the Hamiltonian with respect to $\lambda$ (expressed in terms of matrix elements $\mel{m}{\dlambda H}{n}$) and the energy gap between eigenstates $E_m - E_n$. It is not too hard to see that when the energy gaps are very large, these terms can be neglected. However, let us try to derive a more concrete and quantitative measure for `slowness'.

    First, we can go back to the intermediate result from Eq.~\eqref{eq:coefficient_exact} and write it out as:
    \begin{equation}\label{eq:perturbative_adiabatic}
        \dlambda c_m = \sum_n c_n \braket{m}{\dlambda n}e^{i\dotlambda^{-1}(\theta_m - \theta_n)}.
    \end{equation}
    Since we want to focus on the RHS terms where $m \neq n$, we can remove the $m = n$ term by a change of variables:
    \begin{equation}\label{eq:changing_variables}
        d_m = c_m e^{\int_0^{\lambda}\braket{m}{\dlambda m} \dlambda} = c_m e^{-i\gamma_m}
    \end{equation}
    and then, using Eq.~\eqref{eq:perturbative_adiabatic}, we find
    \begin{equation}\label{eq:n_neq_m_terms}
        \begin{aligned}
            \dlambda d_m &= - \sum_n c_n \braket{m}{\dlambda n}e^{i\dotlambda^{-1}(\theta_m - \theta_n)}e^{-i\gamma_m} + c_m \braket{m}{\dlambda m} e^{-\gamma_m} \\
            &= - \sum_{n \neq m} d_n \braket{m}{\dlambda n} e^{-i(\gamma_m - \gamma_n)} e^{i\dotlambda^{-1}(\theta_m - \theta_n)} \\
            \Rightarrow{} e^{i\gamma_m} \dlambda(c_m e^{-i\gamma_m}) &= - \sum_{n \neq m} c_n \braket{m}{\dlambda n} e^{i \gamma_n} e^{i\dotlambda^{-1}(\theta_m - \theta_n)}
        \end{aligned}
    \end{equation}
    Now all that is left is integration, which leads to:
    \begin{equation}\label{eq:adiabatic_coefficients_dyanmics}
        c_m(1)e^{-i\gamma_m} = c_m(0) - \int_0^1 \sum_{n \neq m} c_n \braket{m}{\dlambda n} e^{i\dotlambda^{-1}(\theta_m - \theta_n)} e^{i(\gamma_n - \gamma_m)} d \lambda.
    \end{equation}

    In the above, we can see that when the integral on the RHS is 0, we recover the result in Eq.~\eqref{eq:c_adiabatic}. The intuition is that when the integral is sufficiently small, the adiabatic condition is valid and the system will follow the instantaneous eigenstate. Since the integral is made up of a sum of terms of the same form, we can focus on determining the bound on one of them. We can represent the integral as:
    \begin{equation}\label{eq:adiabatic_integral_main}
        I_{n \neq m}(1) = \int_0^1 c_n \frac{\mel{m}{\dlambda H}{n}}{E_m - E_n} e^{i\dotlambda^{-1}(\theta_m - \theta_n)} e^{i(\gamma_n - \gamma_m)} d \lambda,
    \end{equation}
    where we used the result from Eq.~\eqref{eq:hamiltonian_derivative}. It may be simplified significantly by using the fact that:
    \begin{equation}
        \begin{aligned}\label{eq:simplifying_adiabatic_integral}
            \dlambda \left( c_n(\lambda) \frac{A_{m,n}(\lambda)}{\omega_{m,n}^2(\lambda)}e^{i\dotlambda^{-1}(\theta_m - \theta_n)} \right) = &\dlambda \left( c_n(\lambda) \frac{A_{m,n}(\lambda)}{\omega_{m,n}^2(\lambda)}\right) e^{i\dotlambda^{-1}(\theta_m - \theta_n)} \\ &- \frac{i}{\hbar \dotlambda} c_n(\lambda) \frac{A_{m,n}(\lambda)}{\omega_{m,n}(\lambda)}e^{i\dotlambda^{-1}(\theta_m - \theta_n)} \\ 
            \Rightarrow c_n(\lambda) \frac{A_{m,n}(\lambda)}{\omega_{m,n}(\lambda)}e^{i\dotlambda^{-1}(\theta_m - \theta_n)} = &-i\hbar\dotlambda \Bigg[ \dlambda \left(c_n(\lambda) \frac{A_{m,n}(\lambda)}{\omega_{m,n}^2(\lambda)} \right) e^{i\dotlambda^{-1}(\theta_m - \theta_n)} \\
            &- \dlambda \left( c_n(\lambda) \frac{A_{m,n}(\lambda)}{\omega_{m,n}^2(\lambda)}e^{i\dotlambda^{-1}(\theta_m - \theta_n)} \right) \Bigg],
        \end{aligned}
    \end{equation}
    where we have used $A_{m,n}(\lambda) = \mel{m(\lambda)}{\dlambda H(\lambda)}{n(\lambda)}e^{-i(\gamma_m(\lambda) - \gamma_n(\lambda))}$ and $\omega_{m,n}(\lambda) = E_m(\lambda) - E_n(\lambda)$. This result can now be inserted into Eq.~\eqref{eq:adiabatic_integral_main}, leading to:
    \begin{equation}\label{eq:final_adiabatic_integral}
        \begin{aligned}
            I_{n \neq m}(1) = \: &i\hbar \dotlambda \left[ c_n(\lambda) \frac{A_{m,n}(\lambda)}{\omega_{m,n}^2(\lambda)} e^{-\frac{i}{\hbar \dotlambda}\int_0^{\lambda}\omega_{m,n}(\lambda')d \lambda'}  \right]_0^1 \\ 
            \quad &-i\hbar \dotlambda \int_0^1 \frac{d}{d\lambda'} \left(c_n(\lambda) \frac{A_{m,n}(\lambda)}{\omega_{m,n}^2(\lambda)} \right) e^{-\frac{i}{\hbar \dotlambda}\int_0^{\lambda'}\omega_{m,n}(\lambda'')d \lambda''} \\
            \approx &-i\hbar \dotlambda \left[ c_n(1) \frac{A_{m,n}(1)}{\omega_{m,n}^2(1)}e^{-\frac{i}{\hbar \dotlambda}\int_0^{1}\omega_{m,n}(\lambda')d \lambda'} - c_n(0) \frac{A_{m,n}(0)}{\omega_{m,n}^2(0)} \right] \\
            = &-i\hbar \dotlambda c_n(1) \frac{A_{m,n}(1)}{\omega_{m,n}^2(1)}e^{-\frac{i}{\hbar \dotlambda}\int_0^{1}\omega_{m,n}(\lambda')d \lambda'},
        \end{aligned}
    \end{equation}
    where the last line is a consequence of the assumption that the that the system starts in the eigenstate $m$, and thus at $\lambda = 0$, the coefficient $c_{n \neq m} = 0$. As for the disappearing integral on the second line, this is due to the fact that $\dotlambda = \frac{1}{\tau}$ and at long times $\tau \rightarrow \infty$, when the adiabatic condition is supposed to hold, the integrand will oscillate so fast that it will effectively vanish\cite{kahane_generalizations_1980}. 

    The term we're left with can effectively be bounded from above, since both exponential terms $e^{-\frac{i}{\hbar \dotlambda}\int_0^{1}\omega_{m,n}(\lambda')d \lambda'}$ and $e^{-i(\gamma_m(\lambda) - \gamma_n(\lambda))}$ (which has been absorbed into $A_{n,m}(\lambda)$) have a maximal value of 1. The same goes for $c_n(1)$. This leaves us with a bound on the remaining quantities:
    \begin{equation}\label{eq:adiabatic_criterion}
        \max _{n,m}\left[\max_{\lambda} \left| \frac{\hbar \dotlambda \mel{m(\lambda)}{\dlambda H(\lambda)}{n(\lambda)}}{(E_m(\lambda) - E_n(\lambda))^2} \right| \right] \ll 1, \quad m \neq n,
    \end{equation}
    which is exactly the adiabatic condition, as required. 

    To illustrate the point more clearly, we can look back to the example Hamiltonian of Eq.~\eqref{eq:rotating_spin_H}, where the energy gap between the two eigenstates $\ket{\psi_1(t)}$ and $\ket{\psi_2(t)}$ is a constant: $E_{\psi_1} - E_{\psi_2} = 2$, and so are the matrix elements $\mel{\psi_1}{\dot{H}}{\psi_2} = \mel{\psi_2}{\dot{H}}{\psi_1} = \frac{\pi}{2 \tau}$. The dependence on $\tau$ of the off-diagonal matrix elements of $\dot{H}$ make the results of Fig.~\ref{fig:bloch_rotating_spin} immediately clearer: as $\tau$ increases (and hence the evolution is slower), the non-adiabatic component of Eq.~\eqref{eq:adiabatic_criterion} decreases proportionately to it. More details on the example and the derivation can be found in Appendix \ref{app:rotating_spin_hamiltonian}.

     In practice, it is not immediately obvious how the quantity stated in Eq.~\eqref{eq:adiabatic_criterion} relates to, say, the fidelity of the final state with respect to the desired state or how large $\tau$, the evolution time, has to be in order to lead to a fidelity of some magnitude. While it is possible to find these bounds, the proof is quite lengthy and not necessary for the purposes of this thesis, so instead we will refer you to \cite{reichardt_quantum_2004, childs_lecture_2008} for more details.
    
    \section{The adiabatic gauge potential}\label{sec:2.2_AGP}

    The previous section introduced quantum adiabaticity and presented some intuition for non-adiabatic effects due to fast driving times. In this section, we would like to establish the deeply related concept of the adiabatic gauge potential (\acrref{AGP}) \cite{kolodrubetz_geometry_2017}, a key player in the subject matter of this thesis and a fascinating mathematical object in its own right. While the \acrref{AGP} has primarily been studied in the context of suppressing non-adiabatic effects \cite{sels_minimizing_2017, claeys_floquet-engineering_2019}, as will be its central role in this thesis, in recent years it has also been shown to be a potential probe for quantum chaos \cite{pandey_adiabatic_2020} and has been proposed for the study of thermalisation \cite{nandy_delayed_2022}. This is a consequence of the fact that quantum chaos, as often defined in the literature, manifests itself through exponential sensitivity of the eigenstates to infinitesimal perturbations that are generated by the \acrref{AGP}.

    \subsection{The moving frame Hamiltonian}

    In Section \ref{sec:2.1.1_proof_adiabatic_theorem} we spent some time working in the instantaneous eigenbasis of the Hamiltonian where it is diagonalised, à la Eq.~\eqref{eq:instantaneous_schroedinger}. For a general Hamiltonian, it is possible to go to this `moving frame' picture by rotating the Hamiltonian via some unitary $U$ so that it becomes diagonal at each point in time. If we start with some arbitrary Hamiltonian $H(\lambda)$ in a `lab frame' (i.e.~one that is viewed from an external, fixed frame of reference) that depends on time through the parameter(s) $\lambda(t)$, it can be diagonalised through $\Tilde{H} = U^{\dagger}(\lambda)H(\lambda)U(\lambda)$, where $\Tilde{\cdot}$ implies that we are now in the basis of the moving frame. In general, whenever the tilde symbol appears above an operator throughout this section, it means that we are working in this new, co-moving basis: $\Tilde{\cdot} = \adj{U}\cdot U$.

    We can also view the quantum system evolving under the Hamiltonian in this moving frame picture: $\ket*{\Tilde{\psi}} = U^{\dagger}\ket{\psi}$, which is equivalent to expanding the wave function in the instantaneous basis (or moving frame) exactly as was done in Eq.~\eqref{eq:adiabatic_basis_expansion}. Given this new basis, rewriting the Schr\"{o}dinger equation reveals:
    \begin{equation}\label{eq:moving_frame_schrodinger}
        \begin{aligned}
            i \hbar \frac{d \ket*{\Tilde{\psi}}}{dt} &= i \hbar \Big( \frac{d \adj{U}}{dt}\ket*{\psi} + \adj{U} \frac{d \ket{\psi}}{dt} \Big) \\
            &= i \hbar \dotlambda \frac{\partial \adj{U}}{\partial \lambda}\ket{\psi} + \adj{U} H\ket{\psi}\\
            &= \dotlambda \Big(i\hbar \frac{\partial \adj{U}}{\partial \lambda}U \Big)\ket*{\Tilde{\psi}} + \adj{U} H U \ket*{\Tilde{\psi}} \\
            &= \Big(\Tilde{H} - \dotlambda\Tilde{\AGP{\lambda}}\Big) \ket*{\Tilde{\psi}},
        \end{aligned}
    \end{equation}
    where the operator $\Tilde{\AGP{\lambda}}$ is the \emph{adiabatic gauge potential} with respect to the parameter $\lambda$ in the moving frame of the Hamiltonian $H$. From the above, we can see that it can be expressed as:
    \begin{equation}\label{eq:AGP_in_terms_of_U}
        \Tilde{\AGP{\lambda}} = i\hbar \adj{U}\dlambda U.
    \end{equation}
    
    The name `gauge potential' refers to operators that are generators of continuous unitary translations in parameter space \cite{kolodrubetz_geometry_2017} of some unitary transformation $U$ and generally takes the form of a derivative operator. In fact, the name originates from quantities under which the physics is invariant. For example, the gauge potential responsible for translations in space is just the momentum operator $p = i\hbar \partial_x$. This can be illustrated in the case of the simple 1D harmonic oscillator with a moving potential centered on $x_0(t)$:
    \begin{equation}\label{eq:harmonic_oscillator}
        H(x_0) = \frac{p^2}{2m} + \frac{1}{2}m \omega^2 (x - x_0)^2,
    \end{equation}
    which can be diagonalised with the transformation $U(x_0) = e^{-i p x_0/\hbar}$. Then the gauge potential is simply:
    \begin{equation}
        \Tilde{\AGP{x}} = i\hbar e^{i p x_0/\hbar} \partial_x e^{-i p x_0/\hbar} = p.
    \end{equation}
    
    In this thesis we will restrict ourselves to the specific case of adiabaticity where the transformation $U(\lambda)$ explicitly takes a wavefunction in an arbitrary basis to the adiabatic or instantaneous basis. This is a non-trivial transformation in practice, as it corresponds to a diagonalisation of the system Hamiltonian at each instantaneous moment in time. The complexity of $U(\lambda)$ and its consequences will become apparent as the chapter progresses.
    
    As we find in Eq.~\eqref{eq:moving_frame_schrodinger}, the wavefunction in the moving frame basis evolves under a combination of a diagonal Hamiltonian $\Tilde{H}$ and some additional term proportional both to the speed at which the parameter $\lambda$ varies and the \acrref{AGP}. At this point we can simplify things by applying the inverse unitary operation in order to return to the lab frame basis: $\Tilde{H} - \dotlambda\Tilde{\AGP{\lambda}} \xrightarrow{U \{ \cdot \} \adj{U}} H - \dotlambda \AGP{\lambda}$. This transformation can be used to demonstrate that we can think of the AGP in the lab frame as nothing more than the derivative operator: $\AGP{\lambda} = i\hbar\dlambda$. To see this, take any quantum state written in some basis, \@e.g.~$\ket{\psi} = \sum_n \psi_n \ket{n}$. Then in the moving frame basis we have:
    \begin{equation}\label{eq:AGP_basis_change}
            \ket{\psi} = \sum_n \psi_n \adj{U}(\lambda) \ket{n} = \sum_{\Tilde{n}} \Tilde{\psi}_n (\lambda) \ket*{\Tilde{n}(\lambda)},
    \end{equation}
    where $\Tilde{\psi}_n (\lambda) = \sum_n \adj{U}(\lambda) \psi_n = \braket{\Tilde{n}(\lambda)}{\psi}$ and the dependence on $\lambda$ enters into the basis vectors through the rotation $U(\lambda)$. We can investigate the matrix elements of $\AGP{\lambda}$ in both bases:
    \begin{equation}\label{eq:AGP_as_derivative}
        \begin{aligned}
            \mel{n}{\Tilde{\AGP{\lambda}}}{m} &= \mel{n}{i\hbar \adj{U}\dlambda U}{m} \\
            &= i\hbar \mel{\Tilde{n}(\lambda)}{\dlambda}{\Tilde{m}(\lambda)} \\
            &= \mel{\Tilde{n}(\lambda)}{\AGP{\lambda}}{\Tilde{m}(\lambda)}
        \end{aligned}
    \end{equation}
    where the last two lines are simply the statement that in the lab frame $\AGP{\lambda} = i\hbar \dlambda$.

    \subsection{Matrix elements of the AGP}

    What we have seen so far is that when we try to solve the Schr\"{o}dinger equation for a quantum system evolving under a time-dependent Hamiltonian in the basis of the moving frame, \@i.e.~in the basis where the time-dependent Hamiltonian is diagonalised, we find that the evolution happens under a `decorated' Hamiltonian composed of the diagonalised $\Tilde{H}$ and an additional operator generally known as the adiabatic gauge potential. We found that in the lab frame, it is the derivative operator with respect to the time-dependent parameters driving the Hamiltonian. What remains is to link this to our discussion of adiabaticity and the adiabatic condition of Section \ref{sec:2.1_adiabatic_theorem}.

    Let us return to the matrix elements of the lab frame \acrref{AGP} and see what they are in the adiabatic basis of Eq.~\eqref{eq:instantaneous_schroedinger}, which is the eigenbasis of the instantaneous Hamiltonian as it varies in time. The first thing to notice is that the diagonal elements of the \acrref{AGP} are very familiar:
    \begin{equation}\label{eq:AGP_diagonal}
        \mel{n(\lambda)}{\AGP{\lambda}}{n(\lambda)} = i\hbar \mel{n(\lambda)}{\dlambda}{n(\lambda)}.
    \end{equation}
    The terms on the RHS are something known as the Berry connections and they look familiar because they are the integrands of the geometric phase (Eq.~\eqref{eq:geometric_phase}) that we found when deriving the adiabatic condition. Earlier, we saw that the geometric phase is related to the geometry or curvature of the parameter space of the adiabatic Hamiltonian and the \acrref{AGP} contains information about this geometry. 

    In order to understand what the off-diagonal elements of the AGP are, we can make use of the fact that in the instantaneous Hamiltonian basis $\mel{m}{H}{n} = 0$ for $m \neq n$. Differentiating with respect to the parameter $\lambda$ gives:
    \begin{equation}\label{eq:differentiating_adiabatic_offdiagonals}
       \begin{aligned}
           \mel{\dlambda m}{H}{n} + \mel{n}{\dlambda H}{n} + \mel{n}{H}{\dlambda m} &= 0 \\
           E_n \braket{\dlambda m}{n} + E_m \braket{m}{\dlambda n} + \mel{n}{\dlambda H}{n} &= 0 \\
           (E_m - E_n) \braket{m}{\dlambda n} + \mel{n}{\dlambda H}{n} &= 0 \\
           \frac{-i}{\hbar} (E_m - E_n) \mel{m}{\AGP{\lambda}}{n} + \mel{n}{\dlambda H}{n} &= 0 \\
           \mel{m}{\AGP{\lambda}}{n} &= i\hbar \frac{\mel{m}{\dlambda H}{n}}{(E_n - E_m)},
       \end{aligned} 
    \end{equation}
    where, since we're working in the adiabatic basis, all eigenstates, eigenenergies and the operators depend on $\lambda$. We can now see that $\AGP{\lambda}$ is Hermitian and the final line is familiar: the off-diagonal elements of the \acrref{AGP} are proportional to the non-adiabatic contribution we derived back in Eq.~\eqref{eq:final_adiabatic_integral}. The full operator in the adiabatic basis is:
    \begin{equation}\label{eq:AGP_adiabatic_basis}
        \AGP{\lambda} = i\hbar\Big(\sum_n \braket{n}{\dlambda n} \dyad{n} + \sum_{m \neq n} \ket{m} \frac{\mel{m}{\dlambda H}{n}}{(E_n - E_m)} \bra{n} \Big).
    \end{equation}

    The outcome of this section then, is the revelation that this (initially mysterious) operator known as the \acrref{AGP} is deeply linked to the notion of adiabaticity in quantum systems: its diagonal terms are related to the geometry of the parameter space of adiabatic dynamics while its off-diagonals elements describe the non-adiabatic eigenstate deformations experienced by a state when it is driven by a time-dependent Hamiltonian. It is useful to note that in the final line of Eq.~\eqref{eq:moving_frame_schrodinger} we found that the Schr\"{o}dinger equation corresponding to the evolution of the instantaneous eigenstates is:
    \begin{equation}\label{eq:schrodinger_moving_frame_2}
        i \hbar \frac{d \ket*{\Tilde{\psi}}}{dt} = \Big( \Tilde{H} - \dotlambda \AGP{\lambda} \Big) \ket*{\Tilde{\psi}},
    \end{equation}
    where now it is not difficult to find how each of these operators contributes to the results of Section \ref{sec:2.1.1_proof_adiabatic_theorem}. The moving frame or instantaneous Hamiltonian generates the dynamical phase factor in Eq.~\eqref{eq:dynamical_phase}, the diagonal elements of the \acrref{AGP} produce the geometric phase factor given by Eq.~\eqref{eq:geometric_phase} and the off-diagonal elements of \acrref{AGP} are responsible for the non-adiabatic transitions out of the eigenstates which we upper bounded in Section \ref{sec:2.1.2_adiabatic_condition}. For a more detailed proof of how to derive the adiabatic theorem starting from Eq.~\eqref{eq:schrodinger_moving_frame_2}, refer to \cite{petiziol_accelerated_2020}.
    
    \section{Counterdiabatic Driving}\label{sec:2.3_CD}

    Having done all this work to characterise and understand the \acrref{AGP} and the adiabatic theorem, we now come to several important questions, starting with \emph{why do we care}? What is it about adiabatic dynamics that makes them important? Why do we want to quantify non-adiabatic transformations or understand what generates them? The answer, at least for the most part, is simple: adiabatic processes are useful. The ability to drive a time-dependent Hamiltonian while remaining in a given eigenstate can be used to prepare interesting quantum states \cite{dimitrova_many-body_2023}, to solve combinatorics problems encoded in quantum systems \cite{ebadi_quantum_2022, pichler_quantum_2018} or even to synthesise effective ramps and quantum logic gates \cite{pelegri_high-fidelity_2022} among many other applications. While there are several ways to achieve these goals, adiabaticity is a comparatively well-understood and general approach, which lends itself broadly to implementation and analysis.
    
    The most natural way of exploiting adiabatic protocols is by adhering to the adiabatic condition. However, as is often the case when it comes to the control of quantum systems, nothing is quite that simple. In practice, changing a Hamiltonian slowly enough to satisfy Eq.~\eqref{eq:adiabatic_criterion} leads to the system being overwhelmed by decoherence. Furthermore, as system sizes get larger, the energy gaps between the instantaneous eigenstates tend to get smaller, requiring slower and slower driving, making adiabatic protocols unscalable. While the adiabatic condition is not impossible to adhere to in specific cases where simple or highly structured systems are considered, in order to have any hope of pushing quantum technologies beyond their current limits, it is necessary to move beyond the adiabatic limit. The result is that we need to find ways to achieve the same results as adiabatic processes but without requiring the prohibitively long driving times that are demanded by Eq.~\eqref{eq:adiabatic_criterion}.
    
    Our analysis of the adiabatic condition has given us a clue as to how we might be able to achieve fast driving without the eigenstate deformations that result from it. Returning to Eq.~\eqref{eq:adiabatic_states}, we may focus our attention on the fact that our goal is simply to have the system follow the eigenstates of the instantaneous Hamiltonian. The approach that aims to do exactly this was first developed independently by Demirplak and Rice \cite{demirplak_adiabatic_2003} and Berry \cite{berry_transitionless_2009}. It began as the observation that one can attempt to reverse-engineer a Hamiltonian that drives the instantaneous eigenstates exactly. Recall from Eq.~\eqref{eq:adiabatic_states} that in the case that we have adiabatic evolution, the instantaneous eigenstates evolve as $\ket{\psi(\lambda)} = e^{i \dotlambda^{-1} \theta_m(\lambda)}e^{i \gamma_m(\lambda)} \ket{m(\lambda)}$ with the dynamical phase $\theta_m$ and geometric phase $\gamma_m$ defined in Eq.~\eqref{eq:dynamical_phase} and Eq.~\eqref{eq:geometric_phase} respectively. If we want to find a Hamiltonian $H_{\rm t-less} (\lambda)$ (transitionless) that drives these states exactly, we can pick a unitary $R(\lambda)$ such that:
    \begin{equation}
        \begin{aligned}
            i\hbar \dotlambda \partial_{\lambda} R(\lambda) &= H_{\rm t-less} (\lambda) R(\lambda), \\
            \Rightarrow H_{\rm t-less} (\lambda) &= i\hbar \dotlambda (\partial_{\lambda} R(\lambda)) \adj{R}(\lambda).\label{eq:H_t_less}
        \end{aligned}
    \end{equation}
    It turns out this unitary is just:
    \begin{equation}\label{eq:transitionless_unitary}
        R(\lambda) = \sum_m e^{i \dotlambda^{-1} \theta_m(\lambda)}e^{i \gamma_m(\lambda)} \dyad{m(\lambda)}{m(0)},
    \end{equation}
    so the transitionless Hamiltonian can be expressed as (from Eq.~\eqref{eq:H_t_less}):
    \begin{equation}
        \begin{aligned}\label{eq:transitionless_hamiltonian}
            \begin{aligned}
                H_{\rm t-less}(\lambda) = \: &i \hbar \dotlambda \sum_m \Bigg[ \Big(- \frac{i E_m}{\dotlambda \hbar} - \braket{m(\lambda)}{\dlambda m(\lambda)}\Big) e^{i \dotlambda^{-1} \theta_m(\lambda)}e^{i \gamma_m(\lambda)} \dyad{m(\lambda)}{m(0)} \\
                &+ e^{i \dotlambda^{-1} \theta_m(\lambda)}e^{i \gamma_m(\lambda)} \dyad{\dlambda m(\lambda)}{m(0)} \Bigg] e^{-i \dotlambda^{-1} \theta_m(\lambda)}e^{-i \gamma_m(\lambda)} \dyad{m(0)}{m(\lambda)} \\
                 = &\sum_m \ket{m} E_m \bra{m} + i\hbar \dotlambda \sum_m (\dyad{\dlambda m}{m} - \braket{m}{\dlambda m} \dyad{m}),
            \end{aligned}
        \end{aligned}
    \end{equation}
    where in the last line the dependence on $\lambda$ has once again been removed from the eigenstates $\ket{m}$ and the eigenenergies $E_m$ noting that all terms for $\lambda = 0$ have been cancelled out. In order to analyse the equation more easily, we rewrite it in terms of two separate components:
    \begin{equation}\label{eq:transitionless_h0_h1}
        H_{\rm t-less}(\lambda) = H_0(\lambda) + H_1(\lambda),
    \end{equation}
    where 
    \begin{equation}
        \begin{aligned}
            H_0 &= \sum_m E_m \dyad{m}, \\
            H_1 &= i\hbar \dotlambda \sum_m (\dyad{\dlambda m}{m} - \braket{m}{\dlambda m} \dyad{m}).
        \end{aligned}
    \end{equation}

    What the above equation shows is that if we can engineer the Hamiltonian $H_{\rm t-less} (\lambda)$, it is possible to drive the system arbitrarily fast, as it will always follow the instantaneous eigenstates. This might seem like a strange statement, but it becomes a lot simpler when we consider that the term $H_1$ looks quite familiar: it is nothing more than the negation of the \acrref{AGP} component in Eq.~\eqref{eq:schrodinger_moving_frame_2}. To see this, let us recall what happens to states driven by the Hamiltonian $H_0$ by returning to Eq.~\eqref{eq:schrodinger_moving_frame_2}:
    \begin{equation}
    i \hbar \frac{d \ket*{\psi(\lambda)}}{dt} = \Big( H_0 - \dotlambda \AGP{\lambda} \Big) \ket*{\psi(\lambda)}.
    \end{equation}
As previously, recall that the additional \acrref{AGP} term scaled by $\dotlambda$ is responsible for the non-adiabatic effects experienced by a system as it gets driven in a time-dependent fashion. From Eq.~\eqref{eq:AGP_diagonal}, we know that the \acrref{AGP} operator can be expressed as
	\begin{equation}
	\begin{aligned}
		\mel{m(\lambda)}{\AGP{\lambda}}{m(\lambda)} &= i\hbar \mel{m(\lambda)}{\dlambda}{m(\lambda)} \\
		&= i\hbar \sum_m (\dyad{\dlambda m}{m} - \braket{m}{\dlambda m} \dyad{m}),
	\end{aligned}
	\end{equation}
which looks remarkably like $H_1$ from our transitionless Hamiltonian, without the $\dotlambda$ scaling factor\@i.e.~$H_1 = \dotlambda \AGP{\lambda}$. Putting these two ideas together, we find that the effective transitionless Hamiltonian $H_{\rm t-less}(\lambda)$ driving the state is just
    \begin{equation}\label{eq:AGP_cancelling_out}
    	H_{\rm t-less}(\lambda) = \Big( H_0 - \dotlambda \AGP{\lambda} \Big)  + \dotlambda \AGP{\lambda} = H_0,
    \end{equation}
    as expected. In the equation above, the effective Hamiltonian in the moving frame is simply the diagonalized version of the driving Hamiltonian $H(\lambda)$ in the lab frame, which drives the instantaneous eigenstates perfectly. This is the idea behind \emph{counterdiabatic driving} or \acrref{CD}. The name, unsurprisingly, stems from the fact that the additional `counterdiabatic' term $+ \dotlambda\AGP{\lambda}$ is added in order to `counter' the non-adiabatic or `diabatic' effects that arise in the effective Hamiltonian throughout the system's evolution. We will note that the second term in $H_1$ is often neglected in constructing the counterdiabatic drive, as it does not contribute to excitations out of the desired eigenstate(s), although it does contribute to a rescaling of their energies. In many applications of adiabatic processes, such a rescaling is not relevant, thus it may be omitted. 
    
    With all this in mind, we can explicitly define the counterdiabatic Hamiltonian:
    \begin{equation}\label{eq:CD_Hamiltonian}
    \HCD(\lambda) = H(\lambda) + \dotlambda \mathcal{A}_\lambda.
    \end{equation}
   If $\HCD$ is known and can be engineered, it is possible to drive a quantum system arbitrarily fast with no deformations associated with non-adiabatic effects. However, if this seems too good to be true, that's because in general it is. The first clue is in the form of the \acrref{AGP} in Eq.~\eqref{eq:AGP_adiabatic_basis}, which implies that in order to know this \acrref{CD} Hamiltonian, we'd need to not only know the full eigenspectrum of the lab frame Hamiltonian for each value of $\lambda$ throughout the protocol, but also to be able to engineer such terms to arbitrary precision in the lab. Furthermore, the off-diagonal elements of $\AGP{\lambda}$, as alluded to earlier, are proportional to the inverse of the energy gaps in the system $(E_n - E_m)^{-1}$. These can become exponentially small as the system size increases, making them diverge or become undefined in the thermodynamic limit \cite{kolodrubetz_geometry_2017, jarzynski_geometric_1995}. In chaotic systems, the \acrref{AGP} cannot be local because no local operator can distinguish many-body states with arbitrary small energy difference \cite{dalessio_quantum_2016}. What all of this really implies is that it is impossible - or at least impractical - to attempt to implement the exact counterdiabatic Hamiltonian given by Eq.~\eqref{eq:CD_Hamiltonian} in the general case, barring some very simple and small systems. This makes \acrref{CD} in its basic form quite impractical in the general case, although several experiments on small and relatively simple systems have demonstrated its effectiveness \cite{bason_high-fidelity_2012}.

    \subsection{Brief interlude: the waiter and the glass of water}

    It may seem like our inability to know or implement the exact \acrref{CD} Hamiltonian of Eq.~\eqref{eq:CD_Hamiltonian} in the general case brings us back to square one in trying to speed up adiabatic protocols. We will show in the next section that it turns out this is not the case at all. However, before we dive back into the math, we can take a moment to illustrate the concept of \acrref{CD} with a classical analogy which is often used in this circumstance \cite{sels_minimizing_2017}, and not only elucidates what we have talked about so far, but also gives some intuition for how we might overcome the practical problems associated with the exact \acrref{AGP}. Furthermore, it sets the stage nicely for the rest of the chapter.
    
    The story goes something like this: imagine that you are a waiter tasked with carrying a glass of water on a tray from the bar to some table on the other side of a rather large restaurant. As you begin to walk, while holding the tray perfectly level with the ground, your acceleration induces a force on the glass which causes it to wobble and the water to splash around. Ideally you would like to stop the water from spilling, so at this point you have two options: either to (a) walk slowly enough so as to minimize the force that is destabilizing the glass or else (b) suitably counteract it by, e.g.~tilting the tray. 

    You may already see where we are going with this. In the analogy, we can view the stable, upright state of the glass full of water as the ground state of some quantum system. The moving waiter then embodies the time-dependent Hamiltonian driving the system from this initial state, where we can model their changing coordinates as they move through the bar via the abstract parameter(s) $\lambda$. Just as in the case of the adiabatic condition of Eq.~\eqref{eq:adiabatic_criterion}, the probability of the glass tipping over depends on both the acceleration and direction of the waiter (the $\dlambda H$ term) as well as how inherently stable the glass is due to\@e.g.~a heavier bottom or more viscous liquid (the energy gap between the ground state and the nearest excited state). In this picture, the two methods the waiter can use to stabilize the glass during transport are analogous to (a) following the adiabatic condition by minimizing the speed at which the Hamiltonian is deformed or (b) applying some suitable control technique, such as counterdiabatic driving, to counteract the non-adiabatic force that appears as a consequence of their fast movement.
    
    This example is not only useful for gaining intuition about adiabaticity and \acrref{CD}, but can also be used to bring attention to several interesting observations. The first is that by including a counterdiabatic component, the waiter introduces a new degree of freedom - a tilt - which would otherwise not show up anywhere in the process or the start/end points of the journey of the glass. Secondly, from the point of view of someone standing by the wayside (the lab frame), the glass is nowhere near standing upright throughout the counterdiabatic tilt, rather it is in some highly excited state, while from the perspective of the glass (the moving frame) it is quite stable and generally close to the instantaneous ground state, as can be garnered from looking at Equations \eqref{eq:AGP_cancelling_out} and \eqref{eq:CD_Hamiltonian} which represent the two perspectives.
    
The most important observation, however, which springboards us into the next section of this chapter, is precisely one which answers the question: \emph{how stable is the glass throughout the waiter's counterdiabatic journey}? We cannot assume, in any realistic scenario, that the waiter has perfect knowledge of the movement of every molecule of water in the glass and can control their movements to such high precision that they instantly counteract even the smallest deviation from the perceived ground state. In fact, it is far more likely that the waiter has very limited ability to tilt the tray as well as only the roughest, low-resolution model of the ways in which the glass wobbles. The result is that far from implementing an `exact' \acrref{CD} Hamiltonian as in Eq.~\eqref{eq:CD_Hamiltonian}, the waiter produces only some high-level approximation of the $\dotlambda \AGP{\lambda}$ term. And yet, barring extreme circumstances, they manage to quickly and safely transport the glass from bar to table. 

    \section{The approximate counterdiabatic drive}\label{sec:2.4_approximate_CD}

    Taking inspiration from the waiter story, we might imagine that a similar idea will hold true for \acrref{CD} protocols in the quantum setting. Why try to derive and implement the exact Hamiltonian of Eq.~\eqref{eq:CD_Hamiltonian}, when some rough version will cancel out most of the non-adiabatic effects? Even in our derivation of the adiabatic condition in Sec.~\ref{sec:2.1.2_adiabatic_condition} we upper-bounded the terms responsible for the unwanted transitions out of the eigenstate rather than trying to work with the full expression. It is this exact philosophy that is the backbone of the rest of this section, where we explore the different ways in which the \acrref{AGP} -- and thus the counterdiabatic drive -- has been approximated, and what are the advantages and drawbacks of each approach. 
 
    \subsection{Local counterdiabatic driving}\label{sec:2.4.1_LCD}

    The first method we will explore was developed by Sels and Polkovnikov in \cite{sels_minimizing_2017}: a variational minimization approach which we will refer to throughout this thesis as local counterdiabatic driving or \acrref{LCD}. Taking inspiration from the story of the waiter in the previous section, we can imagine constraining our counterdiabatic degrees of freedom in some way due to physical restrictions. In the case of the waiter, this might be related to the reaction time and physical capabilities of the human body, while in the case of quantum systems such degrees of freedom are generally best expressed as operators which may be implemented in some physical system. In the case of many-body systems, engineering arbitrary highly non-local many-body operators is hard and experimentally one tends to have access to and control of only a limited set of physical operators. With this in mind, it makes sense to focus our approximation of the \acrref{AGP} to operators that are highly local or at least physically realisable, so that we could actually \emph{implement} them when the time comes.

    The task of not only finding a viable approximation of the \acrref{CD}, but also restricting it to a specific set of operators is not an easy one. Luckily, there is some structure in the \acrref{AGP} that we can exploit in order to write it in a slightly different form. We start by differentiating the eigenenergies of the instantaneous basis Hamiltonian:
    \begin{equation}\label{eq:energy_derivative}
        \begin{aligned}
            \frac{dE}{dt} &= \frac{d}{dt}\ev*{\Tilde{H}}{\Tilde{\psi}} \\
            &= \mel*{\partial_t \Tilde{\psi}}{\Tilde{H}}{\Tilde{\psi}} + \mel*{\Tilde{\psi}}{\partial_t \Tilde{H}}{\Tilde{\psi}} + \mel*{\Tilde{\psi}}{\Tilde{H}}{\partial_t \Tilde{\psi}} \\
            &= \frac{i}{\hbar}\ev*{(\Tilde{H} - \dotlambda \Tilde{\AGP{\lambda}})\Tilde{H}}{\Tilde{\psi}} + \dotlambda \ev*{\dlambda \Tilde{H}}{\Tilde{\psi}} - \frac{i}{\hbar}\ev*{(\Tilde{H} - \dotlambda \Tilde{\AGP{\lambda}})\Tilde{H}}{\Tilde{\psi}}\\
            &= \dotlambda \ev*{\dlambda \Tilde{H}}{\Tilde{\psi}} - \frac{i}{\hbar} \dotlambda \ev*{\comm*{\Tilde{\AGP{\lambda}}}{\Tilde{H}}}{\Tilde{\psi}} \\
            \Rightarrow i\hbar F_{\lambda} &= \comm{\AGP{\lambda}}{H} - i \hbar \dlambda H, \\
            \Rightarrow \comm{\AGP{\lambda}}{H} &= i\hbar \left( F_{\lambda} + \dlambda H\right),
        \end{aligned}
    \end{equation}
    where we use the result from Eq.~\eqref{eq:moving_frame_schrodinger} that $i\hbar \partial_t \ket*{\Tilde{\psi}} = (\Tilde{H} - \dotlambda \Tilde{\AGP{\lambda}})\ket*{\Tilde{\psi}}$ and
    \begin{equation}\label{eq:generalized_force_operator}
        F_{\lambda} = - \sum_n \dlambda E_n(\lambda) \dyad{n(\lambda)},
    \end{equation}
    is the generalised force operator \cite{kolodrubetz_geometry_2017,sels_minimizing_2017,jarzynski_generating_2013}. 

    It turns out that this result can be used to quantify how close some arbitrary operator is to the \acrref{AGP}. In order to do this, we first define an \emph{ansatz} Hermitian operator for $\AGP{\lambda}$ which acts on the same Hilbert space and which we denote $\approxAGP$. Then, we can define an operator $G_{\lambda}$:
    \begin{equation}\label{eq:G_operator}
        G_{\lambda}(\approxAGP) = \dlambda H + \frac{i}{\hbar} \comm{\approxAGP}{H}.
    \end{equation}
    We can see that when $\approxAGP = \AGP{\lambda}$, \@i.e.~when our guess - or approximation - for the \acrref{AGP} is exactly correct, then $G_{\lambda}(\AGP{\lambda}) = - F_{\lambda}$. This fact essentially allows us to reformulate the problem of trying to determine the \acrref{AGP} into one of minimization of distance between the operators $G_{\lambda}(\approxAGP)$ and $-F_{\lambda}$ with respect to the ansatz $\approxAGP$. 

    There are several options for a distance metric between two operators, each providing different information about their properties . However, for our purpose, the task can be simplified simply by noticing that in the case where the ansatz is exact, $G_{\lambda}$ has no off-diagonal elements, or $\comm{H}{G_{\lambda}(\AGP{\lambda})} = 0$. Thus a way to minimise the distance between $G_{\lambda}(\approxAGP)$ and $-F_{\lambda}$ is simply to minimise its Hilbert-Schmidt norm, as this fully and efficiently captures the desired properties of the operator $G_{\lambda}(\AGP{\lambda})$. Let us express this norm as an \emph{action}\cite{kolodrubetz_geometry_2017} associated with the \acrref{AGP}:
    \begin{equation}\label{eq:agp_action}
        \mathcal{S}(\approxAGP) = \Tr[G^2_{\lambda}(\approxAGP)],
    \end{equation}
    which is minimised whenever $\approxAGP$ satisfies:
    \begin{equation}
        \left. \frac{\delta \mathcal{S}(\approxAGP)}{\delta \approxAGP} \right\vert_{\approxAGP = \AGP{\lambda}} = 0 \quad \Rightarrow \quad \left[ H, \dlambda H + \frac{i}{\hbar}\comm{\AGP{\lambda}}{H} \right] = 0.
    \end{equation}

    This all leads to a relatively simple recipe for finding a local, physically realisable counterdiabatic drive. To do this, we can choose a set of operators $\{\mathcal{O}_{\rm LCD}\}$ which satisfy the constraints of our physical system. We can then define an approximate \acrref{AGP} in the basis of these operators as:
    \begin{equation}\label{eq:LCD_basis}
        \approxAGP = \sum_j \alpha_j (\lambda) \mathcal{O}_{\rm LCD}^{(j)},
    \end{equation}
    where the index $j$ indicates the $j^{\rm th}$ operator in the basis and the coefficients $\alpha_j(\lambda)$ describe the continuous schedule of the counterdiabatic drive. Once we choose a set of operators $\{\mathcal{O}_{\rm LCD}\}$, we can think of them as a fixed parameter, and the minimisation procedure consists of minimising the resulting action $\mathcal{S}(\approxAGP)$ with respect to the coefficients $\alpha_j(\lambda)$.

    To make this clearer, let us return to the rotating spin Hamiltonian from Eq.~\eqref{eq:rotating_spin_H}. In order to simplify things, we can rewrite it with a change in parameters, taking $\lambda(t) = \frac{\pi t}{2 \tau}$:
    \begin{equation}\label{eq:rotating_spin_H_lambda}
        H(\lambda) = -\cos(\lambda)\sx - \sin(\lambda)\sz.
    \end{equation}

    Since this is such a simple example, the only operators we could possibly include in the basis for our approximate \acrref{AGP} are single-spin operators. While any of the single-spin Pauli operators $\{ \sx, \sy, \sz\}$ are viable choices here, we note that it is not hard to see from $\Tilde{\AGP{\lambda}} = i\hbar \adj{U} \dlambda U$ that if the Hamiltonian $H(\lambda)$ is real, the counterdiabatic term should be purely imaginary, as follows from the fact that a real Hamiltonian can always be diagonalised by a real orthogonal matrix $U$. If a real operator is elected as the ansatz in this case, we will find that the coefficients $\alpha_j$ of such operators will be equal to $0$. This leaves us with a single degree of freedom that could act as the basis of $\approxAGP$, which is $\sy$:
    \begin{equation}
        \approxAGP = \alpha(\lambda) \sy.
    \end{equation}
    In fact, as there are no other operators in this basis that could fit the description of being both a single-spin operator \emph{and} imaginary, we expect that this ansatz should, for the correct $\alpha$, be equal to the exact \acrref{CD}.

    \begin{wrapfigure}{R}{0.4\textwidth}
    \centering
    \includegraphics[width=0.9\linewidth]{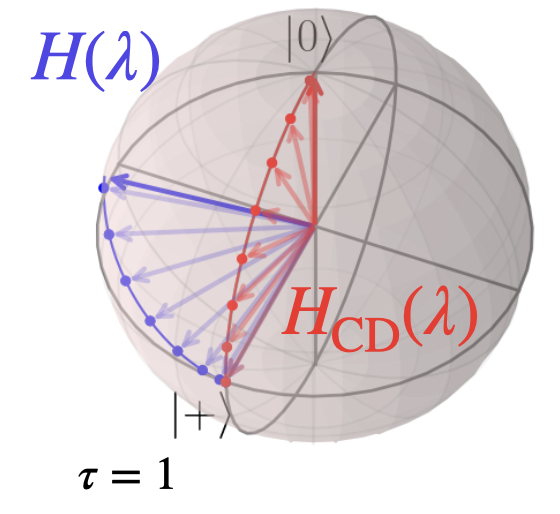} \caption[Rotating spin Bloch sphere illustration with counterdiabatic driving]{State of the rotating spin starting in state $\ket{+}$ driven without \acrref{CD} as in the Hamiltonian of Eq.~\eqref{eq:rotating_spin_H_lambda} (blue) and with \acrref{CD} as given by Eq.~\eqref{eq:rotating_spin_HCD} (red) for total driving time $\tau = 1$.}\label{fig:rotating_CD}
    \end{wrapfigure}
    
    All that remains is to find $G_{\lambda}(\approxAGP)$ and to minimize the corresponding action $\mathcal{S}(\approxAGP)$ with respect to the driving coefficient $\alpha$. Once this process is complete (see Appendix~\ref{app:rotating_spin_hamiltonian} for details), we find that
    \begin{equation}\label{eq:rotating_spin_alpha}
        \alpha(\lambda) = -\frac{\sin^2(\lambda) + \cos^2(\lambda)}{2(\sin^2(\lambda) + \cos^2(\lambda))} = - \frac{1}{2},
    \end{equation}
    meaning that the counterdiabatic Hamiltonian can be written simply as:
    \begin{equation}\label{eq:rotating_spin_HCD}
        \begin{aligned}
            \HCD(\lambda) &= H(\lambda) + \dotlambda \alpha(\lambda) \sy \\
            &= -\cos(\lambda)\sx - \sin(\lambda)\sz - \frac{\pi}{4 \tau} \sy,
        \end{aligned}
    \end{equation}
    where we have used the fact that $\dotlambda = \pi/2\tau$. In Fig.~\ref{fig:rotating_CD}, we can see that even at very fast driving times, the rotating spin does not stray from the plane of rotation when the \acrref{CD} is applied. We can compare this to Fig.~\ref{fig:bloch_rotating_spin}, where similar dynamics without the application of a \acrref{CD} drive were only achieved at around $500$ times longer driving speeds.
    
    In this case, it turns out that the counterdiabatic term is constant as a result of the choice of basis and $H(\lambda)$. In general, however, this is not the case and the \acrref{CD} term depends on $\lambda$ through the coefficients of the lab frame Hamiltonian. Furthermore, while for this example only one operator was needed to describe the full \acrref{CD}, the number of such possible operators for a many-body system grows exponentially with system size, meaning that restricting to a highly-local and physically realisable basis is quite a sizeable reduction in the true number operators in the full \acrref{AGP}. One may yet only hope that the exact gauge potential has significant support only over a small, finite subset of all the possible relevant operators that could be implemented \cite{lawrence_numerical_2024}.
    
    \subsection{Nested commutator expansion}\label{sec:2.4.2_nested_commutators}

    The \acrref{LCD} approach is particularly useful in the case where one wants to implement a \acrref{CD} approximation constrained by some very limited, pre-determined set of operators, but it says absolutely nothing about what the operators should be when no constraints are imposed. A useful question to ask is whether or not there is any way to know what the operator basis of the approximate \acrref{CD} should be prior to performing the optimisation. This is useful not only in the case of determining the form of the \acrref{CD} in order to implement it, but also as a general tool in characterising non-adiabatic effects.

    In this section we will focus on an approach developed in \cite{claeys_floquet-engineering_2019}, where it was found that the \acrref{AGP} to some $\ell^{\rm th}$ order can be extracted from a series of nested commutators:
    \begin{equation}\label{eq:nested_commutator_AGP}
        \Bar{\AGP{\lambda}}^{(\ell)} = i\hbar \sum_{k=1}^\ell \alpha_k(\lambda) \underbrace{[H(\lambda),[H(\lambda),...[H(\lambda)}_{2k-1},\dlambda H(\lambda)]]],
    \end{equation}
    where the coefficients $\alpha_k(\lambda)$ are used in a similar manner as \acrref{LCD}. The minimisation procedure outlined in section \ref{sec:2.4.1_LCD} can be implemented to determine the coefficients $\alpha_k(\lambda)$ for all orders of the nested commutator expansion. On the other hand, one might choose to instead reparameterise and use a different set of coefficients: one $\alpha$ for each orthogonal operator that is obtained in the nested commutator expansion after a chosen number of commutations. The primary difference between the two approaches is merely the parameterisation of the approximate counterdiabatic drive. A more fine-grained parameterisation, with a larger number of coefficients $\alpha$, is liable to give a better approximation of the drive. In the original work \cite{claeys_floquet-engineering_2019}, $\alpha_k(\lambda)$ is used rather than a different parameterisation, as it allows one to easily determine how to engineer a Floquet Hamiltonian that implements the given counterdiabatic drive. This is due to a similarity in the structure of the high-frequency expansion of the Floquet Hamiltonian and the nested commutator expansion described above.
    In the limit of $\ell \rightarrow \infty$, the expression in Eq.~\eqref{eq:nested_commutator_AGP} should represent the exact \acrref{AGP}, although there is no guarantee of convergence prior to this point as during each iteration of the commutations, the set of operators that are obtained need not be orthogonal to the previous set. This, however, may not be an issue in practice, as due to previously discussed reasons relating to a difficulty in implementation, we generally only wish to obtain a simple approximation of the counterdiabatic terms.
    As noted in \cite{claeys_floquet-engineering_2019}, there are several ways to motivate this form of the \acrref{AGP}, \@e.g.~by noticing that such commutator terms appear in the Baker-Campbell-Hausdorff (BCH) expansion in the definition of a (properly regularized) \cite{jarzynski_geometric_1995} \acrref{AGP} for a fixed $\lambda$:
    \begin{equation}\label{eq:regularised_AGP}
        \AGP{\lambda} = \lim_{\epsilon \rightarrow 0^{+}} \int_0^{\infty} dt e^{-\epsilon t} \left( e^{-iH(\lambda) t}\dlambda H(\lambda)e^{iH(\lambda) t} + F_{\lambda} \right),
    \end{equation}
    where $F_{\lambda}$ is defined in Eq.~\eqref{eq:generalized_force_operator}. From the BCH expansion, we can find
    \begin{equation}
        e^{-iH t}\dlambda H e^{iH t} = \sum_{k = 0}^{\infty} \frac{(-it)^k}{k!} \underbrace{[H,[H,...[H}_{k},\dlambda H]]],
    \end{equation}
    where even-order commutators contribute to $F_{\lambda}$ and odd-order commutators to $\AGP{\lambda}$.
    
    To gain more intuition for Eq.~\eqref{eq:nested_commutator_AGP}, one can try to evaluate it in the instantaneous eigenbasis of $H(\lambda)$:
    \begin{equation}\label{eq:nested_commutator_matrix_elements}
        \begin{aligned}
            \mel*{m}{\Bar{\AGP{\lambda}}^{(\ell)}}{n} &= i\hbar \sum_{k=1}^\ell \alpha_k(\lambda) \mel*{m}{\underbrace{[H(\lambda),[H(\lambda),...[H(\lambda)}_{2k-1},\dlambda H(\lambda)]]]}{n} \\
            &= i\hbar \Bigg[ \sum_{k=1}^\ell \alpha_k(\lambda)(E_m - E_n)^{2k - 1} \Bigg] \mel{m}{\dlambda H}{n},
        \end{aligned}
    \end{equation}
    where we can see that the term we obtain at the end looks very similar to the matrix elements we got in deriving the \acrref{AGP} in Eq.~\eqref{eq:differentiating_adiabatic_offdiagonals}. In the case of the nested commutator expansion then, the use of the variational \acrref{LCD} approach in determining the coefficients $\alpha_k$ is equivalent to trying to approximate the factor $(E_m - E_n)^{- 1}$ in the exact \acrref{AGP} via a power-series approximation:
    \begin{equation}
        \alpha_{\lambda}^{(\ell)}(\omega_{mn}) = \sum_{k=1}^\ell \alpha_k \omega_{mn}^{2k - 1},
    \end{equation}
    where $\omega_{mn} = (E_m - E_n)$. While this shows that the nested commutator approximation wouldn't work in regimes where the energy gap is exponentially small or exponentially big (\@i.e.~where $\omega_{mn} \rightarrow 0$ or $\omega_{mn} \rightarrow \infty$), this turns out to not be an issue in practice. In the limit of very large energy gaps, the term $\mel{m}{\dlambda H}{n}$ decays exponentially meaning that the contribution from these elements to the \acrref{AGP} is negligible anyway. As the energy gaps close, the \acrref{AGP} elements become undefined and generally in speeding up adiabatic processes, one only cares about suppressing transitions across some energy gap $\Delta$. In that case, as long as $\omega_{mn} \geq \Delta$, the approximation does its job in the \acrref{CD} protocol.
    
    While the nested commutator expansion does not appear in later chapters, it forms a backbone in the research on practical approximations of counterdiabatic protocols and was used extensively behind the scenes for many of the results presented in the thesis. As such, we found it prudent to include. For further reading on how one might combine the results from the rest of the thesis and the nested commutator method to great success, we refer the reader to \cite{lawrence_numerical_2024}.

\chapter{Quantum Optimal Control}\label{chap:3_Quantum_Optimal_control}
\epigraph{``Neo, sooner or later you’re going to realize, just as I did, that there’s a difference between knowing the path and walking the path."}{Morpheus, \emph{The Matrix (1999)}}

The future of quantum technologies depends on our ability to control quantum systems with precision and accuracy. It is a key factor in the realisaton of, for example, quantum computers \cite{ball_software_2021}, communication systems \cite{omran_generation_2019} and quantum sensors \cite{le_robust_2021}, as well as being necessary in the exploration and understanding of fundamental physics. The research field which concerns itself with such control problems is generally known as Quantum Optimal Control Theory (\acrref{QOCT}) \cite{koch_quantum_2022,glaser_training_2015} and its primary objective is the development of techniques which allow for the construction and analysis of strategies, primarily electromagnetic field shapes, that manipulate quantum dynamical processes in the most efficient and effective way possible in order to achieve certain objectives. Common control objectives in the quantum setting can range from state preparation \cite{zhang_when_2019} and quantum gate synthesis \cite{pelegri_high-fidelity_2022}, to protection against decoherence \cite{rooney_decoherence_2012} and entanglement generation \cite{omran_generation_2019}.

While the field of quantum optimal control is vast and would take an entire book to summarize \cite{dalessio_quantum_2016}, this chapter aims to give a broad overview of the topic highlighting its structure, mechanisms, and practical applications, in particular with respect to the methods that are relevant to the rest of the work presented in this thesis. As such, in Sec.~\ref{sec:3.1_structure_quantum_control}, we will begin by exploring the general structure of optimal control problems in detail and showing how an abstract goal can be transformed into a quantitative formula that guides us toward a desired outcome satisfying a given control objective. First, we will discuss the mathematical structure of optimal control problems (Sec.~\ref{sec:3.1.1_mathematical_structure}) followed by an overview and examples of analytical (Sec.~\ref{sec:3.1.2_analytic_optimisation}) and numerical (Sec.~\ref{sec:3.1.3_numerical_optimisation}) methods for finding solutions to said problems, with a focus on methods that will be relevant to the rest of the content in this thesis. Sec.~\ref{sec:3.2_Quantum_optimal_control} will review how optimal control is adapted in the quantum setting and the main idea behind \acrref{QOCT}, while Sec.~\ref{sec:3.3_qoct_methods} will focus on specific methods used for constructing and optimising driving pulses with quantum systems in mind.

\section{The structure of optimal control problems}\label{sec:3.1_structure_quantum_control}

The idea of an optimal control problem is simple: envision a target you want to achieve, cast it into some form of quantitative or abstract mathematical formula and then use this formula to derive the `best' path to get to said objective. There may be many paths to achieve the target and there may be many metrics to determine what `best' means. The aim of the first part of this section is thus to broadly cover the mathematical structure of optimal control problems and to try and convey an idea of what an optimal path \emph{is} and how one might quantify its optimality. Later in the chapter, we will delve more into practical questions of controllability and the process of optimisation, \@i.e.~once an optimal control problem is constructed, how could one go about finding the solution to it. We will cover both analytical methods in Sec.~\ref{sec:3.1.2_analytic_optimisation} and numerical approaches in Sec.~\ref{sec:3.1.3_numerical_optimisation} focusing on a select few optimisation algorithms which will be relevant to further chapters of this thesis. 

\subsection{Mathematical structure}\label{sec:3.1.1_mathematical_structure}

In general, an optimal control problem is composed of a set of state functions $X: \R \rightarrow \R^n$, and a set of time-dependent control functions $U:\R \rightarrow \R^m$ and the optimal control problem consists of finding $x \in X$ and $u \in U$ that minimise some functional $C: X \cross U \rightarrow \R$ such that the constraint
\begin{equation}\label{eq:control_ODE}
    \dot{x} = f(x, u),
\end{equation}
is satisfied almost everywhere. This is a very abstract description and just about any control problem can be expressed as a special case of this formulation \cite{dalessandro_introduction_2021}. To gain more intuition, we can imagine a more concrete example where, \@e.g.~$U$ and $X$ are sets of continuous functions on the interval $[0, \tau]$ satisfying $x(0) = x_0$. In this scenario, $\tau$ could be a time interval during which we want to drive the system from an initial state $x_0$ to a final state $x_f$ using the control function $u(t)$, $t \in [0, \tau]$. The choice of functional $C$ would have to capture the desired outcome of the protocol: that the state of the system after the driving $x(\tau)$ be equal to the target $x_f$. This can be done by choosing a distance metric that depends only on the drive $u$ and is minimised when $x(\tau) = x_f$, \@~e.g.
\begin{equation}\label{eq:example_cost_func}
    C(u) = \norm{x(\tau)  - x_f},
\end{equation}
where $\norm{\cdot}$ represents some norm on the space $X$.

	The functional $C$ is often referred to in literature as the \emph{cost} or \emph{loss} function \cite{wald_statistical_1950} as it encodes the quality of the final protocol with respect to the desired outcome of the protocol. In that sense, we can imagine adding constraints to the problem that may increase the `cost' of the protocol output if they are not satisfied to some degree. For example, Eq.~\eqref{eq:example_cost_func} can be modified to include additional terms:
\begin{equation}\label{eq:example_cost_func2}
    C(u) = \gamma \norm{x(\tau)  - x_f}^2 + \int_0^{\tau} \norm{u(t)}^2 dt,
\end{equation}
where $\gamma$ is a penalty term on the final state that scales its importance relative to the additional second term, which is analogous to the cost in the energy required to achieve the final state. This updated cost function can be read as introducing a competition between the quality of the final state and the amount of energy expended to get it there, mediated by the value of $\gamma$.

\begin{figure}[t]
\centering
\includegraphics[width=0.9\linewidth]{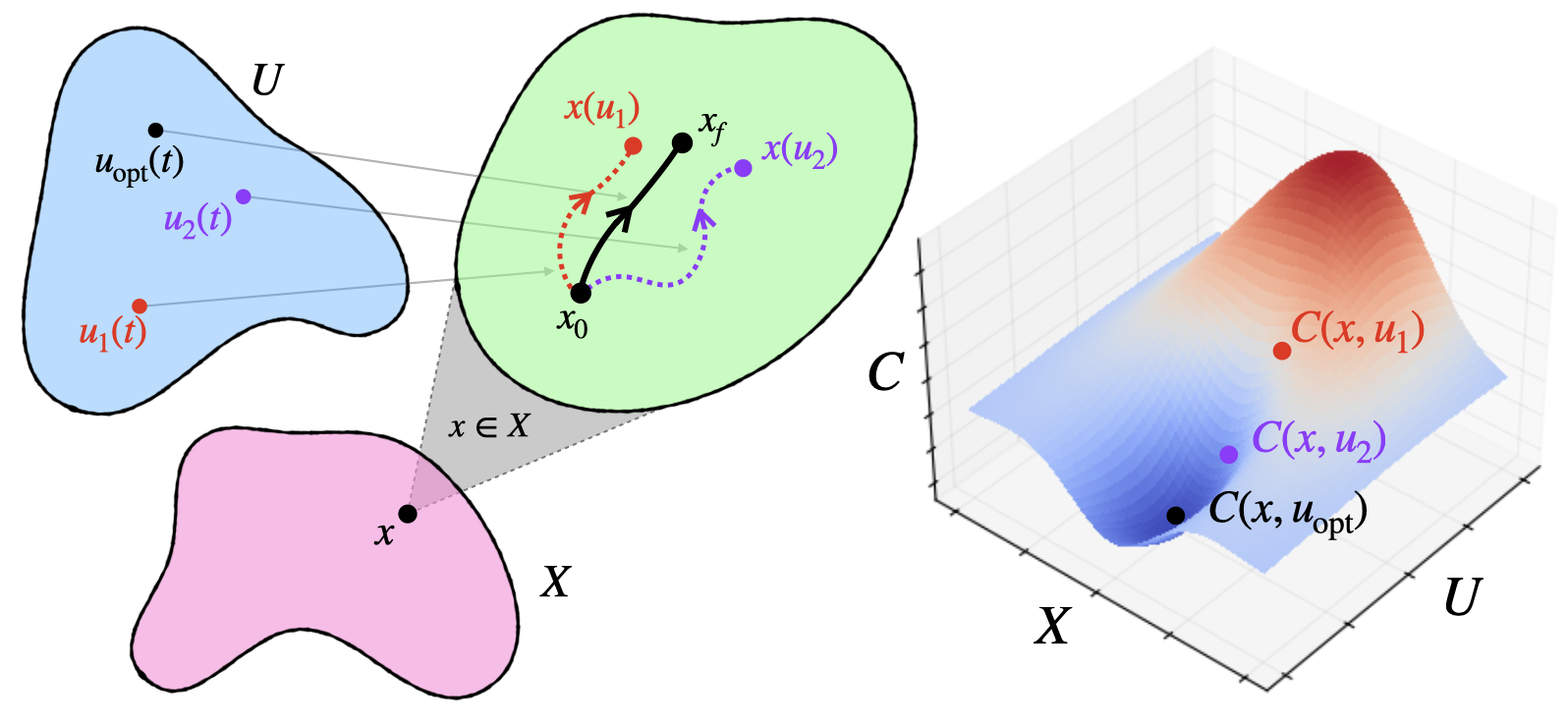} \caption[Illustration of optimal control problem structure]{Illustration of the Mayer-type optimal control problem: when an initial value of the system state $x_0$ is fixed, the choice of control function $u \in U$ and the requirement of satisfying Eq.~\eqref{eq:control_ODE} determine $x$ uniquely. The task is then to find $u_{\rm opt}$ such that the functional $C(x, u_{\rm opt})$ is minimised.}\label{fig:optimal_control}
\end{figure}

There are primarily three different types of problem structures in optimal control centering on different constraints and targets: \emph{Mayer-type}, \emph{Lagrange-type} and their combination, \emph{Bolza-type} problems \cite{dalessandro_introduction_2021}. In this thesis, we will mostly focus on Mayer-type problems, particularly in Ch.~\ref{chap:4_COLD} and Ch.~\ref{chap:6_Applications_fidelity}. In Mayer-type problems, the initial state is specified $x(0) = x_0$ and the cost function is of the form
\begin{equation}\label{eq:mayer_costfunc}
    C(u) = \phi(x(\tau), \tau),
\end{equation}
with $\phi$ a smooth function and $\tau$ the total time of the protocol. These two constraints and the requirement given by Eq.~\eqref{eq:control_ODE} define the state function $x$ uniquely and the problem is then to determine a control function $u$ on the appropriate set $[0, \tau]$ which minimises Eq.~\eqref{eq:mayer_costfunc}. We note that the expression in Eq.~\eqref{eq:mayer_costfunc} is quite general and can include multiple types of `constraints', e.g.~as a linear superposition. In Mayer-type problems, a specific target state can be defined in the cost function as a constraint, which is the case in Eq.~\eqref{eq:example_cost_func} and this is illustrated in Fig.~\ref{fig:optimal_control}. However, this need not be the case as target states can be made implicit by having the cost function target some property of the state instead, like Euclidean distance from the initial state in the case of real vectors over Cartesian coordinates. 

From the above, we can view Mayer-type problems as being concerned primarily with the final state of the system and not its path. Lagrange-type problems, on the other hand, put focus on the behaviour of the system throughout the control trajectory and they encompass cost functions of the type
\begin{equation}\label{eq:lagrange_type_costfunc}
    C(u) = \int_0^{\tau} L(x, u, t) dt,
\end{equation}
where $L$ is a smooth function. This type of cost function is applicable, for example, in cases where one wants to minimise the expenditure of some path-dependent resource during the control procedure, or where a path-dependent quantity is easier to optimise over than a target state quantity. This type of optimisation is something that will become relevant in Ch.~\ref{chap:5_cd_as_costfunc} and Ch.~\ref{chap:7_higher_order_agp}. 

The most general type of problem is the Bolza-type problem, which combines both Mayer and Lagrange in a way that puts emphasis both on the target state of the optimal control and the trajectory that a system takes to get there:
\begin{equation}\label{eq:bolza_tyoe_costfunc}
    C(u) = \phi(x(\tau), \tau) + \int_0^{\tau} L(x, u, t) dt,
\end{equation}
where $\phi$ and $L$ are smooth functions given in Eq.~\eqref{eq:mayer_costfunc}. A great example of Bolza-type problems is the cost function given by Eq.~\eqref{eq:example_cost_func2}, which comprises a competition between distance to a target state and the energy expended to drive the system to said state. 

Apart from identifying the basic anatomy of control problems in terms of $X$, $U$ and $C$, there is a myriad of additional information about their mathematical structure that can help to analyse and thus solve them. For example, it might be useful to identify if, for a particular optimal control problem, the system in question is \emph{controllable}\cite{dirr_lie_2008, fleming_optimal_1975} \@i.e.~can any initial state be transformed into any desired target state. Equally, it might be useful to study the related concept of \emph{reachable sets}\cite{vom_ende_reachability_2020, fleming_optimal_1975}, which are sets containing all the states that an initial state can be driven to by the set of control functions $U$. It is not hard to see how the concept of controllabilty relates to reachability: a system is controllable if its reachable set contains all target states. In the case of Mayer-type problems, for example, it might be sensible to define a reachable set parameterised by the final evolution time $\tau$ such that it contains all possible states that can be obtained by the system during a driving time $\tau$. Finally, it would be remiss not to mention the concept of \emph{necessary conditions for optimality}\cite{mangasarian_sufficient_1966}, which focus on determining what formal conditions need to be satisfied for a specific control $u \in U$ to be optimal. Generally, this involves perturbing an assumed optimal control $u$ by some small parameter $\epsilon$ giving $u^{\epsilon}$ and then imposing the constraint that
\begin{equation}\label{eq:optimality_condition}
    C(u^{\epsilon}) - C(u) \geq 0,
\end{equation}
which is then considered the necessary condition for optimality. The most basic of these optimality conditions is the Pontryagin maximum principle or \acrref{PMP} \cite{boltyanski_nonclassical_1999} (see Appendix \ref{app:PMP}), which states that for an optimal control problem, the optimal control and state trajectories should maximize a specific function which combines the system dynamics, the control inputs, and the Lagrange multipliers which encode the constraints of the control problem.

\subsection{Analytical optimisation}\label{sec:3.1.2_analytic_optimisation}

While the first part of optimal control is the construction of the problem, the second part is the search for a solution. The methods used to do this can generally be classified either as analytical or numerical approaches. While both are widely used in optimal control theory, this thesis will largely only focus on the latter, as such we will be brief in introducing the former. 

Analytical optimal control techniques are those that leverage mathematical rigor and formalism to derive solutions or insights, as opposed to relying primarily on numerical simulations, heuristics, or experimentation. They provide a theoretical foundation for understanding the properties and solutions of optimal control problems and are closely related to the discussion in Sec.~\ref{sec:3.1.1_mathematical_structure}. They can allow for a complete geometric understanding of the control problem leading to, for example, knowledge of the structure of a solution or even some proof about a global optimum. For a given set of constraints they might even be used to derive time limits of state transformations, \@i.e.~the concept of reachability. An example of analytical methods is the aforementioned \acrref{PMP}, which provides information about the optimal solution via a set of differential equations. A different analytical control theory tool, the Hamilton-Jacobi-Bellman equation \cite{yong_dynamic_1999}, provides a way to find the optimal protocol via dynamic programming\cite{sniedovich_dynamic_2010}. We note here that the concepts introduced in the previous chapter may be viewed as a form of analytical optimisation - computation of the \acrref{AGP} (Sec.~\ref{sec:2.2_AGP}) analytically provides information about the non-adiabatic effects a system experiences given a certain path through the Hamiltonian parameter space and \acrref{CD} (Sec.~\ref{sec:2.3_CD}) then provides an optimal control protocol that drives a system via a desired trajectory. The cost function can be viewed as some measure of the magnitude of non-adiabatic effects generated by the \acrref{AGP} -- something we will explore in more detail in Ch.~\ref{chap:5_cd_as_costfunc}.

The trouble with analytical approaches, despite the commonplace rigorous guarantees of optimality and the scope of information they provide about the system, trajectory and structure of the control problems and their solutions, is that they are very difficult to scale up and quite inflexible to complex problem constraints. Once again, the concept of \acrref{CD} from the last chapter provides and excellent example of this problem, since an exact counterdiabatic drive may get exponentially more difficult to compute as the system scale or complexity increases. As such, analytical approaches are generally reserved for special cases, when problems have low dimensionality and simple structures with a cost function that is generally linear in the arguments. Many real-world control systems require more complexity and flexibility than can be afforded by analytical methods.

\subsection{Numerical optimisation}\label{sec:3.1.3_numerical_optimisation}

To overcome the drawbacks of analytical approaches, many optimal control problems are instead solved using numerical optimisation methods. These are generally algorithmic, iterative  techniques which explore the cost function landscape step-by-step in order to converge to a minimum value. Numerical methods, as a general rule, do not offer the same analysis or guarantees of optimality that analytical methods do. Their iterative nature may lead to a dependence of the outcome on the initial conditions of the algorithm, such as an initial guess for an optimal solution from which the iterations proceed or the bounds on the search space. Despite these drawbacks, however, numerical methods tend to be far more popular than analytical ones simply due to their flexibility and applicability. Where analytical approaches fail, the only way forward is often a numerical method.

A general numerical optimisation technique consists of an initialisation step, a series of search steps and a termination step. These can be summarised as follows:
\begin{enumerate}
    \item \emph{Initialisation}: set up the necessary constraints of the optimal control problem, such as bounds on the solution space or an initial guess for the optimal solution.
    \item \emph{Search}: Perform some iterative search steps (deterministic or stochastic) with the goal of converging to the minimum of the cost function. What constitutes a single step varies massively between different techniques.
    \item \emph{Termination}: Return a solution after some condition is satisfied. This can be a convergence criterion based on the change in the cost function value between steps or a limit on the number of search steps that the algorithm is allowed to perform.
\end{enumerate}

The simplicity of these three components leaves a lot of room for creativity and over the years many numerical optimisation algorithms and techniques have been developed to deal with different constraints and topologies of various cost function landscapes. It would take an entire book \cite{nocedal_numerical_2006} to cover the various categories and subcategories that exist within the field, so we will restrict ourselves to exploring a few key classifications of the structure of numerical optimisation methods. 

One of the more broad ways to classify numerical optimisation methods is into the categories of \emph{gradient-based} methods and \emph{gradient-free} methods. Gradient-based methods, as the name implies, make use of gradient information (the first derivative of the cost function) to guide the search for an optimal solution. These methods are often efficient and converge rapidly when the cost function is smooth and differentiable. A popular example of a gradient-based method is the gradient descent algorithm, which iteratively adjusts the solution in the direction opposite to the gradient, as this direction is likely the steepest decrease in the cost function value. A typical gradient descent protocol might look like:
\begin{equation}\label{eq:gradient_descent}
    \ubb_{n + 1} = \ubb_n - \mu \grad_{\ubb} C(\ubb_n),
\end{equation}
where $n$ denotes the current iteration of the algorithm, $\grad_{\ubb} C(\ubb_n)$ is the derivative of the cost function $C$ with respect to the control parameters $\ubb$ and $\mu$ is generally known as the `learning rate' or `step size' and its job is to control the resolution at which the algorithm traverses the cost landscape. Larger $\mu$ might lead to faster convergence but it might also mean overshooting the cost function minimum, so adjusting its value is often a heuristic that requires some experimentation. One way that many popular optimisers, such as ADAM \cite{kingma_adam_2015}, overcome this issue is by implementing a variable, adaptive value of $\mu$ throughout the optimisation process. Other examples of gradient-based methods include Newton's method and quasi-Newton methods \cite{suli_introduction_2003}, which employ information about the second derivative to guide the search and provide faster convergence as well as a myriad of other approaches including stochastic methods \cite{bottou_tradeoffs_2007}. 

Gradient-free methods, on the other hand, do not require gradient information, making them suitable for optimization problems where the cost function is, \@e.g.~ discontinuous, non-differentiable, or its gradient is difficult or expensive to compute. Examples of gradient-free methods include particle swarm optimization \cite{bonyadi_particle_2017}, the Nelder-Mead method, which we will explore in more detail in Sec.~\ref{sec:3.1.3.1_Nelder_Mead} as well as the Powell method of Sec.~\ref{sec:3.1.3.2_Powell}. These methods often rely on trial and error, random sampling, or mimicking natural phenomena like evolutionary mechanisms\cite{vikhar_evolutionary_2016} to explore the solution space. As in the case of gradient-based approaches, there is a veritable zoo of methods under this umbrella. As the rest of this thesis we will deal almost exclusively with gradient-free methods, we will provide examples of how these techniques look in the next couple of sections.

Apart from the gradient-information, another key way to classify optimisation algorithms is either as \emph{local} or \emph{global}. Local optimization methods are designed to find a local minimum, which is a solution that is better than all other feasible solutions in its vicinity in the landscape of the cost function. They are typically efficient at converging to the local minimum, but they provide no guarantee of finding the global minimum if the cost function is non-convex \@i.e.~the local minimum is not automatically also the global minimum. Both Nelder-Mead and Powell are local methods.

Global optimization methods, on the other hand, aim to find a global optimum, which is the best solution among all feasible solutions, not just those in a local neighborhood. These methods typically employ a strategy to explore the entire solution space, either deterministically or stochastically, to avoid getting trapped in a local optimum. As a result of this larger scope, global optimization methods are generally more computationally intensive than local methods. An example of global optimisation that we will explore in more detail in Sec.~\ref{sec:3.1.3.3_dual_annealing} is Dual-Annealing, which combines generalized simulated annealing\cite{tsallis_generalized_1996}, a global search algorithm, with local optimisers in order to find an optimal solution. Global methods are often used when the optimization problem is complex, non-convex, or the global solution is significantly better than any local solution.

Finally, in numerical optimal control we can make a distinction between open-loop and closed-loop optimisation, particularly when referring to the real-life use or experiments on a given system:
\begin{itemize}
    \item \emph{Open-loop} approaches calculate the control sequence ahead of time and apply it to the system irrespective of the system's actual behavior during the protocol. 
    \item \emph{Closed-loop} methods actively adjust the control strategy based on the current and past states of the system (see Fig.~\ref{fig:quantum_optimal_control}).
\end{itemize}
The closed loop approach is more resilient to uncertainties and disturbances but requires real-time computation or pre-computed feedback laws. In this thesis, the focus will be exclusively on open-loop approaches, as closed-loop methods require access to live experimental data which was not available in the case of the methods explored in later chapters. However, it is important to acknowledge that the results obtained in open-loop optimisations may not reflect the realistic, complex response a physical system might have to a specific control protocol, given that the model we use may not include the full details of the physical system.

In the following sections we give examples of some common numerical optimisation methods that were used to obtain the results presented in this thesis.

\subsubsection{Nelder-Mead}\label{sec:3.1.3.1_Nelder_Mead}

A frequently used gradient-free optimiser is the Nelder-Mead (or downhill-simplex) method \cite{nelder_simplex_1965} developed by J. Nelder and R. Mead in 1965. It is referred to as a \emph{direct search} or \emph{pattern search} approach and it is a gradient-free local method, making it generally quite efficient, but not guaranteed to converge to a global optimum of the cost function. Direct search methods work by varying each optimisable parameter by some small stepsize from the current minimum in each direction and computing the cost function at the updated value. The change that leads to the largest decrease in the cost function value is taken as the new minimum. Once no such variation leads to an improvement, the stepsize is halved and the process is repeated until some convergence criterion is satisfied.

\begin{figure}[t]
\centering
\includegraphics[width=\linewidth]{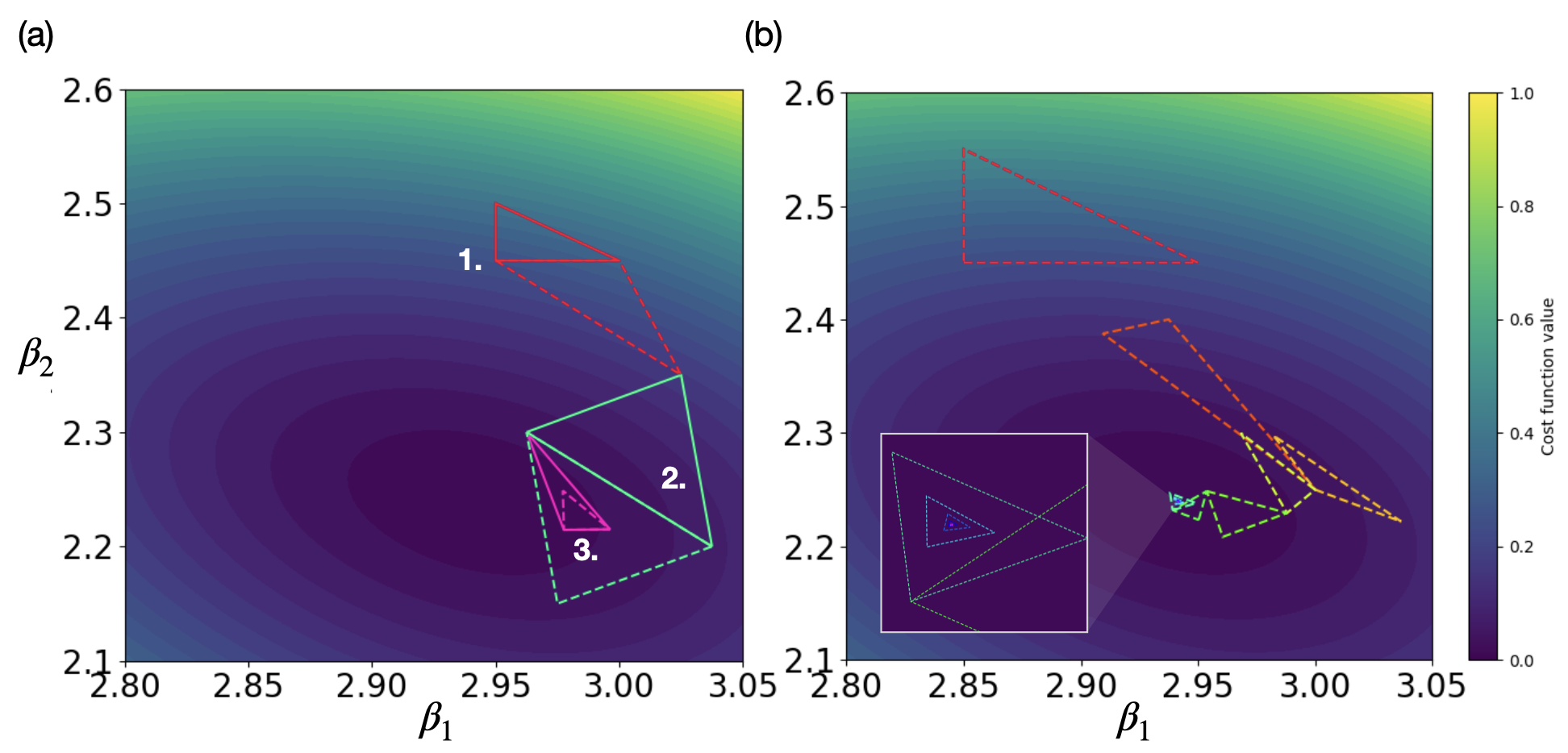} \caption[Visualising the Nelder-Mead optimisation algorithm.]{Illustration of the Nelder-Mead algorithm for a cost function parameterised by two parameters $\beta_1$ and $\beta_2$.  The simplex for a 2-dimensional landscape is a triangle which transforms continuously seeking out a lower cost function value in the landscape as described in the main text. (a) Examples of three of the transformations that the simplex performs during the optimisation: 1. expansion, 2. reflection and 3. contraction. Solid lines indicate the simplex prior to the transformation while dashed lines indicate its state after. (b) A plot of every third step of the algorithm, with the changing color of the triangle, starting from red and ending with bright purple, indicating the iterations. The inset shows a magnification of the final steps of the algorithm. The cost function landscape is represented by the contour plot and the aim of the Nelder-Mead algorithm in this case is to find its local minimal value (dark purple, as indicated by the colorbar).}\label{fig:nelder_mead}
\end{figure}

The way this direct search approach is adapted in Nelder-Mead is by constructing simplices, which are geometric objects that generalise triangles in lower and higher dimensions. For a cost function dependent on $n$ parameters, Nelder-Mead constructs an $n$-dimensional simplex. For $n=0$ this is a point, for $n=1,2,3$ a line segment, triangle and tetrahedron respectively and then higher-dimensional versions as $n$ increases. Thus a simplex has $n+1$ vertices for $n$ parameters. 

The vertices of this simplex then traverse the cost function landscape according to the Nelder-Mead algorithm in order to converge to some minimum value.  In most of the search steps, the primary change is to shift the highest point of the simplex (\@i.e.~where the cost function value is largest) through the opposite face of the simplex, moving to a point with a lower cost function value. These steps are known as \emph{reflections} and they are designed to preserve the volume of the simplex, ensuring it remains non-degenerate. Whenever possible, the method will \emph{expand} the simplex along a particular direction, which allows it to take bigger steps in search of a minimum. When the simplex encounters a region that can be thought of as a `valley floor' in the cost function landscape, it \emph{contracts} its dimensions orthogonal to the valley, so that it can slide down. See Fig.~\ref{fig:nelder_mead} (a) for a visual reference. In situations where the simplex has to navigate through a narrow passage, it shrinks itself in all directions, wrapping itself around its best (lowest) point, enabling it to continue its search for the minimum. The whole process is illustrated for a simple example in Fig.~\ref{fig:nelder_mead} (b).

This description of the Nelder-Mead method only outlines the basic idea that was first developed in the original 1965 paper. Many variations and improvements have been developed in the years since and the actual implementations vary. In general, the Nelder-Mead approach is simple to understand and implement, as well as being quite efficient and flexible. However, it often suffers from convergence issues, being both likely to return a sub-optimal local minimum and to get stuck without converging far longer than necessary, undoing any efficiency it otherwise promised. Furthermore, the simplex method doesn't scale well in higher dimensions, making it less effective as the number of parameters increases.

\subsubsection{Powell's method}\label{sec:3.1.3.2_Powell}

Another approach from the gradient-free, local optimiser crowd is Powell's method, first developed by Michael J. D. Powell in 1964 \cite{powell_efficient_1964}. The algorithm is known as a \emph{conjugate-direction} approach, not to be confused with the more common conjugate-gradient approach \cite{hestenes_methods_1952}, although the two are related as the latter can be viewed as a specialisation of the former. 

The basis of Powell's method relies on the idea of conjugate vectors or conjugate directions. Two vectors $\ubb$ and $\boldsymbol{v}$ are said to be conjugate with respect to some positive semidefinite matrix $A$ if $\ubb^T A \boldsymbol{v} = 0$. A set of conjugate directions, thus, is a set of vectors that are pairwise conjugate. Furthermore, one can make the observation \cite{brent_algorithms_2002} that the function
\begin{equation}
    f(\xbb) = \xbb^T A \xbb - 2\boldsymbol{b}^T \xbb + c
\end{equation}
for some positive semidefinite matrix $A$, $\boldsymbol{b} \in \R^n$ and $c \in \R$ has a minimum at the point $\xbb = \sum_{i=1}^n \beta_i \ubb_i$ in the space spanned by the set of conjugate vectors $\{ u_j \}_{j = 1,...,n}$ with
\begin{equation}
    \beta_i = \frac{\ubb_i^T\boldsymbol{b}}{\ubb_i^T A \ubb_i}.
\end{equation}
This minimum can be calculated efficiently just through evaluating the cost function, without needing explicit access to $A$, $\boldsymbol{b}$ or $c$. This property allowed Powell to develop a simple but powerful gradient-free approach, which can be summarised in the following bit of pseudocode.
\begin{algorithm}
\caption{Powell's Method}
\begin{algorithmic}[1]
\Procedure{Powell}{}
\State Initialise the method with ansatz solution $\ubb_0 \in \R^m$ and $n \leq m$ conjugate search vectors $\{\xbb_1, ..., \xbb_n \}$. If none are provided, use columns of the $m$-dimensional identity matrix.
\For{$i = 1,...,n$}
\State Compute $\beta_i$ to minimise $f(\ubb_{i - 1} + \beta_i \xbb_i)$
\State Define $\ubb_i \gets \ubb_{i - 1} + \beta_i \xbb_i$
\EndFor
\For{$i = 1,...,n-1$}
\State $\xbb_{i} \gets \xbb_{i+1}$
\EndFor
\State $\xbb_n \gets (\ubb_n - \ubb_0)$
\State Compute $\beta$ to minimize $(f(\ubb_0  + \beta \xbb_n))$
\State $\ubb_0 \gets \ubb_0 + \beta \xbb_n$
\EndProcedure
\end{algorithmic}
\end{algorithm}

In other words, the algorithm is initialised with a guess for a solution and a set of conjugate directions. It then proceeds to find a minimum along each direction and shifts to a new point along a superposition of their minima, adding the vector along in which it shifted to the list of conjugate direction vectors and removing the first vector in the list before starting the next step.

\begin{figure}[t]
\centering
\includegraphics[width=\linewidth]{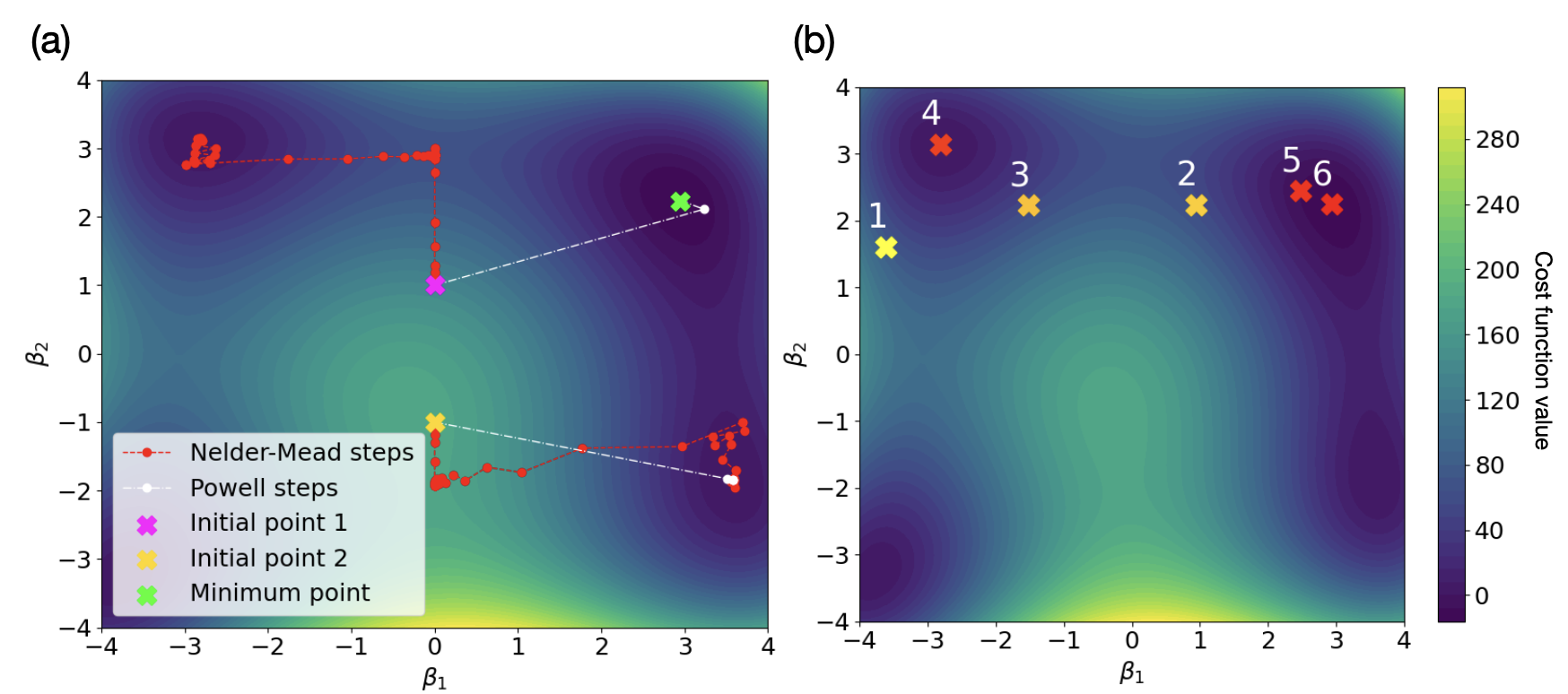} \caption[Visualising optimisers in action]{An illustration of the optmisation strategy of several numerical optimisation methods. (a) Local optimisers: steps of the Nelder-Mead method and Powell's method when instantiated in two different locations of the loss function landscape. The global mimimum is illustrated by a green cross. (b) The local minima (crosses) visited by the Dual-Annealing algorithm in the order indicated by numbered labels. The colour of the crosses ranges from yellow (high cost function value) to red (low cost function value).}\label{fig:optimisers}
\end{figure}

The Powell method is far more complex than is presented here, in particular due to the fact that several extra steps are usually added in order to guarantee convergence and and additional features to help optimise it. Additionally, the minimisation procedure of steps 4 and 11 in the pseudocode is highly non-trivial and can be achieved via several different algorithms like Brent's method\cite{brent_algorithms_2002}. It has guarantees of being very efficient in convex optimisation problems and excels in high-dimensional spaces, unlike Nelder-Mead. A plot of the search steps of the two methods in Fig.~\ref{fig:optimisers}(a) shows how they compare in terms of number of steps taken and accuracy in finding the optimum of some non-convex loss function. Importantly, given the more complicated nature of the steps in Powell's method, the fact that it requires fewer steps to converge to a solution does not necessarily make it more efficient.

\subsubsection{Dual-annealing}\label{sec:3.1.3.3_dual_annealing}

Unlike both Nelder-Mead and Powell's method, dual-annealing is a \emph{global} optimization algorithm, meaning that its primary goal is to find a global minimum of the function. It is also a stochastic method, since rather than following a pre-defined set of rules or procedures, it employs probabilistic transitions or decisions during the search. This added randomness can help the algorithm escape local optima and explore the solution space more broadly, however it also adds to the computational complexity of such approaches. As mentioned earlier, global optimisation algorithms tend to be far less efficient than local ones, but this is the price that needs to be paid when solutions obtained in local minima are simply not enough and the cost function landscape is highly non-convex. 

What is particularly interesting about dual-annealing is that it combines Generalized Simulated Annealing (\acrref{GSA}) \cite{tsallis_generalized_1996}, a global search algorithm, with a choice of \emph{local} optimiser that refines the solution once the global search is done. This is important because global algorithms, including \acrref{GSA}, are often good at locating the vicinity of the global minimum (the basin) but not necessarily the minimum itself. 

The \acrref{GSA} part of dual-annealing function is, unsurprisingly, a generalisation of the simulated annealing algorithm \cite{kirkpatrick_optimization_1983} inspired by the annealing process of metallurgy which causes a molten metal to reach its crystalline state which is the global minimum in terms of thermodynamic energy. In simulated annealing, the cost function is treated as the energy function of a molten metal and one or more artificial temperatures are introduced and gradually cooled. In \acrref{GSA}, this presents itself as a series of probabilistic jumps across the cost function landscape that depend on an artificial temperature parameter which decreases as the search progresses.

More concretely, at each step of the search, the algorithm generates a trial jump in the cost function space from the current temporary solution to a new point. This is done by sampling from a modified Cauchy-Lorentz distribution over the cost function space. The distribution peaks around the current temporary solution and its scale parameter (a variable that controls its spread) is a function of the artificial temperature $T_{q_v}$. Thus, the higher the temperature, the more likely it is that the trial jump will be larger, taking the solver further away from its current location in the cost function space. The $q_v$ parameter can be set to different values in order to speed up or slow down the cooling process. 

Once the trial jump has been generated, it is either accepted or rejected based on the cost function value at the new point as compared to the current point. If the new point is better (\@i.e.~the jump is `downhill', towards a lower energy), then the jump is accepted. If, on the other hand, the jump is worse or `uphill', it might still be accepted with some probability based on a parameterised Metropolis algorithm \cite{chib_understanding_1995}, where the probability of acceptance is calculated roughly as
\begin{equation}
	P = \exp{-\Delta C / T_{q_v}},
\end{equation}
with $\Delta C$ the change in the cost function value from the previous trial solution to the new one and $T_{q_v}$ the artificial temperature. This allows for the algorithm to potentially escape local minima. If a jump is accepted, the search then continues in a similar manner from the new point and the temperature parameter is decreased, reducing the probability of the next generated jump being far away from the current point and the potential of accepting a jump to a `worse' value. Once the temperature reaches 0, the system `freezes' and can only transition to states with lower `energy' or cost function value in its immediate vicinity until it reaches a local minimum.

The dual-annealing algorithm proceeds by first using \acrref{GSA} to identify a `basin' in the cost function landscape and then using the best solution so far as an initial guess for a local optimisation algorithm like Nelder-Mead or Powell's method to refine the solution. The local search is generally called when the artificial temperature decreases below some pre-defined value and once the local search is done, the whole process restarts again while keeping track of the current best solution. The entire algorithm terminates when some convergence criterion is satisfied. Usually this is when some number of search iterations or cost function evaluations is reached, or there is no more improvement to the solution below some tolerance. This process is illustrated in Fig.~\ref{fig:optimisers}(b), where the dual-annealing algorithm returns points 1, 2 and 4 as minima detected during the annealing stages with 3 and 5 corresponding to minima detected during the local searches. 

The verdict regarding dual-annealing, with respect to the local optimisers that we addressed previously, is that it is more powerful and can lead to better solutions, given that it has a better ability to explore the cost function landscape. Many real-world cost function landscapes are non-convex, high-dimensional and consist of many local minima. In such cases local optimisers, by virtue of only searching locally, will always get trapped in a local minimum if initialised near one. If there are many such local minima, then a local optimiser is highly likely to consistently not find the global optimum/minimum. A global optimiser like dual-annealing, on the other hand, may be initialised anywhere on the landscape and will hop around the entire solution space due to its stochastic nature, ignoring locally good solutions in order to occasionally move `up the hill'. This allows it to converge to a global optimum rather than a local one with far higher certainty than any local optimiser.

However, dual-annealing is also more computationally expensive than local methods, as can be made obvious by the fact that local search is merely a subroutine of the algorithm. The constant hopping around the landscape generally requires far more iterations than an algorithm like Nelder-Mead. Ultimately, the choice of which approach to use comes down to having knowledge about the cost function landscape as well as trial-and-error. The use of a global optimiser may be overkill when the cost function landscape lends itself well to local methods (e.g.~when it is highly convex) and each evaluation of the cost function is expensive. If, however, locally optimal solutions are not enough, then global methods are by far the best option.

\section{Quantum optimal control}\label{sec:3.2_Quantum_optimal_control}

\begin{figure}[t]
\centering
\includegraphics[width=0.8\linewidth]{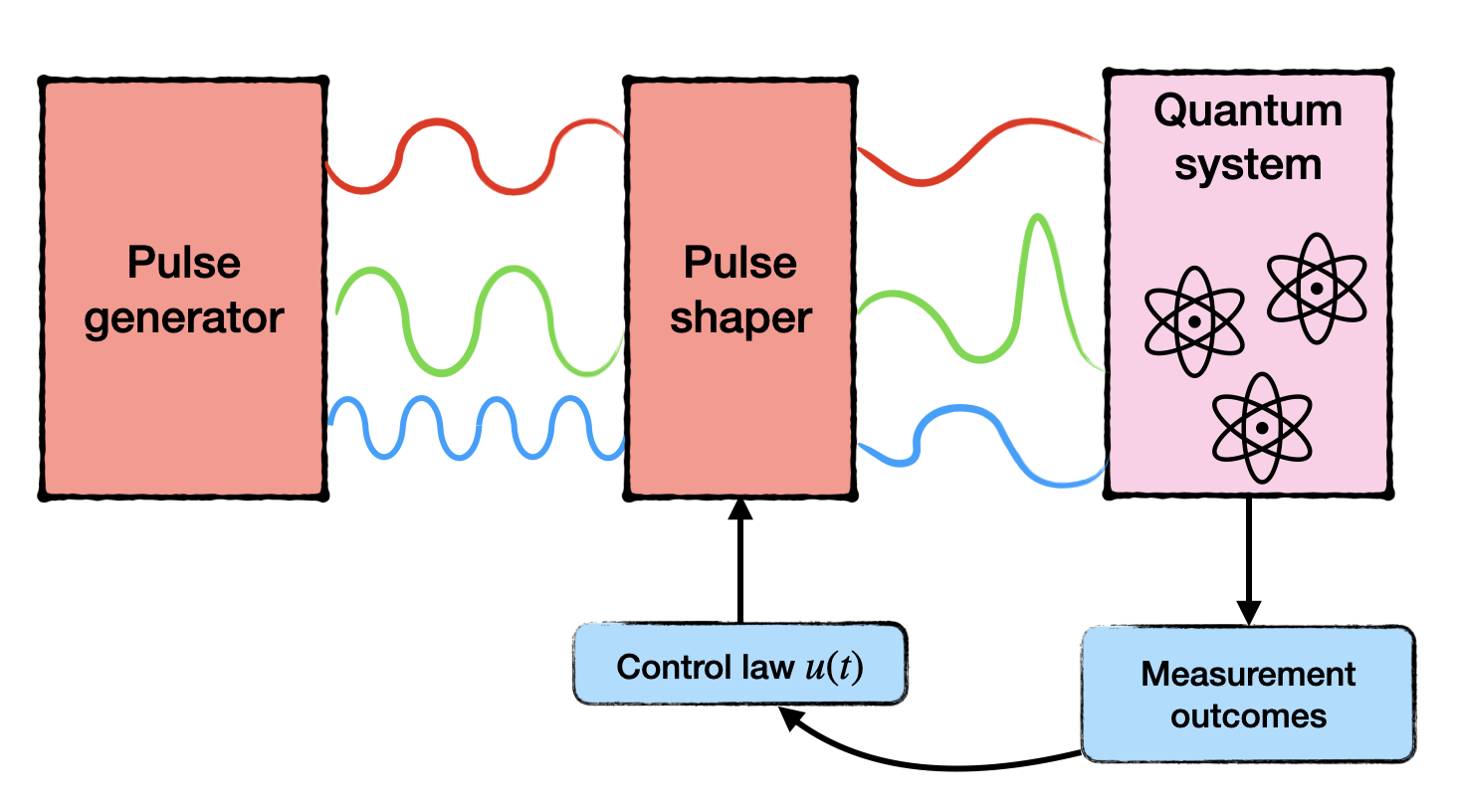} \caption[Schematic diagram of open-loop quantum optimal control]{A sketch of a quantum optimal control closed-loop set-up. A quantum system is directly controlled by a set of electromagnetic pulses which are shaped according to a set of control functions $\ubb(t)$ that are optimised based on feedback from the information obtained through measurements of the system.}\label{fig:quantum_optimal_control}
\end{figure}

We've now established that the broad goal of optimal control theory is the design of protocols and strategies which optimise the behaviour of some abstract control system with respect to some abstract target. Quantum optimal control theory (\acrref{QOCT}), rather predictably, does this in the setting where the abstract system is a quantum system. Very broadly then, \acrref{QOCT} concerns itself with the design and analysis of control fields (usually electromagnetic fields) that manipulate quantum dynamical processes at the atomic or molecular scale in the best way possible, as illustrated in Fig.~\ref{fig:quantum_optimal_control}. In this chapter, we will broadly cover the basics of \acrref{QOCT}, starting with how the mathematical structure discussed in Sec.~\ref{sec:3.1.1_mathematical_structure} can be adapted to the quantum setting and ending with detailed descriptions of \acrref{CRAB} (Sec.~\ref{sec:3.3.1_CRAB}) and \acrref{GRAPE} (Sec.~\ref{sec:3.3.2_GRAPE}), popular \acrref{QOCT} methods which will be relevant to later work presented in this thesis. As the content of later chapters will focus on closed systems, that will be the perspective we will take with respect to \acrref{QOCT}. More concretely, we will focus on cases where the generator of transformations of a quantum system is primarily modelled as the Hamiltonian as opposed to, \@e.g.~a Liouvillian, but a similar, if generally more complex, analysis holds in the case of open systems.  

Returning to the material covered in Sec.~\ref{sec:3.1.1_mathematical_structure}, we can now add more structure to the abstract notions of system, control function and cost function. In the quantum setting, the set of state functions $X$ often takes the form of a set of quantum states, be they complex vectors, density matrices or operators. The set of control functions $U$ is usually represented by a set of functions of parameterised Hamiltonians. It is common to decompose a control Hamiltonian into two components: the time-dependent `drive' part and the time-independent `drift' part. The time-dependent part can then be further decomposed into a set of $N_k$ operators $\mathcal{O}_{\rm opt} = \{\mathcal{O}_{\rm opt}^{(k)}\}_{k=1,...,N_k}$, such that the full control Hamiltonian reads:
\begin{equation}\label{eq:optimal_control_H}
    H(\ubb(t)) = H_0 + \sum_{k = 1}^{N_k} u_k(t) \mathcal{O}_{\rm opt}^{(k)},
\end{equation}
where $H_0$ is the so-called `drift' Hamiltonian which drives the free evolution of the system and the additional terms made up of control functions $u_k(t) \in \ubb(t)$ driving the corresponding operators $\mathcal{O}_{\rm opt}^{(k)}$.

Given this, we can describe a general quantum optimal control problem in analogy to Eq.~\eqref{eq:control_ODE} as one where the aim is to solve the Schr\"{o}dinger equation:
\begin{equation}\label{eq:optimal_control_schroedinger}
    i \hbar \partial_t \ket{\psi(t)} = H(\ubb(t))\ket{\psi(t)},
\end{equation}
with the constraint of starting in a state from a set of initial states $\ket{\psi_0} \in \boldsymbol \Psi_0$ and while minimising some cost function that targets a set of final states $\ket{\psi_T} \in \boldsymbol \Psi_T$. We note that while we are using wavefunctions as representations for the states of the control system, it is actually quite common in \acrref{QOCT} to instead work in the operator picture, where the Schr\"{o}dinger equation is
\begin{equation}\label{eq:operator_schroedinger_eq}
    \partial_t \mathcal{O}(t) = -i H(\ubb(t))\mathcal{O}(t),
\end{equation}
where $\mathcal{O}$ is an operator on some pre-defined Hilbert space. A useful constraint in this setting is to take the initial state of the operator to be the identity $\mathcal{O}(0) = \mathds{1}$. The choice of wave mechanics or matrix mechanics depends on the specific \acrref{QOCT} problem at hand, although questions of \@e.g.~controllability are usually best-solved with operators rather than state vectors. For example, if we can show that the set of possible matrices that can be obtained for system \eqref{eq:operator_schroedinger_eq} is the set of all the unitary matrices (with the rank of the system Hilbert space), then the system can theoretically be steered to any arbitrary state and thus it is controllable. 

The choice of cost function in the quantum setting is generally informed by the desired properties of the target state(s) combined with considerations for what information can be extracted from the system and other constraints. For example, when the aim of the optimisation is to prepare a single, well-defined quantum state $\ket{\psi_T}$ with high accuracy, then the most informative cost function is:
\begin{equation}\label{eq:costfunc_fidelity}
    C_{\rm F}(\tau, \ubb) = 1 - F(\tau, \ubb) = 1 - \abs{\braket{\psi(\tau, \ubb)}{\psi_T}}^2,
\end{equation}
where $F(\tau, \ubb)$ is the fidelity of the final state $\ket{\psi(\tau, \ubb)}$ with respect to the desired target $\ket{\psi_T}$. Here $\ket{\psi(\tau, \ubb)}$ is generated by driving an initial state $\ket{\psi_0} \in \boldsymbol \Psi_0$ for a time $\tau$ via the time-dependent Hamiltonian $H(\ubb(t))$. If, on the other hand, the target state need only be a ground state of some Hamiltonian $H_T$, then it might be far more convenient to use the final system energy, which encompasses several degenerate states (rather than a single unique target state) and in such cases often provides better convergence:
\begin{equation}\label{eq:costfunc_energy}
    C_{\rm E}(\tau, \ubb) = \mel{\psi(\tau, \ubb)}{H_T}{\psi(\tau, \ubb)}.
\end{equation}
Finally, should one be interested only in a specific property of the final state, like its entanglement, then the cost function might look something like:
\begin{equation}\label{eq:costfunc_entanglement}
    C_{\rm S}(\tau, \ubb) = -S[\ket{\psi(\tau, \ubb)}],
\end{equation}
where $S[\cdot]$ is some appropriate measure of entanglement. As well as being informed by the set of target states, the cost function may include further constraints, like the total power of the driving fields in the Hamiltonian. This is analogous to the cost function in Eq.~\eqref{eq:example_cost_func2}, which in the quantum case might look something like:
\begin{equation}\label{eq:costfunc_constraint}
    C(\tau, \ubb) = C_{\rm F}(\tau, \ubb) + \sum_{k = 1}^{N_k} \int_0^{\tau} \abs{u_k(t)}^2 dt,
\end{equation}
which can be read as an optimisation for final state fidelity with the added constraint that the time-integrated flux of the driving fields is minimised. As in the case of abstract quantum control, the inclusion of constraints need not be additive - it could be multiplicative or take on some completely different form, depending on what behaviour is expected from the loss function landscape. Additive constraints are often easy to understand when building a problem: if a constraint adds positively to the cost function value and one wants to minimise the cost function, then optimisation will lead to a reduction in whatever component corresponds to the constraint. 

\section{Quantum optimal control methods}\label{sec:3.3_qoct_methods}

Both analytical (Sec.~\ref{sec:3.1.2_analytic_optimisation}) and numerical (Sec.~\ref{sec:3.1.3_numerical_optimisation}) methods have been developed for the optimal control of quantum systems in recent decades. Analytical methods in \acrref{QOCT} generally deal with questions of necessary conditions for controllability \cite{schirmer_complete_2001} or reachability of states, \@e.g.~exploring quantum speed limits \cite{khaneja_time_2001, hegerfeldt_driving_2013, poggi_quantum_2013}. Analytical methods \emph{can} be used to find solutions to quantum optimal control problems rather than just classify their structure, but the Achilles' heel of analytical approaches remains a general inability to deal with complex systems. The volatile and often exponentially complex nature of quantum systems means that numerical approaches tend to be the preferred method for actually determining solutions to \acrref{QOCT} problems. 

Numerical methods in \acrref{QOCT} tend to consist of the development and analysis of iterative algorithms focused on optimising pulses for quantum systems. This can be done by constructing a mathematical description of the pulse, including parameters that control its shape and which can then be numerically optimised. Most numerical methods under the umbrella of \acrref{QOCT} involve a classical optimiser, like those discussed in Sec.~\ref{sec:3.1.3_numerical_optimisation}, as a subroutine in the approach which finds the optimal values for the pulse parameters. In this section we will explore two of the more broadly used numerical approaches in quantum optimal control, \acrref{CRAB} and \acrref{GRAPE}. 

\subsection{Chopped random-basis quantum optimization (CRAB)}\label{sec:3.3.1_CRAB}

The ``Chopped random-basis quantum optimization" or \acrref{CRAB} method is a quantum optimal control method first introduced in \cite{doria_optimal_2011, caneva_chopped_2011} which revolves around the construction of a truncated randomized basis of functions for the control fields of a quantum system. It was originally developed for quantum many-body systems whose time evolution can be efficiently simulated by time-dependent density matrix renormalization group (tDMRG)\cite{white_density_1992, schollwock_density-matrix_2005, schollwock_density-matrix_2011}. It was believed that such systems were mostly intractable for control optimization using gradient-based algorithms \cite{brif_control_2010}, although such potential limitations have been overcome in more recent work\cite{jensen_approximate_2021}. \acrref{CRAB} provides a way to reduce the space of search parameters, making the optimisation process more efficient, while retaining access to a large solution space through the added randomisation component.

The key idea is to expand the control pulse $\ubb(t)$ in some truncated basis of dimension $N_k$: 
\begin{equation}\label{eq:basic_CRAB}
    u(t) = \sum_{i = 1}^{N_k} c_i u_i(t),
\end{equation}
where the cost function landscape is spanned by the coefficients $c_i$, that need to be optimised over using numerical optimisation methods like those described in Sec.~\ref{sec:3.1.3_numerical_optimisation}. Generally this basis is made up of trigonometric functions, since their behaviour is considered easy to understand, although it could be any basis that spans the space of admissible controls \@e.g.~something like the generalized Chebyshev polynomials. A choice of basis can further be enhanced or modified by a shape function $g(t)$ that fixes the pulse to some initial and/or final value:
\begin{equation}\label{eq:trigonometric_CRAB}
    u(g(t),t) = \sum_{i = 1}^{N_k/2} c_i \frac{\cos{\omega_i t}}{g(t)} + \sum_{i = N_k/2 + 1}^{N_k} c_i \frac{\sin{\omega_i} t}{g(t)}.
\end{equation}

Importantly, the key to expanding the solution space in order to find better pulses using the \acrref{CRAB} approach lies in the randomisation of the frequencies $\omega_i$. During each optimisation process, the $\omega_i$ are chosen randomly around the principal harmonics within some interval $[0, \omega_{\rm max}]$, allowing the pulse shapes to be more diverse and more complex than by simply keeping them fixed at a given value. The optimisation process can then be parallelised, with several optimisation instances running simultaneously exploring several different sets of random frequencies and the optimal solution can be picked from the final outcomes of all optimisations. 

The \acrref{CRAB} approach lends itself very easily to the incorporation of additional features and constraints like the shape function. For example, it is quite easy to start with a trial pulse, say $f(t)$, which cannot be expanded efficiently or exactly in the chosen basis and to dress it according to
\begin{equation}\label{eq:trial_pulse_CRAB}
    u(t) = f(t)\Bigg(1 + \sum_{i = 1}^{N_k} c_i u_i(t)\Bigg).
\end{equation}
Another particularly useful alteration to the basic \acrref{CRAB} procedure is what is known as `dressed' \acrref{CRAB} or dCRAB \cite{rach_dressing_2015}, which in a similar vein aims to iteratively re-dress solutions obtained from previous optimisations with new sets of basis functions added onto the existing solution. These \emph{super-iterations} $j$ can be modelled as
\begin{equation}\label{eq:dCRAB}
    u^j(t) = c_0^j u^{j-1}(t) + \sum_{i = 1}^{N_k} c_i^j u^j_i(t),
\end{equation}
where $u^j_i(t)$ are new basis functions and $u^{j-1}(t)$ is the pulse obtained from a previous, $(j-1)^{\rm st}$ optimisation. The coefficient $c_0^j$ can be seen as shifting the solution in the direction of the previous solution pulse while $\{ c_i^j\}_{i = 1, ..., N_k}$ move it in new search directions $u^j_i(t)$. This is, in fact, very similar to Powell's optimisation method which we covered in Sec.~\ref{sec:3.1.3.2_Powell}, wherein a finite set of search directions in cost function space is continuously updated with linear combinations of their optima. dCRAB can be seen as doing the same but with updates sampled from an infinite-dimensional search space. This iterative approach is useful in  avoiding local minima and in exploring a far larger search space, avoiding the hard constraint of a finite set of basis functions for each optimisation. 

There are several key advantages in the \acrref{CRAB} approach that have led to its widespread use in the \acrref{QOCT} community. For one, the randomization of the control field basis allows for a more comprehensive exploration of the control landscape, which can lead to the discovery of better solutions. It also offers relatively quick convergence as the number of optimisable parameters is usually small when compared to other approaches (such as \acrref{GRAPE}, which we will explore in the next section). Finally, it is very flexible: the basis functions can be altered and constraints can be incorporated quite easily, whether they concern the physical implementation (\@e.g.~the shaping function) or the efficiency of the optimisation itself.

\subsection{Gradient Ascent Pulse Engineering (GRAPE)}\label{sec:3.3.2_GRAPE}

The ``Gradient Ascent Pulse Engineering" (GRAPE) algorithm is yet another widely used \acrref{QOCT} numerical method. It was first developed in order to design pulse sequences in NMR spectroscopy \cite{khaneja_optimal_2005} and has since been iterated upon and improved a number of times as well as being integrated into several optimal control packages \cite{de_fouquieres_second_2011, chen_iterative_2022, machnes_comparing_2011, johansson_qutip_2013}. As the name suggests, it is a gradient-based optimisation method and while initially it was used primarily for the preparation of specific target states, its powerful flexibility has since lent itself to many other applications in the setting of quantum technologies, like the optimisation of quantum logic gates \cite{motzoi_optimal_2011, anderson_accurate_2015}.

The key idea behind \acrref{GRAPE} is to replace continuous control functions, like \@e.g.~those used in the basis functions of \acrref{CRAB}, with piecewise constant control amplitudes $u_j(t_k)$, each applied to the control system at time $t_k \in [0, \tau]$ for a time interval $\Delta t$, where $\tau$ is the total evolution time. One may view this as discretizing the time-evolution of the system into $N_m$ slices of time $\Delta t = t_{k + 1} - t_k$. These slices need not all be of equal size, but for simplicity let us work in the setting where they are, meaning that $\tau = N_m \Delta t$.

At this point we can pause and notice that since the control amplitudes $u_j(t_k)$ are piecewise constant for all time intervals, they can be treated as a set of parameters that can be optimised using a numerical optimisation algorithm. This gives $N_j \cross N_m$ total parameters to optimise, as each $j^{\rm th}$ pulse will be made up of $N_m$ time-steps. Given this relatively large number of parameters, the original \acrref{GRAPE} algorithm includes an analysis of how to compute the gradient of the cost function with respect to each $u_j(t_k)$ in order to implement gradient-based optimisation methods like gradient-descent (Eq.~\eqref{eq:gradient_descent}). Recalling the form of the quantum control Hamiltonian from Eq.~\eqref{eq:optimal_control_H}, the propagator for the time-evolution of the quantum system using \acrref{GRAPE} during a single time step $\Delta t$ at time $t_k$ is
\begin{equation}
    U_k(\Delta t) = \exp{- i \Delta t \Big( H_0 + \sum_{j=1}^{N_j} u_j(t_k) \mathcal{O}_{\rm opt}^{(j)} \Big)}
\end{equation}
for some drift component of the Hamiltonian $H_0$ and some basis of control operators $\{\mathcal{O}_{\rm opt}^{(j)}\}_{j = 1,...,N_j}$. The full evolution of the system can thus be captured by the product of operators (with dependence on $\Delta t$ removed):
\begin{equation}
    U(\tau) = U_{N_m} U_{N_m - 1} ... U_{2} U_{1},
\end{equation}
such that for some initial state $\rho_0$ (where we are now working with density matrices rather than state vectors), the final evolved state can be written as
\begin{equation}\label{eq:grape_rho_evolved}
    \begin{aligned}
        \rho(\tau) &= U(\tau)\rho_0 \adj{U}(\tau) \\
        &= U_{N_m} ... U_{1} \rho_0 \adj{U}_{1} ... \adj{U}_{N_m}
    \end{aligned}
\end{equation}

In order to derive a way to compute the gradient of the cost function with respect to the parameters, it is necessary to define a cost function. In this case we will use the overlap of the final state $\rho(\tau)$ with respect to some target state $\rho_T$, a density matrix version of Eq.~\ref{eq:costfunc_fidelity}:
\begin{equation}
    C(\ubb) = \Tr{\rho_T^{\dagger}\rho(\tau)},
\end{equation}
where $\ubb$ in this case is the set of all $N_j \cross N_m$ parameters to be optimised $\ubb: \{ u_j(t_k)\}_{j = 1, ..., N_j}^{k = 1,...,N_m}$. Using Eq.~\ref{eq:grape_rho_evolved} and the cyclic property of the trace we can write
\begin{equation}\label{eq:grape_costfunc_rewrite}
    \begin{aligned}
        C(\ubb ) &= \Tr{\rho_T^{\dagger}U_{N_m} ... U_{1} \rho_0 \adj{U}_{1} ... \adj{U}_{N_m}} \\
        &=  \Tr{\adj{U}_{k+1} ... \adj{U}_{N_m} \rho_T U_{N_m} ... _{k+1} U_{k} ... U_{1} \rho_0 \adj{U}_{1} ... \adj{U}_{j}} \\
        &= \Tr{\Lambda_k \rho_k},
    \end{aligned}
\end{equation}
where $\Lambda_k = \adj{U}_{k+1} ... \adj{U}_{N_m} \rho_T U_{N_m} ... _{k+1}$ and $\rho_k = U_{k} ... U_{1} \rho_0 \adj{U}_{1} ... \adj{U}_{j}$. 

In order to calculate the gradient of $C(\ubb)$ with respect to each parameter $u_j(t_k)$, we first investigate what happens to $U_k$ when we perturb each parameter by some small amount $\delta u_j(t_k)$. To first order in $\delta u_j(t_k)$ we get
\begin{equation}
    \delta U_k = -i \delta u_j(t_k) U_k \int_0^{\Delta t} U_k(t') \mathcal{O}_{\rm opt}^{(j)} U_k(-t') dt'.
\end{equation}
Then, for small $\Delta t$ (\@i.e.~when it is much smaller than the norm of the control Hamiltonian), we find that the integral in the expression above can be approximated as the average value of the integrand, leading to:
\begin{equation}\label{eq:grape_costfunc_gradient}
    \frac{\delta C(\ubb)}{\delta u_j(t_k)} = - \Tr{\Lambda_k \Big(i\Delta t \comm{\mathcal{O}_{\rm opt}^{(j)}}{\rho_k}\Big)}.
\end{equation}

Using this, it is now possible to implement gradient-based numerical optimisation algorithms in order to find optimal values of $\ubb$, in the vein of gradient-descent from Eq.~\ref{eq:gradient_descent}. The method has been improved upon after the initial algorithm was first published, \@e.g.~in \cite{machnes_comparing_2011} in order to include information about second-derivatives of the cost function, allowing for more complex gradient-based optimisation like quasi-Newton methods (see discussion in Sec.~\ref{sec:3.1.3_numerical_optimisation}). Recent years have also seen improvements similar to those of dCRAB in the case of the \acrref{CRAB} algorithm, where an iterative optimisation procedure is applied on top of the basic \acrref{GRAPE} algorithm \cite{chen_iterative_2022}. It would be pertinent to mention, that there are very similar approaches to constructing \acrref{GRAPE}-type pulses out in the literature known as Krotov schemes \cite{reich_monotonically_2012}. The key difference between \acrref{GRAPE} and Krotov is in the update step in the iterative optimisation procedure: where the \acrref{GRAPE} algorithm updates all control parameters in a single iteration at once, the Krotov-based methods do so sequentially. Furthermore, Krotov-based approaches are only well-defined in the continuum limit \@i.e.~where the time step $\Delta t \ll 1$, which need not be the case for many implementations of \acrref{GRAPE}. In fact, Krotov approaches have been shown to be monotonically convergent in that limit \cite{morzhin_krotov_2019}, so they can suffer in the face of discretisation, something that is a roadblock for some implementations. \acrref{GRAPE} escapes this fate and, despite in many ways being less sophisticated, can be far more efficient in more restrictive settings with large time steps far away from the continuum limit.

Ultimately, \acrref{GRAPE} is a simple and powerful approach for constructing a control pulse, but it can suffer from the large number of parameters that need to be optimised. In more simple settings, where the cost function is smooth and convex, it is a very powerful tool, as on top of the high degree of control over the exact shape of the pulse, it offers a gradient-based method level of convergence. Gradient-based methods, as discussed in Sec.~\ref{sec:3.1.3_numerical_optimisation}), are efficient and converge rapidly given convexity guarantees, regardless of the number of parameters that describe the control function. There has been a lot of work done in recent years analyzing the topological and mathematical properties of quantum control landscapes, including their smoothness \cite{chakrabarti_quantum_2007, rabitz_surprising_2023,dong_quantum_2022}, although these only apply to problems where the cost function is ``well behaved" - \@i.e.~is some polynomial of the final state vector or matrix which itself evolves continuously on a smooth manifold.  However, there is no reason to expect that the cost function landscape will be particularly smooth nor convex in any specific instance, meaning the gradient information obtained in the \acrref{GRAPE} algorithm may not be useful. Furthermore, the gradient evaluation step can be quite computationally intensive. At the end of the day, one can always construct a \acrref{GRAPE}-type pulse and optimise the many parameters using, for example, a global optimiser like dual-annealing from Sec.~\ref{sec:3.1.3.3_dual_annealing}, but given how high-dimensional the problem might be due to the many parameters involved, this can be a very computationally intensive task.

It is useful to compare \acrref{GRAPE} and \acrref{CRAB}, as each offers a different set of advantages and disadvantages. The effectiveness of \acrref{CRAB}, for example, relies a lot on the choice of basis functions used in constructing the pulse, but the number of parameters to be optimised is generally far lower than that of \acrref{GRAPE}. Both offer a lot of flexibility in terms of incorporating constraints and using different numerical optimisers, although \acrref{CRAB} generally does not include a systematic way to compute cost function gradients, leaving it subject to gradient-free methods. 
\part{Optimising approximate counterdiabatic driving}\label{part:COLD}

\chapter{Counterdiabatic optimised local driving}\label{chap:4_COLD}

\epigraph{I feel a need… a need for speed.}{LT Pete "Maverick" Mitchell, \emph{Top Gun, 1986}}

In Ch.~\ref{chap:2_adiabaticity} we established that adiabatic evolution of a quantum system requires timescales that scale with the inverse of the energy gap, without which it experiences non-adiabatic excitations out of its instantaneous eigenstate(s). This presents a problem, as the results of adiabatic dynamics - \@i.e.~the production of the set of adiabatic eigenstates of the final Hamiltonian after the system evolution - is useful in many applications of quantum technologies \cite{dimitrova_many-body_2023, campo_more_2014, ebadi_quantum_2022}, but the timescales this requires are often difficult to achieve due to decoherence and other physical constraints. 

The dual motivation of implementing adiabatic evolution and doing so \emph{fast} has led to the development of a number of methods and approaches under the umbrella of \acrref{STA} \cite{guery-odelin_shortcuts_2019, torrontegui_chapter_2013}, with a universal \acrref{STA} approach being provided by \acrref{CD} \cite{berry_transitionless_2009, demirplak_adiabatic_2003}, introduced in detail in Sec.~\ref{sec:2.3_CD}. However, as established in Sec.~\ref{sec:2.3_CD}, exact \acrref{CD} is often difficult to derive and even more difficult to implement in an experimental setting \cite{meier_counterdiabatic_2020, ban_counter-diabatic_2014}, leading to the development of approximate methods such as \acrref{LCD} \cite{sels_minimizing_2017} and the truncated nested-commutator approach \cite{claeys_floquet-engineering_2019} which were discussed in detail in Sec.~\ref{sec:2.4.1_LCD} and Sec.~\ref{sec:2.4.2_nested_commutators} respectively. Apart from the already mentioned techniques, many other approaches \cite{saberi_adiabatic_2014, campbell_shortcut_2015, whitty_quantum_2020} have been developed which aim to bypass the inherent complexity of the exact \acrref{CD}, either in the case of small systems or ones which have scaling transformations \cite{del_campo_shortcuts_2013, deffner_classical_2014, deng_superadiabatic_2018}. These approximate methods all have their advantages and drawbacks when applied to particular adiabatic processes, owing both to their approximate nature and the practical aspects of their implementation.  

In this chapter we will present a new method for speeding up adiabatic processes: Counterdiabatic Optimised Local Driving (\acrref{COLD}), which was first developed with the goal of improving upon the results of \acrref{LCD} while retaining the advantages that it offers. Namely: \acrref{COLD} is a method that, given a time-dependent Hamiltonian and a set of physical constraints for the system that is being driven, can be used to construct an approximate counterdiabatic protocol that performs optimally for the given set of constraints on the Hamiltonian and the system. It does this by combining \acrref{LCD} and optimal control, which we explored in detail in Ch.~\ref{chap:3_Quantum_Optimal_control}. What we generally mean by `physical constraints' in this case arises from what can be implemented in an experiment: types of quantum operators, the range of magnitudes that each pulse driving an operator can take on or any other physical constraint, like the topology of the physical system. Optimality in this case is also understood to be the ability to drive a system from some initial state to some target state with minimal loss. While \acrref{LCD} can be used to implement an approximate \acrref{CD} protocol built out of restricted, physically realisable operators, \acrref{COLD} does this via finding an optimal path for the system, such that the approximate counterdiabatic drive is maximally effective in suppressing non-adiabatic effects.

We will begin the chapter by introducing the \acrref{COLD} method in detail. Then, in Sec.~\ref{sec:4.2_COLD_QOCT}, we will explore exactly what part \acrref{QOCT} plays in the new method. This chapter lays the groundwork for the method of \acrref{COLD}, while in Part \ref{part:applications} of the thesis we will present and analyse the results of its numerical implementation in various physical systems.

\section{Counterdiabatic driving and optimal control}\label{sec:4.1_COLD}

Let us begin by explicitly setting the stage for the problem that we want to solve. Given a Hamiltonian $H(\lambda)$, which depends on time via the parameter $\lambda(t)$, and a system prepared in an eigenstate of $H(\lambda_0)$, where $\lambda_0 = \lambda(0)$ (often this is the ground state, but it need not be), our task is to vary the parameter $\lambda$ from its initial value $\lambda_0$ to some final value $\lambda_f = \lambda(\tau)$ during a duration of time $\tau$ such that at the end of the process, the system is in the corresponding eigenstate of $H(\lambda_f)$. More precisely, if \@e.g.~the system starts in the ground state of $H(\lambda_0)$, after the evolution it should be in the ground state of $H(\lambda_f)$. This can be done quite reliably, as per the discussion of Ch.~\ref{chap:2_adiabaticity}, as long as the instantaneous eigenstates of the Hamiltonian driving the system are not degenerate throughout the evolution and the driving happens slowly enough (see Sec.~\ref{sec:2.1.2_adiabatic_condition}). However, bearing in mind that such slow evolution is generally not accessible, our primary goal is to achieve this result while keeping $\tau$ small \@i.e.~making the evolution as fast as possible while still achieving the desired outcomes.

As already mentioned in the introduction to this Chapter, one way to achieve this task is by using \acrref{CD} (Sec.~\ref{sec:2.3_CD}). That is, for a given $H(\lambda)$, it may be possible to derive and implement an exact counterdiabatic Hamiltonian from Eq.~\eqref{eq:CD_Hamiltonian} which suppresses all non-adiabatic effects experienced by the system due to fast driving. Exact \acrref{CD} could, in this way, keep a system in the instantaneous eigenstate of $H(\lambda)$ during arbitrarily short driving times (within the geometric speed limit \cite{bukov_geometric_2019}), but exact \acrref{CD} is not generally accessible for an arbitrary Hamiltonian \cite{kolodrubetz_geometry_2017} and often requires highly non-local operators. 

The next best thing to try, then, might be an approximate \acrref{CD} method like \acrref{LCD}. As discussed in Sec.~\ref{sec:2.4.1_LCD}, \acrref{LCD} not only allows one to variationally approximate the full \acrref{CD}, thus suppressing some of the losses associated with non-adiabatic effects, but it also gives one the freedom to choose the basis of operators for the approximation, making it very attractive in experimental settings where only a limited set of physical operators are available, e.g. when there is no local control of subsystems and only global pulses are available. If an ansatz is not forthcoming, it is also possible to use the ideas presented in Sec.~\ref{sec:2.4.2_nested_commutators} to build up a local operator basis which contributes to the full \acrref{CD} via the nested commutator approach \cite{geier_floquet_2021} and to then use the variational method of \acrref{LCD} in order to construct a counterdiabatic schedule made up of a physically implementable subset of that basis. 

The \acrref{LCD} method is powerful, but it is not without its faults. The primary disadvantage of such an approach is that the counterdiabatic drive being implemented will always be an \emph{approximation} unless the ansatz basis is fully representative of the exact \acrref{CD}. In cases where the approximation is a poor one, the \acrref{LCD} technique might not offer any suppression of errors at all. One solution to this would simply be to expand the ansatz basis in order to access more degrees of freedom in describing the \acrref{CD}, but this would be counter to the idea of only requiring a physically implementable set of operators as part of the approximation in order to make it useful in an experimental setting. Another solution would be to use a different method entirely to achieve the same result by, for example, taking a page out of optimal control theory as covered extensively in Ch.~\ref{chap:3_Quantum_Optimal_control}. It is not obvious, however, that a switch in tactics would lead to an improvement or what the complexity of designing a new approach might be. As discussed earlier, optimal control pulses can be constructed in a multitude of different ways, many of which have structure that may be completely ineffective for suppressing non-adiabatic effects. In the case of more flexible control pulses like \acrref{GRAPE}, which might offer a larger solution space, what we often run into is an issue of efficiency as the number of control parameters increases very quickly. 

This is where we come to the new method, \acrref{COLD}, which was developed with the aim of retaining the advantages of \acrref{LCD} while improving upon its results. The approach begins with the observation that any counterdiabatic schedule will depend on the driving path of the original Hamiltonian for which it is constructed, as discussed extensively in Ch.~\ref{chap:2_adiabaticity}. Namely, if we write a Hamiltonian $H(\lambda)$ as a sum of $N_H$ operators $\{ \mathcal{O}_{\rm H}^{(i)} \}_{i = 1,...,N_H}$ each scaled by a $\lambda$-dependent coefficient $h_i(\lambda) \in \hbb$ (note that we include constant functions here too instead of treating them as time-independent):
\begin{equation}
    H(\lambda, \hbb) = \sum_{i = 1}^{N_H} h_i(\lambda) \mathcal{O}_{\rm H}^{(i)},
\end{equation}
then the \acrref{CD} drive can be expressed as a sum of operators $\{\mathcal{O}_{\rm CD}^{(j)}\}_{j = 1,...,N_{\rm CD}}$ which are scaled by functions $\alpha_{j}(\lambda, \hbb)$ and the rate of change in the parameters $\dotlambda = \frac{d \lambda}{dt}$. That is to say the form of the counterdiabatic drive is a function of the time-dependent parameter $\lambda$, the operators $\mathcal{O}_{\rm H}^{(i)}$ and their $\lambda$-dependent coefficients. Returning to Eq.~\eqref{eq:CD_Hamiltonian}, we can now write the counterdiabatic Hamiltonian as:
\begin{equation}\label{eq:H_cd_operators}
    \begin{aligned}
        H_{\rm CD} &= H(\lambda, \hbb) + \dotlambda \AGP{\lambda} \\
        &= \sum_{i = 1}^{N_H} h_i(\lambda) \mathcal{O}_{\rm H}^{(i)} + \sum_{j = 1}^{N_{\rm CD}} \dotlambda \alpha_{j}(\lambda, \hbb) \mathcal{O}_{\rm CD}^{(j)},
    \end{aligned}
\end{equation}
where $\mathcal{O}_{\rm CD} = \{\mathcal{O}_{\rm CD}^{(j)}\}_{j = 1,...,N_{\rm CD}}$ is an operator basis of the adiabatic gauge potential $\AGP{\lambda}$ (\acrref{AGP}) which was introduced at length in Sec.~\ref{sec:2.2_AGP}. We can even see how this relationship comes about by looking at the matrix elements of the \acrref{AGP} in Eq.~\eqref{eq:AGP_adiabatic_basis}, which are a function of the instantaneous eigenenergies of $H(\lambda, \hbb)$ and the matrix elements of $\dlambda H(\lambda, \hbb)$, all of which can be written as functons of $\lambda$ and $\hbb$. Note, that in any finite system the operator basis of \acrref{AGP} will be finite.

In this setting, \acrref{LCD} is a way to variationally find the coefficients $\alpha_j$ for a given subset of the full \acrref{AGP} basis $\mathcal{O}_{\rm LCD} \subset \mathcal{O}_{\rm CD}$ which minimise the operator distance between the generalised adiabatic force from the exact \acrref{AGP} and the force generated by the approximate \acrref{AGP}. In the case where the ansatz is the full basis set $\mathcal{O}_{\rm CD}$, one should recover the exact \acrref{AGP} using the variational approach.

The reason for expressing the counterdiabatic Hamiltonian in this way is to make the dependence of the coefficients $\alpha_j$ on the functions $\hbb$ and $\lambda$ explicit. As mentioned at the start of this section, the philosophy of \acrref{COLD} begins with the observation that the form of the counterdiabatic drive will depend on the path of the Hamiltonian in the parameter space of its coefficients, \@i.e.~if we change either $\hbb$ or $\lambda$ (or both), the form of the counterdiabatic drive will change via the functions $\alpha_j$. A different way of looking at it is to view each specific set of parameters $(\hbb, \lambda)$ as defining a new time-dependent Hamiltonian with its own instantaneous eigenbasis that generates different non-adiabatic effects for a finite evolution time (see, \@e.g.~Eq.~\eqref{eq:moving_frame_schrodinger} and the discussion surrounding the \acrref{AGP}). For clarity, we note that the set of operators $\mathcal{O}_{\rm CD}$ is not modified by varying $\hbb$ or $\lambda$, but could be affected instead if the operators in the set $\mathcal{O}_{\rm H}$  are changed.

Let us return to the problem stated at the beginning of this section: our aim is to drive a system which started in an eigenstate of $H(\lambda_0)$ to the corresponding eigenstate of $H(\lambda_f)$ in the shortest amount of time possible. It is important to note that the problem statement does not say anything about the state for any other value of $\lambda$ throughout the evolution, although in the case where exact \acrref{CD} is implemented, the system should follow the instantaneous eigenstates of $H$ throughout the full dynamics. This more relaxed condition means that, as long as the time-dependent Hamiltonian driving the system matches up with the problem Hamiltonian at the start and end of the dynamics and the system is in the correct eigenstate at those two points, the path that it takes between them is not particularly important barring any other constraints. This observation can now allow us to finally introduce \acrref{COLD}.

\subsection{The method}

In Sec.~\ref{sec:3.2_Quantum_optimal_control}, we delved into the many ways in which driving pulses for quantum systems can be systematically constructed and modified or optimised in order to achieve particular goals. In fact, it is possible to use \acrref{QOCT} to speed up adiabatic protocols too, something that has been studied extensively, with the resulting methods generally grouped under the umbrella of \acrref{STA} \cite{guery-odelin_shortcuts_2019, torrontegui_chapter_2013}. We could imagine casting our original problem of driving a system to a particular eigenstate of $H(\lambda_f)$ as simply an optimisation problem with a cost function focused on state fidelity given by Eq.~\eqref{eq:costfunc_fidelity}. 

In the case of \acrref{COLD}, the first step is to construct a control Hamiltonian in the vein of Eq.~\eqref{eq:optimal_control_H} with the constraints that it be equal to $H(\lambda_0)$ at $t=0$ and $H(\lambda_f)$ at $t=\tau$:
\begin{equation}\label{eq:COLD_optimal_control}
    H_{\betabb}(\lambda, \hbb, \betabb) = \sum_{i = 1}^{N_H} h_i(\lambda) \mathcal{O}_H^{(i)} + \sum_{k = 1}^{N_k} \beta_k(\lambda) \mathcal{O}_{\rm opt}^{(k)}.
\end{equation}
Here, $\beta_k(\lambda) \in \betabb$ are the control functions which can be constructed and optimised using the methods described in Sec.~\ref{sec:3.2_Quantum_optimal_control} and $\{\mathcal{O}_{\rm opt}^{(k)}\}_{k = 1, ..., N_k}$ are controllable operators, which can be a subset of $\mathcal{O}_{\rm H}$ or introduce a new degree of freedom to the Hamiltonian, as long as the constraints that $H_{\betabb}(\lambda_0) = H(\lambda_0)$ and $H_{\betabb}(\lambda_f) = H(\lambda_f)$ are satisfied. 

In the second step of defining \acrref{COLD}, we return to \acrref{LCD} and the observation that our approximate counterdiabatic drive will depend on the coefficients of the Hamiltonian. If we are given an ansatz set of operators $\{\mathcal{O}_{\rm LCD}^{(j)}\}_{j = 1, ..., N_{\rm LCD}}$ and use them to variationally determine the approximate \acrref{CD} protocol for the control Hamiltonian $H_{\betabb}$ of Eq.~\eqref{eq:COLD_optimal_control}, the resulting Hamiltonian will look something like this:
\begin{equation}\label{eq:COLD_Hamiltonian}
    \begin{aligned}
        H_{\rm COLD}(\lambda, \hbb, \betabb) &= H_{\betabb}(\lambda, \hbb, \betabb) + H_{\rm LCD}(\lambda, \hbb, \betabb) \\
        &= \sum_{i = 1}^{N_H} h_i(\lambda) \mathcal{O}_{\rm H}^{(i)} + \sum_{k = 1}^{N_k} \beta_k(\lambda) \mathcal{O}_{\rm opt}^{(k)} + \sum_{j = 1}^{N_{\rm LCD}} \dotlambda \alpha_{j}(\lambda, \hbb, \betabb) \mathcal{O}_{\rm LCD}^{(j)},
    \end{aligned}
\end{equation}
which is the \acrref{COLD} Hamiltonian. Note that if the set $\mathcal{O}_{\rm opt}$ is not a subset of $\mathcal{O}_{\rm H}$, then the operators in $\mathcal{O}_{\rm CD}$ and consequently $\mathcal{O}_{\rm LCD}$ may be different than in the case when only driving the bare Hamiltonian. 

All that is left now is the third \acrref{COLD} step, which is the optimisation of the coefficients $\beta_k(\lambda)$ using \acrref{QOCT} methods presented in Sec.~\ref{sec:3.2_Quantum_optimal_control}. A natural, though not exclusive, cost function for this process would be the final state fidelity from Eq.~\eqref{eq:costfunc_fidelity} with respect to the desired eigenstate of $H(\lambda_f)$. We will provide a more detailed discussion of the optimal control component of \acrref{COLD} in the next section.

\begin{figure}[t]
    \centering
    \includegraphics[width=0.8\linewidth]{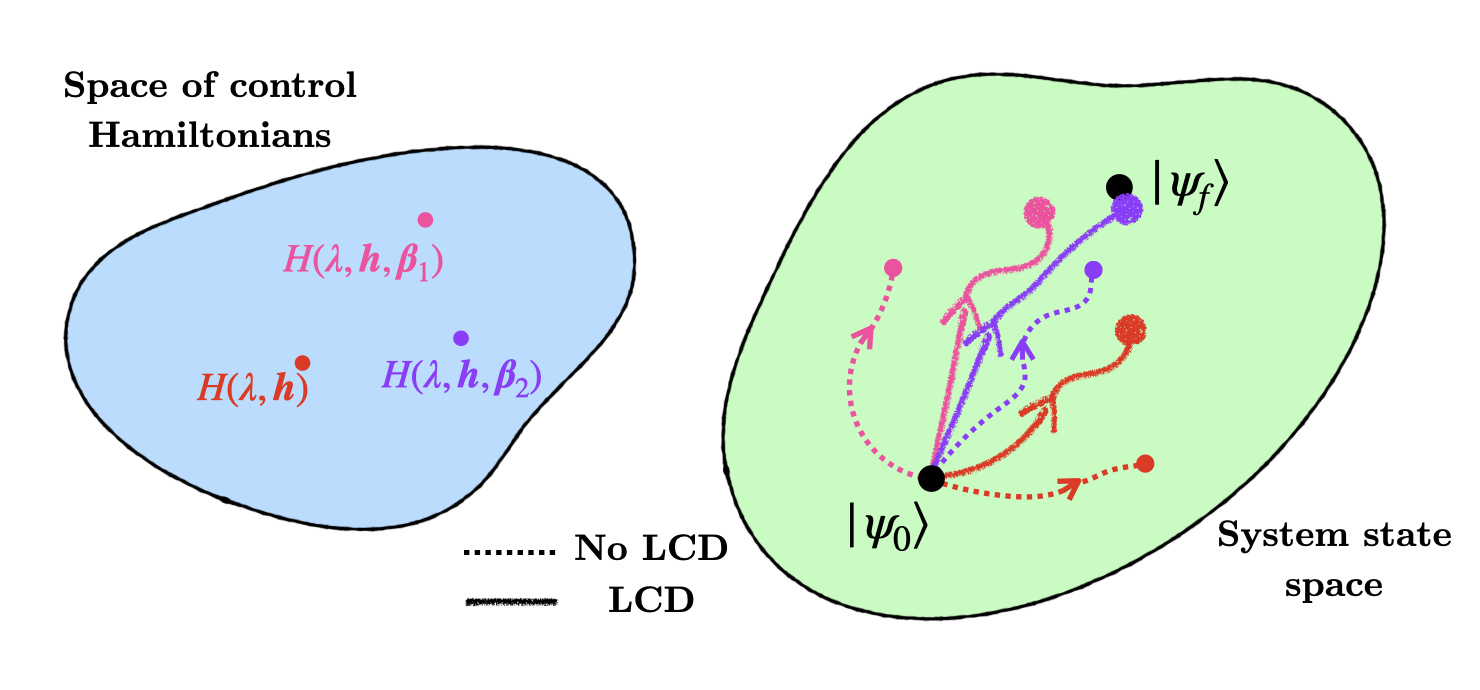} \caption[A diagrammatic illustration of the COLD method]{A diagrammatic illustration of the COLD method. (left) The set of control Hamiltonians with control parameters $\betabb$, with each point within the shape representing a different instance of $\betabb$. In the case where the control amplitude is $0$ throughout the evolution, we recover the bare Hamiltonian $H(\lambda, \hbb)$ (red point). (right) The system state space with $\ket{\psi_0}$ the eigenstate of $H(\lambda_0)$ that the system is prepared in and $\ket{\psi_f}$ the corresponding eigenstate of $H(\lambda_f)$, which is the target. The dotted (no \acrref{LCD} drive) and chalky (\acrref{LCD} drive included) directed lines show how the system is driven across state space for a fixed total time $\tau$ in the case of each control Hamiltonian indicated on the blue shape. \acrref{COLD} essentially allows one to use optimisation in order to find a path via the value of $\betabb$ which will lead to the result which is closest to the final state when \acrref{LCD} is applied, \@e.g.~going from the red path to the purple. Note that in cases where \acrref{LCD} is equivalent to the exact \acrref{CD} the arrows in the state space will lead from the initial state to the final state exactly, although the control parameter might change the shape of the path that they take to get there.}\label{fig:COLD_illustration}
\end{figure}

The addition of the control pulse prior to applying \acrref{LCD} makes it such that the \acrref{CD} coefficients $\alpha_j$ are now functions of $\betabb$, meaning that varying $\betabb$ will change the shape of the approximate counterdiabatic drive. This is illustrated in Fig.~\ref{fig:COLD_illustration} with changing paths in state space of the driven system. We take the set of operators $\mathcal{O}_{\rm LCD}$ to be fixed, since in a practical scenario this set would depend on physical constraints of the system for which it is implemented. However, the relative contribution of each operator to the exact counterdiabatic pulse (and thus its effectiveness at suppressing non-adiabatic effects) is governed by the path of the Hamiltonian. By adding a control term, we can now optimise the system evolution to follow a path which allows the truncated counterdiabatic drive to maximally suppress non-adiabatic effects. Fig.~\ref{fig:COLD_illustration} illustrates how the application of \acrref{LCD} will modify any path in system state space to end up closer to the target state than in its absence, but for particular paths it will get much closer. Examples of the effectiveness of \acrref{COLD} in different systems are demonstrated and analysed in more detail in Ch.~\ref{chap:6_Applications_fidelity}, where we explore how \acrref{COLD} holds up against each of its components on their own -- \acrref{LCD} and \acrref{QOCT}.

\section{Optimal control toolbox}\label{sec:4.2_COLD_QOCT}

With optimal control being one of the two key components of \acrref{COLD}, we will now revisit the content covered in Ch.~\ref{chap:3_Quantum_Optimal_control}, linking it to the way one might go about constructing control pulses in the \acrref{COLD} setting. In Sec.~\ref{sec:3.3_qoct_methods} we covered ``Chopped Randomised Basis" (\acrref{CRAB}) and ``Gradient Ascent Pulse Engineering" (\acrref{GRAPE}), two quantum optimal control methods that offer very flexible yet powerful approaches to constructing and optimising control pulses. Consequently, we can use them in the setting of \acrref{COLD} too. In our original work presented in Ref.~\cite{cepaite_counterdiabatic_2023}, three separate techniques for constructing optimal control pulses were implemented which will be the focal point of Part \ref{part:applications} of the thesis:
\begin{itemize}
    \item `Bare' pulses, which are functions composed of a Fourier basis where each basis function is scaled by an optimisable coefficient, similar to those given by Eq.~\ref{eq:trigonometric_CRAB}. The name `bare' is used to distinguish them from \acrref{CRAB} as they do not include any randomisation component.
    \item COLD-CRAB pulses, which are like the bare version but with the inclusion of randomisation in the frequencies of the basis functions used to construct the pulse, as discussed in Sec.~\ref{sec:3.3.1_CRAB}.
    \item COLD-GRAPE pulses, wherein the optimisable function is constructed using the \acrref{GRAPE} approach of parameterised piecewise constant time slices, as was expanded upon in detail in Sec.~\ref{sec:3.3.2_GRAPE}. 
\end{itemize}

To illustrate, a `bare' pulse, which will make a return often in the next part of the thesis, is the function
\begin{equation}\label{eq:bare_pulse}
    f(\lambda, \betabb) = \sum_{k=1}^{N_k} \beta^k \sin (2 \pi k \lambda),
\end{equation}
which fulfills the boundary conditions of $H(\lambda_0)$ and $H(\lambda_f)$ for $\lambda_0 = 0$ and $\lambda_f = 1$. The parameters $\beta_k$ for each frequency $k$ can be optimised using a numerical approach from Sec.~\ref{sec:3.1.3_numerical_optimisation}. This is a very simple way to construct a control pulse, but the simplicity is its appeal: it describes a continuous function with very few parameters and a solution can often be easily analysed once it is found.

The case of COLD-CRAB is more self-explanatory, as it is just an implementation of the \acrref{CRAB} algorithm, which was discussed in detail in Sec.~\ref{sec:3.3.1_CRAB}, to construct the control pulse for \acrref{COLD}. In the numerical results presented in the next part of the thesis, the most common implementation is taking the bare pulse that defined above in Eq.~\eqref{eq:bare_pulse} and randomising the principal frequencies $\omega_k = 2 \pi k$ of the trigonometric functions. This is done by drawing parameters $r_k$ from a uniform random distribution $r_k \in [-0.5,0.5]$ at each optimisation instance of the pulse and replacing $k \rightarrow k(1+r_k)$. This makes optimisation more complex - as already discussed in Sec.~\ref{sec:3.3.1_CRAB} - but it also often allows for far better results without any increase in optimisable parameters. Depending on the computational resources at hand, especially parallelisation, COLD-CRAB is a far better option than just the bare pulse in terms of results.

Finally, COLD-GRAPE is exactly what it says on the tin - if the tin were Sec.~\ref{sec:3.3.2_GRAPE}. In this case the optimisable pulse is built up out of piecewise constant control amplitudes, as in the original \acrref{GRAPE} algorithm and these are optimised once again using numerical methods from Sec.~\ref{sec:3.1.3_numerical_optimisation}. The method generally requires more optimisation parameters and thus is computationally intensive, but it also removes the need to choose a good pulse basis, like in the bare and COLD-CRAB cases. It is possible, but not necessary, to use the gradient information of the cost function that was provided in the original \acrref{GRAPE} paper \cite{khaneja_optimal_2005}. As will become clearer in Ch.~\ref{chap:7_higher_order_agp} however, some of the cost function landscapes we have to deal with in the case of \acrref{COLD} are highly non-convex and as such gradient-based optimisation techniques generally do not work well. 

One computational issue to address in using \acrref{GRAPE} for \acrref{COLD} is that the coefficients $\alpha_j$ of the \acrref{LCD} pulses generally have a dependence on $\dlambda \betabb$ due to the \acrref{AGP} operator being a function of the matrix elements of $\dlambda H$. These are not well-defined for a pulse constructed out of piecewise constant amplitudes masquerading as a continuous function. The way to get around this issue is to use spline interpolation \cite{noauthor_spline_nodate}, which is a method used to interpolate between the piecewise components and recover a continuous pulse which can then be used to calculate the derivatives $\dlambda \betabb$. 

These three methods are by no means the only way to construct optimal pulses for \acrref{COLD} and, as discussed in Sec.~\ref{sec:3.2_Quantum_optimal_control}, the field of quantum optimal control is vast \cite{koch_quantum_2022}. A consideration that is specific to \acrref{COLD} is the inclusion of constraints in the cost function on the \acrref{LCD} pulse as well as the control drive, as it can diverge \@e.g. across phase transitions \cite{hatomura_controlling_2021}, something that makes experimental implementation difficult and goes against the philosophy of \acrref{COLD} as a method. Ultimately, both the \acrref{LCD} pulse and the control drive should be designed with the goal of making them useful in an experimental setting and the optimal control component needs to reflect this. 
\chapter{Adiabatic gauge potential as a cost function}\label{chap:5_cd_as_costfunc}

\epigraph{Always remember, however, that there’s usually a simpler and better way to do something than the first way that pops into your head.}{\emph{Donald Knuth}}

In the previous chapter we presented the \acrref{COLD} method, which combines \acrref{LCD} (Sec.~\ref{sec:2.4.1_LCD}) and quantum optimal control (Sec.~\ref{sec:3.2_Quantum_optimal_control}) in order to speed up adiabatic quantum processes while minimising transitions out of the instantaneous eigenstates. The strategy of \acrref{COLD} is largely concerned with implementing optimal control in order to modify the path of the time-dependent Hamiltonian in parameter space in a way that maximises the effectiveness of a given \acrref{LCD} drive in driving a system to a target eigenstate of the adiabatic Hamiltonian. The optimal control component of \acrref{COLD} is thus constructed around optimising for the final state of the system after evolution: whether by assessing its fidelity with respect to some target state or a property like entanglement.

In this chapter we will take a slightly different but complementary perspective on combining \acrref{LCD} and optimal control by asking the question of what happens when, instead of optimising for a particular target state, we use only information about the counterdiabatic drive - or rather, the \acrref{AGP} from Sec.~\ref{sec:2.2_AGP} - as the optimal control cost function. Since the \acrref{AGP} contains information about the non-adiabatic effects experienced by a system, it is reasonable to believe that this information can be extracted and its analysis can be useful in designing optimal fast driving schedules for adiabatic protocols. 

We will begin the chapter with a brief motivation behind using the counterdiabatic pulse as a metric for optimising fast adiabatic processes in Sec.~\ref{sec:5.1_motivation} and then in Sec.~\ref{sec:5.2_designing_costfunc_hocd} we will explore several different ways in which the \acrref{CD} pulse can be transformed into an optimal control cost function. While this chapter will introduce the theory behind the idea, Ch.~\ref{chap:7_higher_order_agp} will present the numerical simulation results obtained using the ideas in this chapter.

\section{Motivation}\label{sec:5.1_motivation}

Returning to the key ideas behind \acrref{CD}, we may recall that the exact \acrref{CD} pulse is comprised of the \acrref{AGP} operator $\AGP{\lambda}$ scaled by the rate of change of $\lambda(t)$ (expressed as $\dotlambda$) in the adiabatic Hamiltonian, as given by Eq.~\eqref{eq:CD_Hamiltonian}. The \acrref{AGP} is the generator of adiabatic deformations between quantum eigenstates, and its off-diagonal elements are responsible for transitions between the instantaneous (or adiabatic) eigenstates. Put another way, the Frobenius norm of the \acrref{AGP} is the distance between nearby adiabatic eigenstates \cite{pandey_adiabatic_2020, nandy_delayed_2022}. Thus, there are two components comprising the non-adiabatic effects experienced by a system driven at finite time by a time-dependent Hamiltonian: the rate of change of the parameters $\dotlambda$ and the \acrref{AGP} operator, which can be quantified \@e.g.~via the magnitude of $\dotlambda$ or the norm of the adiabatic gauge potential. 

In the design of fast adiabatic protocols such as the techniques under the umbrella of \acrref{STA}, we are competing against $\dotlambda$, since the aim in general is speed rather than adherence to the adiabatic condition. This leaves us with minimising $\AGP{\lambda}$, or else, as in the case of \acrref{CD}, suppressing or mitigating its effects. To whit, \acrref{LCD} does this by implementing an operator that approximately suppresses the \acrref{AGP} for a given Hamiltonian. \acrref{COLD} then aids in this endeavour by modifying the Hamiltonian and thus the corresponding \acrref{AGP} operator in a way that allows for a given \acrref{LCD} protocol to perform better. The optimal control component of \acrref{COLD} in \cite{cepaite_counterdiabatic_2023} and throughout Ch.~\ref{chap:6_Applications_fidelity} is implemented with the target state fidelity (Eq.~\eqref{eq:costfunc_fidelity}) as a cost function, which is due to the fact that the primary application of adiabatic protocols in often (though not always \cite{pelegri_high-fidelity_2022}) state preparation. The use of fidelity as a cost function, however, necessitates access to the wavefunction of the final prepared state, something that becomes difficult in the case of large or highly correlated systems, and may result in a highly non-convex or complex cost function landscape (see, for example, Fig.~\ref{fig:ghz_contours}). Furthermore, fidelity is not a useful cost function in practice, particularly in the case where the target state is unknown, which is common in \@e.g.~applications of adiabatic protocols to Hamiltonians whose ground states encode solutions to combinatorics problems \cite{ebadi_quantum_2022, albash_adiabatic_2018}. 

A solution to the problem of fidelity as a cost function might be the use of a different metric for optimising the Hamiltonian path. We have established that $\AGP{\lambda}$ contains information pertaining to non-adiabatic losses experienced by a driven system. Thus, a natural approach to optimising the path of the Hamiltonian would be to minimise the \acrref{AGP} operator (we will discuss the details of this in the next section), as this should in principle minimise losses associated with non-adiabaticity. The main advantages of such an approach would be the fact that  the minimisation should be far more efficient than any attempt to compute the full system evolution, assuming we have access to the \acrref{AGP} as a function of the Hamiltonian path in parameter space. In the case of \acrref{LCD}, this is a valid assumption. Furthermore, this process would require no knowledge of the system wavefunction at any point, removing the drawbacks discussed earlier concerning the efficiency of implementing the fidelity cost function.

\section{Designing a cost function around the counterdiabatic pulse}\label{sec:5.2_designing_costfunc_hocd}

There are several ways to define a metric for a time-dependent control pulse and in this thesis we will explore two in particular. In order to do this, we will first return to Ch.~\ref{chap:4_COLD}, where we expressed the \acrref{CD} pulse as a sum of operators $\{\mathcal{O}_{\rm CD}^{(j)}\}_{j = 1, ..., N_{\rm CD}}$ which are scaled by coefficients $\alpha_j(\lambda, \hbb)$ with $\lambda$ a function of time and $\hbb$ the $\lambda$-dependent coefficients of the Hamiltonian $H(\lambda)$. In the \acrref{LCD} case, this sum is truncated to a set of operators $\mathcal{O}_{\rm LCD} \subset \mathcal{O}_{\rm CD}$ which are themselves scaled by a different set of coefficients $\alpha_j^{\prime}(\lambda, \hbb)$. Expressed this way, we can view the operators as static, with the $\lambda$- and $\hbb$-dependent coefficients $\alpha_j$ and $\alpha_j^{\prime}$ encoding the shape of the pulse and containing the information about non-adiabatic effects for given values of $\lambda$ and $\hbb$. If we include an optimal control component parameterised by functions $\beta_k(\lambda) \in \betabb$ as in Eq.~\eqref{eq:COLD_optimal_control} in the case of \acrref{COLD}, then the counterdiabatic coefficients become dependent on $\betabb$ too, allowing us to optimise them by varying the control functions.

Given this form for the counterdiabatic pulses, we can choose two types of metrics: (i) one which looks at the whole pulse, whether in the case of the exact \acrref{AGP} or some truncation obtained using \acrref{LCD}, and (ii) one which instead only picks out an extremum of the pulse, like its maximum amplitude. In the first case, a natural option would be the time-integrated absolute value of the $\alpha_j$ coefficients, which in the optimal control community is often referred to as the ``control effort"\cite{petersen_control_1987}:
\begin{equation}\label{eq:COLD_costfunc_integral}
    C_{\rm I}(\tau, \betabb) = \sum_j^{N_I} \int_{0}^{\tau} dt^{\prime} |\alpha_j^{\prime}(\lambda(t^{\prime}), \hbb, \betabb)|,
\end{equation}
where, as before, $\tau$ is the total driving time and here we take the sum over $N_I$ coefficients. When the quantity $C_I(\tau, \betabb)$ is minimised, naturally the contribution of the operators scaled by the coefficients in the sum will be reduced. When it is $0$, then these operators will not contribute at all to the non-adiabatic losses experienced by the driven system. It should be noted that understanding the physical meaning behind non-zero values of $C_I$ is quite non-trivial, although it is natural to expect that the larger it is, the more losses the system experiences. 

In the case where we want to minimise the exact \acrref{AGP}, $N_I = N_{CD}$. On the other hand, when it comes to \acrref{LCD} it may be fruitful to minimise $\alpha_j^{\prime}$ corresponding to operators which are not actually applied to the system. Say one has access to a set of operators with a non-zero contribution to the exact \acrref{CD} with the ability to implement only a subset, \@e.g.~they were obtained via the nested commutator approach from Sec.~\ref{sec:2.4.2_nested_commutators}. Then it is sensible to optimise for a path which minimises the contribution of the subset of operators that \emph{cannot} be implemented. This should, in theory, reduce their contribution to the non-adiabatic losses. In this way, an \acrref{LCD} protocol can be used to both suppress non-adiabatic losses by the application of an approximate counterdiabatic drive and to optimise for a Hamiltonian path which reduces the non-suppressed losses experienced by the system.

The second option for a cost function which uses nothing but the counterdiabatic pulse coefficients is one which instead minimises some extremum of the entire pulse, such as the maximal absolute amplitude reached by the pulse throughout the evolution, which we will write as:
\begin{equation}\label{eq:COLD_costfunc_maximum}
    C_{\rm A}(\tau, \betabb) = \max_{t^{\prime} \in [0,\tau]} \left( \sum_j^{N_I} | \alpha_j(\lambda(t^{\prime}), \hbb, \betabb)|\right).
\end{equation}

Intuitively, if the integral cost function is $0$, then this second approach does not provide any more information as the two will be equivalent. However, what this cost function captures is whether or not the path of the system for a given $\betabb$ and $\tau$ experiences any critical point in its evolution where the non-adiabatic effects are maximised, for example in the case of closing gaps between the instantaneous eigenstates. This cost function may be useful in cases where one wants to avoid such closing gaps within the system evolution and no path optimally suppresses all non-adiabatic effects. In particular, it might be used in \acrref{LCD} as a means to avoid cases where the \acrref{CD} operators \emph{which are not countered via a counterdiabatic drive} are responsible for most of the avoided transitions out of the instanteous eigenstate. For example, should one only be able to apply a counterdiabatic drive to a single subsystem of many, then it would make sense to aim to create a driving protocol for the full system which minimises the non-adiabatic effects experienced by every other system that cannot have them be suppressed in such a way. In other words, if you can't counter the losses, find a way to not generate them in the first place. 
\part{Applications of COLD}\label{part:applications}

\chapter{Optimising for properties of the state}\label{chap:6_Applications_fidelity}

\epigraph{In theory, theory and practice are the same. In practice, they are not.}{Unknown}

In Ch.~\ref{chap:4_COLD} we introduced \acrref{COLD}, a new method for speeding up adiabatic processes while suppressing non-adiabatic losses. In this chapter, we will investigate how such a method might perform for different adiabatic protocols in various physical systems via results from numerical simulations. 

There are a number of parameters that can be varied in each instance of applying COLD, including different ways to construct the control pulse (Sec.~\ref{sec:4.2_COLD_QOCT}), physical constraints placed on the system, and the operator basis used for the \acrref{LCD}, among others. Furthermore, it is important to compare the effects of \acrref{COLD} against either of its two components: \acrref{LCD} and quantum optimal control, which have been implemented with the same goal as \acrref{COLD} in the past \cite{sels_minimizing_2017, glaser_training_2015, guery-odelin_shortcuts_2019}. We made the claim in Ch.~\ref{chap:4_COLD} that \acrref{COLD} should outperform either approach simply by construction and here we will demonstrate this in practice.

We will begin with an example of a two-spin annealing process in order to illustrate the \acrref{COLD} approach in detail on a simple toy example. This will also serve as a good test bed for the variational method for deriving an \acrref{LCD} pulse and constructing an optimal control pulse. We will then proceed to illustrate how \acrref{COLD} can be applied for a series of different example Hamiltonians and systems, starting with the Ising spin chain in Sec.~\ref{sec:5.2_Ising_chain}, then a case of population transfer in a synthetic lattice via an adiabatic rapid passage (\acrref{ARP}) protocol in Sec.~\ref{sec:5.3_synthetic} and finally in preparing maximally entangled states in a system of frustrated spins in Sec.~\ref{sec:6.4_ghz_states}. We will demonstrate how driving amplitude constraints affect the performance of \acrref{COLD} and other approaches in Sec.~\ref{sec:6.2.1_restricting_amps} and how different optimisation cost functions can inform the results in Sec.~\ref{sec:6.4.1_t3} where we will use entanglement as an optimisation metric instead of final state fidelity.

\section{Two-spin annealing}\label{sec:5.1_2spin_annealing}

\begin{figure}[t]
    \centering
    \includegraphics[width=0.8\linewidth]{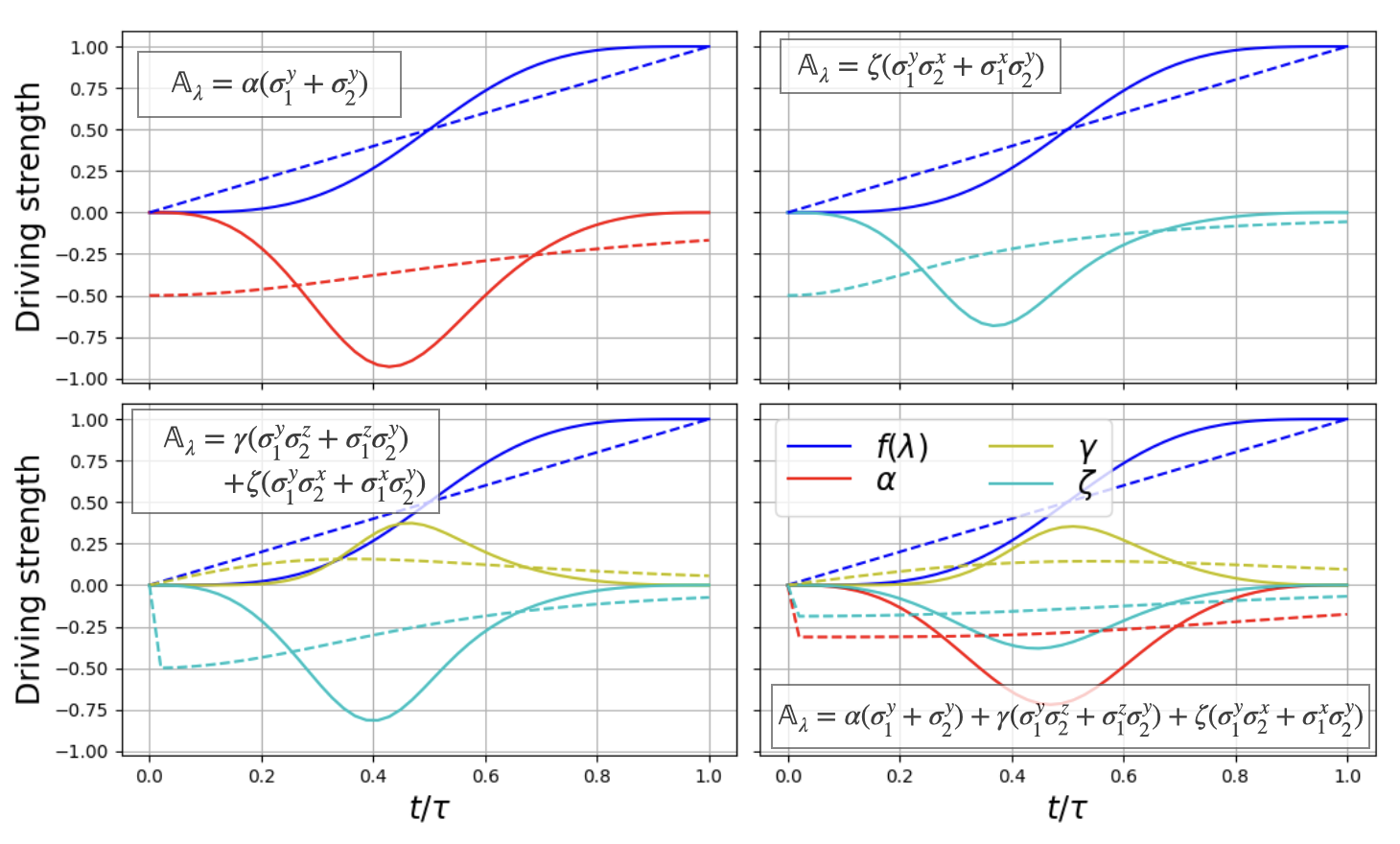} \caption[Driving strengths of LCD for different Hamiltonian protocols.]{Figure illustrating the functional forms of the drives in the two-spin annealing protocol of Eq.~\eqref{eq:two_spin_hamiltonian}. The dark blue plots indicate the functional form of $X(\lambda)$ scaled by $(2h_0)^{-1}$ (such that $X(\lambda) = 2 h_0 f(\lambda)$ where $f(\lambda)$ is the function being plotted). The solid plots are for the case given in Eq.~\eqref{eq:lambda_func1}, where $f(\lambda) = \sin^2\left(\frac{\pi}{2} \sin^2 \left( \frac{\pi}{2} \lambda \right) \right)$, while the dashed plots describe the case where a linear drive is used instead: $f(\lambda) = \lambda$, remembering that $\lambda = \frac{t}{\tau}$. Each set of axes above shows the resulting drives for different choices of \acrref{LCD} ansatz, as indicated by the inset equations. Note that the case where $\approxAGP = \gamma(\sz_1\sy_2 + \sy_1\sz_2)$, the resulting drive is zero throughout the protocol and hence it is not plotted above.}\label{fig:lcd_drives}
\end{figure}

To showcase and explore the use of \acrref{COLD} in a relatively simple setting we will consider a two spin quantum annealing problem with Hamiltonian
\begin{equation}\label{eq:two_spin_hamiltonian}
H_0(\hbb, \lambda) = J(\lambda) \sz_1 \sz_2 + Z(\lambda) ( \sz_{1} + \sz_{2}) +  X(\lambda) (\sx_{1} + \sx_{2}),
\end{equation}
where the operator subscripts denote the index of the spin on which they act, $H_0$ is parameterised by the functions $\hbb = \{J(\lambda), Z(\lambda), X(\lambda)\}$, with the $\lambda(t) = \frac{t}{\tau}$ term encoding the time-dependence and where $J(\lambda) = -2J_0$ and $z(\lambda) = -h_0$ are constant functions. The added transverse field allows us to explore a larger operator basis for the \acrref{LCD}. For this example we use
\begin{equation}\label{eq:lambda_func1}
X(\lambda) = 2 h_0 \sin^2\left(\frac{\pi}{2} \sin^2 \left( \frac{\pi}{2} \lambda \right) \right),
\end{equation}
where we note that since $\lambda(0) = 0$ and $\lambda(\tau) = 1$, the transverse field is tuned from $0$ to $2h_0$ as $t$ goes from $0$ to $\tau$. The choice of function is rather arbitrary, but in general we want driving functions which are smooth, in particular near the beginning and end of the protocol, since they lead to a more smooth,  well-behaved counterdiabatic term which tends to 0 at the beginning and end of the driving time without any sudden jumps in value, making its experimental implementation more amenable. This is a consequence of the \acrref{LCD} drive's dependence on the derivatives of the functions in set $\hbb$. Fig.~\ref{fig:lcd_drives} illustrates the difference in the resulting \acrref{LCD} drives when using the smooth version of $X(\lambda)$ given above and a linear protocol. We can see that in the case of a linearly increasing drive, the \acrref{LCD} driving terms are generally non-zero at the start and end of the protocol and may exhibit sudden jumps in value due to the larger derivative of the driving function at certain instances during the driving procedure. 

We consider the case where $J_0/h_0 = 0.5$, meaning that the initial ground state of the system is in the $\ket{\uparrow \uparrow}$ state and the ground state at $t = \tau$ should be a superposition of all the symmetric states. As per the discussion in Sec.~\ref{sec:2.4.1_LCD}, since $H_0$ is a real-valued Hamiltonian, the single-spin LCD operators should be fully imaginary and thus given by the following ansatz for the adiabatic gauge potential:
\begin{equation}\label{eq:LCD1st}
\approxAGP^{(1)}(\lambda, \hbb) = \alpha(\lambda, \hbb) (\sy_1 + \sy_2),
\end{equation}
which we will indicate as `first-order' or \acrref{FO} \acrref{LCD}, referring to the fact that these are the most local spin operators and denoting this fact with the superscript $(1)$. In this case, we know that the \acrref{LCD} pulse is of the same form for both $\sy_1$ and $\sy_2$ due to the symmetry of the Hamiltonian. Thus, we only require one coefficient $\alpha$ for both operators. In future sections where this is not the case, we will instead differentiate between global and local \acrref{LCD} pulses. In the case of local \acrref{LCD}, unique pulses may be required for each separate operator to capture the counterdiabatic effects more faithfully, but for the sake of simplicity and the commonplace difficulty of highly precise local control of systems in expreiment, we will generally approximate them with a single coefficient.

Using the methods described in \cite{sels_minimizing_2017} and summarised in Sec.~\ref{sec:2.4.1_LCD}, we can determine the form of the coefficient $\alpha$ using a variational approach, which we will set out in detail here to illustrate the method. For the given $H_0$, and $\approxAGP$, the first step is to find the operator $G_{\lambda}$ from Eq.~\eqref{eq:G_operator} (setting, in true Physics fashion, $\hbar = 1$):
\begin{equation}
    \begin{aligned}
        G_{\lambda}(\approxAGP^{(1)}, H_0) &= \dlambda H_0 + i \comm{\approxAGP^{(1)}}{H_0} \\
        &= \dlambda {X} (\sx_{1} + \sx_{2}) - 2 \alpha J (\sx_1\sz_2 + \sz_1\sx_2) - 2 \alpha Z (\sx_{1} + \sx_{2}) \\
        &+ 2 \alpha X (\sz_{1} + \sz_{2}),
    \end{aligned}
\end{equation}
where the $\lambda$-dependence is omitted and the only non-constant function is $X(\lambda)$, making it the only non-zero contribution to $\dlambda H_0$. We then obtain the action from Eq.~\eqref{eq:agp_action} defined as $\mathcal{S} = \Tr[G^2_{\lambda}]$:
\begin{equation}
    \frac{1}{8} \Tr[G^2_{\lambda}] = 4 \alpha^2 X^2 + (\dlambda X - 2 \alpha Z)^2 + 4 \alpha^2 J^2,
\end{equation}
which can be minimised with respect to the coefficient $\alpha$ in order to find the \acrref{LCD} pulse:
\begin{equation}
    \begin{aligned}
        \frac{\partial \mathcal{S}}{\partial \alpha} &= 8X^2\alpha - 4Z(\dlambda X - 2 \alpha Z) + 8 J^2 \alpha = 0 \\
        & \Rightarrow \alpha = \frac{1}{2} \frac{Z \dlambda X}{X^2 + Z^2 + J^2},
    \end{aligned}
\end{equation}
where we find, as expected, that the \acrref{LCD} pulse is a function of $\lambda$ and the Hamiltonian coefficients $\hbb$ (and their derivatives with respect to $\lambda$). For completeness, the full \acrref{LCD} Hamiltonian, recalling Eq.~\eqref{eq:CD_Hamiltonian}, then reads:
\begin{equation}\label{eq:FO_LCD_H}
    \begin{aligned}
        H_{\rm LCD}(\hbb, \lambda) &= H_0(\hbb, \lambda) + \dotlambda \alpha(\hbb, \lambda) (\sy_{1} + \sy_{2}),
    \end{aligned}
\end{equation}
where the counterdiabatic pulse is simply the (approximate) \acrref{AGP} scaled by the rate of change in the time-dependent parameter $\lambda$.

We can do the same as above for the next most local, two-spin operators, which we will refer to as `second-order' or \acrref{SO} \acrref{LCD}. They too should be imaginary and, due to the Hamiltonian symmetry, can be cast into two groups with two different \acrref{LCD} coefficients $\gamma$ and $\zeta$:
\begin{equation}\label{eq:twospin_so_lcd}
        \approxAGP^{(2)}(\hbb, \lambda) = \gamma(\hbb, \lambda) (\sx_1 \sy_2 + \sy_1 \sx_2) + \zeta(\hbb, \lambda) (\sz_1 \sy_2 + \sy_1 \sz_2).
\end{equation}
These can be solved for in a similar vein to the method for $\alpha$, although in this case the minimisation of $\mathcal{S}(\approxAGP^{(2)})$ will happen separately for both $\gamma$ and $\zeta$, giving a coupled set of equations which can be solved numerically. Since the whole system is only made up of two spins, the two orders of \acrref{LCD} ansatz are enough to characterise the \acrref{AGP} fully as they contain all completely imaginary orthogonal operators in the Pauli basis. Thus, solving the coupled set of equations
\begin{equation}\label{eq:two_spin_coupled_eqs}
        \begin{pmatrix}
        2(X^2 + Z^2 + J^2) & - 2JX & 4JZ \\ 
        -XJ & X^2 + 4Z^2 & -3XZ \\ 
        4JZ & -6ZX & 2J^2 + 2Z^2 + 8X^2
        \end{pmatrix} 
        \begin{pmatrix}
            \alpha \\
            \gamma \\
            \zeta
        \end{pmatrix} = 
        \begin{pmatrix}
            Z \dlambda X \\
            0 \\
            J \dlambda X
        \end{pmatrix}
\end{equation}
for the coefficients $\alpha$, $\gamma$ and $\zeta$ should give the exact \acrref{AGP} operator 
\begin{equation}\label{eq:two_spin_exact_AGP}
    \AGP{\lambda}(\hbb, \lambda) = \alpha(\hbb, \lambda) (\sy_1 + \sy_2) + \gamma(\hbb, \lambda) (\sx_1 \sy_2 + \sy_1 \sx_2) + \zeta(\hbb, \lambda) (\sz_1 \sy_2 + \sy_1 \sz_2)
\end{equation}
for the Hamiltonian $H_0(\hbb, \lambda)$. Given any other $\hbb$ and $\lambda$, the equations above can be modified quite easily by changing the form of the action correspondingly. 

\begin{figure}[t]
    \centering
    \includegraphics[width=0.8\linewidth]{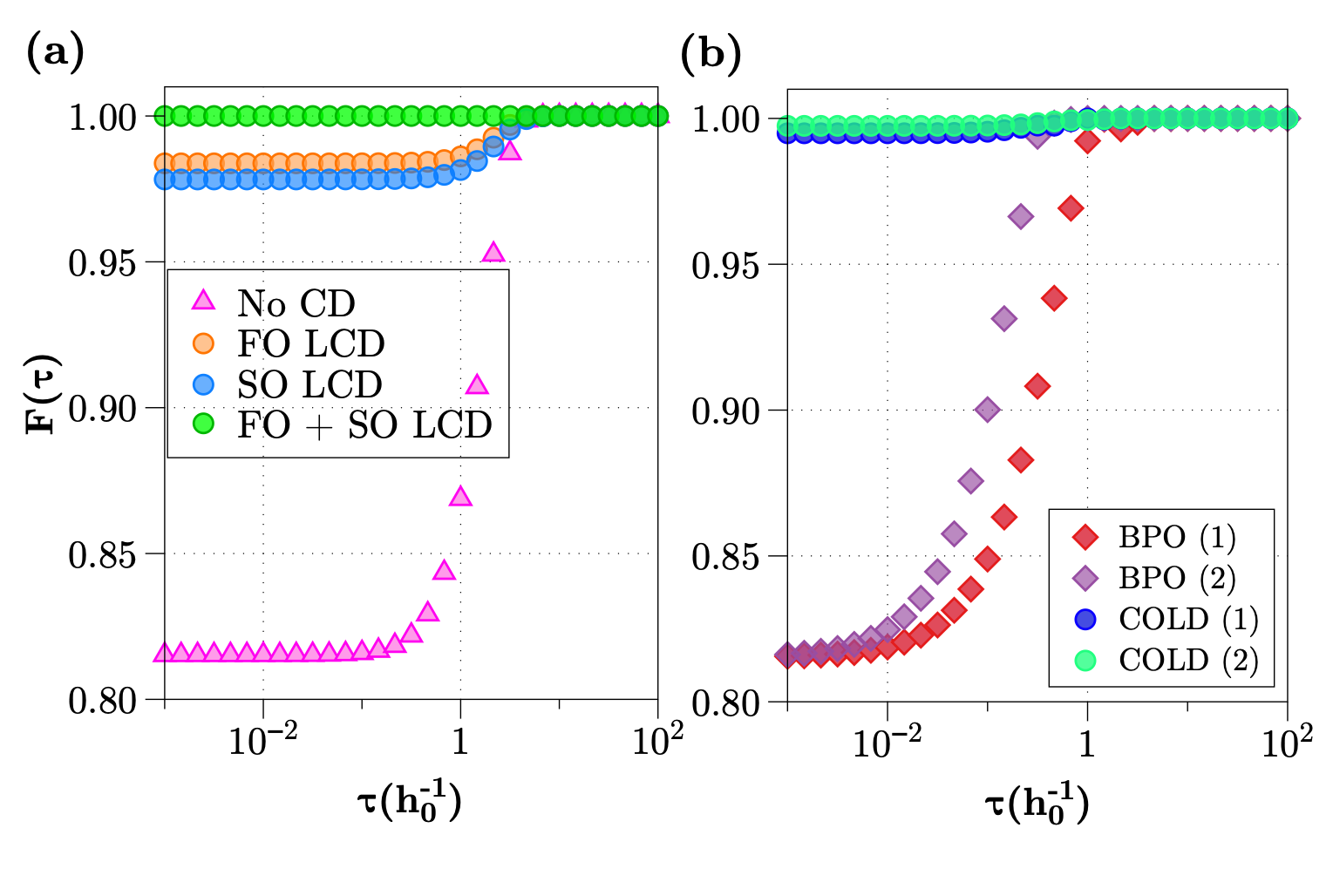} \caption[COLD applied to two-spin annealing]{Optimisation of the annealing protocol for two spin Hamiltonian given by Eq.~\eqref{eq:two_spin_hamiltonian} and with parameters as described in the main text.  (a) Final fidelities of the annealing protocol with triangles (pink) representing the case where no \acrref{CD} is applied and circles showing the case of \acrref{FO} \acrref{LCD} (orange), \acrref{SO} \acrref{LCD} (blue) as well as the combination of \acrref{FO} and \acrref{SO} \acrref{LCD} (green). (b) Final fidelities achieved when using the optimal control method \acrref{BPO} ($N_k = 1$: red diamonds, $N_k = 2$: purple diamonds) and \acrref{COLD} ($N_k = 1$: blue circles, $N_k = 2$: aquamarine circles) with \acrref{FO} \acrref{LCD} operators as described in the text.}\label{fig:twospin_fidelities}
\end{figure}

The different approaches are demonstrated in Fig~\ref{fig:twospin_fidelities}(a) via numerical simulations of the system evolution for different total evolution times $\tau (h_0^{-1})$. We compare the final evolved state $\ket{\psi(\tau)}$ of the system with the ground state of $H_0(\lambda = 1)$ denoted by $\ket{\psi_{GS}}$ by computing their fidelity:
\begin{equation}
    F(\tau) = \left|\braket{\psi_{GS}}{\psi(\tau)}\right|^2.
\end{equation}
The results are computed for the case where only the bare Hamiltonian $H_0(\hbb, \lambda)$ from Eq.~\eqref{eq:two_spin_hamiltonian} drives the system and they are then compared to \acrref{FO} \acrref{LCD} (Eq.~\eqref{eq:FO_LCD_H}), \acrref{SO} \acrref{LCD} (the solution to Eq.~\eqref{eq:two_spin_coupled_eqs} with $\alpha$ set to $0$) and the exact counterdiabatic drive (Eq.~\eqref{eq:two_spin_exact_AGP}). The results show that at fast evolution times ($\tau < 1 h_0^{-1}$) when no counterdiabatic drive is applied the final state remains far from the ground state of $H_0(\lambda = 1)$ and only begins to approach the target as the evolution time is increased, exactly as one might expect given the adiabatic condition (Sec.~\ref{sec:2.1.2_adiabatic_condition}). In the case of both \acrref{FO} and \acrref{SO} \acrref{LCD}, however, the system approaches the desired ground state with close to unit fidelity even at very short driving times several orders of magnitude faster. As expected, when the exact \acrref{CD} is applied including all single- and two-spin imaginary operators, the system reaches the desired state with unit fidelity at arbitrarily short driving times, as in the rotating spin example we discussed in Ch.~\ref{chap:2_adiabaticity} and Appendix~\ref{app:rotating_spin_hamiltonian}. There is a curious phenomenon, wherein for all pulses barring exact \acrref{CD} (\acrref{FO} + \acrref{SO} \acrref{LCD}), the system fidelity is flat for some time before rising to 1 at around the same time. A natural interpretation of this is that around driving time $\tau = 1 h_0^{-1}$ the system enters the adiabatic regime. Prior to this point, it is likely that in each of the different cases (\acrref{BPO}, \acrref{FO} and \acrref{SO} \acrref{LCD}), there is a different fundamental limit to the controllability of the system that depends on what types of external drives are applied, as the fidelity plateaus at different values for each method regardless of driving time. For a more in-depth discussion on how this might relate to the quantum speed limit and the way that the \acrref{LCD} affects the spectrum of the Hamiltonian, we refer the reader to Ref.~\cite{abah_energetic_2019}.

At this point all we have done is to implement the \acrref{LCD} method for a simple example where the exact \acrref{CD} can be easily derived. This was done primarily to illustrate how the \acrref{LCD} pulse is constructed and how it can be used to significantly speed up system evolution while driving it close to the target state by suppressing a large proportion of non-adiabatic losses. With these components explored in depth, we can now finally introduce \acrref{COLD} in a practical setting.

As the goal of \acrref{COLD} is to improve \acrref{LCD} in restricted settings, the most natural approach to demonstrate it in a simple example like this is to take the \acrref{FO} \acrref{LCD} ansatz as a fixed approximation for the counterdiabatic drive and to ignore \acrref{SO} terms. This is a realistic scenario, as even two-spin operators like those found in the \acrref{SO} \acrref{LCD} anstaz of Eq.~\eqref{eq:twospin_so_lcd} are generally difficult to engineer in physical systems and thus even in this simple case it is unlikely that the exact \acrref{CD} pulse could be implemented, though its functional form is known.

Revisiting Ch.~\ref{chap:4_COLD}, we find that the first step of \acrref{COLD} is to construct a control pulse for the Hamiltonian. Ideally, this is something that can be easily controlled in an experimental setting and for the given $H_0$, we can imagine introducing a $\lambda$-dependent control for the coupling, transverse or longitudinal fields. For simplicity and pedagogy, we can introduce a `bare' pulse (Sec.~\ref{sec:4.2_COLD_QOCT}) control component to the Hamiltonian which drives the $(\sz_1 + \sz_2)$ operators (the choice is arbitrary - we may as well have picked  $(\sx_1 + \sx_2)$ or any other operator form the available degrees of freedom - though the results would change depending on that choice) and obeys the constraint of being $0$ at the beginning and end of the driving time. This gives a control Hamiltonian:
\begin{equation}\label{eq:COLD_twospin_controlH}
    \begin{aligned}
        H_\beta(\betabb, \hbb, \lambda) &= H_0(\hbb, \lambda) + \sum_{k=1}^{N_k} \beta_k(\lambda) (\sz_{1} + \sz_{2}) \\
        &= H_0(\hbb, \lambda) + \sum_{k=1}^{N_k} c_k \sin (\pi k \lambda) (\sz_{1} + \sz_{2}),
    \end{aligned}
\end{equation}
where $\beta_k(\lambda) \in \betabb$, $\beta_k(\lambda) = c_k \sin (\pi k \lambda)$, the value $N_k$ denotes the total number of control functions and the parameters $c_k$ can be optimised using a numerical optimal control method introduced in Sec.~\ref{sec:3.1.3_numerical_optimisation}. In this case, we will implement Powell optimisation (Sec.~\ref{sec:3.1.3.2_Powell}) as it is an efficient, gradient-free method that heuristically appears to avoid local minima in the cost function space better than Nelder-Mead (Sec.~\ref{sec:3.1.3.1_Nelder_Mead}), although both can be used given the relatively simple control problem at hand. 

Since the control Hamiltonian includes additional non-trivial $\lambda$-dependent components that $H_0$ did not, we need to re-derive the updated \acrref{FO} \acrref{LCD} pulse for the control Hamiltonian $H_{\beta}$, which is not particularly difficult if we follow the earlier recipe:
\begin{equation}\label{eq:twospin_alpha_COLD}
    \alpha(\betabb, \hbb, \lambda) = \frac{1}{2} \frac{(Z + f_{\rm opt}^{N_k}) \dlambda X - (\dlambda f_{\rm opt}^{N_k}) X}{(Z + f_{\rm opt}^{N_k})^2 + X^2 + J^2},
\end{equation}
where $f_{\rm opt}^{N_k} = \sum_{k=1}^{N_k} \beta_k$ is the full control pulse constructed out of the $N_k$ components. The total \acrref{COLD} Hamiltonian with \acrref{FO} \acrref{LCD} is then:
\begin{equation}\label{eq:twospin_COLD_Ham}
    H_{\rm COLD}(\betabb, \hbb, \lambda) = H_0(\hbb, \lambda) + f_{\rm opt}^{N_k}(\betabb, \lambda)(\sz_1 + \sz_2) + \dotlambda \alpha(\betabb, \hbb, \lambda)(\sy_1 + \sy_2).
\end{equation}

All that remains is to optimise the parameters $c_k$ with Powell's optimisation algorithm. As our goal is to drive the system to the ground state of $H_0(\lambda = 1)$, we can implement the fidelity cost function from Eq.~\eqref{eq:costfunc_fidelity} as a metric for the optimisation. Fig.~\ref{fig:twospin_fidelities}(b) shows the results of numerical simulations in the same regime as for the optimisation-free case, comparing evolution under the optimal control Hamiltonian from Eq.~\eqref{eq:COLD_twospin_controlH}, an approach we call `Bare Powell Optimisation' or \acrref{BPO} and the \acrref{COLD} Hamiltonian (Eq.~\eqref{eq:twospin_COLD_Ham} with a \acrref{FO} \acrref{LCD} pulse included. Even for a single optimisable parameter, \acrref{COLD} achieves $\sim 99.5\%$ fidelity at arbitrarily short times, while \acrref{BPO} remains stuck below $\sim 85\%$ until $\tau \sim 0.1h_0^{-1}$. Both approaches show some improvement with an added control parameter ($N_k = 2$), with the \acrref{COLD} result starting at $\sim 99.8\%$ fidelity even at short times. Neither approach, however, shows any noticeable improvement when adding more control parameters beyond this. 

Before moving on to the next section, we will note that this example is very simple and largely pedagogical. It may be possible to improve the results of both \acrref{BPO} and \acrref{COLD} with more sophisticated optimal control techniques, but it shows that even in the simplest case, \acrref{COLD} is a powerful method. If reliable access to the operators making up the exact \acrref{CD} is available, then it is better to implement \acrref{CD} rather than attempting optimal control. However, as this is almost never the case given the complexity and non-locality of the exact \acrref{AGP}, \acrref{COLD} is the best way to make the most out of the limited counterdiabatic capacity available.

\section{Ising chain}\label{sec:5.2_Ising_chain}

A more complex and widely studied example system that we can apply \acrref{COLD} to is the one-dimensional Ising spin chain for $N$ spins in the presence of a transverse and longitudinal field. The Ising model is an often-studied model in quantum mechanics and its ground states can be used to encode solutions to many combinatorics problems when we extend to arbitrary spin-spin couplings \cite{mohseni_ising_2022, ebadi_quantum_2022, pichler_quantum_2018}. As such, studying annealing protocols for Ising Hamiltonians with arbitrary connectivity is of particular interest. While in this section we will look only at the restricted case of the Ising chain, in Appendix~\ref{app:arbitrary_ising_derivation} we derive the coupled equations for \acrref{FO} and \acrref{SO} \acrref{LCD} coefficients for a model with arbitrary $\sz\sz$ connectivity between the spins.

The Ising chain is described by the Hamiltonian
\begin{equation}\label{eq:ising_chain_hamiltonian}
    H_0(\hbb,\lambda) = J(\lambda) \sum_{j}^{N-1} \sz_j \sz_{j+1} + Z_0\sum_j^N \sz_j + X(\lambda) \sum_j^N \sx_j,
\end{equation}
where once again $\lambda(t) = t/\tau$, $\hbb = \{ J(\lambda), Z(\lambda), X(\lambda) \}$ with constant functions $J(\lambda) = -J_0$, $Z(\lambda) = Z_0$ and
\begin{equation}
    X(\lambda) = X_0 \sin^2\left(\frac{\pi}{2} \sin^2 \left( \frac{\pi}{2} \lambda \right) \right).
\end{equation}
In this section, all results are obtained using $J_0 = 1$, $Z_0 = 0.02J_0$ and $X_0 = 10J_0$.

Many of the steps in this section will be rehashed from the two spin example as the approach is very similar. In this case, we will only focus on the \acrref{FO} \acrref{LCD} terms, which are the single-spin $\sy$ operators applied to each spin in the chain, in the same vein as in the two spin example  of Eq.~\eqref{eq:LCD1st} due to the Hamiltonian being real:
\begin{equation}\label{eq:ising_fo_agp}
    \approxAGP^{(1)}(\hbb, \lambda) = \alpha(\hbb, \lambda) \sum_{j}^N\sy_j,
\end{equation}
where using the variational \acrref{LCD} approach we find that 
\begin{equation}
    \alpha = \frac{1}{2} \frac{Z \dlambda X}{X^2 + Z^2 + 2(1 - 1/N)J^2},
\end{equation}
with the $N$-dependent factor in the denominator is a consequence of the edge effects of the chain which disappear in the case of a ring. 

\begin{figure}[t]
    \centering
    \includegraphics[width=\linewidth]{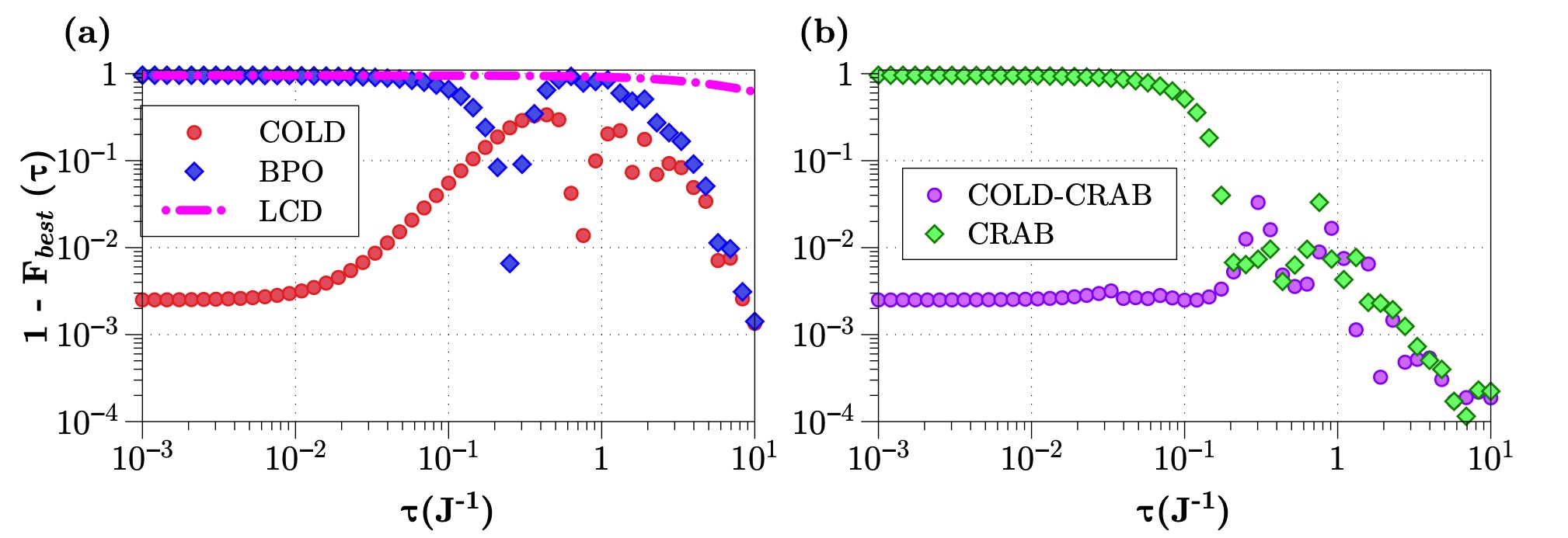} \caption[Applying COLD and COLD-CRAB to the Ising chain for 5 spins without constraints on the driving amplitudes.]{Final state fidelities for the Ising spin chain example described in the text for $N=5$ spins. In (a) we plot the best $(1 - F(\tau))$ obtained from 500 optimisations for different evolution times $\tau$ in the case where only \acrref{FO} \acrref{LCD} is applied (pink dash-dot line), as well as \acrref{BPO} (Eq.~\eqref{eq:ising_chain_BPO_H}, blue diamonds) and \acrref{COLD} with \acrref{FO} \acrref{LCD} (red circles) with a bare pulse control as described in the text. (b) shows implementations of \acrref{CRAB} and \acrref{COLD}-\acrref{CRAB}, with the best result again chosen from 500 optimisations, each with a different randomised frequencies in the trigonometric basis as described in the text. All control pulses use $N_k = 1$ optimisable parameter. Reprinted with permission from \cite{cepaite_counterdiabatic_2023}. Copyright 2023, American Physical Society.} \label{fig:ising_unconstrained}
\end{figure}

In order to implement \acrref{COLD} we need to once again construct a control pulse. In this case, we will start with a similar naive bare pulse as in the two spin case from the previous section
\begin{equation}\label{eq:ising_chain_BPO_H}
    \begin{aligned}
        H_{\beta}(\betabb, \hbb, \lambda) &= H_0(\hbb, \lambda) + \sum_k^{N_k} \beta_k(\lambda) \sum_j \sz_j \\
        &= H_0(\hbb, \lambda) + \sum_k^{N_k} c_k \sin(\omega_k \lambda) \sum_j \sz_j,
    \end{aligned}
\end{equation}
with $\omega_k = 2\pi k$ the $k$th principal frequency. In the case where no \acrref{LCD} is added to this Hamiltonian, we will again refer to the method as \acrref{BPO} as the numerical optimisation is carried out using Powell's method.

To construct the \acrref{COLD} Hamiltonian, the only additional step is to include the \acrref{LCD} pulse which we will restrict to single-spin terms as in Eq.~\eqref{eq:ising_fo_agp}:
\begin{equation}
    H_{\rm COLD}(\betabb, \hbb, \lambda) = H_0(\hbb, \lambda) + f_{\rm opt}^{N_k}(\betabb, \lambda)\sum_j \sz_j + \dotlambda \alpha(\betabb, \hbb, \lambda)\sum_j \sy_j,
\end{equation}
where again $f_{\rm opt}^{N_k} = \sum_k^{N_k} \beta_k(\lambda)$ represents the full control pulse made up of $N_k$ functions. 

As well as the bare pulse, however, in this case we will also implement the \acrref{CRAB} algorithm which was first introduced in Sec.~\ref{sec:3.3.1_CRAB} and compare it to the naive approach. The \acrref{COLD} algorithm with a \acrref{CRAB}-type pulse is thus referred to as \acrref{COLD}-\acrref{CRAB}. In our case, the inclusion of \acrref{CRAB} simply necessitates adding a randomised component to the naive basis functions $\beta_k(\lambda)$. Namely, the pulse $f_{\rm opt}^{N_k}$ is replaced by $f_{\rm CRAB}^{N_k}$ where the principal frequencies $k$ are modified as $k \rightarrow k(1+r_k)$, with $r_k$ drawn from a uniform random distribution $r_k \in [-0.5,0.5]$ such that:
\begin{equation}\label{eq:ising_crab_pulse}
    f_{\rm CRAB}^{N_k} = \sum_k^{N_k} c_k \sin(2 \pi k(1+r_k)\lambda).
\end{equation}
Since each optimisation instance for the \acrref{CRAB} algorithm will implement a slightly different pulse for the same number of control parameters owing to the randomised component, it is liable to lead to cost functions that have both better and worse minima than those of the naive pulse, meaning that the optimisation needs to be carried out many times in order to be sure of exploring as much of the solution space as possible. This added complexity is, as already discussed in Sec.~\ref{sec:3.3.1_CRAB}, what generally makes \acrref{CRAB} a better approach in terms of results obtained and a more difficult one due to the computational overhead required for the optimisation.  

In Fig.~\ref{fig:ising_unconstrained}(a) we plot the results (note that we now plot infidelity rather than fidelity) for the \acrref{FO} \acrref{LCD}, \acrref{BPO} and \acrref{COLD} for $N=5$ spins, observing that unlike in the two spin case, the \acrref{LCD} approach for this set of operators no longer shows a significant speed up, with final state fidelities remaining low even at long times. In fact, for all plotted times, it shows barely a $1\%$ improvement in fidelity over the bare Hamiltonian, which is not plotted as it overlaps with the \acrref{LCD} line. The \acrref{BPO} approach, on the other hand, appears to perform better at longer times with a sharp increase in fidelity around what is likely a natural timescale for the system, where it dips below the results produced by \acrref{COLD}. This is likely the phenomenon of a `magic time', as identified in \cite{couvert_optimal_2008, kiely_fast_2021}. The \acrref{COLD} approach, as before, performs really well at short driving times, where it is orders of magnitude better than either \acrref{LCD} or \acrref{BPO}. At longer times, this advantage wanes whether due to the optimisation process as multiple minima emerge in the cost function landscape or due to the fact that the local non-adiabatic effects are no longer the main source of losses. 

Fig.~\ref{fig:ising_unconstrained}(b) shows the results when using \acrref{CRAB} as well as \acrref{COLD}-\acrref{CRAB} in the same setting, with both performing far better at longer times than their naive \acrref{BPO} and \acrref{COLD} counterparts in plot (a) respectively, but remaining at similar levels of performance at short times, likely due to something more fundamental like the geometric speed limit \cite{bukov_geometric_2019}. This is definitely an argument for using something more sophisticated like \acrref{COLD}-\acrref{CRAB} as a general rule, as long as the computational resources are available for many and/or parallel optimisation instances.

While in Fig.~\ref{fig:ising_unconstrained} we only explore one optimisable parameter $N_k = 1$ and a system size of $N=5$ spins, the resulting advantage of \acrref{COLD} over bare optimisation scales with system size and we find that, at least in the case of the naive pulse, increasing the number of parameters does not seem to make too much of a difference. These results and more discussion can be found in Appendix~\ref{app:ising}.

\subsection{Restricting the driving amplitudes}\label{sec:6.2.1_restricting_amps}

While it is all well and good to talk about practical protocols implementing only local \acrref{LCD} operators with control drives that can be accessed by real experiments, one thing that we have so far failed to mention and which is not hard to observe from the form of the \acrref{CD} drive in Eq.~\eqref{eq:CD_Hamiltonian}, is that the amount of power required for the counterdiabatic pulse at short times scales with the speed of the changing Hamiltonian due to the $\dotlambda$ coefficient present in the counterdiabatic term. This means that while both \acrref{LCD} and \acrref{COLD} may lead to really high final state fidelities at very short driving times, they might also do this at the cost of impossibly high power requirements for the drives that implement the counterdiabatic component in either approach. An extensive discussion of this phenomenon and its consequences can be found in \cite{campbell_trade-off_2017, funo_universal_2017, santos_superadiabatic_2015}.

We find, (details in Fig.~\ref{fig:ising_maxamp} in Appendix~\ref{app:ising}), that this is indeed what happens in the Ising chain case: as the total time of the evolution is reduced, the maximum amplitude reached by the \acrref{LCD} pulse increases, leading to two orders of magnitude in difference between the power requirements at $\tau = 10^{-3}J_0^{-1}$ and $\tau = 10^{-1}J_0^{-1}$. As one of the goals of \acrref{COLD} as a method is to be practically implementable, this result is inherently counterproductive, although it may still provide some insights into the diabatic effects experienced by a system.

\begin{figure}[t]
    \centering
    \includegraphics[width=\linewidth]{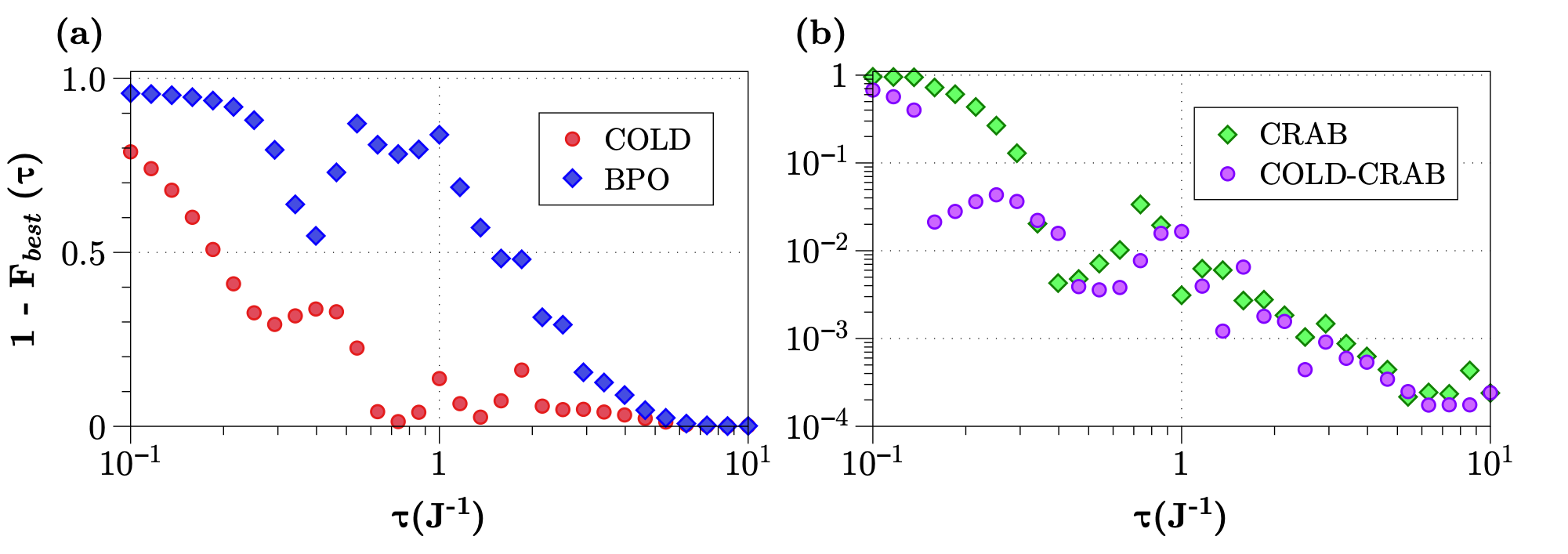} \caption[Applying COLD and COLD-CRAB to the Ising chain for 5 spins with constrained driving amplitudes.]{Optimisation of the constrained annealing protocol for the Ising model for $N=5$ spins with a maximum amplitude limit on each term in the Hamiltonian of Eq.~\eqref{eq:ising_chain_hamiltonian} of $X_0 = 10J_0$. (a) shows a comparison between BPO (blue diamonds). In (b) the comparison is between CRAB (green diamonds) and COLD-CRAB (purple circles). The plotted best results are obtained from 200 optimisations for each method. Reprinted with permission from \cite{cepaite_counterdiabatic_2023}. Copyright 2023, American Physical Society.}\label{fig:ising_constrained}
\end{figure}

In order to solve this issue, we can include penalty terms in the cost function used to optimise the pulses for all of the different approaches: \acrref{BPO}, \acrref{COLD}, \acrref{CRAB} and \acrref{COLD}-\acrref{CRAB}. These penalty terms behave as constraints on the behaviour of the optimised pulse, and they may also include constraints on the \acrref{LCD} pulse that is included in \acrref{COLD}. If we take our original fidelity cost function from Eq.~\eqref{eq:costfunc_fidelity}, we can modify it by including terms that are conditioned on the maximum amplitude of a given pulse:
\begin{equation}
    C_F^{\rm const}(\hbb, \betabb, \lambda) = 1 - F(\hbb, \betabb, \lambda) + \sum_m \Lambda_m(\hbb, \betabb, \lambda),
\end{equation}
where $\Lambda_m(\hbb, \betabb, \lambda) = 0$ if the $m^{\rm th}$ constraint is satisfied and $\Lambda_m(\hbb, \betabb, \lambda) \gg \min(1 - F(\hbb, \betabb, \lambda))$ otherwise. We can implement such constraints for all of the driving amplitudes for all of the coefficients of each Hamiltonian, \@e.g.~$\alpha$, $f_{\rm opt}^{N_k}$ and $f_{\rm CRAB}^{N_k}$. In Fig.~\ref{fig:ising_constrained}, this is done for the case where any of the drives exceeds the maximum amplitude of any of the bare Hamiltonian $H_0$ coefficients, which in our case if when $X(\lambda = 1) = X_0 = 10J_0$. We find that, while the resulting fidelities are greatly reduced in the \acrref{BPO} and \acrref{COLD} cases, the \acrref{COLD} approach outperforms the plain optimal control method. Far better results are obtained using \acrref{CRAB} and \acrref{COLD}-\acrref{CRAB}, which require far more computational resources but, as expected, allow a lot more flexibility within the constraints to obtain good final state fidelities. 

The results presented here are supplemented further in Ch.~\ref{chap:7_higher_order_agp}, where we investigate how to use the ideas from Ch.~\ref{chap:5_cd_as_costfunc} in order to implement \acrref{CD}-based cost functions to optimise \acrref{COLD} and \acrref{BPO} for the Ising spin chain. Additional plots for different system sizes and information on the variances between result outcomes in different optimisation instances can be found in Appendix~\ref{app:ising}.

\section{Transport in a synthetic lattice}\label{sec:5.3_synthetic}

Let us turn our sights to a completely different type of system for a moment and explore how \acrref{COLD} might perform. The efficient transfer of states between opposite ends of a lattice is an important protocol that could have future applications in the settings of quantum computation and simulation due to its promise of efficient transport of information \cite{lang_topological_2017}. This objective is often tackled in the setting of ultracold atoms in optical lattices. While the problem can be tuned to be a single-particle system and the analytical solutions of the corresponding instantaneous Schr\"odinger equation are known \cite{hatsugai_chern_1993,hugel_chiral_2014}, the efficient evolution for state transfer is not straight-forward due to the states being largely delocalised across the lattice throughout the transfer. 

In implementing \acrref{CD}, this delocalisation of states implies the requirement for the exact \acrref{AGP} operator to be highly delocalised too, which presents a practical difficulty. While such terms can be generated via the interactions of the atoms with cavity modes \cite{landig_quantum_2016,keller_phases_2017} or from dipolar interactions \cite{baranov_ultracold_2002, trefzger_ultracold_2011}, the most tractable option remains that of approximate methods like \acrref{LCD} \cite{kiely_fast_2021, coopmans_optimal_2022}.

Recently, \acrref{LCD} was successfully applied to improve an adiabatic rapid passage (\acrref{ARP}) protocol for population transfer across a synthetic lattice \cite{meier_counterdiabatic_2020}. In this realisation, population transfer was achieved in a synthetic tight-binding lattice of laser coupled atomic momentum states. We will consider the same problem as in \cite{meier_counterdiabatic_2020} but with the improvement that can be gained by \acrref{COLD}. 

The system is described by the Hamiltonian on $N$ lattice sites
\begin{equation}\label{eq:lattice_hamiltonian}
    H_0(\hbb, \lambda) = - \sum_n^N J_n(\lambda)(c_n^{\dag}c_{n+1} + H.c.) + \sum_n V_n(\lambda) c_n^{\dag}c_n,
\end{equation}
where $\hbb = \{ J_n(\lambda), V_n(\lambda)\}_{n = 1, ..., N}$, $J_n(\lambda)$ is the $\lambda$-dependent tunnelling that describes the nearest-neighbour coupling, $V_n(\lambda)$ is the on-site energy offset with respect to neighbouring sites and $c_n^{\dag}$ ($c_{n}$) is the creation (annihilation) operator on a given lattice site $n$. In the \acrref{ARP} protocol, the population gets moved from one end of the lattice to the other by linearly ramping the lattice from a positive tilt to a negative tilt via
\begin{equation} \label{eq:J_lattice}
    \begin{aligned}
        J_n(\lambda) &= J_0(0.1 + \lambda) \\
        V_n(\lambda) &= n V_0 (1 - 2\lambda),
    \end{aligned}
\end{equation}
where $V_0 = 4J_0$ is the initial site energy slope, $J_0$ is the characteristic tunnelling scale of the lattice and $\lambda(t) = \frac{t}{\tau}$ as previously.

In \cite{meier_counterdiabatic_2020}, the first order \acrref{LCD} is constructed by decomposing the tunneling into two $\lambda$-dependent components:
\begin{equation}\label{eq:tunneling}
    J_n(\lambda) \rightarrow J_{n, \mathrm{CD}}(\hbb, \lambda) e^{-i\phi_{n, \mathrm{CD}}(\hbb, \lambda)},
\end{equation}
where
\begin{equation}\label{eq:J_cd}
    \begin{aligned}
    J_{n, \mathrm{CD}}(\hbb, \lambda) &= \sqrt{J_n(\lambda)^2 + (\alpha_n(\hbb, \lambda)/\tau)^2}, \\
    \phi_{n, \mathrm{CD}}(\hbb, \lambda)  &= \arctan\left(-\frac{J_n(\lambda)\tau}{\alpha_n(\hbb, \lambda)}\right),
    \end{aligned}
\end{equation}
and the $\alpha_n(\lambda)$ terms correspond to the \acrref{LCD} coefficients. They can be found by solving a set of linear equations
\begin{equation}
    \begin{aligned}
        &-3(J_n J_{n+1})\alpha_{n+1} + (J_{n-1}^2 + 4J^2_n + J_{n+1}^2)\alpha_n \\ &- 3(J_n J_{n-1})\alpha_{n-1} + (V_{n+1} - V_n)^2 \alpha_n \\ &= -\partial_{\lambda}J_n (V_{n+1} - V_{n}).
    \end{aligned}
\end{equation}

\begin{figure}[t]
    \centering
    \includegraphics[width=\linewidth]{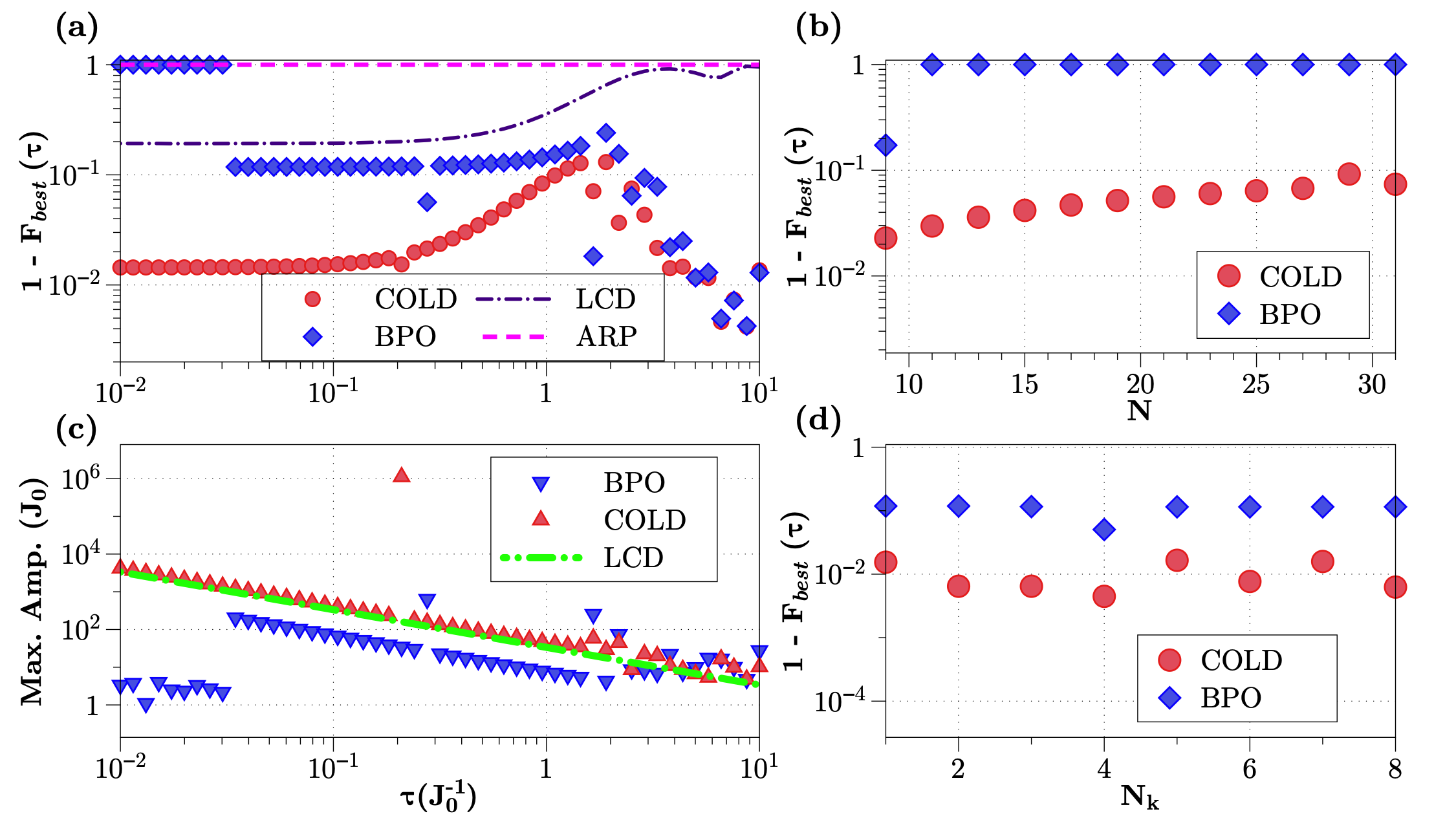} \caption[COLD plots for ARP transport in a synthetic lattice]{Optimisation of state transfer in a synthetic lattice.  In (a) we compare the infidelities obtained via the bare \acrref{ARP} protocol (pink dashed line) and \acrref{FO} \acrref{LCD} implemented in \cite{meier_counterdiabatic_2020} (purple dash-dot line) to \acrref{BPO} (blue diamonds) and \acrref{COLD} (red circles). (c) Maximum amplitude of the tunneling term at each driving time for \acrref{LCD} (green dash-dot line) as given by Eq.~\eqref{eq:tunneling} as well as \acrref{COLD} (red triangles) which includes additional control parameters (Eq.~\eqref{eq:synthetic_control_funcs}) and BPO (blue triangles) which omits the modifications due to CD but retains the control terms. In both (a) and (c) we simulate $N = 7$ lattice sites and use $N_k = 1$ parameter for optimisation of BPO and COLD.  (b) Scaling of fidelities with increasing number of lattice sites (where $N_k = 1$) for both COLD (red circles) and BPO (blue diamonds). (d) does the same for the number of parameters while keeping $N=7$.  Note that both (b) and (d) are simulated for driving time $\tau = 0.5 J^{-1}$ and the best fidelities are obtained across 500 optimisations. Reprinted with permission from \cite{cepaite_counterdiabatic_2023}. Copyright 2023, American Physical Society.}\label{fig:Synthetic}
\end{figure}

In order to implement COLD we once again have to include a control component into the Hamiltonian:
\begin{equation}\label{eq:ising_control_nocrab}
    \begin{aligned}
        H_{\beta}(\hbb, \betabb, \lambda) = H_0(\hbb, \lambda) + \sum_n^N f_{\rm opt}^{N_k}(\betabb, \lambda) (c_n^{\dag}c_{n+1} + H.c.),
    \end{aligned}
\end{equation}
where the function $f_{\rm opt}^{N_k} = \sum_k^{N_k} \beta_k(\lambda)$ is the same as in Eq.~\eqref{eq:ising_chain_BPO_H}. This control pulse can be viewed as modifying the tunneling terms:
\begin{equation}\label{eq:synthetic_control_funcs}
    \begin{aligned}
        J_n(\lambda) \rightarrow J_n^{\rm opt}(\betabb, \lambda) &= J_n(\lambda) + f_{\rm opt}^{N_k}(\betabb, \lambda), \\
        \Rightarrow J_{n, \rm CD}(\betabb, \hbb, \lambda) &= \sqrt{J_n^{\rm opt}(\betabb,\hbb, \lambda)^2 + (\alpha_n(\hbb, \betabb, \lambda)/\tau)^2} \\
        \Rightarrow \phi_{n, \mathrm{CD}}(\betabb, \hbb, \lambda) &= \arctan\left(-\frac{J_n^{\rm opt}(\betabb, \lambda)\tau}{\alpha_n(\betabb, \hbb, \lambda)}\right),
    \end{aligned}
\end{equation} 
and the control functions are optimised using Powell's method as before by minimizing with respect to the fidelity of the final state, where the population has been fully transferred to the opposite lattice site. 

Now that all of our ducks are in a row, we first consider a system size of $N=7$ sites which was successfully experimentally probed in \cite{meier_counterdiabatic_2020}, where final state fidelities of $0.75$ were achieved for $\tau = 1$ms with a final tunnelling strength of $J/\hbar = 1/2\pi kHz$ (equivalent to $\tau \sim 1 J_0^{-1}$ in our units). We initially confirm the breakdown of \acrref{ARP} in this setting for fast times, and the success of the \acrref{LCD} protocol at short times, as shown in Fig.~\ref{fig:Synthetic} (a) and found in \cite{meier_counterdiabatic_2020}. Implementing \acrref{BPO} on its own manages to enhance the achievable fidelities at intermediate times of $\tau > 0.03 J^{-1}$. However, eventually, as observed in all scenarios in this work, \acrref{BPO} becomes stuck in the initial state at fast evolution times. Implementing the \acrref{COLD} protocol achieves an order of magnitude improvement in the fidelity over \acrref{LCD}. This is also plotted in Fig.~\ref{fig:Synthetic}(a).

It may be that \acrref{COLD} achieves this advantage by pumping power into the tunnelling term, as discussed in the Ising chain example, but we can see in Fig.~\ref{fig:Synthetic}(c) that the maximum amplitude of the tunnelling term tracks that of \acrref{LCD}. A key issue for experiments is the maximum amplitude achievable by a driving term and with this result we can stipulate that \acrref{COLD} is likely to be feasible in the same regimes as \acrref{LCD} in this synthetic lattice system, but with far higher resulting fidelities. There is single outlier at intermediate times as indicated by the single point peaking in maximum amplitude in Fig.~\ref{fig:Synthetic}(c), this is the exception to the rule, where the optimisation has found a marginally higher fidelity (see the corresponding point in Fig.~\ref{fig:Synthetic}(a)) by pumping in more power. The maximal amplitude plot may also explain the discontinuous jump in infidelities for the \acrref{BPO} case that happens at very short times: there is also a discontinuity in the maximal amplitude reached by the drive for the same points. This can hint either at some controllability limit, wherein a minimal amount of time is required for the \acrref{BPO} Hamiltonian to shift the population, or more simply this is a consequence of a local optimiser not being able to find a global minimum which may lead to a higher fidelity and a higher driving amplitude. As the total driving time increases, the cost function landscape changes and likely reveals more optimal minima to the optimiser. 

Furthermore, we explore the infidelities with increasing system size for both \acrref{BPO} and \acrref{COLD} in Fig.~\ref{fig:Synthetic}(b). While both protocols show a decreasing fidelity with increasing system size as expected, \acrref{COLD} does not suffer from getting stuck in the initial state, which is what happens in the \acrref{BPO} case as infidelities go to unity for larger systems in Fig.~\ref{fig:Synthetic}(b). This is the same mechanism as for the short driving times in Fig.~\ref{fig:Synthetic}(a). We also find, as plotted in Fig.~\ref{fig:Synthetic}(d), that increasing the number of optimisable parameters $N_k$ in the control pulse does not contribute to an improvement in the results either for \acrref{BPO} or \acrref{COLD}. This means that, at least in this very simple control setting, there is no reason to expect that \acrref{BPO} will outperform \acrref{COLD} by simply adding more complexity to the control pulse.

It is important to acknowledge once again that it may be possible to achieve better results for both the control Hamiltonian and \acrref{COLD} with the use of more sophisticated optimal control methods like \acrref{CRAB}/\acrref{COLD}-\acrref{CRAB} or a global optimiser instead of Powell's method. Any of these methods might prove to be better, but they are also far more computationally intensive and the results presented already show significant imrpovements over \acrref{LCD} or the bare Hamiltonian. In the case where such a protocol is to be implemented in practice, it would be advantageous to explore more refined control methods than those presented here.

\section{Preparing GHZ states in a system of frustrated spins}\label{sec:6.4_ghz_states}

Multipartite entanglement is a powerful resource for quantum computing and more broadly in quantum technologies as a whole, offering unique capabilities for information processing \cite{nielsen_quantum_2010}, secure communication \cite{bostrom_deterministic_2002}, high-precision measurements \cite{kim_heisenberg-limited_2022}, and understanding the foundations of quantum mechanics \cite{einstein_can_1935}. An example of such highly entangled states is the GHZ (Greenberger–Horne–Zeilinger) state \cite{greenberger_bells_1990} on $N > 1$ spins:
\begin{equation}\label{eq:GHZ_state}
    \ket{\rm GHZ} = \frac{1}{\sqrt{2}} (\ket{0}^{\otimes N} + \ket{1}^{\otimes N}),
\end{equation}
here written in the $0,1$ qubit basis.

It is possible to prepare such states in a system of frustrated spins (see Fig.~\ref{fig:ghz_mainfig}(a)) for odd $N > 1$ via an annealing protocol. The Hamiltonian describing such a system is
\begin{equation}\label{eq:ghz_hamiltonian}
    H_0(\hbb, \lambda) = J(\lambda) \Big( \sum_{j}^{N-1} \sz_j \sz_{j+1} + \sum_{j}^{N-2} \sz_j \sz_{j+2} \Big) + h(\lambda) \Big( \sum_j^N (\sz_j + \sx_j) \Big).
\end{equation}
where $\hbb = \{ J(\lambda), h(\lambda) \}$ with $\lambda = \frac{t}{\tau}$, $J(\lambda)= - J_0$ and
\begin{equation}
    h(\lambda) = - h_0 \left(1 - \sin^2\left(\frac{\pi}{2} \sin^2 \left( \frac{\pi}{2} \lambda \right) \right)\right).
\end{equation}
The parameters used for the results in this section are $J_0 = 1$ and $h_0 = 10J_0$, meaning the spins start close to the $\ket{\downarrow^{\otimes N}}$ state. 

In this case, we will apply both \acrref{FO} \acrref{LCD} and \acrref{SO} \acrref{LCD} ans\"{a}tze. The former is the same one as used previously for the Ising chain case, Eq.~\eqref{eq:ising_fo_agp}, while the latter will consist of additional operators between next-nearest-neighbour spins to reflect the geometry of the bare Hamiltonian:
\begin{equation}\label{eq:ghz_so_lcd}
    \begin{aligned}
        \approxAGP^{(2)}(\hbb, \lambda) &= \gamma(\hbb, \lambda) \Big( \sum_{j}^{N-1} \sx_j \sy_{j+1} + \sum_{j}^{N-2} \sy_j \sx_{j+2} \Big) \\
        &+ \zeta(\hbb, \lambda) \Big( \sum_{j}^{N-1} \sz_j \sy_{j+1} + \sum_{j}^{N-2} \sy_j \sz_{j+2} \Big).
    \end{aligned}
\end{equation}
As $H_0$ is simply the Ising Hamiltonian with added couplings between spins $j$ and $j+2$, its \acrref{LCD} coefficients can be found by using the coupled equations in Appendix~\ref{app:arbitrary_ising_derivation} and setting the unused operator coefficients to $0$. 

What remains is to construct the optimal control pulse, which in this case will be done using the \acrref{GRAPE} method from Sec.~\ref{sec:3.3.2_GRAPE}. The control Hamiltonian is
\begin{equation}\label{eq:ghz_control}
    H_{\beta}(\betabb, \hbb, \lambda) = H_0(\hbb, \lambda) + f^{N_k}_{\rm GRAPE}(\betabb, \lambda) \sum_j^N \sz_j,
\end{equation}
where the pulse $f^{N_k}_{\rm GRAPE}$ is comprised of $N_k$ time slices $\Delta \lambda = \lambda_{k+1} - \lambda_k$, such that $N_k \times \Delta \lambda = 1$. During each slice $\lambda_k$, a piecewise constant control amplitude $\beta_k(\lambda_k) = c_k$, is applied to the control system, with $c_k$ denoting the optimisable parameter which represents the amplitude of the pulse in the interval $[\lambda_k, \lambda_{k+1})$. Thus, the pulse comprises of $N_k$ optimisable parameters for $N_k$ time slices. In order to enforce the constraints that the control pulse vanish at the beginning and end of the protocol and to make the pulse more smooth, we also apply a shaping function, which acts during each time interval $[\lambda_k, \lambda_{k+1})$ as
\begin{equation}
        f_{\rm shape}(\beta_k, \lambda_k) = c_k \tanh (\kappa \theta(\lambda_k)) \tanh (- \kappa \theta(\lambda_k - \tau)),
\end{equation}
with $\theta(\lambda) = \sin \frac{\pi}{2} \lambda$ and $\kappa = 30$ an offset parameter. As discussed in Sec.~\ref{sec:4.2_COLD_QOCT}, we use spline interpolation to calculate the derivatives of the control drive when they are required to obtain the \acrref{LCD} drives. The resulting function requires more parameters than the chopped basis we chose to use in previous examples, but it also allows for more flexibility in the final shape of the drive. Furthermore, due the increased number of parameters and search space, instead of Powell optimisation as in previous examples we choose to instead implement dual annealing, first presented in detail in Sec.~\ref{sec:3.1.3.3_dual_annealing}. Dual annealing is a global optimiser and while computationally more costly, it is generally far better in the case of a complex parameter space with multiple minima, which is what may be expected in this case (see Fig.~\ref{fig:ghz_contours} later in the thesis). In this case, instead of referring to the optimised control Hamiltonian of Eq.~\eqref{eq:ghz_control} as \acrref{BPO}, we will instead dub it `bare dual-annealing' or \acrref{BDA}. 

As well as implementing the \acrref{GRAPE} pulse for all spins as in Eq.~\eqref{eq:ghz_control}, as this is a more interesting and complex system than those encountered previously, we will include a separate protocol where three separate control pulses are used: one for each corner spin (see Fig.~\ref{fig:ghz_mainfig}(a)) and one for all the spins in-between. The reason for this choice of pulses is because it is often easier to independently address the edges of a lattice in practical implementations of such protocols, rather than attempting to have local control of all of the spins individually. We will include a `-C' suffix to each method in order to differentiate the corner approach from that of a global control pulse. The resulting control Hamiltonian will thus have $3 \times N_k$ optimisable parameters, which is a very large search space, making the optimisation process very computationally expensive when compared to all of the previous examples discussed in this Chapter. 

In Fig.~\ref{fig:ghz_mainfig}(c) we plot the resulting final state fidelities at different total driving times $\tau$ for a 5 spin frustrated system. We implement \acrref{BDA}, \acrref{FO} and \acrref{SO} \acrref{COLD} as well as their corner-optimised versions while optimising the pulse parameters for final state fidelity with respect to the GHZ state of Eq.~\eqref{eq:GHZ_state}. We observe that \acrref{FO} \acrref{COLD} is not particularly effective at short driving times and does not move the system out of its initial state (see the density matrix plots in (b)), regardless of whether or not separate control is applied to the corner spins. This is very likely due to the fact that the $\sy$ terms making up \acrref{FO} \acrref{LCD} are only a small contribution to the full counterdiabatic drive and thus we need to look to higher-order ans\"{a}tze to see improvements. Notably, the bare control protocol \acrref{BDA} does not fare any better, refusing to budge from the initial state at very small $\tau$. \acrref{SO} \acrref{COLD}, on the other hand, shows a five-fold improvement over the first order when a global optimisable drive is applied and up to a further two more orders of magnitude improvement when the corner spins are driven separately at short times ($\tau = 0.001J_0^{-1}$).  We run optimisations for larger systems at time $\tau = 0.1J^{-1}$ and find that this advantage is retained even with increasing system size, as plotted in Fig.~\ref{fig:ghz_mainfig}(d). This is a large improvement over a recent result presented in \cite{sun_optimizing_2022}, where optimal control was used to directly determine \acrref{SO} \acrref{LCD} coefficients for preparing the GHZ state on an Ising chain rather than creating a control Hamiltonian. At 10 spins the final state fidelity for $\tau = 1J_0^{-1}$ obtained in their paper was 0.18, while we reach a fidelity of 0.72 for 15 spins when using corner optimisation at $\tau = 0.1J_0^{-1}$. 

\subsection{Tripartite GHZ entanglement}\label{sec:6.4.1_t3}

\begin{figure}[t!]
    \centering
    \includegraphics[width=\linewidth]{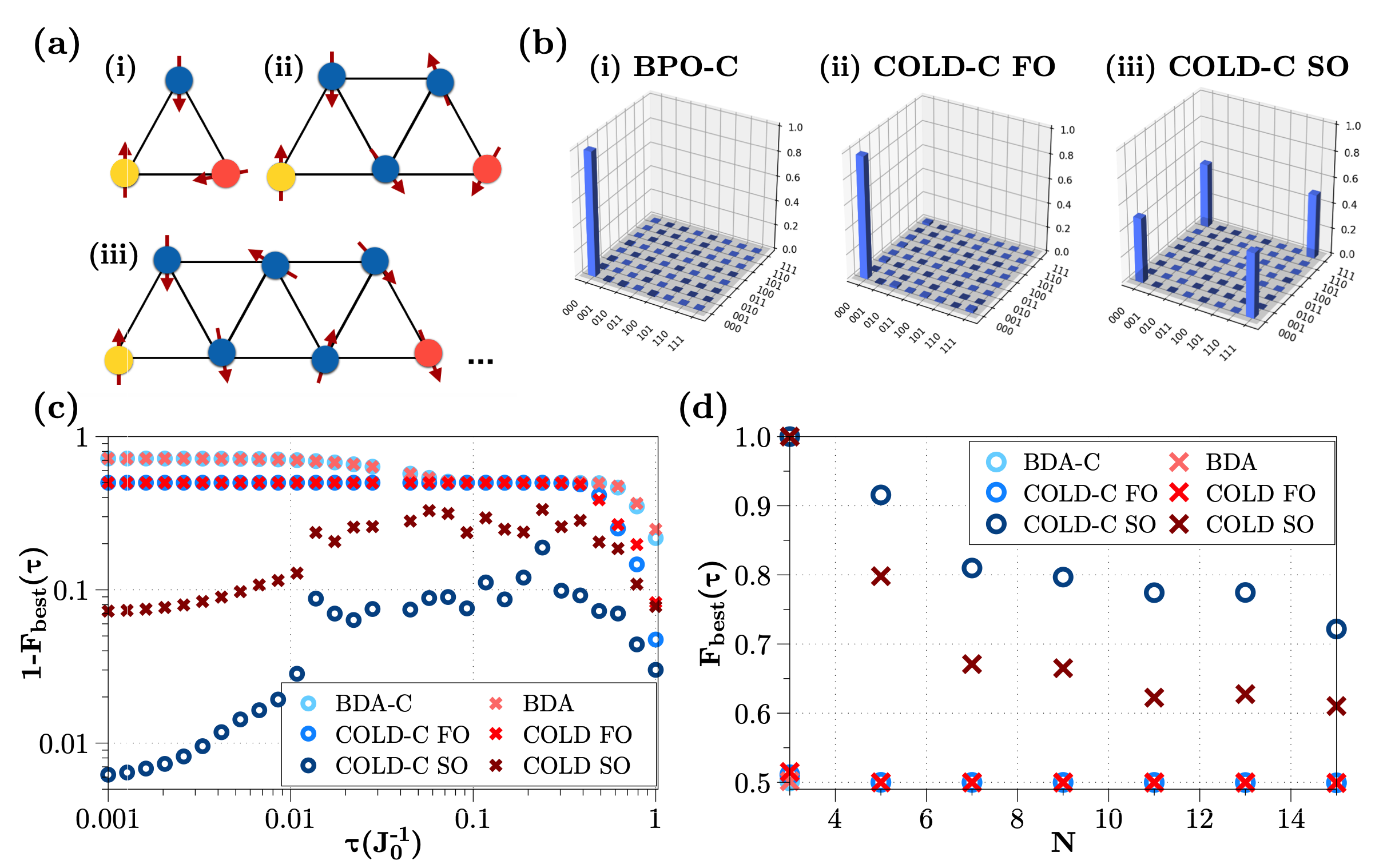} \caption[Preparation of GHZ states in a system of frustrated spins.]{GHZ state preparation in systems of frustrated spins. Spins are arranged in triangular formations as depicted in (a) for (i) 3, (ii) 5 and (iii)7 spins, with spins on the vertices and edges representing couplings.  In the case of corner optimisation, three separate optimisable drives are applied: one for the yellow corner spin, one for the red corner spin and a third drive for all of the blue spins in-between. (b) Density matrix plots of the final state of a 3 spin triangle after an evolution time $\tau = 0.1J_0^{-1}$ when optimised using (i) BDA-C, (ii) \acrref{FO} \acrref{COLD}-C and (iii) \acrref{SO} \acrref{COLD}-C. (c) Final fidelities of the GHZ state on 5 spins for optimised global drive (red crosses) and locally driven corner spins (blue rings). (d) Final fidelities at driving time $\tau = 0.1J_0^{-1}$ for different systems sizes $N$. In the global case we use $N_k = 10$ and in the corner case the total parameters are $3 \times N_k = 30$. The plots are for best results of 5 optimisations for each data point and the dual-annealing search space was bounded in the range $[-50, 50]$ for all parameters. Reprinted with permission from \cite{cepaite_counterdiabatic_2023}. Copyright 2023, American Physical Society.} \label{fig:ghz_mainfig}
\end{figure}

While throughout this chapter we have focused on using final state fidelity as a cost function for the optimisation, in the case of GHZ state preparation, the goal is often the maximal entanglement of such states. Therefore, we can imagine using a cost function which maximises some entanglement metric like that in Eq.~\eqref{eq:costfunc_entanglement}, rather than the final state fidelity. This can be done in cases where several maximally entangled states are targeted, or where a single target state maximises entanglement. In the latter case, the advantage may lie in the fact that entanglement as a metric might lead to a smoother and more convex cost function landscape.

The GHZ state exhibits a particular type of entanglement: when one of the subsystems is measured, the rest are no longer entangled and collapse into a product state. This is different to the other canonical type of multipartite entanglement exhibited by the W state \cite{cabello_bells_2002}, which for 3 spins can be written as:
\begin{equation}\label{eq:W_state}
    \ket{W} = \frac{1}{\sqrt{3}}(\ket{001} + \ket{010} + \ket{100}).
\end{equation}

While measuring the entanglement of a multipartite system is not quite as simple as in the bipartite case, there exists a notion of entanglement for a system of three spins: namely, the three-tangle, first introduced in Ref.~\cite{coffman_distributed_2000}, which is maximised when a three spin system exhibits maximal GHZ-type entanglement and minimised for W-type entanglement and product states. The three-tangle is a very efficient metric and can be expressed as
\begin{equation}\label{eq:3-tangle}
	\begin{aligned}
		T_3(\ket{\psi}) &= 4 \left| d_1 - 2 d_2 + 4 d_3 \right|, \\
		d_1 &= c^2_{000}c^2_{111} + c^2_{001}c^2_{110} + c^2_{010}c^2_{101} + c^2_{011}c^2_{100}, \\
		d_2 &= c_{000}c_{001}c_{110}c_{111} + c_{000}c_{010}c_{101}c_{111} + c_{000}c_{011}c_{100}c_{111} \\
		 &+ c_{001}c_{010}c_{101}c_{110} + c_{001}c_{011}c_{100}c_{110} + c_{010}c_{011}c_{100}c_{101} \\
		d_3 &= c_{000}c_{110}c_{101}c_{011} + c_{100}c_{010}c_{001}c_{111},
	\end{aligned}
\end{equation}
where $c_{ijk}$ represents the complex coefficient of the state $\ket{ijk}$ of the three spin state $\ket{\psi}$. 

\begin{figure}[t]
    \centering
    \includegraphics[width=\linewidth]{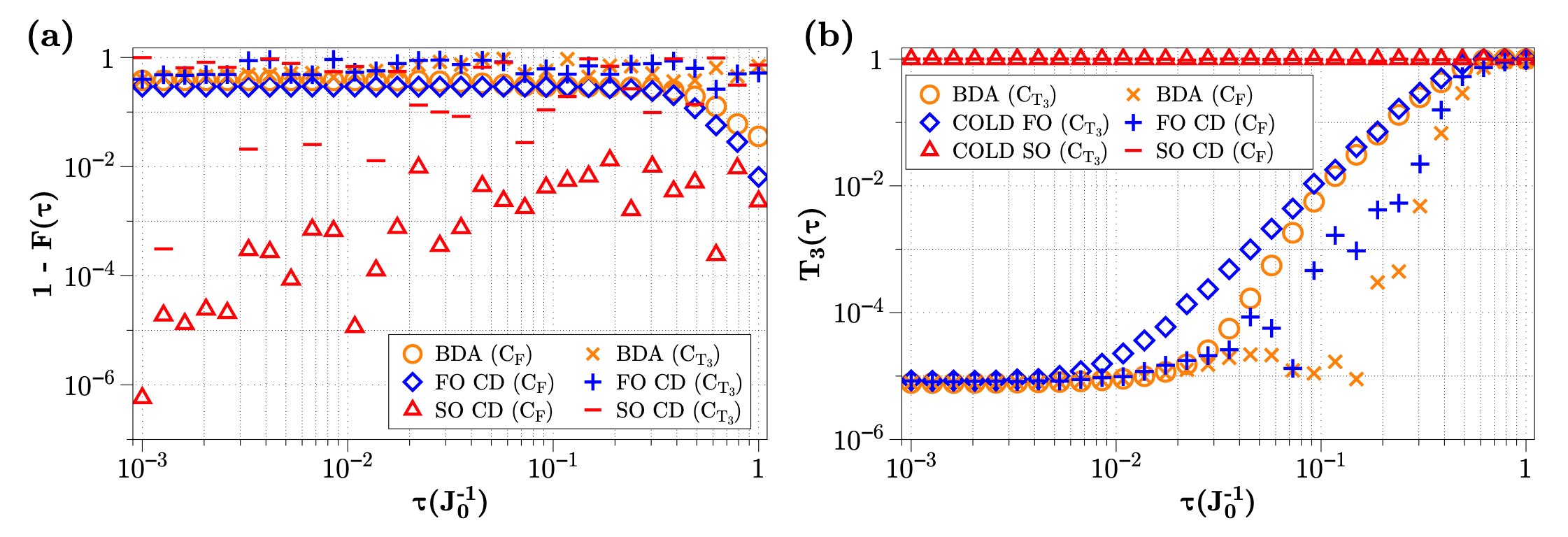} \caption[Preparing 3-spin GHZ states using the 3-tangle as a metric.]{Plots of $T_3$ (Eq.~\eqref{eq:3-tangle}) and fidelity $(1 - F)$ with respect to the GHZ state of the final state prepared after optimising a GRAPE pulse with $N_k = 6$ parameters for each value of $\tau$. In (a) we plot final state fidelity when optimisation is performed using $C_{\rm F}$ and $C_{T_3}$ and in (b) we plot the three-tangle from Eq.~\eqref{eq:tangle_costfunc} for the same cost functions. In both cases, the result is from a single optimisation for each data point and the dual-annealing search space bounds are $[-50,50]$ for all optimisable parameters.}\label{fig:tangle_v_fidelity}
\end{figure}

Fig.~\ref{fig:tangle_v_fidelity}(a), shows the results of the final state fidelity when optimising with $C_{\rm F}$ and optimising solely for the three tangle, using cost function
\begin{equation}\label{eq:tangle_costfunc}
    C_{T_3}(\betabb, \tau) = 1 - T_3(\ket{\psi_f(\betabb, \tau)}),
\end{equation}
where $\ket{\psi_f(\betabb, \tau)}$ is the final state obtained throughout the evolution during time $\tau$ and with optimal controls $\betabb$. In Fig.~\ref{fig:tangle_v_fidelity}(b) we investigate the values of $T_3$ obtained when using the same two cost functions. Both plots show results in a three spin system like that of Fig.~\ref{fig:ghz_mainfig}(a)(i) with only a global control pulse and dual-annealing optimisation. We find that the value of the three tangle and hence the amount of entanglement in the system begins to increase prior to any noticeable improvement in fidelity in the case of \acrref{BDA} and \acrref{FO} \acrref{COLD}, while, as expected, given the much higher fidelities obtained when using \acrref{SO} \acrref{COLD}, the entanglement is maximised for even for very small $\tau$. This is an interesting result, as it indicates that \acrref{SO} counterdiabatic operators are required to be able to speed up entanglement generation. Even when \acrref{FO} terms are applied, it takes long evolution times to generate entanglement. We find that even when maximising for entanglement, the results for final $T_3$ values are not very different to those obtained when optimising for final state fidelity. This is again an indication that neither \acrref{FO} terms nor the plain control Hamiltonian of Eq.~\eqref{eq:ghz_control} are enough to generate entanglement quickly and that \acrref{SO} terms are necessary for this purpose. The much lower fidelities obtained with $C_{T_3}$ are unsurprising, as the three-tangle is a measure of GHZ-type entanglement, which is maximised for several states including the orthogonal state to that presented in Eq.~\eqref{eq:GHZ_state}. This means that maximising for entanglement in this case will not always lead to final states that have high fidelity with respect to the canonical GHZ state presented at the start of this section. An interesting future line of work may be to investigate how \acrref{LCD} or \acrref{COLD} impact the generation of different types of entanglement like W-type entanglement.

\paragraph{Summary} This section contains quite a lot of material, so a brief summary may be in order to place it firmly in the context of what follows in the next chapter. So far we have presented several numerical implementations of the \acrref{COLD} approach for different physical systems. In each case, we optimised the control pulse using either the fidelity cost function of Eq.~\eqref{eq:costfunc_fidelity} with respect to the target ground state of the final Hamiltonian, or a measure of the final state entanglement. We showed that \acrref{COLD} outperforms both the \acrref{LCD} method, for the same set of ansatz operators in the approximation of the counterdiabatic drive, and the control pulse on its own. The next section will revisit several of the systems that were presented here, with the aim of foregoing optimisation with respect to fidelity and instead optimising for properties of the approximate \acrref{CD} components instead, as was discussed in Ch.~\ref{chap:5_cd_as_costfunc}.
\chapter{Higher order AGP as a cost function}\label{chap:7_higher_order_agp}

\epigraph{`Fast' was a word particularly associated with tortoises because they were not it.}{Terry Pratchett, \emph{Pyramids (1989)}}

In Ch.~\ref{chap:5_cd_as_costfunc} we discussed the idea of using the \acrref{AGP} operator and its approximations in order to construct cost functions for the optimisation of Hamiltonian paths in parameter space. There are several reasons why one might expect this to be a good idea: for example, a path in the Hamiltonian parameter space which minimises the \acrref{AGP} norm should, in principle, also minimise the non-adiabatic effects experienced by the system when it is driven along that path. Furthermore, such cost functions should be very efficient to compute once they are written in the correct functional form, giving them an advantage over the fidelity cost function that we used in most examples of Ch.\ref{chap:6_Applications_fidelity}. Computing the final state fidelity is an approach which suffers from increasing complexity and inefficiency with growing system sizes due to requiring access to the system dynamics along the entire path of the evolution.

In this chapter we will motivate the idea of \acrref{AGP}-based cost functions with numerical results, investigating two types of cost functions presented in Ch.~\ref{chap:5_cd_as_costfunc}: minimisation of absolute integrals of the \acrref{LCD} coefficients and minimisation of their maximal amplitudes. We will do this for three different example systems which we covered in the previous chapter: two-spin annealing in, the Ising spin chain, and finally the preparation of maximally entangled GHZ states in systems of frustrated spins. The last example will demonstrate a situation where this new approach might not be optimal and we will discuss the reasons behind this, with the goal of understanding regimes in which one might want to implement the new method. 

\section{Return to two-spin annealing}\label{sec:7.1_two_spin_ho}

In Sec.~\ref{sec:5.1_2spin_annealing} we investigated the \acrref{COLD} protocol in the case of a two-spin annealing protocol described by the Hamiltonian from Eq.~\eqref{eq:two_spin_hamiltonian}, where the system starts close to the state $\ket{\uparrow\uparrow}$ and is driven towards a superposition of all the symmetric states. We will return to this simple example in order to illustrate how the integral and maximum amplitude cost functions from Sec.~\ref{chap:5_cd_as_costfunc}, given by Eq.~\eqref{eq:COLD_costfunc_integral} and Eq.~\eqref{eq:COLD_costfunc_maximum} respectively, behave when used to optimise the Hamiltonian path in parameter space for \acrref{COLD}.

In Ch.~\ref{chap:6_Applications_fidelity} we presented the results of optimisation using properties of the final system state as metrics for success. What we wish to demonstrate here is the use of different cost functions in the optimisation of the parameters $c_k$ from Eq.~\eqref{eq:COLD_twospin_controlH}. As discussed in Ch.~\ref{chap:5_cd_as_costfunc}, once we have the \acrref{AGP} or \acrref{LCD} operators expressed as functions of the control Hamiltonian, they can be used to construct cost functions that can be evaluated very efficiently, regardless of the scale or complexity of the driven system. This is important, because while the fidelity cost function $C_{\rm F}$ from Eq.~\eqref{eq:costfunc_fidelity} that we used in the previous chapter is very effective -- given that it evaluates exactly how close we are to the true goal of the optimisation -- its efficiency scales very poorly with increasing system size. After all, having access to the final state fidelity requires a calculation of the complete system dynamics, as well as full knowledge of the target state ahead of time. 

In this example, we will use the same control Hamiltonian as the two-spin example from the previous chapter given by Eq.~\eqref{eq:COLD_twospin_controlH} as well as the same \acrref{FO} and \acrref{SO} operator ans\"{a}tze for the \acrref{LCD} operators (Eq.~\eqref{eq:LCD1st} and Eq.~\eqref{eq:twospin_so_lcd} respectively). We found previously that the coefficient $\alpha(\betabb, \hbb, \lambda)$ which drives the \acrref{FO} \acrref{LCD} terms and the coefficients $\gamma(\betabb, \hbb, \lambda)$ and $\zeta(\betabb, \hbb, \lambda)$ driving the \acrref{SO} \acrref{LCD} terms can be found by solving the coupled set of equations given in Eq.~\eqref{eq:two_spin_coupled_eqs}. We noted that these three coefficients and the operators they drive are enough to describe the exact \acrref{CD} pulse for any given parameters $(\betabb, \hbb, \lambda)$ in the case of two spins.

\begin{figure}[t]
    \centering
    \includegraphics[width=\linewidth]{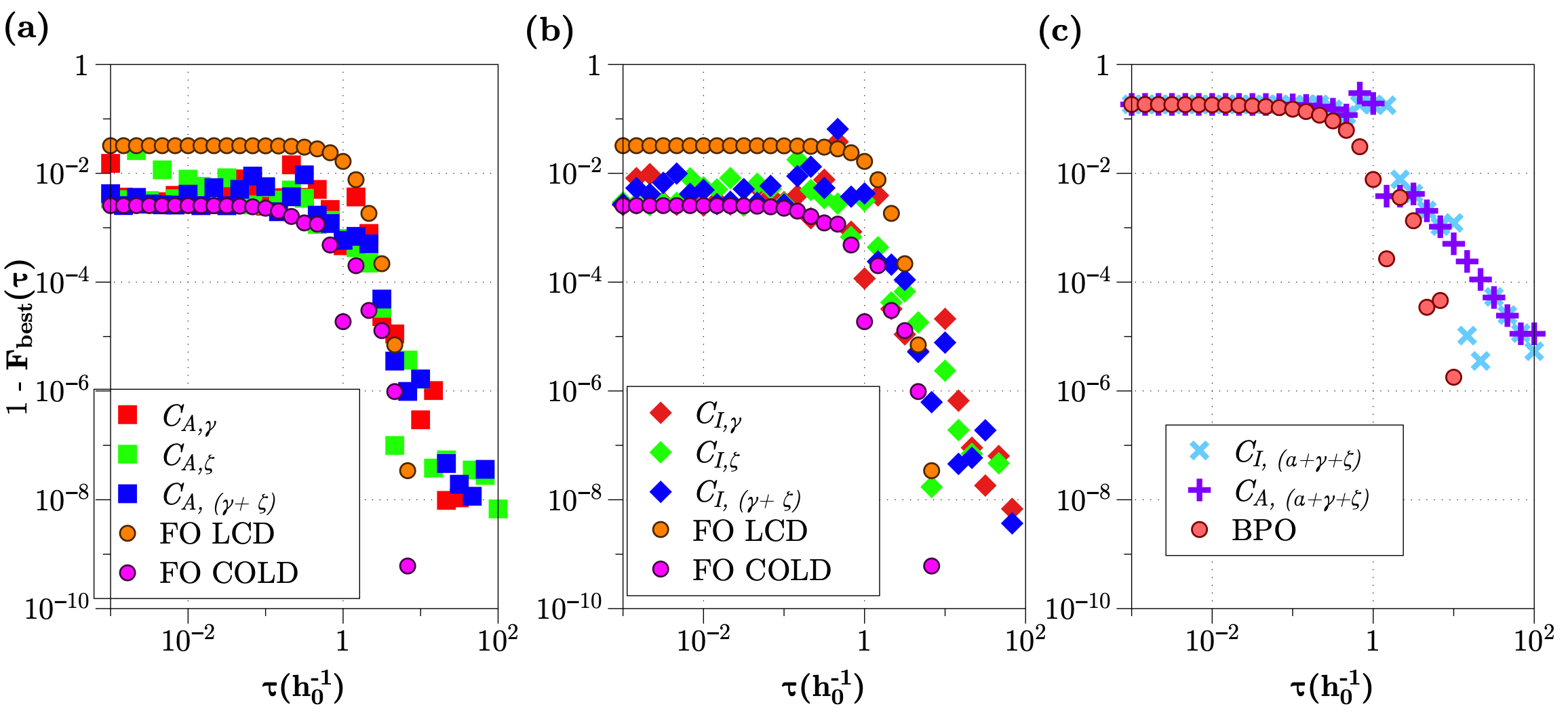} \caption[Two-spin annealing plots of fidelity versus driving time when minimising different orders of LCD.]{Two-spin annealing plots of fidelity versus driving time when optimising using different cost functions. In (a)-(b), we show the results for \acrref{FO} \acrref{LCD} (orange circles) and \acrref{FO} \acrref{COLD} (pink circles) from Fig.~\ref{fig:twospin_fidelities} for comparison. We do the same in (c) for \acrref{BPO} (red circles). Then, in (a) we plot the resulting final state fidelities at different driving times $\tau$ obtained when applying \acrref{FO} \acrref{COLD} to control Hamiltonians from Eq.~\eqref{eq:COLD_twospin_controlH} with parameters optimised using the maximum amplitude cost function $C_{\rm A}$. The results are shown when optimising for $\gamma$ coefficients (red squares, $C_{\rm A, \gamma}$), $\zeta$ coefficients (green squares, $C_{\rm A, \zeta}$) and their sum (blue squares, $C_{\rm A, (\gamma + \zeta)}$). The same is done in (b) for the integral cost function $C_{\rm I}$, where we plot results in the case of optimising $\gamma$ coefficients (red diamonds, $C_{\rm I, \gamma}$), $\zeta$ coefficients (green diamonds, $C_{\rm I, \zeta}$) and their sum (blue diamonds, $C_{\rm I, (\gamma + \zeta)}$). In (c), we plot the resulting final state fidelities when \acrref{BPO} is applied with a control pulse optimised using $C_{\rm I, (\alpha + \gamma + \zeta)}$ (light blue crosses) and $C_{\rm A, (\alpha + \gamma + \zeta)}$ (purple pluses). In all cases, we use $N_k = 1$. The optimisation is done 10 times for each data point with the best final fidelity plotted. We use the Powell optimisation method from Sec.~\ref{sec:3.1.3.2_Powell} for the minimisation.}\label{fig:twospin_scatter}
\end{figure}

We know that applying the exact \acrref{CD} pulse made up of all of the \acrref{FO} and \acrref{SO} \acrref{LCD} terms returns unit fidelity regardless of driving time or any optimisable parameters, as shown in Fig.~\ref{fig:twospin_fidelities}. We also found, plotted in the same Figure, that applying \acrref{FO} \acrref{LCD} to the problem without any control pulse performed worse than applying \acrref{COLD} with \acrref{FO} terms, where a control pulse is included and the control parameters are optimised for final state fidelity. Here we will also aim to optimise the control pulse as before, but in this case we will use a series of cost functions constructed in a similar manner to Eq.~\eqref{eq:COLD_costfunc_integral} and Eq.~\eqref{eq:COLD_costfunc_maximum}. First, we define the cost functions which use the magnitudes of the coefficient integrals as:
 \begin{equation}\label{eq:twospin_costfunc_int}
    \begin{aligned}
        C_{\rm I, \gamma}(\tau, \betabb) &= \int_{0}^{\tau} dt^{\prime} |\gamma(\lambda(t^{\prime}), \hbb, \betabb)|,\\
        C_{\rm I, \zeta}(\tau, \betabb) &= \int_{0}^{\tau} dt^{\prime} |\zeta(\lambda(t^{\prime}), \hbb, \betabb)|,\\
        C_{\rm I, (\gamma + \zeta))}(\tau, \betabb) &= \int_{0}^{\tau} dt^{\prime} \left(|\gamma(\lambda(t^{\prime}), \hbb, \betabb)| + |\zeta(\lambda(t^{\prime}), \hbb, \betabb)|\right), \\
        C_{\rm I, (\alpha + \gamma + \zeta))}(\tau, \betabb) &= \int_{0}^{\tau} dt^{\prime} \left(|\alpha(\lambda(t^{\prime}), \hbb, \betabb)| + |\gamma(\lambda(t^{\prime}), \hbb, \betabb)| + |\zeta(\lambda(t^{\prime}), \hbb, \betabb)|\right),
    \end{aligned}
\end{equation}
where the subscript $I$ denotes an integral-based cost function and the letters $\gamma$ and $\zeta$ simply refer to which pulse coefficient is being minimised for the optimisation process. We then do the same for the maximum amplitude of the pulses, using the subscript $A$ to denote `amplitude':
\begin{equation}\label{eq:twospin_costfunc_amp}
    \begin{aligned}
        C_{\rm A, \gamma}(\tau, \betabb) &= \max_{t^{\prime} \in [0,\tau]} | \gamma(\lambda(t^{\prime}), \hbb, \betabb)|,\\
        C_{\rm A, \zeta}(\tau, \betabb) &= \max_{t^{\prime} \in [0,\tau]} | \zeta(\lambda(t^{\prime}), \hbb, \betabb)|,\\
        C_{\rm A, (\gamma + \zeta)}(\tau, \betabb) &= \max_{t^{\prime} \in [0,\tau]} \left(| \gamma(\lambda(t^{\prime}), \hbb, \betabb)| + |\zeta(\lambda(t^{\prime}), \hbb, \betabb)| \right),\\
        C_{\rm A, (\alpha + \gamma + \zeta)}(\tau, \betabb) &= \max_{t^{\prime} \in [0,\tau]} \left(| \alpha(\lambda(t^{\prime}), \hbb, \betabb)| + |\gamma(\lambda(t^{\prime}), \hbb, \betabb)| + |\zeta(\lambda(t^{\prime}), \hbb, \betabb)| \right),
    \end{aligned}
\end{equation}
which are used to optimise the parameters $\betabb$ by minimising the maximum amplitude of the given \acrref{LCD} coefficients. We recall that the minimisation of these quantities is not necessarily directly related to the maximisation of final state fidelity, as in each case the system still needs to take some path in finite time from an initial to a final state while experiencing some amount of non-adiabatic losses, unless exact \acrref{CD} is implemented. What we expect to change, however, other than the path in parameter space due to varying $\betabb$, is the \emph{structure} of the non-adiabatic effects, as mandated by the constraints imposed by each of the cost functions. For example, minimising the total power of two-body non-adiabatic terms as in $C_{\rm I, (\gamma + \zeta)}$, might inadvertently maximise the effects of the local non-adiabatic terms $\alpha$ or even more non-local operators. This is because the cost function captures nothing about the behaviour of such terms.

The results of optimising for the \acrref{SO} \acrref{LCD} coefficients $\gamma$ and $\zeta$ are plotted in Fig.~\ref{fig:twospin_scatter}(a) and (b). In (a) we show the results of using $C_{\rm A, \gamma}$, $C_{\rm A, \zeta}$ and $C_{\rm A, (\gamma + \zeta)}$ as cost functions for optimising a \acrref{FO} \acrref{COLD} pulse, with \acrref{FO} \acrref{LCD} and \acrref{FO} \acrref{COLD} (optimised using fidelity) from Fig.~\ref{fig:twospin_fidelities} plotted for comparison. We do the same for the respective integral cost functions in (b). The results indicate that while optimising for final state fidelity, not unexpectedly, shows better final state fidelities when \acrref{COLD} is applied, the amplitude and integral cost functions still generally perform better in producing a state close to the target ground state than a naive application of \acrref{FO} \acrref{LCD} with no optimal control. This is a positive result, as it implies that there is a correlation between properties of the \acrref{LCD} coefficients and the final state fidelity in designing optimal control pulses. This is exemplified further in plot (c) of the figure, where we optimise the pulse for a case when no \acrref{LCD} is applied, using a minimisation of the exact \acrref{CD} pulse comprised of all three \acrref{LCD} coefficients. At short evolution times, the optimisation performs as well as the fidelity cost function, while at longer times it begins to lag a little, with a few outliers appearing potentially due to local minima in the cost function landscape. This is to be expected, as minimisation of the exact \acrref{CD} pulse should be equivalent to the minimisation of non-adiabatic effects experienced by the system, and at very short driving times these will probably dominate the loss of final state fidelity, while at longer times there may be several paths in parameter space that are similarly effective. 

\begin{figure}[t!]
    \centering
    \includegraphics[width=\linewidth]{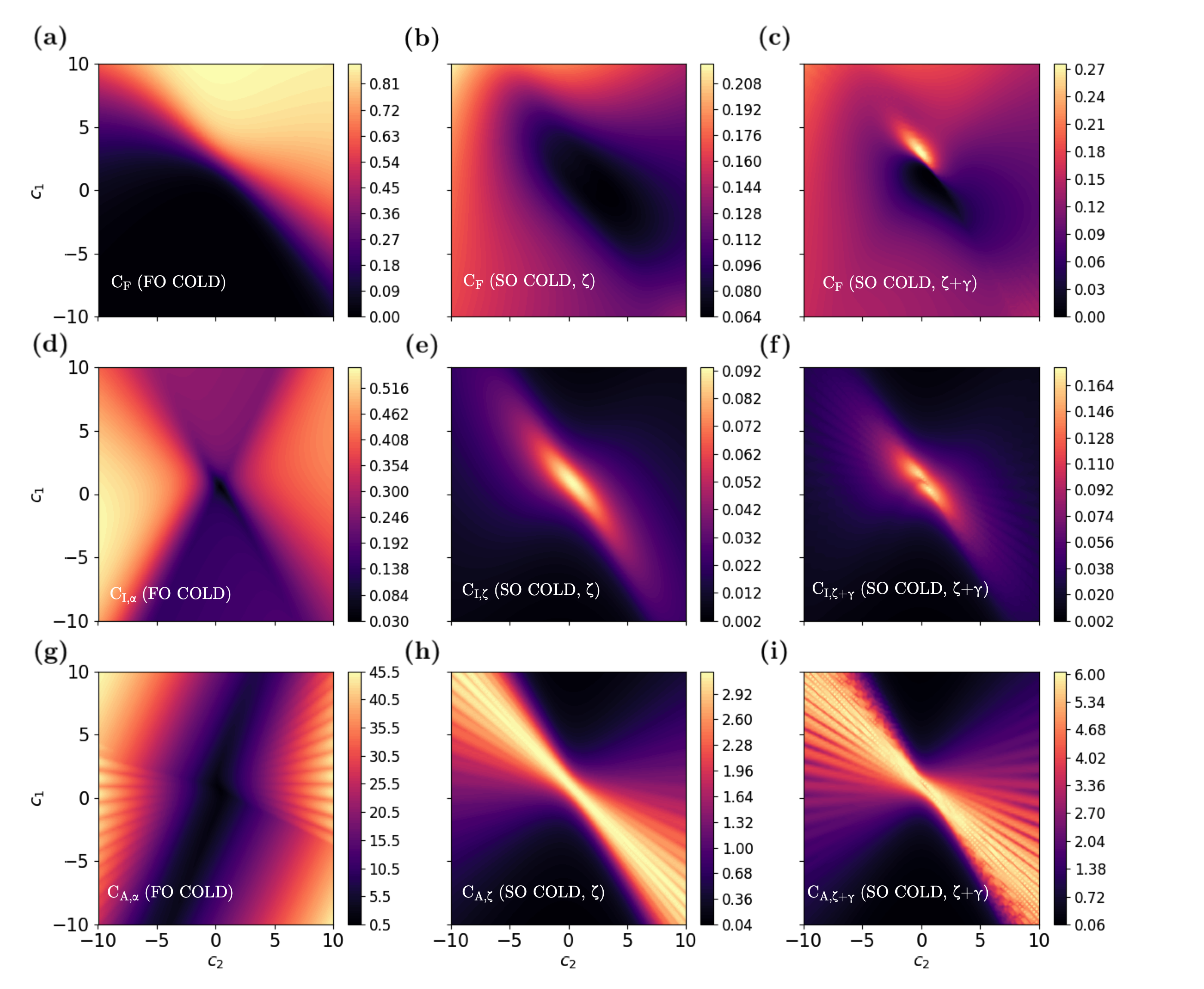} \caption[Two-spin annealing contour plots for final state fidelity and AGP cost function values.]{Contour plots of fidelity and \acrref{AGP} cost function landscapes for two parameters $c_1, c_2 \in [-10, 10]$ (first introduced in Eq.~\eqref{eq:COLD_twospin_controlH}) for two spin annealing, as discussed in the text. All plots are for total evolution time $\tau = 0.1 h_0^{-1}$. (a-c) show plots of fidelity cost function $C_{\rm F}$ values when (a) \acrref{FO} \acrref{COLD} is applied ($\alpha$ terms), (b) \acrref{SO} \acrref{COLD} $\zeta$ terms are applied and (c) both \acrref{SO} \acrref{COLD} terms $\gamma$ and $\zeta$ are applied as discussed in the text. (d-f) show plots of the integral cost function $C_{\rm I}$ values for (d) the coefficient $\alpha$, $C_{\rm I, \alpha}$ (e) the coefficient $\zeta$, $C_{\rm I, \zeta}$ and (f) sum of the coefficients $\gamma$ and $\zeta$, $C_{\rm I, \gamma + \zeta}$. (g-i) do the same for $C_{\rm A}$, where (g) shows a plot of $C_{\rm A, \alpha}$, (h) shows $C_{\rm A, \zeta}$ and (i) is a plot of $C_{\rm A, \gamma + \zeta}$}\label{fig:two_spin_higher_order}
\end{figure}

In order to better understand the relationship between the \acrref{LCD} coefficients and the value of the final state fidelity with respect to the target, in Fig.~\ref{fig:two_spin_higher_order}(a-c) we plot the cost function landscapes for fidelity ($C_{\rm F}$) at $\tau = 0.1h_0^{-1}$ for two parameters $c_1$ and $c_2$ in the range $[-10, 10]$. In (a) we plot the value of $C_{\rm F}$ when \acrref{FO} \acrref{COLD} is applied, which corresponds to the $C_{\rm F}$ landscape for the orange circle plots in Fig.~\ref{fig:twospin_scatter}(a-b). Then, in (b) we plot the values of $C_{\rm F}$ when \acrref{COLD} with only the $\zeta$ terms $\sy_1\sz_2 + \sz_1\sy_2$ is applied. Finally, in (c), we plot $C_{\rm F}$ for  \acrref{COLD} with both of the \acrref{SO} terms $\gamma$ and $\zeta$ applied.  We then do the same for $C_{\rm I, \alpha}$ in (d), $C_{\rm I,\zeta}$ in (e) and $C_{\rm I,(\gamma + \zeta)}$ in (f). This is also done for the respective maximum amplitude cost functions. The $\alpha$ coefficients for (a), (d) and (g) are obtained by solving the coupled equations of Eq.~\eqref{eq:two_spin_coupled_eqs} with $\gamma$ and $\zeta$ set to $0$, while for the plots in (b),(e) and (h) we set $\alpha$ to $0$, and find that the $\gamma$ pulse turns out to be $0$ for all values of $\betabb$ and $\lambda$, hence retaining only values of $\zeta$. In (c),(f) and (i) we solve for the full \acrref{CD} pulse with all coefficients but only plot the \acrref{SO} components. What we find is that while there appears to be some relationship between the maximum and minimum of the \acrref{AGP} cost functions and the final state fidelities, it is not clear cut. Certainly, where the final state fidelity $F(\tau)$ is maximised in (a) (\@i.e.~when the plot shows a minimum value as $C_{\rm F} = 1 - F$), the \acrref{SO} \acrref{LCD} plots (e-f) and (h-i) have small values. However, it appears as though the \acrref{SO} components are also small for values of $c_1$ and $c_2$ which lead to very bad final state fidelities in (a). It is also clear that maximising fidelity for \acrref{FO} \acrref{COLD} is not necessarily equivalent to maximising the \acrref{FO} \acrref{LCD} coefficient. What these plots are intended to illustrate is that while there appears to be some relationship between the various approximations of the \acrref{AGP} and the final state fidelity of the system when \acrref{COLD} of various orders is applied, this relationship need not be clear cut.

We note that there is no contour plot for $C_{\rm F}$ in the case of \acrref{BPO} nor plots of the integral or amplitude functions for the exact \acrref{CD} pulse comprised of all of the \acrref{LCD} coefficients and that is because the optimal parameter values of $c_1$ and $c_2$ were very large (of the order of $1\times10^3 h_0^{-1}$) and varied in the results of Fig.~\ref{fig:twospin_scatter}(c), making it difficult to capture the relevant cost function landscapes visually. The fact that the \acrref{COLD} optimal control pulses require quite low amplitudes even at short driving times could, in fact, also be considered an advantage of the method.

While in this simple example it may be more favourable to implement the fidelity cost function $C_{\rm F}$ given that for two spins it is reasonably efficient to compute the state evolution, in more complex systems this is no longer the case. Iterative optimisation procedures like Powell's method (Sec.~\ref{sec:3.1.3.2_Powell}), which we have been using, may require hundreds or thousands of cost function evaluations for a single optimisation procedure. The fact that we get results which are comparable to $C_{\rm F}$ while using a set of cost functions which become exponentially more efficient to implement as the system size grows is something that can become very useful in such more complex cases.

\section{Return to the Ising spin chain}\label{sec:7.2_ising_ho_lcd}

As in the previous section, here we revisit a system that was already explored in the previous chapter: the Ising spin chain of Sec.~\ref{sec:5.2_Ising_chain}. As in the two-spin case, we are interested in retaining the same parameters as those explored for \acrref{COLD} with the fidelity cost function, changing only the optimisation landscape via the integral and amplitude cost functions outlined in Eq.~\eqref{eq:twospin_costfunc_int} and Eq.~\eqref{eq:twospin_costfunc_amp} respectively. Thus, we use the same bare Hamiltonian and parameters (Eq.~\eqref{eq:ising_chain_hamiltonian}) along with the Powell optimal control pulse from Eq.~\eqref{eq:ising_chain_BPO_H}, parameterised by the control functions $\beta_k \in \betabb$ and by constant control parameters $c_k$. The \acrref{FO} \acrref{LCD} ansatz is still the same as in Eq.~\eqref{eq:ising_fo_agp}, which is a set of local $\sy$ operators on each spin and the \acrref{SO} operators are taken to be all of the nearest neighbour, two-body Pauli terms on $N$ spins, expressed as
\begin{equation}\label{eq:ising_so_lcd_terms}
    \begin{aligned}
        \approxAGP^{(2)}(\betabb, \hbb, \lambda) = & \sum_j^{N-1} \gamma(\betabb, \hbb, \lambda) (\sx_j\sy_{j+1} + \sy_j\sx_{j+1}) \\
        &+ \sum_j^{N-1} \zeta(\betabb, \hbb, \lambda) (\sz_j\sy_{j+1} + \sy_j\sz_{j+1}),
    \end{aligned}
\end{equation}
where $\gamma$ and $\zeta$ are the \acrref{SO} \acrref{LCD} coefficients as in the previous example. In this section we use the same integral and amplitude cost functions as in Eq.~\eqref{eq:twospin_costfunc_int} and Eq.~\eqref{eq:twospin_costfunc_amp}, with the coefficients $\alpha$, $\gamma$ and $\zeta$ as presented above. These can be solved by using the results presented in Appendix~\ref{app:arbitrary_ising_derivation}, as discussed in Sec.~\ref{sec:5.2_Ising_chain}.

The main difference between the two spin example and this one, in particular when considering using \acrref{AGP} constructed cost functions, is that for increasing spin chain lengths we can no longer expect to have access to the exact \acrref{CD} pulse, as the non-adiabatic effects may become delocalised quickly throughout the chain. Even if the effects of the delocalised \acrref{AGP} operators are small, they are not necessarily guaranteed to be non-zero. Thus, we are now operating in a setting where we might not have all of the information about the non-adiabatic effects on the system and must instead contend with \acrref{LCD} approximations of the exact counterdiabatic pulse explicitly.
\begin{figure}[t]
    \centering
    \includegraphics[width=0.8\linewidth]{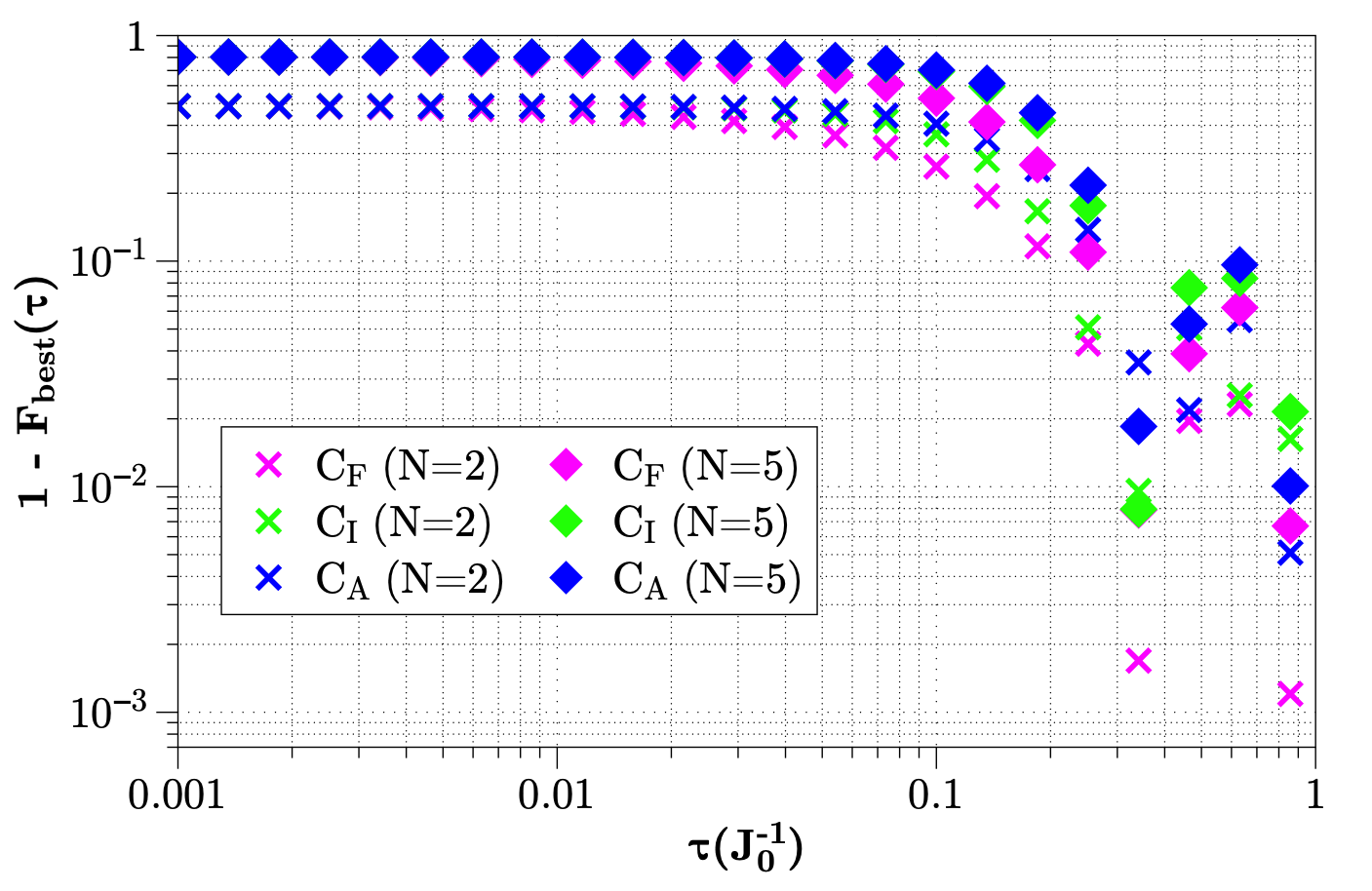} \caption[Plot of final state fidelity for the Ising spin chain for different cost functions and no counterdiabatic component in the implementation.]{Plot of final state fidelities obtained when optimising for the fidelity cost function $C_{\rm F}$ (pink), the integral cost function for \acrref{FO} and \acrref{SO} \acrref{LCD} pulses $C_{\rm I, (\alpha + \gamma + \zeta)}$ (green) as well as the maximum amplitude cost function $C_{\rm A, (\alpha + \gamma + \zeta)}$ (blue). We plot results for two spins (crosses) and five spins (diamonds), optimising separately for both. Results are plotted for $N_k = 1$ and different total driving times $\tau$. Optimisation is performed 10 times for each data point and the lowest obtained value for the cost function is used to compute the fidelity in the case of $C_{\rm A}$ and $C_{\rm I}$.}\label{fig:ising_nocd_higher_order}
\end{figure}

With this in mind, the first thing we explore is whether or not we can use the \acrref{FO} and \acrref{SO} \acrref{LCD} coefficients in the same vein as in Fig.~\ref{fig:twospin_scatter}(c) to optimise the \acrref{BPO} pulse. We plot the results in Fig.~\ref{fig:ising_nocd_higher_order} for the cases of $N=2$ spins and $N=5$ spins, using $C_{\rm F}$, $C_{\rm A, (\alpha + \gamma +\zeta)}$ and $C_{\rm I, (\alpha + \gamma +\zeta)}$. We expect that in the two-spin case we will get a similar result as in Fig.~\ref{fig:twospin_scatter}(c), given that we should be minimising the exact \acrref{CD} pulse in the case of two spins when using both \acrref{FO} and \acrref{SO} in the cost function and this appears to be the case: at short evolution times, the results of the final state fidelity match up regardless of which cost function is used, while at longer times $C_{\rm F}$ begins to perform better, although all cost functions show a similar pattern in final state fidelity scaling with $\tau$. However, in the five spin case, we do not have an explicit reason to expect a similar behaviour unless higher order \acrref{LCD} does not play a big part in the non-adiabatic effects experienced by the system. It turns out, in fact, that this is the case: the behaviour for five spins is comparable to that of two spins: at short times the cost functions are equally as effective, with differences appearing only at longer times. We note further, that as well as the results plotted in Fig.~\ref{fig:twospin_scatter}, which indicate a similar effect of the various cost functions on fidelity, we find the optimised pulse shapes to be quite similar in each case too. This lends further credence to the fact that in the case of the Ising spin chain Hamiltonian, the minima of all cost functions are close to each other in parameter space.

\begin{figure}[t!]
    \centering
    \includegraphics[width=\linewidth]{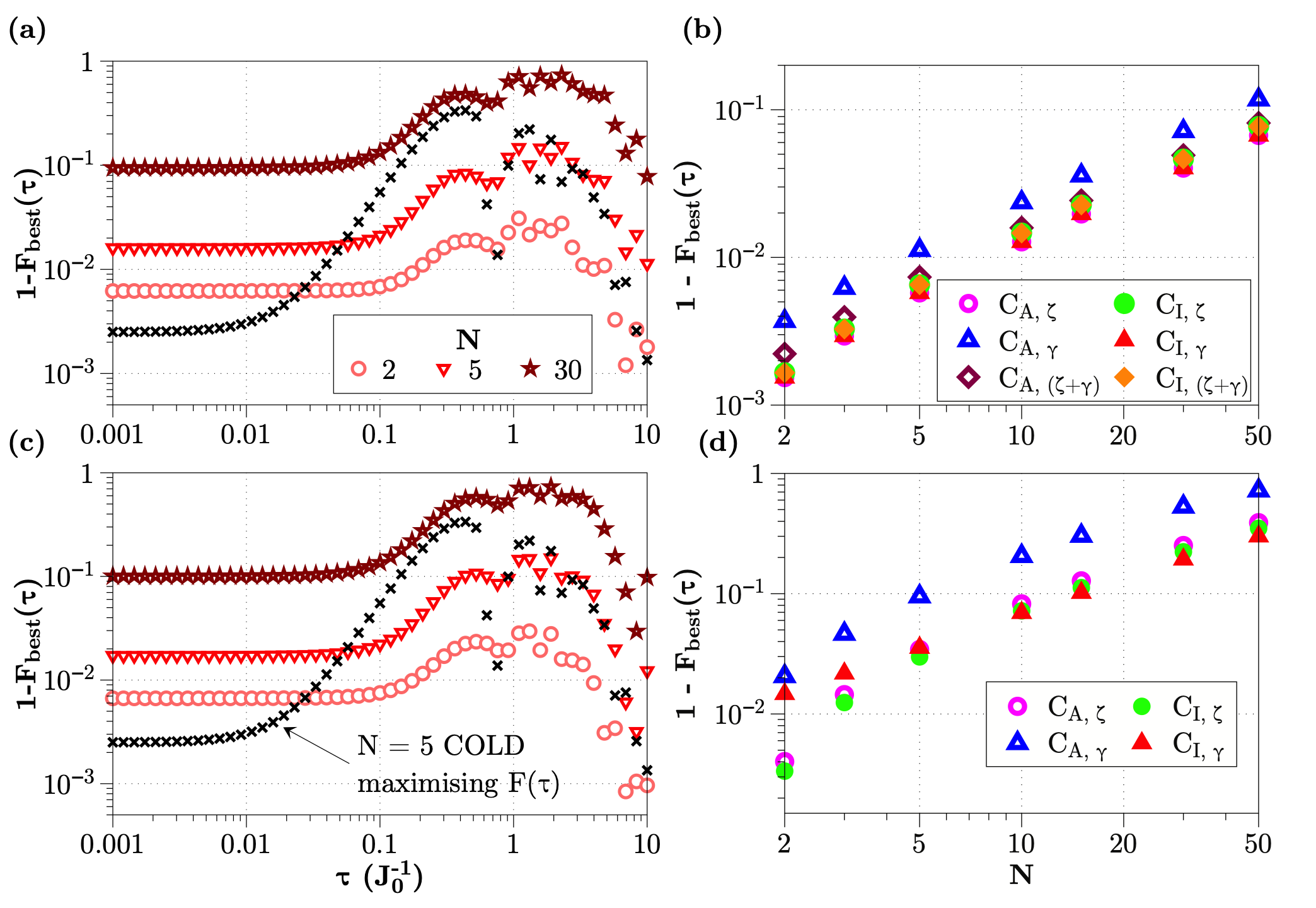} \caption[Plot of final state fidelity for the Ising spin chain for different cost functions with LCD applied.]{Final state fidelities when control parameters are optimised via integral and amplitude cost functions of the \acrref{SO} \acrref{LCD} coefficients for the Ising spin chain with \acrref{FO} \acrref{COLD} applied to the system. In (a) we plot the results obtained when using the integral cost function $C_{\rm I, (\zeta)}$ and in (c) we do the same in the case of maximum amplitude $C_{\rm A, (\zeta)}$ for Ising chains of lengths $N = 2$ (pink circles), $N = 5$ (red inverted triangles) and $N=30$ (dark red stars). We also plot results for \acrref{COLD} optimisation with $C_{\rm F}$ in the case of $N=5$ spins from Fig.~\ref{fig:ising_unconstrained}(a) for comparison (black crosses). We investigate how the different cost functions perform for evolution time $\tau = 0.1J_0^{-1}$ for different system sizes $N$ in the case where (b) \acrref{FO} \acrref{COLD} is applied to the system while minimising a \acrref{SO} component and (d) when applying both \acrref{FO} terms and one of the \acrref{SO} terms. For example, in (d), the pink circles are the result of implementing \acrref{COLD} with \acrref{FO} local $\sy$ terms and $\gamma$ terms $\sx\sy$, $\sy\sx$ while minimising the maximal amplitudes of the $\zeta$ terms $\sz\sy$, $\sy\sz$. For system sizes above $N=10$ we used ITensor\cite{fishman_itensor_2022} MPS calculations which were converged with a truncation level of $10^{-14}$ per time step at each site reaching a maximum bond dimension of $D = 4$. In all cases, a single optimisable parameter is used ($N_k = 1$) and the best optimisation out of 50 (lowest cost function value for each cost function) is used. Reprinted with permission from \cite{cepaite_counterdiabatic_2023}. Copyright 2023, American Physical Society.}\label{fig:ising_minimising_ho_mainplot}
\end{figure}

The other case that we can explore, then, to improve our results, is to see if we could implement \acrref{FO} or partial \acrref{SO} \acrref{COLD} while minimising the other \acrref{LCD} term coefficients. This is presented in Fig.~\ref{fig:ising_minimising_ho_mainplot}, where in (a-b) we implement \acrref{COLD} with \acrref{FO} \acrref{LCD} coefficients while minimising the integral cost function $C_{\rm I, (\zeta)}$ in (a) and the amplitude cost function $C_{\rm A, (\zeta)}$ in (c), both for system sizes of $N = 2, 5, 30$. We also plot the results from Fig.~\ref{fig:ising_unconstrained}(a) which show results when optimising \acrref{COLD} using the final state fidelity cost function $C_{\rm F}$ in the case of $N = 5$ spins for fidelity. What we find is a consistent pattern in the behaviour of the final state fidelity when using either of the two cost functions, with consistently stable final fidelities at shorter driving times and, surprisingly, better final fidelities at longer driving times, at least in the $N=5$ case. We might attribute this to different cost function landscapes due to the different cost functions, which might have more or less optimal minima within reach for a local optimiser like Powell's method (Sec.~\ref{sec:3.1.3.2_Powell}). Regardless, what we do find is that we can get final state fidelities consistently above $90\%$ for a system of $N=30$ spins while using an exponentially more efficient cost function for optimisation. While we use tensor network methods to compute the fidelities of chains with $N=10$ spins and above in all cases, these approaches are still orders of magnitude slower at calculating a single iteration of the $C_{\rm F}$ cost function than in the case of either the integral or amplitude cost functions. We find that this trend is consistent, if slightly shifted depending on which \acrref{AGP}-based cost function is used in plot (b) of Fig.~\ref{fig:ising_minimising_ho_mainplot}, and that it is also consistent when minimising one of the \acrref{SO} coefficients and implementing the other along with the \acrref{FO} terms, at least at relatively short driving times of $\tau = 0.1J_0^{-1}$.

All of these results are not conclusive proof for the advantage of using the integral or amplitude cost functions in place of $C_{\rm F}$, especially if the \acrref{LCD} terms are highly local with respect to the full system size. However, the results presented do indicate a potential advantage, especially in the case of larger systems, of using knowledge about the approximate \acrref{AGP} operator in an optimisation procedure of the control Hamiltonian, assuming the wavefunction of the state is prohibitively difficult to access. In doing so, it is possible to sacrifice some amount of effectiveness for a large gain in efficiency and, possibly, this kind of approach could be used in cases where fidelities are simply not a tractable option in the case of numerical optimisation. Furthermore, as discussed near the end of Sec.~\ref{sec:3.2_Quantum_optimal_control}, this type of optimisation may be more useful in settings where a specific target state may not be the goal. Rather, we may instead desire a particular property of the state, like entanglement. On that note, we move on to the next section.

\section{GHZ states and frustrated spins}\label{sec:7.3_ghz_ho}

Finally, we return to the GHZ state preparation scheme from Sec.~\ref{sec:6.4_ghz_states}. Thus far, both in the two-spin case and in the Ising spin chain case we have seen some evidence for advantage when it comes to using \acrref{AGP}-informed cost functions to optimise the control parameters with respect to the final state fidelity. While we have shown that optimising using the fidelity cost function $C_{\rm F}$ generally guarantees better results, it is also far less efficient for larger systems. Using the integral cost function $C_{\rm I}$ or the amplitude cost function $C_{\rm A}$ may not be as effective, but it is far more efficient and can be used to obtain good results for both \acrref{COLD} and optimal control pulses with no \acrref{CD} added to them. In this case, we will explore an example of a system and control pulse combination for which this option may not be viable, at least given the parameters we are working with.

We will set up the problem in the same way as in Sec.~\ref{sec:6.4_ghz_states}, with the bare Hamiltonian from Eq.~\eqref{eq:ghz_hamiltonian}, as well as a \acrref{GRAPE} control pulse as given by Eq.~\eqref{eq:ghz_control} and the surrounding description. In this case, from Sec.~\ref{sec:6.4.1_t3} we recall that we already attempted implementing a different cost function to $C_{\rm F}$, the aim of which was to maximise a measure of tripartite entanglement in the three spin case: $C_{T_3}$ from Eq.~\eqref{eq:tangle_costfunc}. We will return to both the fidelity and the tangle as measures of success for the final state and thus we will be looking solely at the $N=3$ spin example, both as the simplest possible example to test out new cost functions on and due to the fact that we have a non-trivial tripartite entanglement metric like the three-tangle available.

\begin{figure}[t!]
    \centering
    \includegraphics[width=0.75\linewidth]{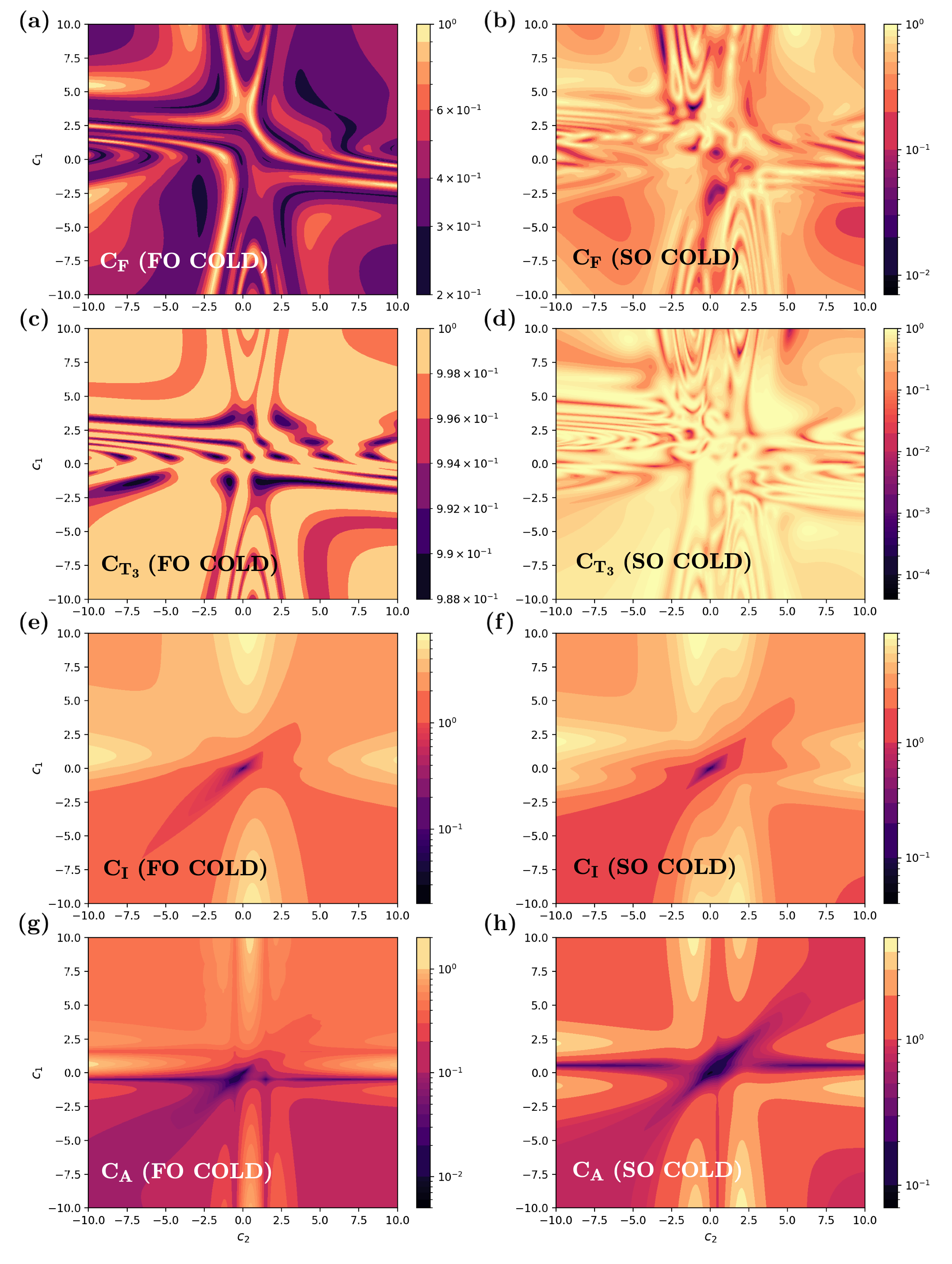} \caption[Contour plots of cost function landscapes for GHZ state preparation in frustrated spin systems (integral cost function).]{Contour plots at $\tau = 0.1 J_0^{-1}$ of different cost function values for GHZ state preparation for parameters $c_1, c_2 \in [-10,10]$ and a \acrref{GRAPE} control pulse. In (a) and (b) we plot $C_{\rm F}$ in the cases where \acrref{FO} and \acrref{SO} \acrref{COLD} is applied respectively. Then, in (c-d) we do the same for $C_{T_3}$, with \acrref{FO} \acrref{COLD} plotted in (c) and \acrref{SO} \acrref{COLD} plotted in (d). (e-h) are then plots of the integral cost function $C_{\rm I}$ values for the same range of parameters. In (e) we plot $C_{\rm I, \alpha^{(1)}}$ when only \acrref{FO} \acrref{LCD} is considered, while in (f) we plot $C_{\rm I, \alpha^{(2)}}$ as described in the text. Then in (g) we plot $C_{\rm I, \gamma}$ and in (h) we plot $C_{\rm I, \zeta}$, corresponding to the \acrref{SO} terms. Note that each plot has its own colour bar, as the colour encodings and the value scaling in each plot is quite different.}\label{fig:ghz_contours}
\end{figure}

The \acrref{FO} \acrref{LCD} terms are defined as previously to be local $\sy$ operators scaled by the coefficient $\alpha(\betabb, \hbb, \lambda)$, while the \acrref{SO} terms are those presented in Eq.~\eqref{eq:ghz_so_lcd}. We will differentiate here between solving for $\alpha$ when only the \acrref{FO} terms are in the \acrref{LCD} ansatz, denoting this case as $\alpha^{(1)}$ and solving for all of the three coefficients $\alpha$, $\gamma$ and $\zeta$ as a combined \acrref{FO} and \acrref{SO} ansatz by using the coupled set of equations presented in Appendix~\ref{app:arbitrary_ising_derivation}, wherein we will refer to the \acrref{FO} coefficient as $\alpha^{(2)}$ instead.

In Fig.~\ref{fig:ghz_contours} we plot the cost function landscapes for $C_{\rm F}$ (a-b), $C_{T_3}$ (c-d) and several integral cost functions $C_{\rm I}$ in (e-h) for a total driving time of $\tau = 0.1J_0^{-1}$ (the results for the maximal amplitude cost functions can be found in Appendix~\ref{app:higher_order_AGP}). The first thing that one might notice when looking at the plots is that while there appears to be some qualitative relationship between the different cost function landscapes, it is certainly not the case that there is a clear overlap between the minimum or maximum values of the integral cost functions and the highest fidelities (lowest values in the contour plot, as $C_{\rm F} = 1 - F$). This is a fact that holds true in the case of the maximal amplitude cost functions in Appendix~\ref{app:higher_order_AGP} too. The $C_{\rm F}$ and $C_{T_3}$ landscapes are highly non-convex, with multiple local minima close to each other, something that may be a consequence of the \acrref{GRAPE} cost function and the very small number of parameters, as this translates to very sudden piecewise shifts in the control pulse (see discussion in Sec.~\ref{sec:4.2_COLD_QOCT}). This may also be a consequence of the degenerate nature of the ground state. It is possible, that the reason there is such a disparity between the final state fidelity and the scale of the \acrref{LCD} coefficients is the small number of parameters in the \acrref{GRAPE} function, but attempts to optimise the \acrref{GRAPE} control pulse with parameter numbers up to $N_k = 12$ using dual-annealing (Sec.~\ref{sec:3.1.3.3_dual_annealing}) with $C_{\rm I}$- and $C_{\rm A}$-type cost functions as in Fig.~\ref{fig:ising_minimising_ho_mainplot} or Fig.~\ref{fig:ising_nocd_higher_order} results in the system barely moving out of its initial state regardless of driving time (not plotted). This is certainly an indication that in a more complex regime with a degenerate ground state where the target is entanglement generation, the \acrref{AGP} cost function approach might not work, or at least needs to be investigated further. Perhaps, as we hope to show in future work, higher orders of the \acrref{LCD} are required in order to capture the non-adiabatic effects associated with entanglement creation.

\part{Conclusion}\label{part:conclusion}

\chapter{Summary}\label{chap:8_Summary}

\epigraph{This is The End; my only friend, The End.}{Jim Morrison,
\emph{``The End", The Doors}}

In this thesis we introduced a new method for speeding up adiabatic quantum protocols while minimising losses due to transitions out of the instantaneous eigenstate: \acrref{COLD}. The new method is comprised of two key components: approximate counterdiabatic driving techniques and quantum optimal control. We discussed the theoretical framework and motivation behind \acrref{COLD}, beginning in Ch.~\ref{chap:2_adiabaticity} with a background introduction to quantum adiabaticity and the losses which arise as a consequence of fast changes in a time-dependent Hamiltonian. We covered how these losses can be described by an operator known as the \acrref{AGP} \cite{kolodrubetz_geometry_2017}, and how the \acrref{CD} technique can be used to suppress the non-adiabatic effects generated by the \acrref{AGP} \cite{berry_transitionless_2009, demirplak_adiabatic_2003}. We explored the reasons why the exact \acrref{CD} pulse is often inaccessible, either in theory or in practice, and introduced several existing methods for constructing an \emph{approximate} counterdiabatic drive. Then, in Ch.~\ref{chap:3_Quantum_Optimal_control}, we covered the theory and methodology involved in optimal control theory, which concerns itself with finding optimal path for a system from some initial state to some final state. We introduced ideas concerning how optimal control theory can be applied in the setting of quantum systems and then we described several popular quantum optimal control methods like \acrref{CRAB} and \acrref{GRAPE}. 

These two background chapters paved the way for \acrref{COLD} in Part~\ref{part:COLD}. This new method is the result of combining an approximate \acrref{CD} method we refer to as \acrref{LCD} \cite{sels_minimizing_2017}, with quantum optimal control techniques. Having outlined how \acrref{LCD} allows one to variationally determine a pulse shape for an ansatz set of physical operators which most closely resembles the exact counterdiabatic drive for a given time-dependent Hamiltonian. The goal of \acrref{LCD} is to suppress as many losses associated with such transitions as possible within the restrictions imposed by the ansatz set of operators. In Ch.~\ref{chap:4_COLD}, we described how \acrref{COLD} can improve upon the \acrref{LCD} approach by using the observation that the non-adiabatic effects experienced by a system driven by a time-dependent Hamiltonian depend on the path of the Hamiltonian through parameter space. We showed how this path can be changed via the implementation of methods from optimal control theory, thus allowing \acrref{COLD} to find a path which maximises the effects of the \acrref{LCD} for a given ansatz set of operators. We then posited, in Ch.~\ref{chap:5_cd_as_costfunc}, that the information about non-adiabatic effects contained in the \acrref{AGP} operator could be used to construct optimisation metrics for the optimal control component of \acrref{COLD}.

Finally, in Part~\ref{part:applications}, we demonstrated how \acrref{COLD} performs by numerically simulating its implementation for various physical systems and time-dependent Hamiltonians. We compared the results to those obtained when using \acrref{LCD} with no optimal control component, as well as to optimal control techniques with no counterdiabatic component. In Ch.~\ref{chap:6_Applications_fidelity}, we focused on using optimal control techniques to target properties of the final state obtained by implementing each method, such as fidelity with respect to a target ground state or amount of entanglement. We showed results for a simple two-spin annealing protocol in order to demonstrate in detail how the \acrref{COLD} approach works. Then, we demonstrated the advantage of using \acrref{COLD} over other approaches in the case of the Ising spin chain, even when the pulse amplitudes of all of the drives involved are constrained to be below some value. This was followed by the case of an \acrref{ARP} protocol for population transport in a synthetic lattice, adapted from \cite{meier_counterdiabatic_2020} wherein only \acrref{LCD} had been implemented. We capped off the chapter with a more complex example, the goal of which was to generate a maximally entangled GHZ state in a system of frustrated spins. We found that \acrref{COLD} showed an advantage in all of these examples and that it could be enhanced with various optimal control techniques like \acrref{CRAB} or \acrref{GRAPE}. In the final example, we discovered that highly local \acrref{LCD} operators cannot generate entanglement through the system at short driving times and that more delocalised pulses might be needed in such systems. We then demonstrated how \acrref{AGP}-informed cost functions, first introduced in Ch.~\ref{chap:5_cd_as_costfunc}, performed for some of the same systems in Ch.~\ref{chap:7_higher_order_agp}. In the case of the two-spin example and the Ising spin chain, we showed that we could implement a far more computationally efficient optimisation protocol than ones which use fidelity as a cost function for finding Hamiltonian paths that minimise non-adiabatic effects. We showed that this could be done in cases where either \acrref{COLD} or only optimal control is implemented. We found, however, that in the case of generating GHZ states, such a cost function did not appear to work as intended, whether due to the complexity of the problem at hand, drawbacks of the \acrref{LCD} approximation or issues with the optimal control. 

\chapter{To boldly go...}\label{chap:9_Future_directions}

\epigraph{Time will explain it all. He is a talker, and needs no questioning before he speaks.}{Euripides}

There is often joy mixed with trepidation in finding that, for all the work that might have already been done, far more remains to be accomplished. This is certainly true in the case of the results presented in this thesis. The \acrref{COLD} method is one that was created with practicality in mind: given a quantum system, a time-dependent Hamiltonian driving it and a set of constraints, like the system controllability or computational resources, it should help one produce an optimal protocol which minimises the non-adiabatic losses experienced by the system while it is driven from an initial eigenstate towards the target as quickly as possible. As the space of systems, Hamiltonians and constraints is vast, merely exploring in which scenarios \acrref{COLD} may or may not have an advantage over the equally vast set of other possible approaches is no small task. However, in this brief chapter, we will discuss several open questions and potential future research directions in a more focused way, including those that arose during the process of constructing and implementing \acrref{COLD}. 

\section{Practical aspects}

The first thing to note is the fact that the field of quantum optimal control is very extensive and that we only explored a few common ways to construct control pulses in this thesis. In general, using a more complex control pulse that has a larger solution space and increased computational resources will almost certainly be a better option than simpler choices, unless there is an informed reason to expect a simpler pulse to do better. In many of the examples in Ch.~\ref{chap:6_Applications_fidelity} and Ch.~\ref{chap:7_higher_order_agp} the control pulse we implemented was the \emph{bare} pulse from Eq.~\eqref{eq:bare_pulse}, which is quite rudimentary. One reason for doing this was to save on time and computational resources, as it required very little of either to implement compared to more complex approaches like \acrref{CRAB} or \acrref{GRAPE}. The other reason was simply the fact that the results obtained using the bare pulse were already enough to demonstrate the functionality and advantages of the method, while also being easier to analyse. In any more focused application of \acrref{COLD}, there would have to be a strong consideration for how a particular choice of optimal control pulse can interact with the constraints of the problem and even the \acrref{LCD} pulse itself, which is a function of the control parameters. A larger gradient in the control pulse could, for example, lead to a spike in the amplitude of the approximate counterdiabatic pulse, due to the $\dlambda H$ matrix elements present in the \acrref{AGP} operator (Eq.~\eqref{eq:AGP_adiabatic_basis}) or else lead to a large non-zero counterdiabatic component at the end of the protocol. 

Apart from the optimal control component, it would be useful to consider the noise present in physical implementations and how that might affect the performance of \acrref{COLD}. The cost function landscapes plotted throughout Ch.~\ref{chap:7_higher_order_agp} give some indication of how smoothly the final state fidelity reacts to a small shift in optimal control parameters. While in some cases, like the two-spin example of Sec.~\ref{sec:7.1_two_spin_ho}, the fidelity cost function is quite smooth, this is absolutely not the case for the GHZ state preparation example in Sec.~\ref{sec:7.3_ghz_ho}. While the highly non-convex nature of the plots might simply be due to the small number of control parameters involved, there is no guarantee that such high susceptibility to parameter values would be avoided in any specific example.

\section{Extensions of COLD}

As well as questions following up from the existing methodology of \acrref{COLD} and the examples covered in this thesis, we might also look forward to new ideas inspired by the content that was presented in previous chapters. The composition of Ch.~\ref{chap:5_cd_as_costfunc} was born, for example, from several observations about the behaviour of different orders of \acrref{LCD} operators in optimised versus un-optimised control pulses (see Appendix~\ref{app:ising} for more details). In a similar vein, we can imagine designing new and better types of cost functions based on information about the non-adiabatic effects experienced by a system rather than just those presented in this thesis. The failure of the approach in the case of GHZ state preparation is certainly a reason to try something different.

We may, for example, opt to optimise the \emph{other} component of the counterdiabatic drive: not the \acrref{AGP} operator, but rather the rate of change of the parameters $\dotlambda$. It may be possible to perform a piecewise optimisation of how fast the changes in the Hamiltonian parameters occur at different critical moments in the system evolution. As discussed in Ch.~\ref{chap:2_adiabaticity}, the `slow' evolution condition for adiabaticity depends on the energy gaps between the instantaneous states. As such, it might be interesting to construct a control pulse which varies $\dlambda$ for each timestep depending on the criticality of the non-adiabatic effects experienced by the system at that point, with the full evolution being constrained to some total evolution time $\tau$.

Beyond these ideas, there may be better approaches to computing the approximate \acrref{CD} components, which is occasionally an arduous task (see \@e.g.~Appendix~\ref{app:arbitrary_ising_derivation}). This is the goal of \cite{lawrence_numerical_2024}, which explores how the structure in certain Hamiltonians like the Ising model can be exploited in order to compute the \acrref{LCD} coefficients for a large numbers of operators very efficiently. Should this be accomplished in a more general setting, \acrref{AGP}-based cost functions may become an even greater resource, as we might be able to better characterise the behaviour of different orders of \acrref{LCD} with respect to each other and the target state. The exact \acrref{AGP} operator may yet have more information for us to use in designing optimal fast adiabatic protocols.

\appendix

\chapter{Rotating spin Hamiltonian}\label{app:rotating_spin_hamiltonian}

In Chap.~\ref{chap:2_adiabaticity} we used the example of a spin rotating in a magnetic field to illustrate adiabatic processes in quantum systems. We considered a spin starting in the $\ket{+}$ state and being rotated from the $x$ direction to the $z$ direction during some total time $\tau$ according to the Hamiltonian in Eq.~\eqref{eq:rotating_spin_H}, which I will reproduce here for convenience:
\begin{equation}\label{eq:rotating_spin_H_lambda_2}
    H(\lambda) = -\cos(\lambda)\sx - \sin(\lambda)\sz,
\end{equation}
with $\lambda(t) = \frac{\pi t}{2 \tau}$. The \acrref{AGP} operator ansatz $\approxAGP$ for this system can be described by the operator $\sy$ scaled by some $\lambda$-dependent coefficient which we will refer to as $\alpha(\lambda)$
\begin{equation}
    \approxAGP = \alpha(\lambda) \sy
\end{equation}
as discussed in the main text. We will now proceed to show how we can arrive at the resulting form of $\alpha$ given in Eq.~\eqref{eq:rotating_spin_alpha} using the \acrref{LCD} method outlined in Sec.~\ref{sec:2.4.1_LCD}.

The first step is to find the operator $G_{\lambda}$ given in Eq.~\eqref{eq:G_operator}:
\begin{equation}
    \begin{aligned}
        G_{\lambda}(\approxAGP) &= \dlambda H + i\comm{\approxAGP}{H} \\
        & = \sin(\lambda)\sx - \cos(\lambda)\sz + 2\alpha(\lambda)\sin(\lambda)\sx - 2\alpha(\lambda) \cos(\lambda)\sz \\
        & = (1 + 2\alpha(\lambda))\sin(\lambda) \sx - (1 + 2\alpha(\lambda))\cos(\lambda)\sz,
    \end{aligned}
\end{equation}
where we have used $\hbar = 1$. This can then be used to define the action
\begin{equation}
    \begin{aligned}
        \mathcal{S}(\approxAGP) &= \Tr\left[G^2_{\lambda}(\approxAGP) \right] \\
        & = 2 (1 + 2\alpha(\lambda))^2 \sin^2(\lambda) + 2 (1 + 2\alpha(\lambda))^2 \cos^2(\lambda) \\
        & = 2 (1 + 2\alpha(\lambda))^2.
    \end{aligned}
\end{equation}
In order to find the form of $\alpha$, we need to find the minimum of the action $\mathcal{S}(\approxAGP)$ with respect to $\alpha$, which can be easily done:
\begin{equation}
    \begin{aligned}
        \frac{\delta \mathcal{S}}{\delta \alpha} &= 8(1 + 2\alpha(\lambda)) \\
        \Rightarrow \alpha(\lambda) &= -\frac{1}{2},
    \end{aligned}
\end{equation}
which is the expected result. 

It can be checked, for this simple example, that $\approxAGP = -\frac{1}{2} \sy$ is the exact \acrref{AGP} operator $\AGP{\lambda}$ by using Eq.~\eqref{eq:AGP_adiabatic_basis} where the matrix elements of the \acrref{AGP} are written out explicitly as:
\begin{equation}
    \begin{aligned}
        \AGP{\lambda} &= i \Big(\sum_n \braket{n}{\dlambda n} \dyad{n} + \sum_{m \neq n} \ket{m} \frac{\mel{m}{\dlambda H}{n}}{(E_n - E_m)} \bra{n} \Big) \\
        & = i \Big(\sum_n \braket{n}{\dlambda n} \dyad{n} + \sum_{m \neq n} \braket{m}{\dlambda n} \dyad{m}{n}\Big)
    \end{aligned}
\end{equation}

In this case, the adiabatic eigenstates of the Hamiltonian $H(\lambda)$ are
\begin{equation}
    \begin{aligned}
        \ket{\psi_1} &= \frac{1}{n_1} \left[(\sec(\lambda) + \tan(\lambda)) \ket{\uparrow} + \ket{\downarrow}\right] \\
        \ket{\psi_2} &= \frac{1}{n_2} \left[(-\sec(\lambda) + \tan(\lambda)) \ket{\uparrow} + \ket{\downarrow}\right],
    \end{aligned}
\end{equation}
where $n_1 = \sqrt{1 + \abs{\sec(\lambda) + \tan(\lambda)}^2}$ and $n_2 = \sqrt{1 + \abs{-\sec(\lambda) + \tan(\lambda)}^2}$ are the normalisation factors. Their derivatives with respect to $\lambda$ are
\begin{equation}
    \begin{aligned}
        \ket{\dlambda \psi_1} &= \frac{1}{n_1^3} \sec(\lambda) \left[(\sec(\lambda) + \tan(\lambda))\ket{\uparrow}  - (\sec(\lambda) + \tan(\lambda))^2 \ket{\downarrow}\right] \\
        \ket{\dlambda \psi_2} &= (2 + 2\sin(\lambda))^{-3/2} \left[(\sin(\lambda) + 1)^2 \ket{\uparrow} + \cos(\lambda)(1 + \sin(\lambda))\ket{\downarrow}\right].
    \end{aligned}
\end{equation}

Evaluating $\braket{\psi_1}{\dlambda \psi_1}$ and $\braket{\psi_2}{\dlambda \psi_2}$ we find that they are equal to $0$, meaning the diagonal elements of $\AGP{\lambda}$ are $0$. Doing the same for the off-diagonals we find 
\begin{equation}
    \begin{aligned}
        \braket{\psi_1}{\dlambda \psi_2} &= \frac{1}{2} \\
        \braket{\psi_2}{\dlambda \psi_1} &= -\frac{1}{2},
    \end{aligned}
\end{equation}
meaning that the exact \acrref{AGP} operator, as found from evaluating its matrix elements is just
\begin{equation}
    \AGP{\lambda} = \mqty(0 & \frac{i}{2} \\ -\frac{i}{2} & 0) = \alpha(\lambda) \sy,
\end{equation}
which is the result obtained previously from the \acrref{LCD} approach. 

\chapter{Pontryagin maximum principle}\label{app:PMP}

This part of the appendix is dedicated solely to introducing the Pontryagin maximum principle or \acrref{PMP}, which, while not used in the main results of the thesis, forms the backbone of analytical optimal control theory, which I discuss at length in Ch.~\ref{chap:3_Quantum_Optimal_control}. In formal terms, the \acrref{PMP} can be defined \cite{dalessandro_introduction_2021} by the following theorem.

\newtheorem{theorem}{Theorem}

\begin{theorem}[\acrref{PMP} for Mayer problems]\label{thm:pmp}
    For fixed final time $\tau$ and free final state assume $u$ is the optimal control and $x$ the corresponding trajectory solution of Eq.~\eqref{eq:control_ODE}. Then, there exists a nonzero vector $\lambda$ solution of the adjoint equations
  \begin{equation}
      \dotlambda^T = - \lambda^T f(x(t), u(t))
  \end{equation}
  with terminal condition
  \begin{equation}
      \lambda^T(\tau) = -\phi(x(\tau))
  \end{equation}
  such that, for almost every $t \in (0, \tau]$, we have
  \begin{equation}\label{eq:pmp_maximisation}
      \lambda^T(t) f(x(t), u(t)) \geq \lambda^T(t) f(x(t), v)
  \end{equation}
  for every $v$ in the set of the admissible values for the control $U$. Furthermore, for every $t \in [0, \tau]$
  \begin{equation}\label{eq:pmp_constant}
      \lambda^T(t) f(x(t), u(t)) = c,
  \end{equation}
  for a constant $c$
\end{theorem}

Using this, one can then define the \emph{optimal control Hamiltonian}:
\begin{equation}
    h(\lambda, x, u) := \lambda^T(t) f(x, u).
\end{equation}
Now we can recast Eqs.~\eqref{eq:pmp_maximisation} and \eqref{eq:pmp_constant}:
\begin{equation}
    \begin{aligned}
        h(\lambda(t), x(t), u(t)) &= c \\
        h(\lambda, x, u) &\geq h(\lambda, x, v),
    \end{aligned}
\end{equation}
The solution will be of the form $u := u(x, \lambda)$ and it can be solved with the system of equations
\begin{equation}
    \begin{aligned}
        \dot{x} &= f(x, u(x, \lambda)), \\
        \dotlambda^T &= -\lambda^T f(x, u(x, \lambda))
    \end{aligned}
\end{equation}
with the boundary conditions $x(0) = x_0$ and $\lambda^T(\tau) = -\phi(x(\tau))$. Every control which is obtained with this procedure satisfies the necessary conditions of optimality and it is a candidate to be the optimal control.

\chapter{Derivation of the CD coefficients for an arbitrary Ising graph}\label{app:arbitrary_ising_derivation}

In the main text, we discuss deriving the local counterdiabatic driving or \acrref{LCD} terms to first and second order for an Ising graph of $N$ spins with arbitrary couplings between them. In this appendix, we will show a full derivation for the coupled set of equations required to determine the coefficients for said terms.

An Ising Hamiltonian for $N$ spins and with both a transverse and longitudinal field and with arbitrary couplings can be written as:
\begin{equation}\label{eq:ising_hamiltonian}
    H(\lambda) = \sum_{i = 1}^{N-1}\sum_{j = i+1}^{N} J_{ij}(\lambda) \sz_i \sz_j + \sum_{i = 1}^{N} \Big( X_i(\lambda) \sx_i + Z_i(\lambda) \sz_i \Big)
\end{equation}
where the coefficients $J_{ij}$ correspond to couplings between spins $i$ and $j$. Systems like this can be viewed as undirected graphs, with each spin corresponding to a vertex and each coupling $J_{ij}$ denoting an edge between the corresponding spins. In the case of a weighted graph, the magnitude of each $J_{ij}$ can be viewed as the weight of the corresponding edge. This type of Hamiltonian, for specific values of $J_{ij}$, $X_i$ and $Z_i$ can be used to describe the two-spin annealing example of Sec.~\ref{sec:5.1_2spin_annealing}, the Ising chain from Sec.~\ref{sec:5.2_Ising_chain} and the frustrated spin model of Sec.~\ref{sec:6.4_ghz_states}. 

The first order \acrref{LCD} ansatz, as stated in the main text, is just single-spin operators:
\begin{equation}\label{eq:spin_agp_1storder}
    \AGP{\lambda}^{(1)} = \sum_{i = 1}^N \alpha_i(\lambda) \sy_i
\end{equation}
and the second order can be split up into 4 separate symmetries of operators:
\begin{equation}\label{eq:spin_agp_2ndorder}
        \AGP{\lambda}^{(2)} = \sum_{i = 1}^{N-1}\sum_{j = i+1}^{N} \Big( \gamma_{ij}(\lambda) \sx_i \sy_j + \Bar{\gamma}_{ij}(\lambda) \sy_i \sx_j + \zeta_{ij}(\lambda) \sz_i \sy_j + \Bar{\zeta}_{ij}(\lambda) \sy_i \sz_j \Big).
\end{equation}

The first order commutators are computed as follows:
\begin{equation}\label{eq:first_order_AGP_commutator}
        i\comm{\alpha_i \sy_i}{H} = 2\alpha_i \Big[ \sum_{j = i+1}^{N} - J_{ij} \Big( \sx_i \sz_j + \sz_i \sx_j \Big) + X_i \sz_i - Z_i \sx_i \Big],
\end{equation}
where I have omitted the dependence on $\lambda$ of the terms. The second order expansions, sadly, look like this:
\begin{equation}
    \begin{aligned}
        i\comm{\gamma_{ij} \sx_i \sy_j}{H(\lambda)} &= 2\gamma_{ij} \Big[ \sum_{k = 1}^{i-1} (J_{ki} \sz_k \sy_i \sy_j - J_{kj} \sz_k \sx_i \sx_j)  + \sum_{k = i + 1}^{j-1} (J_{ik} \sy_i \sz_k \sy_j - J_{kj} \sx_i \sz_k \sx_j) \\ 
        &+ \sum_{k = j + 1}^N (J_{ik} \sy_i \sy_j \sz_k - J_{jk} \sx_i \sx_j \sz_k) + Z_i \sy_i \sy_j + X_j \sx_i \sz_j - Z_j \sx_i \sx_j \Big] \\
        i\comm{\Bar{\gamma}_{ij} \sy_i \sx_j}{H} &= 2\Bar{\gamma}_{ij} \Big[ \sum_{k = 1}^{i-1} (J_{kj} \sz_k \sy_i \sy_j - J_{ki} \sz_k \sx_i \sx_j) + \sum_{k = i + 1}^{j-1} (J_{kj} \sy_i \sz_k \sy_j - J_{ik} \sx_i \sz_k \sx_j) \\ 
        &+ \sum_{k = j + 1}^N (J_{jk} \sy_i \sy_j \sz_k - J_{ik} \sx_i \sx_j \sz_k)+ Z_j \sy_i \sy_j + X_i \sz_i \sx_j - Z_i \sx_i \sx_j \Big] \\ 
        i\comm{\zeta_{ij} \sz_i \sy_j}{H(\lambda)} &= 2\zeta_{ij} \Big[ -\sum_{k = 1}^{i-1} J_{kj} \sz_k \sz_i \sx_j - \sum_{k = i + 1}^{j-1} J_{kj} \sz_i \sz_k \sx_j - \sum_{k = j + 1}^N J_{jk} \sz_i \sx_j \sz_k  \\
        &- J_{ij} \sx_j - X_i \sy_i \sy_j + X_j \sz_i \sz_j - Z_j \sz_i\sx_j \Big] \\
        i\comm{\Bar{\zeta}_{ij} \sy_i \sz_j}{H(\lambda)} &= 2\Bar{\zeta}_{ij} \Big[ - \sum_{k = 1}^{i-1} J_{ki} \sz_k \sx_i \sz_j - \sum_{k = i + 1}^{j-1} J_{ik} \sx_i \sz_k \sz_j 
        - \sum_{k = j + 1}^N J_{ik} \sx_i \sz_j \sz_k \\
        &- J_{ij} \sx_i - X_j \sy_i \sy_j + X_i \sz_i \sz_j - Z_i \sx_i \sz_j \Big]
    \end{aligned}
\end{equation}
Combined, the above commutators along with the coefficients of $\dlambda H$ give the operator $G_{\lambda}(\AGP{\lambda}^{(1,2)})$ for an ansatz \acrref{AGP} constructed from both single- and two-spin operators (as per Eq.~\eqref{eq:G_operator}):
\begin{equation}\label{eq:ising_graph_G_operator}
    \begin{aligned}
        G_{\lambda}(\AGP{\lambda}^{(1,2)}) &= \sum_{i=1}^N \Bigg[ (\dot{X}_i - 2\alpha_i Z_i - 2\sum_{j=1}^{i-1} J_{ji}\zeta_{ji} - 2\sum_{j=i+1}^N J_{ij}\zetabar_{ij})\sx_i \\
        &+ (\dot{Z}_i + 2\alpha_i X_i)\sz_i \Bigg] \\
        &+ \sum_{i = 1}^{N-1} \sum_{j = i+1}^{N} \Bigg[(\dot{J}_{ij} + 2\zeta_{ij} X_j + 2\zetabar_{ij} X_i)\sz_i\sz_j \\
        &+ (2\gamma_{ij}Z_i + 2\gammabar_{ij} Z_j - 2 \zeta_{ij} X_i - 2 \zetabar_{ij} X_j)\sy_i\sy_j \\
        &+ (2\gamma_{ij}Z_j + 2\gammabar_{ij} Z_i)\sx_i\sx_j \\
        &+ (-2\alpha_i J_{ij} + 2\gamma_{ij}X_j - 2\zetabar_{ij} Z_i)\sx_i\sz_j \\
        &+ (-2\alpha_j J_{ij} + 2\gammabar_{ij}X_i - 2\zeta_{ij} Z_j)\sz_i\sx_j \\
        &+ \sum_{k = 1}^{i-1} \Big[ (2\gamma_{ij}J_{ki} + 2\gammabar_{ij} J_{kj})\sz_k\sy_i\sy_j + (2\gamma_{ij}J_{kj} + 2\gammabar_{ij} J_{ki})\sz_k\sx_i\sx_j \\
        &+ (- 2 \zeta_{ij} J_{kj} - 2 \zeta_{kj}J_{ij})\sz_k\sz_i\sx_j \Big] \\
        &+ \sum_{k = i+1}^{j-1} \Big[ (2\gamma_{ij}J_{ik} + 2\gammabar_{ij} J_{kj})\sy_i\sz_k\sy_j + (2\gamma_{ij}J_{kj} \\
        &+ 2\gammabar_{ij} J_{ik})\sx_i\sz_k\sx_j + (- 2 \zetabar_{ij} J_{ik} - 2 \zeta_{ik}J_{ij})\sz_k\sx_i\sz_j \Big] \\
        &+ \sum_{k = j+1}^N \Big[ (2\gamma_{ij}J_{ik} + 2\gammabar_{ij} J_{jk})\sy_i\sy_j\sz_k + (2\gamma_{ij}J_{jk} + 2\gammabar_{ij} J_{ik})\sx_i\sx_j\sz_k \\
        &+ (- 2 \zetabar_{ij} J_{ik} - 2 \zetabar_{ik}J_{ij})\sz_i\sx_j\sz_k \Big] \Bigg].
    \end{aligned}
\end{equation}
In order to find the coupled set of equations that allow us to compute each of the coefficients in the approximate \acrref{AGP} according to the \acrref{LCD} approach, we need to minimise the action $\mathcal{S} = \Tr[G_{\lambda}^2]$ with respect to each of the coefficients. As the Pauli operators and their tensor products are traceless, this means that the action is merely the sum of the squares of all the orthogonal operator coefficients of $G_{\lambda}$. Minimising $\mathcal{S}$ with respect to each $\alpha_i$ gives:
\begin{equation}\label{eq:ising_graph_minimise_alpha}
    \begin{aligned}
        &\alpha_i \Big[2Z_i^2 + 2X_i^2 + \sum_{j = 1}^{i-1}2J_{ji}^2 + \sum_{i+1}^{N}2J_{ij}^2 \Big] \\
        \sum_{j = i+1}^N &\gamma_{ij} \Big[ -2J_{ij}X_j \Big] + \sum_{j = 1}^{i-1} \gammabar_{ji} \Big[ -2J_{ji}X_j \Big] \\
        \sum_{j = i+1}^N &\zetabar_{ij} \Big[ 4J_{ij}Z_i \Big] + \sum_{j = 1}^{i-1} \zeta_{ji} \Big[4 J_{ji}Z_i \Big] \\
        &= Z_i \dot{X}_i - X_i \dot{Z}_i,
    \end{aligned}
\end{equation}
where $i$ is fixed. Fixing $i$ and $j$ and minimising with respect to each $\gamma_{ij}$ gives:
\begin{equation}\label{eq:ising_graph_minimise_gamma}
    \begin{aligned}
        &\alpha_i\Big[ -X_j J_{ij} \Big] + \zeta_{ij}\Big[ -  X_i Z_i\Big] + \zetabar_{ij} \Big[ -2 X_j Z_i \Big] \\
        + \: &\gamma_{ij} \Big[ Z_i^2 + Z_j^2 + X_j^2 + \sum_{k = 1}^{i-1} (J_{ki}^2 + J_{kj}^2) + \sum_{k = i+1}^{j-1} (J_{ik}^2 + J_{kj}^2) + \sum_{k = j+1}^N (J_{ik}^2 + J_{jk}^2) \Big] \\
        + \: &\gammabar_{ij} \Big[ 2 Z_i Z_j + \sum_{k = 1}^{i-1} 2J_{ki}J_{kj} + \sum_{k = i+1}^{j-1} 2J_{ik}J_{kj} + \sum_{k = j+1}^N 2J_{ik}J_{jk} \Big] = 0
    \end{aligned}
\end{equation}
and likewise for each $\gammabar$:
\begin{equation}\label{eq:ising_graph_minimise_gammabar}
    \begin{aligned}
        &\alpha_j\Big[ -X_i J_{ij} \Big] + \zeta_{ij}\Big[ - 2 X_i Z_j\Big] + \zetabar_{ij} \Big[ - X_j Z_j \Big] \\
        + \: &\gammabar_{ij} \Big[ Z_i^2 + Z_j^2 + X_i^2 + \sum_{k = 1}^{i-1} (J_{ki}^2 + J_{kj}^2) + \sum_{k = i+1}^{j-1} (J_{ik}^2 + J_{kj}^2) + \sum_{k = j+1}^N (J_{ik}^2 + J_{jk}^2) \Big] \\
        + \: &\gamma_{ij} \Big[ 2 Z_i Z_j + \sum_{k = 1}^{i-1} 2J_{ki}J_{kj} + \sum_{k = i+1}^{j-1} 2J_{ik}J_{kj} + \sum_{k = j+1}^N 2J_{ik}J_{jk} \Big] = 0.
    \end{aligned}
\end{equation}
Finally, for fixed $i$, $j$, we minimise with respect to $\zeta_{ij}$:
\begin{equation}\label{eq:ising_graph_minimise_zeta}
    \begin{aligned}
        &\alpha_j\Big[ 4 Z_j J_{ij} \Big] + \gamma_{ij}\Big[ - 2 X_i Z_i\Big] + \gammabar_{ij} \Big[ - 4 X_i Z_j \Big] \\
        + \: &\zeta_{ij} \Big[ 2 Z_j^2 + 2X_i^2 + 2X_j^2 + \sum_{k = 1}^{i-1} 2 J_{kj}^2 + \sum_{k = i+1}^{j-1} 2 J_{jk}^2 \Big] + \zetabar_{ij}\Big[ 4 X_i X_j \Big] \\
        + \sum_{k = 1}^{i-1} &\zeta_{kj}\Big[ 2 J_{ij} J_{kj} \Big] + \sum_{k = 1}^{j-1} \zeta_{kj} \Big[2 J_{ij}J_{kj}\Big] \\
        + \sum_{k = i +1}^{j-1} &\zetabar_{jk} 2 J_{ij} J_{jk} + \sum_{k = j +1}^N \zetabar_{jk} 2 J_{ij} J_{jk} = J_{ij} \dot{X}_j - \dot{J}_{ij} X_j
    \end{aligned}
\end{equation}
and with respect to $\zetabar_{ij}$:
\begin{equation}\label{eq:ising_graph_minimise_zetabar}
    \begin{aligned}
        &\alpha_i\Big[ 4 Z_i J_{ij} \Big] + \gamma_{ij}\Big[ - 4 X_j Z_i\Big] + \gammabar_{ij} \Big[ - 2 X_j Z_j \Big] \\
        + \: &\zetabar_{ij} \Big[ 2 Z_i^2 + 2X_i^2 + 2X_j^2 + \sum_{k = i+1}^{j-1} 2 J_{ik}^2 + \sum_{k = j+1}^N 2 J_{ik}^2 \Big] + \zeta_{ij}\Big[ 4 X_i X_j \Big] \\
        + \sum_{k = i+1}^N &\zetabar_{ik}\Big[ 2 J_{ij} J_{ik} \Big] + \sum_{k = j+1}^N \zetabar_{ik} \Big[2 J_{ij}J_{ik}\Big] \\
        + \sum_{k = 1}^{i-1} &\zeta_{ki} 2 J_{ij} J_{ki} + \sum_{k = i+1}^{j-1} \zeta_{ik} 2 J_{ij} J_{ik} = J_{ij} \dot{X}_i - \dot{J}_{ij} X_i.
    \end{aligned}
\end{equation}
armed with this knowledge, we can now explore the non-adiabatic effects generated by one- and two-spin operators on any random time-dependent Ising graph Hamiltonian. These results are particularly relevant in the case of the Ising spin chain from Sec.~\ref{sec:5.2_Ising_chain} and the frustrated spin example from Sec.~\ref{sec:6.4_ghz_states}.

\chapter{Additional plots for the Ising spin chain example}\label{app:ising}

This appendix contains additional information and plots concerning the Ising spin chain example from Sec.~\ref{sec:5.2_Ising_chain}, where we investigate implementations of \acrref{COLD}, \acrref{LCD} and \acrref{BPO} for this particular system. In the plots presented here, all parameters are the same as those discussed in Sec.~\ref{sec:5.2_Ising_chain} unless stated otherwise.

\begin{figure}[t!]
    \centering
    \includegraphics[width=\linewidth]{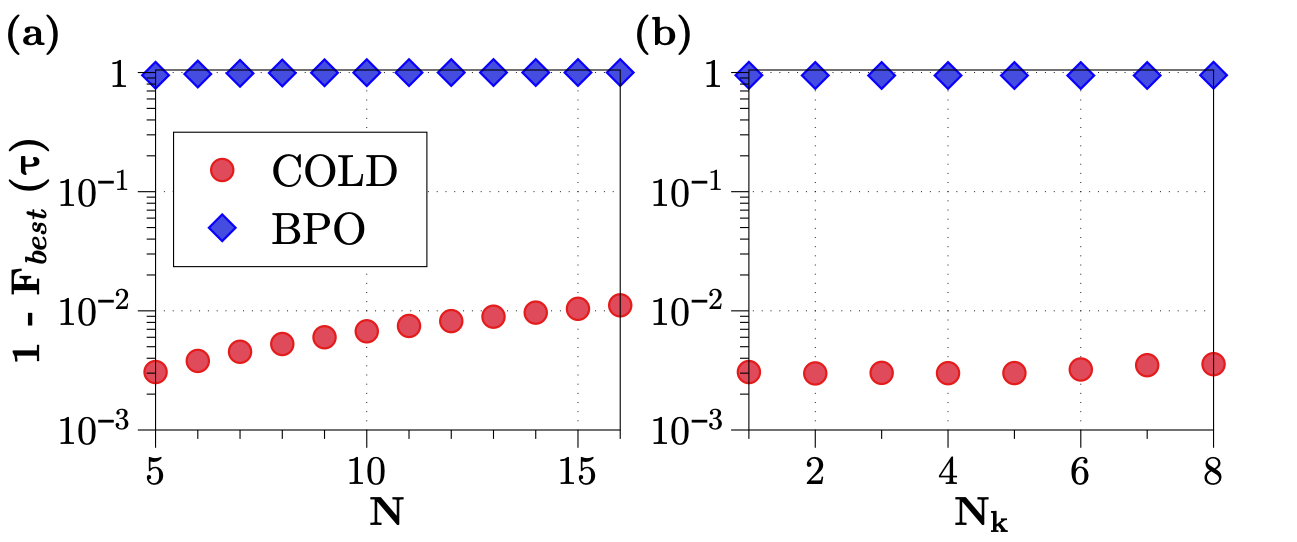} \caption[Plots of how final state fidelities scale using COLD and BPO for different system sizes and optimisable parameters.]{Scaling of fidelities in the annealing protocol for the Ising model with (a) system size $N$ and (b) optimisation parameters $N_k$ at driving time $\tau=10^{-2}J_0^{-1}$. Plots show a comparison between \acrref{BPO} (blue diamonds) and \acrref{COLD} (red circles). Plotted best fidelities are obtained across 500 optimisations. Reprinted with permission from \cite{cepaite_counterdiabatic_2023}. Copyright 2023, American Physical Society.}\label{fig:ising_scalingN}
\end{figure}

As well as the five spin chain which is analysed in detail in the, in Fig.~\ref{fig:ising_scalingN} we also present a plot of how \acrref{FO} \acrref{COLD} and \acrref{BPO} scale with (a) increasing chain lengths and (b) increasing number of control parameters $N_k$ when the bare optimisation pulse from Eq.~\eqref{eq:ising_control_nocrab} is used on $N=5$ spins. We see that the \acrref{COLD} fidelity decreases as a function of the number of spins $N$ but remains very high when compared to \acrref{BPO}, while there appears to be no noticeable improvement for ether \acrref{BPO} or \acrref{COLD} when the number of control parameters $N_k$ of the bare control pulse are increased. This may be a consequence of the way in which the bare control pulse is constructed, as the impact of parameterisation on the result should depend heavily on the type of control pulse used. In the case of \acrref{GRAPE} or \acrref{CRAB} we would not necessarily expect the trend in Fig.~\ref{fig:ising_scalingN}(b) to be replicated.

\begin{figure}[t!]
    \centering
    \includegraphics[width=\linewidth]{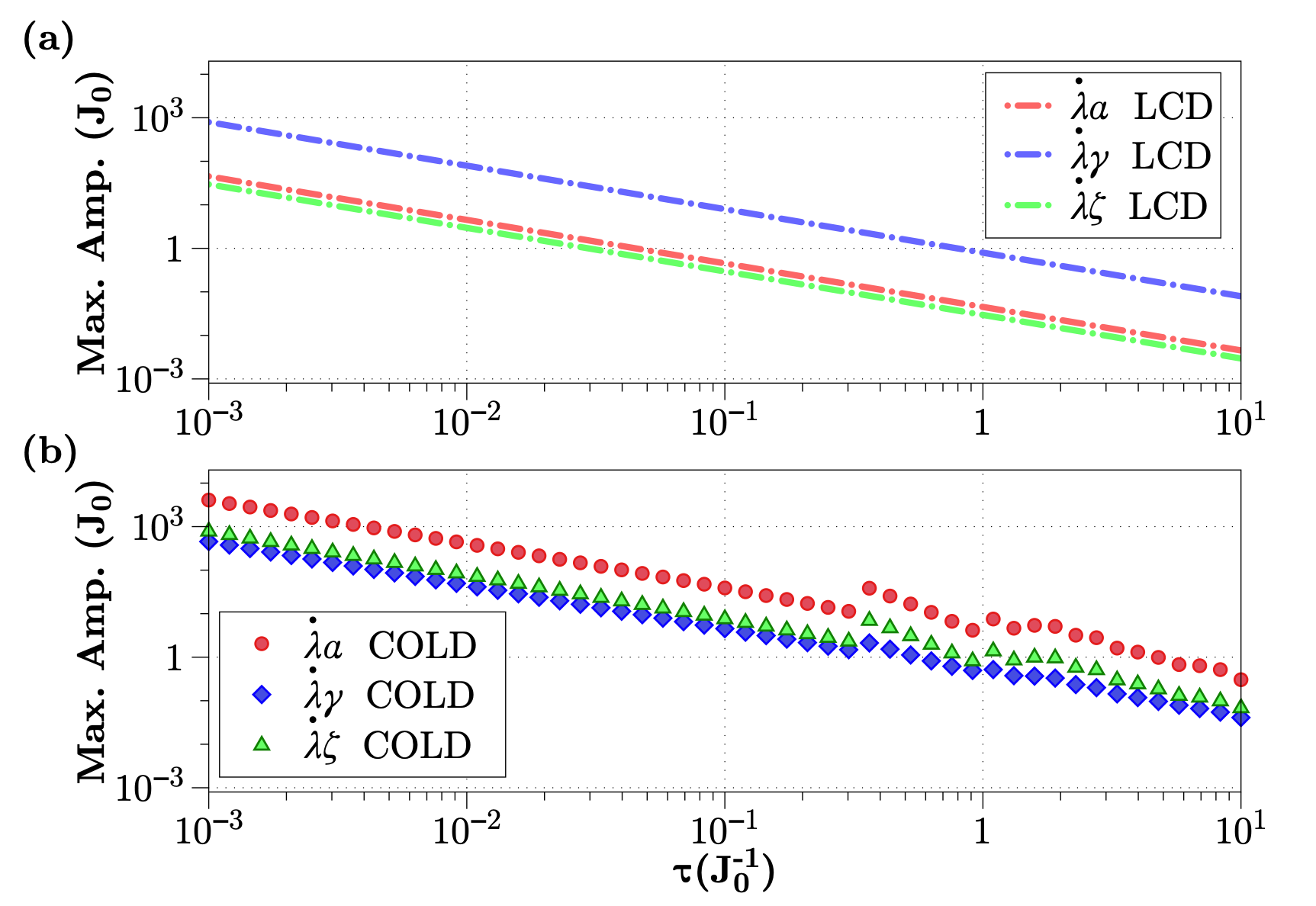} \caption[Plots of maximum amplitudes of LCD drives for the Ising spin chain.]{Maximum amplitudes of \acrref{LCD} terms in the Ising model annealing protocol for (a) \acrref{FO} and \acrref{SO} \acrref{LCD} with no additional optimal control fields and (b) the \acrref{COLD} approach optimised for the best final state fidelity implementing \acrref{FO} LCD as shown in Fig.~\ref{fig:ising_unconstrained}(a). The plot shows the maximum amplitude reached at any point in the drive for different driving times $\tau$ in the case of \acrref{FO} terms $\alpha$ (red circles) and \acrref{SO} terms $\gamma$ (blue diamonds) and $\zeta$ in the case when only the \acrref{FO} terms are applied to the system. Reprinted with changes with permission from \cite{cepaite_counterdiabatic_2023}. Copyright 2023, American Physical Society.}\label{fig:ising_maxamp}
\end{figure}

In Fig.~\ref{fig:ising_maxamp}, we plot the scaling of the \acrref{FO} and \acrref{SO} counterdiabatic terms applied to the five spin Ising chain from Eq.~\eqref{eq:ising_fo_agp} and Eq.~\eqref{eq:ising_so_lcd_terms} with the $\dotlambda$ term included, \@i.e. for the coefficient $\alpha$ we plot the exact \acrref{CD} amplitude of the given \acrref{LCD} operator $\dotlambda \alpha$. In (a) we see the how the different \acrref{LCD} drive amplitudes scale with increasing driving time and in (b) we do the same in the case of the optimised pulses for \acrref{COLD} which were used to obtain the fidelities in Fig.~\ref{fig:ising_unconstrained}(a). Note that while we plot the \acrref{SO} terms for both cases, these were not actually implemented in obtaining the fidelities plotted in the main text. We find, as expected based on the included $\dotlambda$ scaling, that the \acrref{LCD} coefficients decrease linearly with respect to the driving time $\tau$ due to the fact that $\dotlambda = \frac{1}{\tau}$. The coefficients $\alpha$, $\gamma$ and $\zeta$, as expected, stay constant due to their lack of dependence on $\tau$. We see that the \acrref{SO} term $\gamma$ is over an order of magnitude larger than either the \acrref{FO} term $\alpha$ or the other \acrref{SO} term $\zeta$. In (b), however, we find that the \acrref{FO} term $\alpha$ dominates the maximal amplitude for all driving times $\tau$. Furthermore, there is no longer a clean, linear dependence of the counterdiabatic coefficients on driving time, as they are now functions of the control pulse which is optimised for a different set of control parameter values at each driving time $\tau$. The inversion in the strength of the \acrref{SO} and \acrref{FO} \acrref{LCD} terms between the \acrref{COLD} and control-free case shows that in this case \acrref{COLD} implements a dynamical Hamiltonain which is favourable for the applied LCD operators, which are local $\sy$ operators on each spin. This behaviour lends support to the ideas presented in Ch.~\ref{chap:5_cd_as_costfunc} as well as the results in Sec.~\ref{sec:7.2_ising_ho_lcd}, wherein properties of the \acrref{LCD} coefficients are used to optimise the Hamiltonian control pulse.

\chapter{Additional plots on GHZ state preparation using AGP as a cost function}\label{app:higher_order_AGP}

In Sec.~\ref{sec:7.3_ghz_ho}, we considered using \acrref{AGP}-based cost functions, which were introduced in detail in Ch.~\ref{chap:5_cd_as_costfunc}, in GHZ state preparation in a system of $N=3$ frustrated spins. We found that, when using a control pulse constructed from \acrref{GRAPE}, there appears to be no advantage to using \acrref{FO} or \acrref{SO} \acrref{LCD} information in the optimisation process, whether integrals or maximum amplitudes of the operator coefficients. In Fig.~\ref{fig:ghz_contours} in the main text, we plotted the cost function landscapes of the fidelity cost function $C_{\rm F}$, the tangle cost function $C_{T_3}$ and several integral cost functions $C_{\rm I}$ in the case of different \acrref{LCD} coefficients for total driving time $\tau = 0.1J_0^{-1}$ and for two control parameters $c_1, c_2 \in [-10,10]$. The results showed a highly non-convex landscape with respect to the final state fidelity and the three-tangle, which measures the amount of GHZ-type entanglement in a system. There also appeared to be no significant correlation between the minimum and maximum values of the integral cost functions and the quality of the final state, \@i.e.~either the final state fidelity or the amount of entanglement in the final state. 

\begin{figure}[t!]
    \centering
    \includegraphics[width=0.8\linewidth]{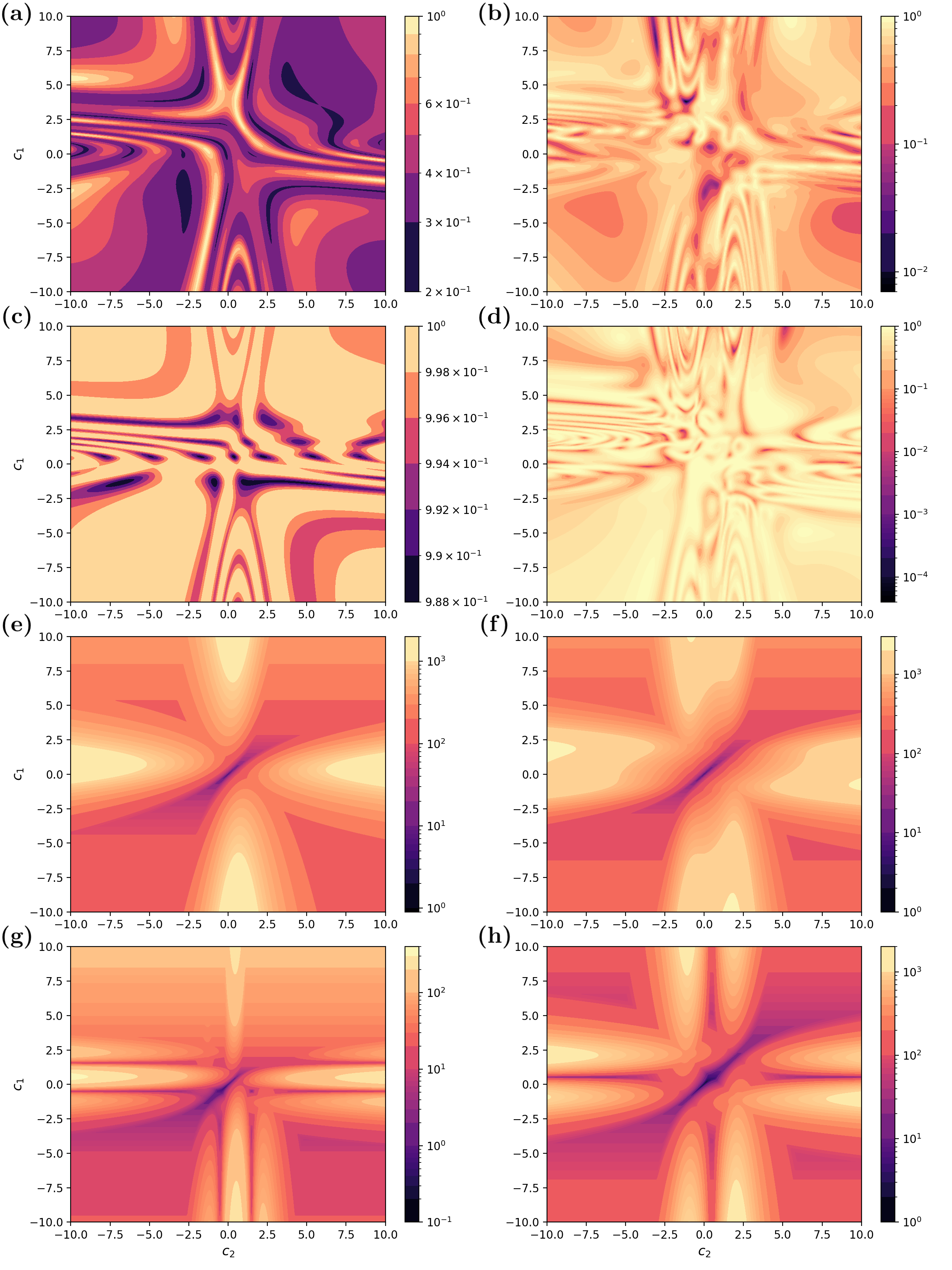} \caption[Contour plots of cost function landscapes for GHZ state preparation in frustrated spin systems (maximum amplitude cost function).]{Contour plots at $\tau = 0.1 J_0^{-1}$ of different cost function values for GHZ state preparation for parameters $c_1, c_2 \in [-10,10]$ and a \acrref{GRAPE} control pulse. In (a) and (b) we plot $C_{\rm F}$ in the cases where \acrref{FO} and \acrref{SO} \acrref{COLD} is applied respectively. Then, in (c-d) we do the same for $C_{T_3}$, with \acrref{FO} \acrref{COLD} plotted in (c) and \acrref{SO} \acrref{COLD} plotted in (d). (e-h) are then plots of the maximum amplitude cost function $C_{\rm A}$ values for the same range of parameters. In (e) we plot $C_{\rm A, \alpha^{(1)}}$ when only \acrref{FO} \acrref{LCD} is considered, while in (f) we plot $C_{\rm A, \alpha^{(2)}}$ as described in the text. Then in (g) we plot $C_{\rm A, \gamma}$ and in (h) we plot $C_{\rm a, \zeta}$, corresponding to the \acrref{SO} terms. Note that each plot has its own color bar, as the color encodings and the value scaling in each plot is quite different.} \label{fig:ghz_contours_max_appendix}
\end{figure}

In Fig.~\ref{fig:ghz_contours_max_appendix} we reproduce the landscapes of $C_{\rm F}$ and $C_{T_3}$ from Fig.~\ref{fig:ghz_contours} and then plot the results for the maximum amplitude cost function $C_{\rm A}$ in plots (e-h) corresponding to the same coefficients as were plotted for the integral cost function in the main text. While there is some minimal difference in the landscapes between $C_{\rm I}$ and $C_{\rm A}$, they broadly follow similar trends and show no correlation with fidelity or entanglement.

\begin{figure}[t!]
    \centering
    \includegraphics[width=0.8\linewidth]{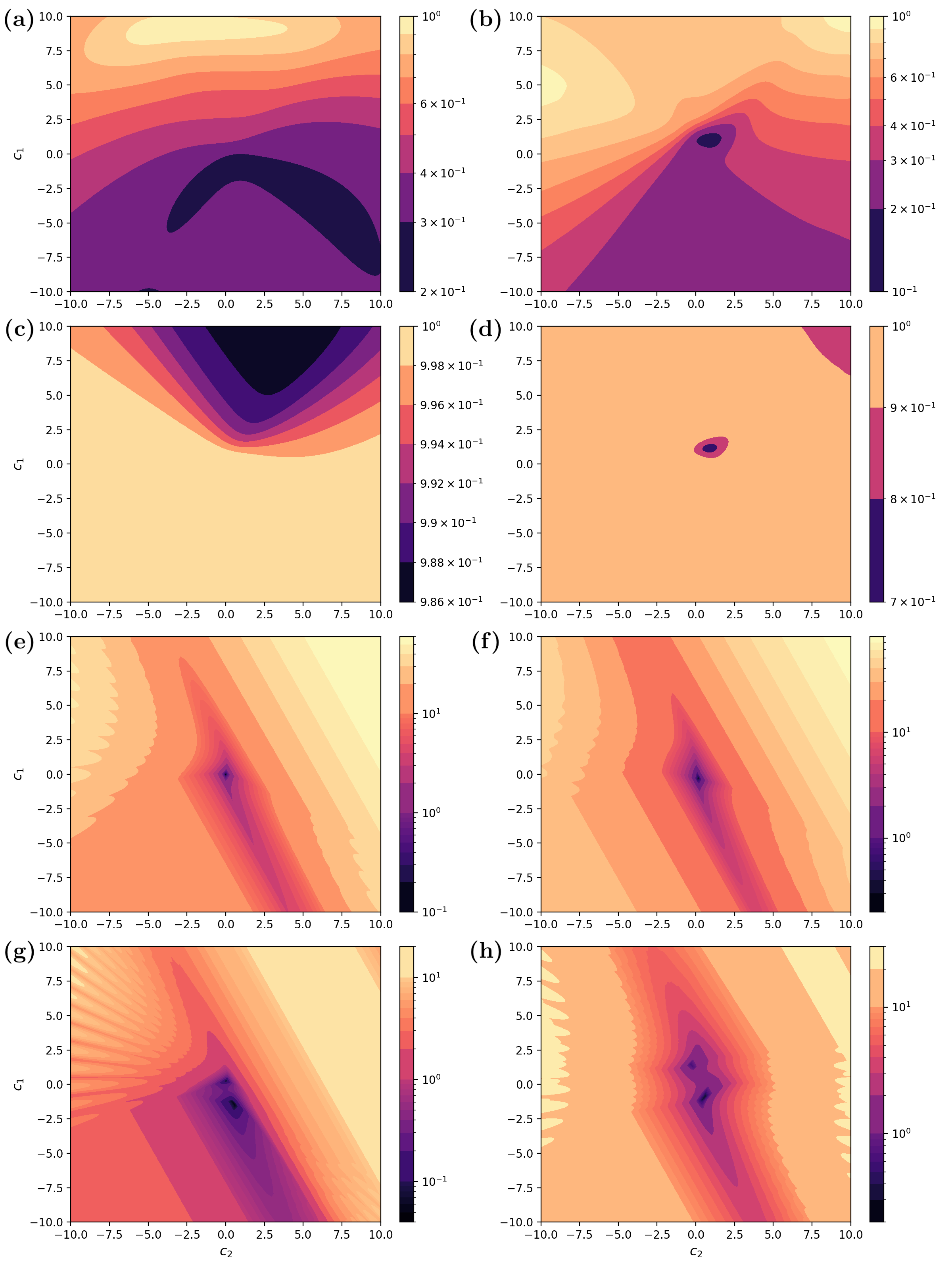} \caption[Contour plots of cost function landscapes for GHZ state preparation in frustrated spin systems (maximum amplitude cost function) using a bare optimisation pulse.]{Contour plots at $\tau = 0.1 J_0^{-1}$ of different cost function values for GHZ state preparation for parameters $c_1, c_2 \in [-10,10]$ and a bare control pulse. In (a) and (b) we plot $C_{\rm F}$ in the cases where \acrref{FO} and \acrref{SO} \acrref{COLD} is applied respectively. Then, in (c-d) we do the same for $C_{T_3}$, with \acrref{FO} \acrref{COLD} plotted in (c) and \acrref{SO} \acrref{COLD} plotted in (d). (e-h) are then plots of the maximum amplitude cost function $C_{\rm A}$ values for the same range of parameters. In (e) we plot $C_{\rm A, \alpha^{(1)}}$ when only \acrref{FO} \acrref{LCD} is considered, while in (f) we plot $C_{\rm A, \alpha^{(2)}}$ as described in the text. Then in (g) we plot $C_{\rm A, \gamma}$ and in (h) we plot $C_{\rm A, \zeta}$, corresponding to the \acrref{SO} terms.}\label{fig:ghz_contours_max_noGRAPE}
\end{figure}

\begin{figure}[t!]
    \centering
    \includegraphics[width=0.9\linewidth]{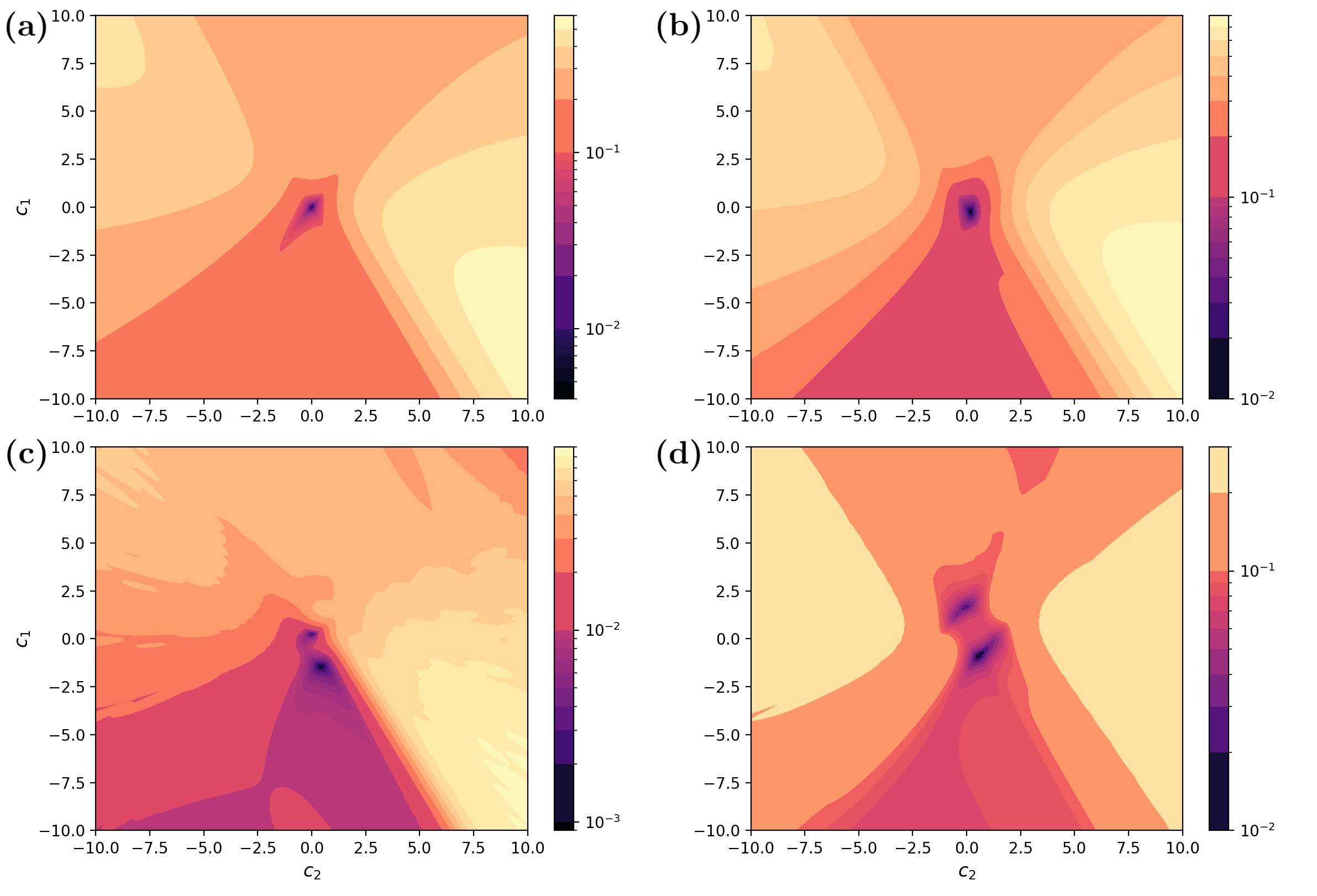} \caption[Contour plots of cost function landscapes for GHZ state preparation in frustrated spin systems (integral cost function) using a bare optimisation pulse.]{Contour plots at $\tau = 0.1 J_0^{-1}$ of different cost function values for GHZ state preparation for parameters $c_1, c_2 \in [-10,10]$ and a bare control pulse. In (a) we plot $C_{\rm I, \alpha^{(1)}}$ when only \acrref{FO} \acrref{LCD} is considered, while in (b) we plot $C_{\rm I, \alpha^{(2)}}$ as described in the text. Then in (c) we plot $C_{\rm I, \gamma}$ and in (d) we plot $C_{\rm I, \zeta}$, corresponding to the \acrref{SO} terms.}\label{fig:ghz_contours_int_noGRAPE}
\end{figure}

To be sure that this failure is not a consequence of constructing the control pulse using the \acrref{GRAPE} algorithm, we also implement a bare control pulse like that described in Eq.~\eqref{eq:ising_control_nocrab} and plot the results in Fig.~\ref{fig:ghz_contours_max_noGRAPE} for the maximum amplitude cost functions $C_{\rm A}$ as well as integral cost functions $C_{\rm I}$ in Fig.~\ref{fig:ghz_contours_int_noGRAPE}. While the cost function landscapes are far smoother, the resulting fidelities and entanglement in the final state are orders of magnitude worse than those using the \acrref{GRAPE} pulse. Furthermore, there once again does not appear to be any advantage to using the \acrref{AGP}-based cost functions. The maximum entanglement (in the given range of parameters) when applying \acrref{SO} terms, for example, as shown in Fig.~\ref{fig:ghz_contours_max_noGRAPE}(d), occurs close to the minimum of the \acrref{SO} pulse integrals and maximal amplitudes (plots (g-h) in Fig.~\ref{fig:ghz_contours_max_noGRAPE} and plots (c-d) in Fig.~\ref{fig:ghz_contours_int_noGRAPE}, which is not what we would expect based on the conjecture that an optimal pulse would maximise the effects of the \acrref{LCD} operators that are being applied. Any attempts at optimisation using $C_{\rm I}$ or $C_{\rm A}$ pulses does not return better results than in the \acrref{GRAPE} case.

\addcontentsline{toc}{chapter}{Bibliography}
\bibliographystyle{util/ieeetran.bst}
\bibliography{references}

\end{document}